\documentclass[reprint,amsmath,amssymb,nofootinbib,aps]{revtex4-1}
\usepackage{subfig}
\usepackage{graphicx}
\graphicspath{ {images/} }
\usepackage{dcolumn}
\usepackage{bm}
\usepackage{amsfonts}
\usepackage{slashed}
\usepackage{mathtools}
\usepackage{color}

\def\be{\begin{equation}}
\def\ee{\end{equation}}
\def\wh{\widehat}
\def\bwt{\begin{widetext}}
\def\ewt{\end{widetext}}
\newcommand{\muT}{\wh{\mu}}
\newcommand{\nuT}{\wh{\nu}}
\newcommand{\Cc}{\mathbb{C}}
\newcommand{\Ss}{\mathbb{S}}
\newcommand{\mT}{\wh{-}}
\newcommand{\pT}{\wh{+}}
\newcommand{\itP}[1]{\wh{#1}}

\begin{document}

	\title{Interpolating Quantum Electrodynamics between Instant and Front Forms}
	\author{Chueng-Ryong Ji}
	\affiliation{Department of Physics, North Carolina State University, Raleigh, North Carolina 27695-8202}
	\author{Ziyue Li}
	\affiliation{Department of Physics, North Carolina State University, Raleigh, North Carolina 27695-8202}
	\author{Bailing Ma}
	\affiliation{Department of Physics, North Carolina State University, Raleigh, North Carolina 27695-8202}
	\author{Alfredo Takashi Suzuki}
	\affiliation{Instituto de F\'{\i}sica Te\'orica-UNESP Universidade Estadual Paulista, Rua Dr. Bento Teobaldo Ferraz, 271 - Bloco II - 01140-070, S\~ao Paulo, SP, Brazil.}
	
	
	\begin{abstract}
		The instant form and the front form of relativistic dynamics proposed by Dirac in 1949 can be linked by an interpolation angle parameter $\delta$ spanning between the instant form dynamics (IFD) at $\delta =0$ and the front form dynamics which is  
now known as the light-front dynamics (LFD) at $\delta =\pi/4$.  We present the formal derivation of the interpolating quantum electrodynamics (QED) in the canonical field theory approach and discuss the constraint fermion degree of freedom which appears uniquely in the LFD.  The constraint component of the fermion degrees of freedom in LFD results in the instantaneous contribution to the fermion propagator, which is genuinely distinguished from the ordinary equal-time forward and backward propagation of relativistic fermion degrees of freedom. As discussed in our previous work, the helicity of the on-mass-shell fermion spinors in LFD is also distinguished from the ordinary Jacob-Wick helicity in the IFD with respect to whether the helicity depends on the reference frame or not. To exemplify the characteristic difference of the fermion propagator between IFD and LFD, we compute the helicity amplitudes of typical QED processes such as $e^+ e^- \to \gamma \gamma$ and $e \gamma \to e \gamma$ and present the whole landscape of the scattering amplitudes in terms of the frame dependence or the scattering angle dependence with respect to the interpolating angle dependence.
Our analysis clarifies any conceivable confusion in the prevailing notion of the equivalence between the infinite momentum frame approach and the LFD.
	\end{abstract}
	
	\maketitle
	
	
	\section{\label{sec:intro}Introduction}
	For the study of relativistic particle systems,
	Dirac~\cite{Dirac} proposed three different forms of the relativistic
	Hamiltonian dynamics in 1949: i.e. the instant ($x^0 =0$), front
	($x^+ = (x^0 + x^3)/\sqrt{2} = 0$), and point ($x_\mu x^\mu = a^2 >
	0, x^0 > 0$) forms.  The instant form dynamics (IFD) of quantum
	field theories is based on the usual equal time $t=x^0$ quantization
	(units such that $c=1$ are taken here), which provides 
	a traditional approach evolved from the non-relativistic dynamics. 
	The IFD makes a close contact with the Euclidean space, developing 
	temperature-dependent quantum field theory, lattice QCD, etc. 
	The equal light-front time $\tau \equiv (t +z/c)/\sqrt{2}=x^+$ quantization yields the front form dynamics, nowadays
	more commonly called light-front dynamics (LFD), which provides an innovative approach to the study of
	relativistic dynamics. The LFD works strictly in the Minkowski space,
	developing useful frameworks for the analyses of deep inelastic scattering (DIS), 
	parton distribution functions (PDFs), deeply virtual Compton scattering (DVCS), generalized parton distributions (GPDs), etc.
	The quantization in the point form ($x^{\mu}x_{\mu}=a^{2}>0, x^{0}>0$) is called radial quantization and this
	quantization procedure has been much used in string theory and conformal field theories~\cite{FHJ-1973}
	as well as in hadron physics~\cite{Glozman,Wagenbrunn,Melde}.
	Among these three forms of relativistic dynamics proposed by Dirac, however, 
	the LFD carries the largest number (seven) of the kinematic (or interaction
	independent) generators leaving the least number (three) of the dynamics generators
	while both the IFD and the point form dynamics carry six kinematic and four dynamic generators 
	within the total ten Poincar\'e generators. Indeed, the maximum number of kinematic generators
	allowed in any form of relativistic dynamics is seven and the LFD is the only one which possesses
	this maximum number of kinematic generators. Effectively, the LFD maximizes the capacity 
	to describe hadrons by saving a lot of dynamical efforts in obtaining the QCD solutions that reflect the full Poincar\'e symmetries. 
	
	To link the LFD with the IFD which has been the traditional approach, we introduce an interpolation angle parameter spanning
between the IFD and LFD. Although we want ultimately to obtain a general formulation for the QCD, we start from the simpler theory to discuss first the bare-bone structure that will persist even in the more complicated theories.
	Starting from the scalar field theory~\cite{Alfredo} to discuss the interpolating scattering amplitude with only momentum degree of freedom, we have extended the discussion to the electromagnetic gauge degree of freedom~\cite{JLS2015} and the on-mass-shell fermion~\cite{LAJ2015}. In particular, we discussed the link between the Coulomb gauge in IFD and the light-front gauge
	in LFD~\cite{JLS2015} and the chiral representation of the helicity spinors interpolating
	between the IFD and the LFD~\cite{LAJ2015}. In this work, we entwine the fermion propagator interpolation 
	with our previous works of the electromagnetic gauge field~\cite{JLS2015} and the helicity spinors~\cite{LAJ2015} and fasten the bolts and nuts necessary to launch the interpolating QED. 
	
	As we have already discussed the prototype of QED scattering processes ``$e\mu\to e\mu$ " and ``$e^+e^-\to \mu^+\mu^-$"
	involving a photon propagator in our previous work~\cite{LAJ2015}, we present in this work 
	the two-photon production amplitude in the pair annihilation of fermion and anti-fermion process 
	``$e^+e^-\to\gamma \gamma$" as well as the Compton scattering amplitude ``$e\gamma \to e\gamma$" involving a fermion propagator.  
	Since the effects of external fermions and bosons have already been studied in our previous works~\cite{JLS2015,LAJ2015}, 
	we will focus on the intermediate fermion propagator in this work. 
	
	To trace the forms of relativistic quantum field theory between
	IFD and LFD, we take the following convention
	of the space-time coordinates to define the interpolation
	angle\cite{Alfredo,JLS2015,LAJ2015,Hornbostel,chad}:
	\begin{align}
	\label{interpolangle}
	\left[ \begin{array}{c}
		x^{\widehat{+}}  \\
		x^{\widehat{-}}
	\end{array}
	\right] = \left[ \begin{array}{cc}\cos\delta & \sin\delta \\
		\sin\delta & -\cos\delta \end{array} \right]\left[\begin{array}{c} x^0 \\ x^3  \end{array}\right],
	\end{align} 
	in which the interpolation angle is allowed to run from 0 through
	$45^\circ$, $0\le \delta \le \frac{\pi}{4}$. The lower index variables $x_{\wh{+}}$ and $x_{\wh{-}}$
	are related to the upper index variables as $x_{\wh{+}}=g_{{\wh{+}\wh{\mu}}}x^{\wh{\mu}}=\mathbb{C}x^{\wh{+}}+\mathbb{S}x^{\wh{-}}$
        and $x_{\wh{-}}=g_{{\wh{-}\wh{\mu}}}x^{\wh{\mu}}=-\mathbb{C}x^{\wh{-}}+\mathbb{S}x^{\wh{+}}$, 
        denoting $\mathbb{C}=\rm{cos} 2\delta$ and $\mathbb{S}=\rm{sin} 2\delta$ and 
        realizing $g_{{\wh{+}\wh{+}}} = -g_{{\wh{-}\wh{-}}}=\rm{cos} 2\delta = \mathbb{C}$ and 
     	$g_{{\wh{+}\wh{-}}} = g_{{\wh{-}\wh{+}}}= \rm{sin} 2\delta = \mathbb{S}$.
		 All the indices with the
	wide-hat notation signify the variables with the interpolation angle
	$\delta$.  For the limit $\delta \rightarrow 0$ we have $x^{\wh+} = x^0$
	and $x^{\wh-} = -x^3$ so that we recover usual space-time coordinates
	although the $z$-axis is inverted while for the other extreme limit, $\delta
	\rightarrow \frac{\pi}{4}$ we have $x^{\wh{\pm}} = (x^0\pm x^3)/\sqrt{2}
	\equiv x^{\pm}$ which leads to the standard light-front coordinates.
	Since the perpendicular components remain the same ($x^{\itP{j}}=x^{j},x_{\itP{j}}=x_{j}, j=1,2$), 
	we will omit the ``\textasciicircum''  notation unless necessary from now on for 
	the perpendicular indices $j=1,2$ in a four-vector.
	Of course, the same interpolation applies to the four-momentum variables too as it applies to all four-vectors.
	The details of the relationship between the interpolating variables and
	the usual space-time variables can be seen in our previous 
	works~Ref.\cite{JLS2015,LAJ2015,Alfredo}.
	
	In Ref.\cite{JLS2015}, we developed the electromagnetic gauge field propagator interpolated between the IFD and the LFD and found that the light-front gauge $A^{+}=0$ in the LFD is naturally linked to the Coulomb gauge $\boldsymbol\nabla \cdot \mathbf{A}=0$ in IFD. We identified the dynamical degrees of freedom for the electromagnetic gauge fields as the transverse photon fields and clarified the equivalence between the contribution of the instantaneous interaction and the contribution from the longitudinal polarization of the virtual photon. Our results for the gauge propagator and time-ordered diagrams clarified whether one should choose the two-term form~\cite{Mustaki1991}
	or the three-term form~\cite{Leibbrandt1984,Srivastava2001,ThreeTermPropagator}
	for the gauge propagator in LFD. There has been a sustained interest and discussion on this issue of the two-term vs. three-term gauge propagator in LFD~\cite{Mantovani2016}.
	Our transverse photon propagator in LFD assumes the three-term form, but the third term cancels the instantaneous interaction contribution. Thus, one can use the two-term form of the gauge propagator for effective calculation of amplitudes if one also omits the instantaneous interaction from the Hamiltonian.
	But if one wants to show equivalence to the covariant theory, all three terms should be kept because the instantaneous interaction is a natural result of the decomposition of Feynman diagrams, and the third term in the propagator is necessary for the total amplitudes to be covariant.
	We also see that the photon propagator was derived according to the generalized gauge that links the Coulomb gauge to light-front gauge and thus the three-term form appears appropriate in order to be consistent with the appropriate gauge.
	
	In Ref.\cite{LAJ2015}, we derived the generalized helicity spinor that links the instant form helicity spinor to the light-front helicity spinor. For a given generalized helicity spinor, the spin direction does not coincide with the momentum direction in general. Thus, we studied how the spin orientation angle $\theta_{s}$ changes in terms of both $\delta$ and the angle $\theta$ that defines the momentum direction of the particle. In particular, the helicity in IFD depends on the reference frame. If the observer moves faster than the positive helicity spinor, then the direction of the momentum becomes opposite to the spin direction and the helicity of the spinor flips its sign.
	In contrast, the helicity defined in LFD is independent of the reference frame. We have detailed the increment of the angle difference $\theta-\theta_s$ with the increment of the interpolation angle $\delta$ in Ref.\cite{LAJ2015}, which bifurcates at a critical interpolation angle $\delta_{c}$. We found this critical interpolation angle $\delta_{c}=\arctan \left( \frac{|\mathbf{P}|}{E} \right)$, where $|\mathbf{P}|$ and $E$ are the magnitude of the three-momentum and the energy of the particle under investigation. 
	The IFD and the LFD belong to separately the two different branches bifurcated and divided out at the critical interpolation angle $\delta_c$. This bifurcation indicates the necessity of the distinction in the spin orientation between the IFD and the LFD and clarifies any conceivable confusion in the prevailing notion of the equivalence between  the LFD and the infinite momentum frame (IMF) approach~\cite{Weinberg} formulated in the IFD.
	
	Now that the spinor has been interpolated between IFD and LFD, we show in this work that the covariant Feynman propagator 
	$\Sigma=\frac{\slashed{q}+m}{q^2-m^2}$
	of the intermediate virtual fermion with the four-momentum $q$ and the mass $m$ 
	can also be decomposed into the two interpolating time-ordered processes, one with the ``forward moving" intermediate fermion
	in the sense that its interpolating longitudinal momentum $q_{\wh{-}}$ is positive, i.e. $q_{\wh{-}}>0$ and the other
	with the ``backward moving" intermediate fermion carrying the opposite sign of $-q_{\wh{-}}$, i.e. $-q_{\wh{-}}<0$.
	The corresponding ``forward" and ``backward" amplitudes are given by
	\begin{align}\label{eqn:Sigma}
	\Sigma_{F} =\frac{1}{2Q^{\wh{+}}}\;\frac{\slashed{Q}_F+m}{q_{\wh{+}}-Q_{F\wh{+}}},
	\Sigma_{B} =\frac{1}{2Q^{\wh{+}}}\;\frac{-\slashed{Q}_B+m}{-q_{\wh{+}}-Q_{B\wh{+}}},
	\end{align}
	where
	\begin{align} \label{eqn:eqn:qb+}
	Q_{F\wh{+}}&=
	\frac{-\mathbb{S}q_{F\wh{-}}+Q^{\wh{+}}}{\mathbb{C}},\\ \label{qb+}
	Q_{B\wh{+}}&=\frac{-\mathbb{S}q_{B\wh{-}}+Q^{\wh{+}}}{\mathbb{C}},
	\end{align}
	and
	\be\label{eqn:q+}
	Q^{\wh{+}}=\sqrt{q_{\wh{-}}^{2}+\mathbb{C} (\mathbf{q}_{\perp}^{2}+m^{2})},
	\ee
	with the 4-momenta $q_F=q$ and $q_B=-q$ which 
	are those of the off-shell fermion and anti-fermion, while $Q_F$ and $Q_B$ are the corresponding on-shell 4-momenta. 
	Only the interpolating energies of the ``forward" and ``backward" moving intermediate fermions, i.e. $Q_{F\wh{+}}$ and $Q_{B\wh{+}}$ are different from $q_F$ and $q_B$, respectively, as given by Eqs.~(\ref{eqn:eqn:qb+}) and (\ref{qb+}).
	In the light-front limit $ \delta\to\frac{\pi}{4} $, i.e., $ \mathbb{C}\to0 $, we get
	\begin{align}\label{eqn:LFlimit-fermion-propagator}
	\Sigma_{F,\delta\to\frac{\pi}{4}} = \frac{\slashed{q}_{on}+m}{q^2-m^2}, \hspace{.3in}
	\Sigma_{B,\delta\to\frac{\pi}{4}} = \frac{\gamma^+}{2q^+},
	\end{align}
	where $ q_{on} $ is the on-shell  momentum 4-vector with its spacial part equal to that of $q$
	while it satisfies the Einstein energy-momentum relationship. Here, $\Sigma_{B,\delta\to\frac{\pi}{4}}$ 
	turns out to be the instantaneous contribution in the light-front propagator. This proves the usual light-front decomposition of
	the fermion propagator given by~\cite{SJChang}
	\begin{equation}\label{eqn:LF-fermion-propagator}
	\frac{1}{\slashed{q} - m} = \frac{\sum_s u(q,s) {\bar u}(q,s)}{q^2 - m^2} + \frac{\gamma^+}{2q^+},
	\end{equation}
	where the numerator $\slashed{q}_{on}+m$ of $\Sigma_{F,\delta\to\frac{\pi}{4}}$ in Eq.~(\ref{eqn:LFlimit-fermion-propagator}) is replaced by the spin sum of the on-shell spinor product $\sum_s u(q,s) {\bar u}(q,s)$.
	
	In the next section, Sec.~\ref{sec:formal}, we present the formal derivation of the interpolating QED. We outline two different derivations of the Feynman rules for $x^{\wh{+}}$-ordered diagrams formulated at any interpolation angle.
	The first approach directly decomposes the covariant Feynman diagram, and the second one utilizes the canonical field theory and the old-fashioned perturbation theory. We notice in particular the constraint fermion degree of freedom which appears uniquely in the LFD, resulting in the instantaneous contribution to the fermion propagator.
	The canonical field theory is studied for the entire range of the interpolation angle $0 \le \delta \le \pi/4$.
	Equations of motion, free fields, gauge condition, momentum and angular momentum tensor are examined, and the Hamiltonian at constant $x^{\wh{+}}$ is found.
	In Sec.~\ref{sec:eess}, we study the $ x^{\wh{+}} $-ordered fermion propagator in more detail. Taking a simple example, the annihilation of fermion and anti-fermion into two scalar particles, we show the characteristic behavior of the amplitudes as the form interpolates between IFD and LFD. Both the collinear and non-collinear cases are discussed examining the 
angular momentum conservation. In Sec.~\ref{sec:calculation}, we present the results for the $ e^+e^-\to \gamma \gamma$ process  and the Compton Scattering $ e\gamma\to e\gamma $. We compute all 16 helicity amplitudes and discuss the frame dependence and/or the scattering angle dependence with respect to the interpolation angle dependence.
For the $e^+ e^- \to \gamma \gamma$ amplitudes, the symmetry between the forward and backward angle dependence is discussed. The limit to the LFD ($\delta=\pi/4$) is analyzed and the comparison with the well-known analytic results from the manifestly covariant calculation is presented. Summary and conclusions follow in Sec.~\ref{sec:conclusion}. In Appendix~\ref{app:position_space}, we derive Eq.(\ref{eqn:Sigma}) and present the fermion propagator in the position space which supplements the discussion 
in Sec.~\ref{sub:scattering_theory}. In Appendix~\ref{app:Hamiltonian}, we present the derivation of interpolating QED Hamiltonian which supplements the discussion in Sec.~\ref{sub:canonical_field_theory}. 
In Appendix~\ref{app:Propagator}, the manifestly covariant fermion propagator is explicitly derived from the sum of the interpolating time-ordered fermion propagators.
In Appendix~\ref{app:ApprantAngle}, we provide the relation between the center of mass scattering angle and the apparent scattering angle in a boosted frame and correspond the angular distributions for the center of mass frame in Sec.~\ref{sec:eess} to the apparent angle distributions in boosted frames. The angular distribution and the frame dependence of the $ e^+e^-\to \gamma \gamma $ helicity amplitudes 
are summarized in in Appendices~\ref{app:BoostedAnnihilation} and \ref{app:PzDepThetaPiOver3}, respectively.\\
	
	\section{Formal derivation of the Interpolation of QED}
	\label{sec:formal}
	In our previous works, we studied in great detail the interpolation of the photon polarization vectors, the gauge propagator and the on-mass-shell helicity spinors.
	In this paper, we complete the interpolation of the QED theory by providing the final piece of the entity: the interpolating fermion propagator.
	The form of this interpolating fermion propagator is derived.
	In subsection \ref{sub:scattering_theory}, we decompose the covariant Feynman diagrams into $x^{\wh{+}}$-ordered diagrams, from which a general set of Feynman rules for any $x^{\wh{+}}$-ordered scattering theory is obtained.
	In subsection \ref{sub:canonical_field_theory}, the canonical field theory approach is studied and the corresponding Hamiltonian for the old fashioned perturbation theory is derived.

	\subsection{Scattering Theory}
	\label{sub:scattering_theory}
	
	Following what Kogut and Soper did in their light-front QED paper~\cite{KS}\footnote{Although Kogut and Soper represented their work in Ref.~\cite{KS} as the QED in the infinite momentum frame, it actually was the formulation of QED in the Light-Front Dynamics (LFD).}, we regard the perturbative expansion of the S matrix in Feynman diagrams as the foundation of quantum electrodynamics.
	In this section, we decompose the covariant Feynman diagram into a sum of $x^{\wh{+}}$-ordered diagrams.
	We shall not be concerned with the convergence of the perturbation series, or convergence and regularization of the integrals in the present work.
	
	\subsubsection{Propagator Decomposition}\label{ssb:propagator_decomposition}
	
	In Ref.~\cite{JLS2015}, we obtained the decomposition of the photon propagator given by
	\begin{widetext}
		\begin{align}
		D_{F}(x)_{\wh{\mu}\wh{\nu}}&= \int \frac{d^{2}\mathbf{q}_{\perp} }{(2\pi)^{3}}\int_{-\infty}^{\infty}d q_{\wh{-}} \wh\Theta(q_{\wh{-}}) \frac{\mathcal{T}_{\wh{\mu}\wh{\nu}}}{2 \sqrt{q_{\wh{-}}^{2}+\mathbb{C} \mathbf{q}_{\perp}^{2}}}
		\left[ \Theta(x^{\wh{+}})e^{-i q_{\wh{\mu}}x^{\wh{\mu}}}+\Theta(-x^{\wh{+}})e^{i q_{\wh{\mu}}x^{\wh{\mu}}} \right] \notag\\
		&+i\delta(x^{\wh{+}})\int \frac{d^{2}\mathbf{q}_{\perp} }{(2\pi)^{3}}\int_{-\infty}^{\infty}d q_{\wh{-}} \frac{n_{\wh{\mu}}n_{\wh{\nu}}}{q_{\wh{-}}^{2}+\mathbb{C} \mathbf{q}_{\perp}^{2}} e^{-i(q_{\wh{-}}x^{\wh{-}}+\mathbf{q}_{\perp}\mathbf{x}^{\perp})}\label{eqn:unified_photon_time_ordered_propagator_QED}
		\end{align}
	\end{widetext}
	where $q_{\wh{+}}=\left(-\mathbb{S} q_{\wh{-}}+\sqrt{q_{\wh{-}}^{2}+\mathbb{C} \mathbf{q}_{\perp}^{2}}\right)/\mathbb{C}$ 
	is the interpolating on-mass-shell energy and 
	the explicit form of $\mathcal{T}_{\wh{\mu}\wh{\nu}}$ is given by 
	\begin{align}\label{eqn:transverse_photon_propagator_numerator_in_radiation_gauge}
	&\mathcal{T}_{\wh{\mu}\wh{\nu}} \equiv \sum_{\lambda=\pm}{\epsilon_{\wh{\mu}}^{*}(\lambda)\epsilon_{\wh{\nu}}(\lambda)} \notag\\ 
	&= -g_{\wh{\mu}\wh{\nu}}+\dfrac{(q\cdot n)(q_{\wh{\mu}}n_{\wh{\nu}}+q_{\wh{\nu}}n_{\wh{\mu}})}{\mathbf{q}_{\perp}^{2}\Cc+q_{\mT}^{2}} 
	- \dfrac{\Cc q_{\wh{\mu}}q_{\wh{\nu}}}{\mathbf{q}_{\perp}^{2}\Cc+q_{\mT}^{2}} - \dfrac{q^{2}n_{\wh{\mu}}n_{\wh{\nu}}}{\mathbf{q}_{\perp}^{2}\Cc+q_{\mT}^{2}}
	\end{align}
	with the obvious familiar notation $q\cdot n = q_{\wh{\mu}}n^{\wh{\mu}}$ and $q^2 = q_{\wh{\mu}}q^{\wh{\mu}}$.
	Here, the polarization vectors $\epsilon^{\wh{\mu}}(p,\pm)$ are explicitly given in Ref.~\cite{JLS2015} and
	$\mathcal{T}_{\wh{\mu}\wh{\nu}}$ given by Eq.~(\ref{eqn:transverse_photon_propagator_numerator_in_radiation_gauge})
	is obtained in the radiation gauge for any interpolating angle, i.e. $A^{\pT}=0$ and 
	$\partial_{\mT}A_{\mT}+\boldsymbol\partial_{\perp}\cdot\mathbf{A}_{\perp}\Cc=0$. As discussed in Ref.~\cite{JLS2015},   
	our interpolating radiation gauge links naturally the Coulomb gauge in IFD ($\Cc=1$) and the light front gauge in LFD ($\Cc=0$). 
	One should also note that $\wh\Theta(q_{\wh{-}})$ in Eq.~(\ref{eqn:unified_photon_time_ordered_propagator_QED}) is 
	the interpolating step function given by
	\begin{align}
	\wh\Theta(q_{\wh{-}}) &= \Theta(q_{\wh{-}}) + (1-\delta_{\mathbb{C} 0})\Theta(-q_{\wh{-}}) \nonumber \\
	&=
	\begin{cases}
	1 \hspace{1cm} &(\mathbb{C}\neq0)\\
	\Theta(q^+) &(\mathbb{C}=0)
	\end{cases}
	\label{eqn:Interpolating_Theta_Function_QED}
	\end{align}
	which was introduced to combine the results of $\mathbb{C}\neq0$ and $\mathbb{C}=0$.
	
	Similarly, the manifestly covariant Klein-Gordon propagator $\Delta_{F}(x)$ in the position space 
	given by 
	\begin{widetext}
		\begin{align}
		\Delta_{F}(x)&\equiv \int\dfrac{d^{4}q}{(2\pi)^{4}} \exp(-i q_{\wh{\mu}} x^{\wh{\mu}})\dfrac{i}{q^{\wh{\mu}}q_{\wh{\mu}}-m^{2}+i\epsilon}\nonumber \\
		&=\int \dfrac{d^{2}\mathbf{q}_{\perp} d q_{\wh{-}} d q_{\wh{+}}} {(2\pi)^{4}}\exp{[-i(q_{\wh{+}}x^{\wh{+}}+q_{\wh{-}}x^{\wh{-}}+\mathbf{q}_{\perp}\cdot\mathbf{x}^{\perp})]}
		\dfrac{i}{\mathbb{C} q_{\wh{+}}^{2}+2 \mathbb{S} q_{\wh{-}}q_{\wh{+}}-\mathbb{C} q_{\wh{-}}^{2}-\mathbf{q}_{\perp}^{2}-m^{2}+i\epsilon} \label{eqn:Klein_Gordon_propagator}
		\end{align}
	\end{widetext}
	can also be obtained by combining the results of $\mathbb{C}\neq0$ and $\mathbb{C}=0$ with the interpolating step function 
	$\wh\Theta(q_{\wh{-}})$:
	\begin{widetext}
		\begin{align}
		\Delta_{F}(x)=\int \dfrac{d^{2}\mathbf{q}_{\perp} }{(2\pi)^{2}}\int_{-\infty}^{\infty}\frac{d q_{\wh{-}}}{2\pi} \wh\Theta(q_{\wh{-}}) \dfrac{1}{2 \sqrt{q_{\wh{-}}^{2}+\mathbb{C} (\mathbf{q}_{\perp}^{2}+m^{2})}}
		\left[ \Theta(x^{\wh{+}})e^{-i q_{\wh{\mu}}x^{\wh{\mu}}}+\Theta(-x^{\wh{+}})e^{i q_{\wh{\mu}}x^{\wh{\mu}}} \right],\label{eqn:Klein_Gordon_time_ordered_propagator}
		\end{align}
	\end{widetext}
	where the value of $ q_{\wh{+}} $ in the exponent is taken to be the interpolating on-mass-shell energy, i.e. 
\begin{equation}
q_{\wh{+}}=\begin{cases}
	\left(-\mathbb{S} q_{\wh{-}}+\sqrt{q_{\wh{-}}^{2}+\mathbb{C} (\mathbf{q}_{\perp}^{2}+m^{2})}\right)/\mathbb{C},\quad\mathrm{for}\ x^{\wh+}>0,\\
	 \left(-\mathbb{S} q_{\wh{-}}-\sqrt{q_{\wh{-}}^{2}+\mathbb{C} (\mathbf{q}_{\perp}^{2}+m^{2})}\right)/\mathbb{C},\quad\mathrm{for}\ x^{\wh+}<0. \notag
	\end{cases}
\end{equation} 
The detailed derivation of Eqs.(\ref{eqn:Klein_Gordon_propagator}) and (\ref{eqn:Klein_Gordon_time_ordered_propagator}) will be given in Appendix~\ref{app:position_space}, where the pole integration is done explicitly.
	
	The result for $\mathbb{C}\neq0$, i.e. $\wh\Theta(q_{\wh{-}}) =1$, in Eq.~(\ref{eqn:Klein_Gordon_time_ordered_propagator})
	can be obtained by noting the two poles for $q_{\wh{+}}$ in Eq.~(\ref{eqn:Klein_Gordon_propagator}) given by
	\begin{align}
	{\cal A}_{\wh+}-i\epsilon'=\left(-\mathbb{S} q_{\wh{-}}+\sqrt{q_{\wh{-}}^{2}+\mathbb{C} (\mathbf{q}_{\perp}^{2}+m^{2})}\right)/\mathbb{C}-i\epsilon',\label{eqn:A_Pole_m}\\
	-{\cal B}_{\wh+}+i\epsilon'=\left(-\mathbb{S} q_{\wh{-}}-\sqrt{q_{\wh{-}}^{2}+\mathbb{C} (\mathbf{q}_{\perp}^{2}+m^{2})}\right)/\mathbb{C}+i\epsilon',\label{eqn:B_Pole_m}
	\end{align}
	where $\epsilon'>0$. In order not to involve any contribution from the arc in the contour integration, we evaluate the $q_{\wh{+}}$ integral in Eq.~(\ref{eqn:Klein_Gordon_propagator}) by closing the contour in the lower (upper) half plane if $x^{\wh{+}}>0$ ($x^{\wh{+}}<0$).
	This produces the desired decomposition for $\Delta_{F}(x)$ with $\wh\Theta(q_{\wh{-}}) =1$ in 
	Eq.~(\ref{eqn:Klein_Gordon_time_ordered_propagator}) given by
	\begin{align}
	\Delta_{F}(x)&=\int \dfrac{d^{2}\mathbf{q}_{\perp} }{(2\pi)^{3}}\int_{-\infty}^{\infty}d q_{\wh{-}} \dfrac{1}{2 Q^{\wh{+}}}\notag\\
	&\times
	\left[ \Theta(x^{\wh{+}})e^{-i q_{\wh{\mu}}x^{\wh{\mu}}}+\Theta(-x^{\wh{+}})e^{i q_{\wh{\mu}}x^{\wh{\mu}}} \right],\label{eqn:Klein_Gordon_time_ordered_propagator1}
	\end{align}
	where we denoted the denominator factor  
	in Eq.~(\ref{eqn:Klein_Gordon_time_ordered_propagator}) by $Q^{\wh{+}}$, i.e.
	\begin{equation}
	Q^{\wh{+}}\equiv\sqrt{q_{\wh{-}}^{2}+\mathbb{C} (\mathbf{q}_{\perp}^{2}+m^{2})}.
	\label{eqn:Q^+}
	\end{equation}
Note here that the integration measure in Eq.~(\ref{eqn:Klein_Gordon_time_ordered_propagator}) is the invariant differential surface element on the mass shell, i.e.
	\begin{equation}
	\int \dfrac{d^{2}\mathbf{q}_{\perp} }{(2\pi)^{3}} \dfrac{d q_{\wh{-}}}{2 Q^{\wh{+}}}=\int\dfrac{d^{4}q}{(2\pi)^{4}}2\pi\delta(q^{2}-m^{2}).\label{eqn:invariant_mass_shell_element}
	\end{equation}
	
	The result for $\mathbb{C} = 0$, i.e. $\wh\Theta(q_{\wh{-}}) = \Theta(q^+)$, in Eq.~(\ref{eqn:Klein_Gordon_time_ordered_propagator})
	can also be obtained by noting the single pole for $q_{\wh{+}}=q^-$ in Eq.~(\ref{eqn:Klein_Gordon_propagator}) given by
	\begin{equation}
	q^-=\frac{\mathbf{q}_{\perp}^{2}+m^{2}}{2q^+} - i \frac{\epsilon}{2q^+}\label{eqn:q-_pole_LF}
	\end{equation}
	which should be taken in the contour integration of the light-front energy $q^-$ variable
	without involving the arc contribution 
	in Eq.~(\ref{eqn:Klein_Gordon_propagator}). Note here that this single pole corresponds to ${\cal A}_{\wh+} $ in Eq.~(\ref{eqn:A_Pole_m}) in the limit of $ \mathbb{C}\to0 $.
	This requires to close the contour in the lower (upper) half plane of the complex $q^-$ space
	if $x^+ > 0 $ $(x^+ < 0)$, as we explained essentially the same procedure for $ \mathbb{C}\neq0 $ case. 
	
	Due to the rational relation between $ q^- $ and $ q^+ $ given by Eq.~(\ref{eqn:q-_pole_LF}), the value of $q^+$ must be positive to keep the $q^-$ pole in the lower half plane for $x^+ > 0$,
	while the value of $q^+$ must be negative to keep the $q^-$ pole in the upper half plane for $x^+ < 0$.
	This leads to the result given by 
	\begin{align}
	\Delta_{F}(x)&=\int \dfrac{d^{2}\mathbf{q}_{\perp} }{(2\pi)^{3}}\int_{0}^{\infty}d q^+ \dfrac{1}{2 q^+}\notag\\
	&\times
	\left[ \Theta(x^{+})e^{-i q \cdot x}+\Theta(-x^{+})e^{i q \cdot x} \right],\label{eqn:Klein_Gordon_time_ordered_propagator0}
	\end{align}
	where $q \cdot x = q^+ x^- + (\frac{\mathbf{q}_{\perp}^{2}+m^{2}}{2q^+})x^- -\mathbf{q}_{\perp}\cdot\mathbf{x}_{\perp}$
	noting $\mathbf{x}^{\perp}=-\mathbf{x}_{\perp}$.
	This result is identical to Eq.~(\ref{eqn:Klein_Gordon_time_ordered_propagator}) for $\Cc = 0$. Thus, our result in Eq.~(\ref{eqn:Klein_Gordon_time_ordered_propagator}) covers both $ \mathbb{C}\neq0 $ and $ \mathbb{C}=0 $ cases together.
	
	As the fermion propagator in the position space can be obtained by 
	\begin{equation}
	S_{F}(x)=(i\partial_{\wh\mu}\gamma^{\wh\mu}+m)\Delta_{F}(x),\label{eqn:covariant_electron_propagator}
	\end{equation}
	we can now use Eqs.~(\ref{eqn:Klein_Gordon_time_ordered_propagator}) and (\ref{eqn:covariant_electron_propagator}) to derive a decomposition for the fermion propagator
	given by
	\begin{widetext}
		\begin{align}
		S_{F}(x)&=\int \dfrac{d^{2}\mathbf{q}_{\perp} }{(2\pi)^{3}}\int_{-\infty}^{\infty}d q_{\wh{-}} \wh\Theta(q_{\wh{-}})\dfrac{1}{2 Q^{\wh{+}}}
		\left[ \Theta(x^{\wh{+}})(\slashed q+m)e^{-i q_{\wh{\mu}}x^{\wh{\mu}}}
		+\Theta(-x^{\wh{+}})(-\slashed q+m)e^{i q_{\wh{\mu}}x^{\wh{\mu}}}\right]\notag\\
		&+i\gamma^{\wh{+}}\int \dfrac{d^{2}\mathbf{q}_{\perp} }{(2\pi)^{3}}\int_{-\infty}^{\infty}d q_{\wh{-}} \wh\Theta(q_{\wh{-}})\dfrac{1}{2 Q^{\wh{+}}}\left[ \delta(x^{\wh{+}}) e^{-i q_{\wh{\mu}}x^{\wh{\mu}}}-\delta(x^{\wh{+}})e^{i q_{\wh{\mu}}x^{\wh{\mu}}}  \right] ,\label{eqn:electron_time_ordered_propagator}
		\end{align}
	\end{widetext}
	where the ``$ \wh+ $'' component of $ q $ takes the corresponding pole values, as mentioned before. 
	Here, 
	the differentiation of $\Theta(x^{\wh{+}})$ and $\Theta(-x^{\wh{+}})$ in Eq.~(\ref{eqn:Klein_Gordon_time_ordered_propagator}) with respect to $x^{\wh{+}}$ gives us two terms: $\delta(x^{\wh{+}})e^{-i q_{\wh{\mu}}x^{\wh{\mu}}}$ and $-\delta(x^{\wh{+}})e^{i q_{\wh{\mu}}x^{\wh{\mu}}}$ in $ \mathbb{C}\neq0 $ case, and these two will cancel each other exactly when an integration with respect to $x^{\wh{+}}$ is performed 
	as we show explicitly in the next subsection, so that they don't contribute to the Feynman rules. 
	Therefore, we can drop them from 
	the decomposition. Thus, when $ \mathbb{C}\neq0 $, the second line in Eq.~(\ref{eqn:electron_time_ordered_propagator}) automatically drops off, and the first line is the whole result. However, in the $ \mathbb{C}=0 $ case, the integration over $q_- = q^+$ (note that $ q_{\wh{-}} $ is just $ q_- $ without hat when $ \mathbb{C}=0 $) goes from $ 0 $ to $ \infty $ instead of $ -\infty $ to $ \infty $ as denoted by the interpolating step function $\wh\Theta(q_{\wh{-}})$. Thus, the two $ \delta(x^+) $ terms resulting from differentiating the $ \Theta(x^+) $ function do not cancel each other, and the term proportional to $ \delta(x^+) $ is left over. This term is the instantaneous contribution unique to the LF. 
	Thus, when we take $ \mathbb{C}=0 $, our fermion propagator result given by Eq.~(\ref{eqn:electron_time_ordered_propagator}) coincides with the LF propagator previously derived by
	Kogut and Soper~\cite{KS}: 
	\begin{widetext}
		\begin{align}
		S_{F}(x)_{\mathrm{LF}}&=\int \dfrac{d^{2}\mathbf{q}_{\perp} }{(2\pi)^{3}}\int_{0}^{\infty} \dfrac{d q^{+}}{2 q^{+}} \left[ \Theta(x^+)(\slashed q+m)e^{-i q\cdot x}
		+\Theta(-x^+)(-\slashed q+m)e^{i q\cdot x} \right]\notag\\
		&+i\delta(x^{+})\gamma^{+}
		\int \dfrac{d^{2}\mathbf{q}_{\perp} }{(2\pi)^{3}}\int_{-\infty}^{\infty} \dfrac{d q^{+}}{2 q^{+}} e^{-iq^{+}x^{-}+i\mathbf{q}_{\perp}\cdot\mathbf{x}_{\perp}}.
		\label{eqn:electron_LF_time_ordered_propagator}
		\end{align}
	\end{widetext}
	Note here that the interpolating wide-hat notations are switched to the usual light front notations.

	\subsubsection{Rules for $x^{\wh{+}}$-ordered Diagrams}
	\label{ssb:Rules_for_x^(pt)-ordered_diagrams}
	\begin{figure}
		\centering
		\subfloat[]{
			\includegraphics[width=0.34\columnwidth]{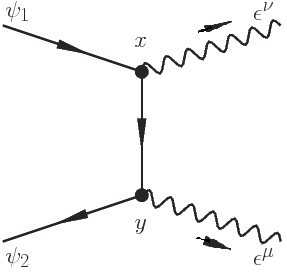}
			\label{fig:annihilation_space}}
		\hspace{20pt}
		\centering
		\subfloat[]{
			\includegraphics[width=0.34\columnwidth]{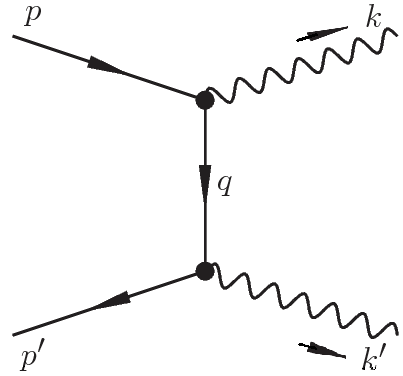}
			\label{fig:annihilation_momentum}}
		\\
		\centering
		\subfloat[]{
			\includegraphics[width=0.34\columnwidth]{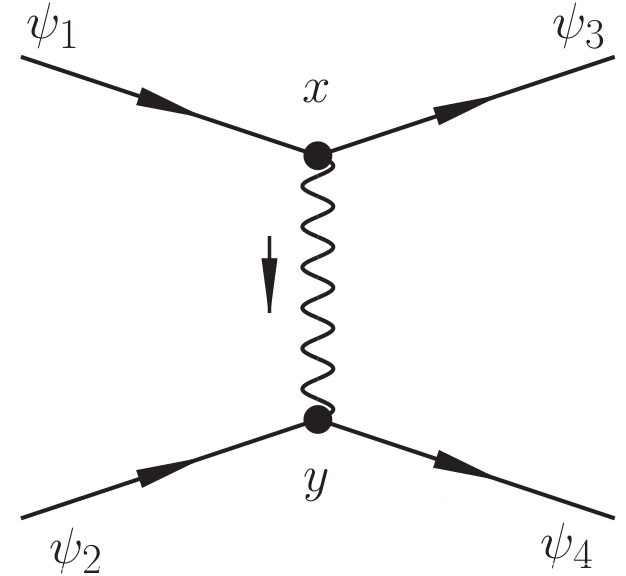}
			\label{fig:scattering_space}}
		\hspace{20pt}
		\centering
		\subfloat[]{
			\includegraphics[width=0.34\columnwidth]{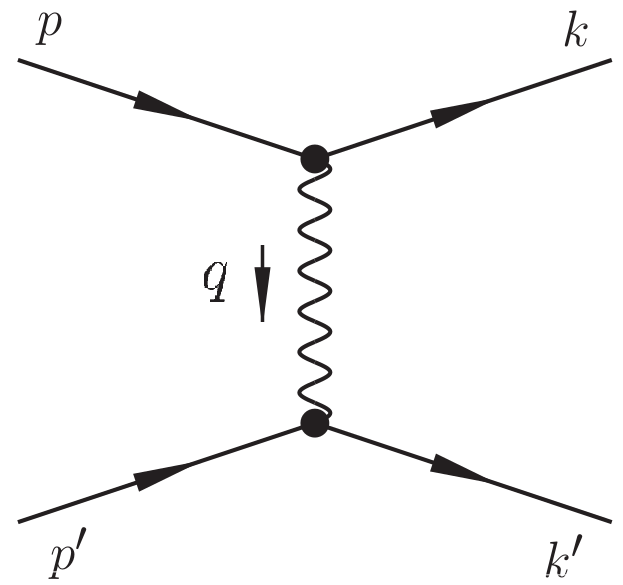}
			\label{fig:scattering_momentum}}
		\caption{\label{fig:covariant_diagrams}
		Lowest-order tree level covariant annihilation diagram in (a) position space and (b) momentum space.
			Lowest-order tree level covariant scattering diagram in (c) position space and (d) momentum space.}
	\end{figure}
	
	To find the rules for $x^{\wh{+}}$-ordered diagrams, we start with the Feynman diagrams in coordinate space.
	The amplitude for diagram shown in Fig.~\ref{fig:annihilation_space} for the process of $e^+e^-\to \gamma \gamma  $ can be written as
	\begin{align}
	i\mathcal{M}=&(-i e)^{2}\int d^{4}x d^{4}y ~\epsilon^{*}_{\wh{\mu}}(y) [\bar{\psi}_{2}(y) \gamma^{\wh{\mu}} \nonumber\\
	&\times S_{F}(y-x) \gamma^{\wh{\nu}} \psi_{1}(x)] \epsilon^{*}_{\wh{\nu}}(x).\label{eqn:electron_propagator_diagram_amplitude}
	\end{align}
	Here, we use the plane wave solution of the Dirac equation for the electron and the charge conjugate plane wave solution for the positron:
	\begin{align}
	\psi_{1}(x)=e^{-i p\cdot x}u(p,s),\label{eqn:electron_wave_function}\\
	\psi_{2}(y)=e^{i p'\cdot  y}v(p',s'),\label{eqn:positron_wave_function}
	\end{align}
	where $p$ and $s$ are the momentum and spin of the fermion.
	The photon wave function is
	\begin{align}
	\epsilon_{\wh{\mu}}(x)=e^{-i k\cdot x}\epsilon_{\wh{\mu}}(k,\lambda),\label{eqn:photon_wave_function}
	\end{align}
	where $\epsilon_{\wh{\mu}}(k,\lambda)$ is the polarization vector with momentum $k$  and helicity $\lambda$, the explicit form of which was given in Ref.~\cite{JLS2015}.
	
	With the change of variables
	\begin{alignat}{2}\label{eqn:xy_to_xT}
	&x\rightarrow x, &\quad&y\rightarrow T=y-x,
	\end{alignat}
	Eq.~(\ref{eqn:electron_propagator_diagram_amplitude}) becomes
	\begin{align}
	i\mathcal{M}=&(-i e)^{2}\int d^{4}x d^{4}T ~e^{i(k'-p')\cdot T}\epsilon^{*}_{\wh{\mu}}(k',\lambda') \left[ \bar{v}(p',s') \gamma^{\wh{\mu}} \right.\notag\\
	&\left. \times S_{F}(T) \gamma^{\wh{\nu}} u(p,s)\right] \epsilon^{*}_{\wh{\nu}}(k,\lambda)e^{i (k+k'-p-p')\cdot x}.
	\label{eqn:electron_propagator_diagram_amplitude_x_T}
	\end{align}
	The $x$ integration immediately gives the total energy-momentum conservation condition.
	After we plug in the decomposed $S_{F}$ given by Eq.~(\ref{eqn:electron_time_ordered_propagator}), 
	we finish the $T^{\wh{+}}$ integration using the following relations
	\begin{align}\label{eqn:M_evaluation_with}
	\int_{-\infty}^{\infty}dT^{\wh{+}}\Theta(T^{\wh{+}})e^{iP_{\wh{+}}T^{\wh{+}}}&=\dfrac{i}{P_{\wh{+}}+i\epsilon},\\
	\int_{-\infty}^{\infty}dT^{\wh{+}}\Theta(-T^{\wh{+}})e^{iP_{\wh{+}}T^{\wh{+}}}&=-\dfrac{i}{P_{\wh{+}}-i\epsilon},
	\end{align}
	where the causality of the relativistic quantum field theory is assured with the $\pm i\epsilon$ factor for the $\pm T^{\wh{+}}$ region, respectively.
	Thus, we get the interpolating energy denominator factor of $\frac{i}{P_{ini \wh{+}}-P_{inter \wh{+}}+i\epsilon}$ for each intermediate state.
	For the momentum assignment shown in Fig.~\ref{fig:annihilation_momentum}, $P_{ini \wh{+}}=p_{\wh{+}}+p'_{\wh{+}}$ is the total ``energy'' of the initial particles, and $P_{inter \wh{+}}$ gives the total ``energy'' of the intermediate particles, which is $k_{\wh{+}}+q_{\wh{+}}+p'_{\wh{+}}$ when $y^{\wh{+}}>x^{\wh{+}}$ and $p_{\wh{+}}-q_{\wh{+}}+k'_{\wh{+}}$ when $y^{\wh{+}}<x^{\wh{+}}$.
	On the other hand, the $dT^{\wh{-}}d^{2}\mathbf{T}^{\perp}$ integration gives straightforwardly $(2\pi)^{3}\delta(P_{\wh{-}}^{\text{in}}-P_{\wh{-}}^{\text{out}})\delta^{2}(\mathbf{P}_{\perp}^{\text{in}}-\mathbf{P}_{\perp}^{\text{out}})$ at each vertex.
	Lastly, the $\delta(T^{+})$ term in Eq.~(\ref{eqn:electron_time_ordered_propagator}) gives an extra instantaneous contribution at the light-front ($\mathbb{C}=0$) and is easy to calculate.
	\begin{figure}
		\centering
		\subfloat[]{\includegraphics[width=0.3\columnwidth]{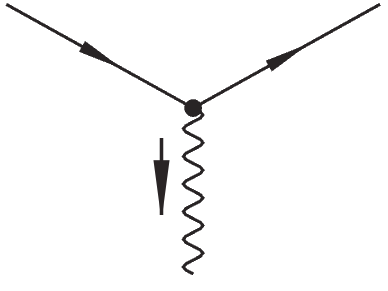}\label{fig:Normal_Vertex}}
		\vspace{0pt}
		\centering
		\subfloat[]{\includegraphics[width=0.3\columnwidth]{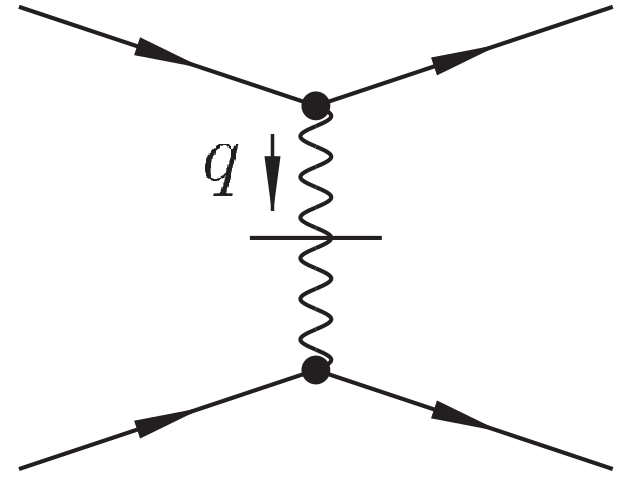}\label{fig:Instantaneous_Photon_Vertex}}
		\vspace{0pt}
		\centering
		\subfloat[]{\includegraphics[width=0.3\columnwidth]{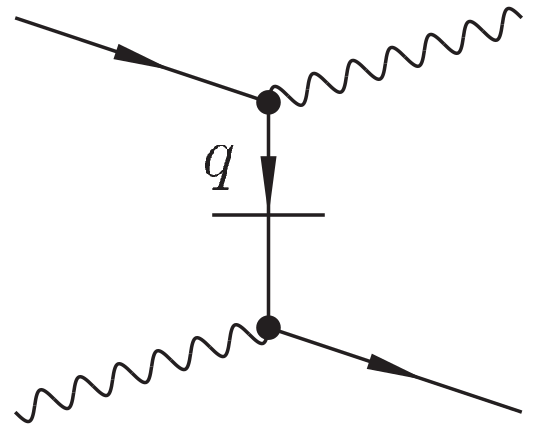}\label{fig:Instantaneous_Electron_Vertex}}
		\caption{\label{fig:Vertices}Vertices that appear in the $x^{\wh{+}}$-ordered diagrams. When $\mathbb{C}\neq0$, only two kinds of vertices (a) and (b) exist. When $\mathbb{C}=0$, all three vertices (a), (b), (c) are present.}
	\end{figure}
	Similar analysis can be done for the process of $ e\mu\to e\mu $ shown in Figs.~\ref{fig:scattering_space} and ~\ref{fig:scattering_momentum}, with the decomposition equation of the photon propagator given by Eq.~(\ref{eqn:unified_photon_time_ordered_propagator_QED}).
	
	After the above analysis, with a little thought, one can summarize and write down the rules for $x^{\wh{+}}$-ordered diagrams 
	as the following:
	\begin{enumerate}
		\item $u(p,s)$, $\bar{u}(p,s)$, $v(p,s)$, $\bar{v}(p,s)$, $\epsilon_{\mu}(p,\lambda)$, and $\epsilon^{*}_{\mu}(p,\lambda)$ for each incoming and outgoing external lines;
		\item $(\slashed{p}+m)=\Sigma_{s}u(p,s)\bar{u}(p,s)$ for electron propagators; $(-\slashed{p}+m)=-\Sigma_{s}v(p,s)\bar{v}(p,s)$ for positron propagators; $\mathcal{T}_{\wh{\mu}\wh{\nu}}\equiv \sum_{\lambda=\pm}{\epsilon_{\wh{\mu}}^{*}(\lambda)\epsilon_{\wh{\nu}}(\lambda)}$ for photon propagators;
		\item $-ie\gamma^{\wh{\mu}}(2\pi)^{3}\delta(P_{\wh{-}}^{\text{in}}-P_{\wh{-}}^{\text{out}})\delta^{2}(\mathbf{P}_{\perp}^{\text{in}}-\mathbf{P}_{\perp}^{\text{out}})$ for each vertex as shown in Fig.~\ref{fig:Normal_Vertex};
		\begin{equation}
		\begin{split}
		-&e^{2}\dfrac{in_{\wh{\mu}}n_{\wh{\nu}}}{q_{\wh{-}}^{2}+\mathbb{C} \mathbf{q}_{\perp}^{2}}(2\pi)^{3}\delta(P_{\wh{-}}^{\text{in}}-P_{\wh{-}}^{\text{out}})\delta^{2}(\mathbf{P}_{\perp}^{\text{in}}-\mathbf{P}_{\perp}^{\text{out}})\\
		&\times\cdots\gamma^{\wh{\mu}}\cdots\gamma^{\wh{\nu}}\cdots \nonumber\label{eqn:instantaneous_photon_vertex}
		\end{split}
		\end{equation}
		for each vertex as shown in Fig.  \ref{fig:Instantaneous_Photon_Vertex}, where $q_{\wh{-}}$, $\mathbf{q}_{\perp}$ are the total momentum transfered;
		\begin{equation}
		-ie^{2}\gamma^{\mu}\gamma^{+}\gamma^{\nu}\frac{1}{2q^{+}}(2\pi)^{3}\delta(P^{+}_{\text{in}}-P^{+}_{\text{out}})\delta^{2}(\mathbf{P}_{\perp}^{\text{in}}-\mathbf{P}_{\perp}^{\text{out}}), \nonumber
		\end{equation}
		for each vertex as shown in Fig.~\ref{fig:Instantaneous_Electron_Vertex} appearing only if $ \mathbb{C}=0 $, i.e. only in LFD, where $q^{+}=k'^{+}-p'^{+}$;
		\item $\frac{i}{P_{ini \wh{+}}-P_{inter \wh{+}}+\i\epsilon}$ for each internal line, where $P_{int \wh{+}}$ and $P_{inter \wh{+}}$ are the sums of energies for the initial and intermediate particles;
		\item an over-all factor of $(2\pi)\delta(P_{\wh{+}}^{in}-P_{\wh{+}}^{out})$ for the interpolating energy conservation;
		\item an integration
		\begin{align}
		\int\frac{d \mathbf{q}_{\perp}}{(2\pi)^{3}}\int_{-\infty}^{\infty}\frac{dq_{\wh{-}}}{2Q^{\wh{+}}} \wh\Theta(q_{\wh{-}})\nonumber
		\end{align}
		for every internal propagating line, with $m$ in Eq.~(\ref{eqn:Q^+}) being the mass of the exchanged particle.
	\end{enumerate}
	
	The rules for $x^{+}$-ordered diagrams on the light front, first derived by Kogut and Soper~\cite{KS}, are reproduced by taking $ \mathbb{C}=0 $ in the above rules. For instance, in rule 6, when $ \mathbb{C}=0 $, the integration limits of $ q_{\wh{-}}=q_-=q^+ $ 
	change to $(0,\infty)$, i.e. 
	\begin{align}
	\int\frac{d \mathbf{q}_{\perp}}{(2\pi)^{3}}\int_{0}^{\infty}\frac{dq^{+}}{2q^{+}} \nonumber
	\end{align}
	for every internal line.
	
	In the next subsection~\ref{sub:canonical_field_theory}, we develop the canonical field theory of quantum electrodynamics in any interpolating angle. And we will see that it reproduces the Feynman rules we obtained here.

	\subsection{Canonical Field Theory}
	\label{sub:canonical_field_theory}
	
	\subsubsection{Equations of Motion}
	\label{ssb:equations_of_motion}
	The Lagrangian density for QED is
	\begin{align}
	\mathcal{L}=-\frac{1}{4}F_{\wh{\mu}\wh{\nu}}F^{\wh{\mu}\wh{\nu}}+\bar{\psi}(i\gamma^{\wh{\mu}}D_{\wh{\mu}}-m)\psi, \label{eqn:QED_Lagrangian}
	\end{align}
	where $D_{\wh{\mu}}=\partial_{\wh{\mu}}+ie A_{\wh{\mu}}$, and $F_{\wh{\mu}\wh{\nu}}=\partial_{\wh{\mu}}A_{\wh{\nu}}-\partial_{\wh{\nu}}A_{\wh{\mu}}$.
	The equations of motion are therefore
	\begin{align}
	\partial_{\wh{\mu}}F^{\wh{\mu}\wh{\nu}}=e J^{\wh{\nu}}=e \bar{\psi}\gamma^{\wh{\nu}}\psi. \label{eqn:gauge_field_EOM} \\
	(i\gamma^{\wh{\mu}}\partial_{\wh{\mu}} - e \gamma^{\wh{\mu}}A_{\wh{\mu}}-m)\psi=0, \label{eqn:fermion_EOM}
	\end{align}
	
	By converting the upper index components into lower index components, Eq.~(\ref{eqn:gauge_field_EOM}) 
	can be written as
	\begin{align}
	&(\mathbb{C}\boldsymbol\partial_{\perp}^{2}+\partial_{\wh{-}}^{2})A_{\wh{+}}\notag\\
	=&(\mathbb{C}\partial_{\wh{+}}+\mathbb{S}\partial_{\wh{-}})\boldsymbol\partial_{\perp}\cdot\mathbf{A}_{\perp}+(\partial_{\wh{+}}\partial_{\wh{-}}-\mathbb{S}\boldsymbol\partial_{\perp}^{2})A_{\wh{-}}-eJ^{\wh{+}}. \label{eqn:A_pT_EOM_step_1}
	\end{align}
	Next, we apply the generalized transverse gauge condition~\cite{JLS2015}:
	\begin{align}\label{eqn:general_gauge_condition_position_space_QED}
	\partial_{\wh{-}}A_{\wh{-}}+\mathbb{C}\boldsymbol\partial_{\perp}\cdot\mathbf{A}_{\perp}=0,
	\end{align}
	and Eq.~(\ref{eqn:A_pT_EOM_step_1}) simplifies to
	\begin{align}
	(\mathbb{C}\boldsymbol\partial_{\perp}^{2}+\partial_{\wh{-}}^{2})(\mathbb{C} A_{\wh{+}}+\mathbb{S} A_{\wh{-}})=(\mathbb{C}\boldsymbol\partial_{\perp}^{2}+\partial_{\wh{-}}^{2})A^{\wh{+}}= -e J^{\wh{+}}\mathbb{C}. \label{eqn:A_pT_EOM}
	\end{align}
	
	From Eqs.  (\ref{eqn:general_gauge_condition_position_space_QED}) and (\ref{eqn:A_pT_EOM}), we see that we can regard $A_{1}$ and $A_{2}$ as the two independent free components, while at any given ``time'' $x^{\wh{+}}$, $A_{\wh{-}}$ can be determined by $A_{1}$, $A_{2}$, and $A_{\wh{+}}$ determined by $A_{1}$, $A_{2}$ and $\psi$.
	We may take the boundary condition, $A_{\wh{-}}(x^{\wh{+}},x^{1},x^{2},+\infty)=-A_{\wh{-}}(x^{\wh{+}},x^{1},x^{2},-\infty)$, which is consistent with the choice made by Kogut and Soper for the light-front QED~\cite{KS}. Then, the solution to Eq.~(\ref{eqn:general_gauge_condition_position_space_QED}) is found as
	\begin{align}
	&A_{\wh{-}}(x^{\wh{+}},x^1,x^2,x^{\wh{-}})\notag\\
	=&-\frac{1}{2}\mathbb{C}\int dx'^{\wh{-}} \epsilon(x^{\wh{-}}-x'^{\wh{-}})\boldsymbol\partial_{\perp}\cdot\mathbf{A}_{\perp}(x^{\wh{+}},x^{1},x^{2},x'^{\wh{-}})\notag\\
	=&\frac{1}{2}\mathbb{C}\int dx'^{\wh{-}} |x^{\wh{-}}-x'^{\wh{-}}|\partial_{\wh{-}}\boldsymbol\partial_{\perp}\cdot\mathbf{A}_{\perp}(x^{\wh{+}},x^{1},x^{2},x'^{\wh{-}}), \label{eqn:A_mT_position_space}
	\end{align}
	using the integration by parts and noting $\epsilon(x) = \frac{d |x|}{dx}$, i.e.
	\begin{equation}
	\epsilon(x)=
	\begin{cases}
	1, &x>0, \\
	-1, &x<0.
	\end{cases}\nonumber\label{eqn:sign_funciton}
	\end{equation} 
	By a simple change of variables $ X^{\wh{+}}\equiv x^{\wh{+}} $, $\mathbf{X}^{\perp}\equiv \mathbf{x}^{\perp}/\sqrt{\mathbb{C}}$, $X^{\wh{-}}\equiv x^{\wh{-}}$, Eq.  (\ref{eqn:A_pT_EOM}) becomes
	\begin{align}
	\bar{\nabla}^{2}A^{\wh{+}}\equiv\left(\dfrac{\partial^{2}}{\partial (X^{i})^{2}}+\dfrac{\partial^{2}}{\partial (X^{\wh{-}})^{2}}\right)A^{\wh{+}}=-eJ^{\wh{+}}\mathbb{C}\quad (i=1,2),
	\label{eqn:transformed_EOM_Laplace}
	\end{align}
	which has the solution
	\begin{align}
	A^{\wh{+}}&= e \int d^{2}\mathbf{X'}^{\perp}dX'^{\wh{-}} \frac{ J^{\wh{+}}(X')\mathbb{C}}{4\pi\sqrt{(\mathbf{X}^{\perp}-\mathbf{X'}^{\perp})^{2}+(X^{\wh{-}}-X'^{\wh{-}})^{2}}},
	\label{eqn:A^pT_position_space}
	\end{align}
	where the argument of $J^{\wh{+}}(X')$ denotes the four-vector 
	$X'^{\wh{\mu}}\equiv(X'^{\wh{+}},X'^1,X'^2,X'^{\wh{-}})=(x'^{\wh{+}},\frac{x'^1}{\sqrt{\mathbb{C}}},\frac{x'^2}{\sqrt{\mathbb{C}}},x'^{\wh{-}})$. In the instant form limit ($\mathbb{C}\rightarrow 1$), $A_{\wh{-}}\rightarrow A^{3}$, $A^{\wh{+}}\rightarrow A^{0}$, $J^{\wh{+}}\rightarrow J^{0}$ and the above solutions given by Eqs.~(\ref{eqn:A_mT_position_space}) and (\ref{eqn:A^pT_position_space}) agree with the instant form results. In the light-front limit ($\mathbb{C}\rightarrow 0$), both $ A_{\wh{-}} $ and $ A^{\wh{+}} $ in Eqs.~(\ref{eqn:A_mT_position_space}) and (\ref{eqn:A^pT_position_space}), respectively, can also be easily shown to be consistent with the light-front gauge $ A^+=0 $ due to the apparent $ \mathbb{C} $ factor in the numerator.
	
	However, we note that both $ A_{\wh{+}}=-\mathbb{S}A_{\wh{-}}/\mathbb{C}+A^{\wh{+}}/\mathbb{C} $ and $ A^{\wh{-}}=\mathbb{S}A^{\wh{+}}/\mathbb{C}-A_{\wh{-}}/\mathbb{C} $ carry overall $ 1/\mathbb{C} $ factor, and thus $ A^- $ in LFD, i.e. the $ \mathbb{C}\to0 $ limit of $A_{\wh{+}}$ or $A^{\wh{-}}$, does not vanish. 
	In fact, the $A_{\wh{+}}$ component satisfies the following constraint equation without containing any time derivatives:
	\begin{equation}
	\bar{\nabla}^{2}(A_{\wh{+}}+\frac{\mathbb{S} A_{\wh{-}}}{\mathbb{C}})=(\mathbb{C}\boldsymbol\partial_{\perp}^{2}+\partial_{\wh{-}}^{2})(A_{\wh{+}}+\frac{\mathbb{S} A_{\wh{-}}}{\mathbb{C}})=-e J^{\wh{+}}, \label{eqn:A_pT_equations}
	\end{equation}
	where the three dimensional Laplace operator reduces to a one dimensional operator when $\mathbb{C}=0$.

	From Eq.~(\ref{eqn:A_mT_position_space}), we can find that the term $-\mathbb{S}A_{\wh{-}}/\mathbb{C}$ in the $\mathbb{C}\to0$ 
	(or $\mathbb{S}\to1$) limit becomes
	\begin{equation}
	\left[-\mathbb{S}A_{\wh{-}}/\mathbb{C}\right] \to
	-\frac{1}{2}\int dx'^{-} |x^{-}-x'^{-}|\partial_{-}\boldsymbol{\partial}_{\perp}\cdot\mathbf{A}^{\perp}(x^{+},x^{1},x^{2},x'^{-}).
	\label{eqn:SAminusoverC}
	\end{equation}
	Also, from Eq.~(\ref{eqn:A^pT_position_space}), we can see that the term $A^{\wh{+}}/\mathbb{C}$ in the $\mathbb{C}\to0$ limit
	becomes
	\begin{equation}
	\left[A^{\wh{+}}/\mathbb{C}\right] \to -\frac{e}{2}\int dx'^{-} |x^{-}-x'^{-}| J^+(x^+,x^1,x^2,x'^{-}),
	\label{eqn:AplusoverC}
	\end{equation}
	where the $ \mathbf{X'}^{\perp} $ integration can be made straightforwardly by realizing 
	the suppression of  $\mathbf{X'}^{\perp}$ component in the light-front ($ \mathbb{C}\to0 $) limit of $J^{\wh{+}}(X')$
	and assigning $Y=X-X'$ to use
	\begin{align}
	&\int d^{2}\mathbf{Y}^{\perp} \frac{1}{4\pi\sqrt{(\mathbf{Y}^{\perp})^{2}+(Y^{\wh{-}})^{2}}} \notag \\
	&= \int d(\mathbf{Y}^{\perp})^2 \frac{1}{4\sqrt{(\mathbf{Y}^{\perp})^{2}+(Y^{\wh{-}})^{2}}} \notag \\
	&= - \frac{1}{2}|Y^{\wh{-}}|
	\label{eqn:transverseint}
	\end{align}
	that becomes $- \frac{1}{2}|x^{-}-x'^{-}|$ in LFD 
	with the current $J^+(x^+,x^1,x^2,x'^-)$ vanishing in the limit $|\mathbf{x}^{\perp}|\to\infty$.
	Combining Eqs.~(\ref{eqn:SAminusoverC}) and (\ref{eqn:AplusoverC}), we thus get the LFD result
	\begin{align}
	&A^{-}(x^{+},x^{1},x^{2},x^{-})\notag\\
	=&-\frac{1}{2}\int dx'^{-} |x^{-}-x'^{-}|\left[ \partial_{-}\boldsymbol{\partial}_{\perp}\cdot\mathbf{A}^{\perp}(x^{+},x^{1},x^{2},x'^{-})\right.\notag\\ &\left.+eJ^{+}(x^{+},x^{1},x^{2},x'^{-})) \right], \label{eqn:A_-_LF_solution}
	\end{align}
	which was also derived in Ref.~\cite{KS} except for some superficial differences in the conventions used.
	Eq.~(\ref{eqn:A_-_LF_solution}) was noted in Ref.~\cite{Zhao} as well.
	
	To simplify the notations and make the derivations easier to follow, we may write
	\begin{align}
	A_{\wh{-}}(x)&=-\mathbb{C}\frac{\boldsymbol\partial_{\perp}\cdot\mathbf{A}_{\perp}(x)}{\partial_{\wh{-}}},  \label{eqn:A_mT_EOM_convinient_form} \\
	A^{\wh{+}}(x)&=-\frac{e J^{\wh{+}}(x)\mathbb{C}}{\mathbb{C}\boldsymbol\partial_{\perp}^{2}+\partial_{\wh{-}}^{2}}, 	\label{eqn:A^pT_EOM_convinient_form}
	\end{align}
	instead of the explicit integral forms shown in Eqs.~(\ref{eqn:A_mT_position_space}) and (\ref{eqn:A^pT_position_space}).
	We also write $A_{\wh{+}}$ as
	\begin{align}
	A_{\wh{+}}(x)&=\mathbb{S}\frac{\boldsymbol\partial_{\perp}\cdot\mathbf{A}_{\perp}(x)}{\partial_{\wh{-}}}-\frac{e J^{\wh{+}}(x)}{\mathbb{C}\boldsymbol\partial_{\perp}^{2}+\partial_{\wh{-}}^{2}}, \label{eqn:A_pT_EOM_convinient_form}
	\end{align}
	which represents $A_{\wh{+}}=-\mathbb{S} A_{\wh{-}}/\mathbb{C}+A^{\wh{+}}/\mathbb{C}$ with $A_{\wh{-}}$ and $A^{\wh{+}}$ given by Eqs.~(\ref{eqn:A_mT_position_space}) and (\ref{eqn:A^pT_position_space}), respectively. 
	Written in this way, Eqs.~(\ref{eqn:A_mT_EOM_convinient_form}) - (\ref{eqn:A_pT_EOM_convinient_form}) also show very clearly, that only the $A_{1}$ and $A_{2}$ components of $A_{\wh{\mu}}$ are dynamical variables.
	
	
	
	For the fermion fields, Eq.~(\ref{eqn:fermion_EOM}) can be written as
	\begin{align}
	&\left[ i\left( \gamma^{\wh{+}}\partial_{\wh{+}}+\gamma^{\wh{-}}\partial_{\wh{-}}+\boldsymbol{\gamma}^{\perp}\cdot\boldsymbol{\partial}_{\perp}\right)\right. \notag\\
	&-\left.e\left( \gamma^{\wh{+}}A_{\wh{+}}+\gamma^{\wh{-}}A_{\wh{-}}+\boldsymbol{\gamma}^{\perp}\cdot\mathbf{A}_{\perp}\right) -m\right] \psi=0, \label{eqn:fermion_EOM_rewritten}
	\end{align}
	where the interpolating gamma matrices satisfy the usual Clifford algebra $\{\gamma^{\wh{\mu}},\gamma^{\wh{\nu}}\}=2g^{\wh{\mu}\wh{\nu}}$ and the interpolating metric is given by 
	\begin{align}\label{eqn:g_munu_interpolation}
	g^{\muT\nuT}
	= g_{\muT\nuT}
	= \begin{pmatrix}
	\Cc & 0  & 0  & \Ss \\
	0          & -1 & 0  & 0 \\
	0          & 0  & -1 & 0 \\
	\Ss & 0  & 0  & -\Cc
	\end{pmatrix}.
	\end{align}
	If $ \mathbb{C}\neq0 $, Eq.~(\ref{eqn:fermion_EOM_rewritten}) contains the interpolating time derivative $\partial_{\wh{+}}$
	and thus all four components of $\psi$ field are dynamical.
	However, if $ \mathbb{C} = 0 $, then one may notice a rather dramatic change of two components of $\psi$ field 
	from being dynamical to the constrained components due to 
	$\{\gamma^{+},\gamma^{+}\}= {\gamma^+}^2 = 0$ as well as  
	$\{\gamma^{-},\gamma^{-}\}= {\gamma^-}^2 = 0$, while $\{\gamma^{+},\gamma^{-}\}=2$.
	This may be shown explicitly by writing Eq.~(\ref{eqn:fermion_EOM_rewritten}) for $\mathbb{C} = 0$,
	\begin{align}
	\label{eqn:LFfermion_EOM_rewritten}
	&\left[ i\left( \gamma^{+}\partial_{+}+\gamma^{-}\partial_{-}+\boldsymbol{\gamma}^{\perp}\cdot\boldsymbol{\partial}_{\perp}\right)\right. \notag\\
	&\left.-e\left( \gamma^{+}A_{+}+\gamma^{-}A_{-}+\boldsymbol{\gamma}^{\perp}\cdot\mathbf{A}_{\perp}\right) -m\right] \psi=0,
	\end{align}
	and splitting $\psi$ into $\psi_+ = P_+ \psi$ and $\psi_- = P_- \psi$ with the projection operators  
	$P_+=\frac{1}{2}\gamma^-\gamma^+ $ and $ P_-=\frac{1}{2}\gamma^+\gamma^- $, i.e.
	\begin{equation}
	\psi=\psi_++\psi_-=P_+\psi+P_-\psi.\label{eqn:projection_psi}
	\end{equation}
	Then, because $\gamma^+P_-=0$, 
	$ \psi_- $ can be determined at any light-front time $x^+$ through the following constraint equation
	\begin{equation}
	2\left( i\partial_--eA_-\right)\psi_-= \left[ \left( i\boldsymbol{\partial}_{\perp}-e\mathbf{A}_{\perp}\right)\boldsymbol{\gamma}^{\perp}+m \right]\gamma^+\psi_+,
	\end{equation}
	which reduces in the light-front gauge $ A_-=A^+=0$ to
	\begin{equation}
	2\left( i\partial_-\psi_-\right) = \left[ \left( i\boldsymbol{\partial}_{\perp}-e\mathbf{A}_{\perp}\right)\boldsymbol{\gamma}^{\perp}+m \right]\gamma^+\psi_+.\label{eqn:constraint_psi_minus}
	\end{equation}
	Thus, the two components of $\psi$ given by $\psi_-$ in LFD become constrained in the sense that 
	the time dependence of $\psi_-$ is provided by the other fields that satisfy the dynamic equation with the light-front time 
	derivative $\partial_+$ such as $ \mathbf{A}_{\perp} $
	and $\psi_+$. No new time-dynamic information can be provided by the constrained field $\psi_-$.
	As done in Ref.~\cite{KS}, we may split this constrained field $\psi_-$ into the ``free" part $\tilde{\psi}_-$ and the 
	``interaction" part $\Upsilon$, i.e. $ \psi_-= \tilde{\psi}_-+\Upsilon $, identifying from Eq.~(\ref{eqn:constraint_psi_minus}):
	\begin{equation}
	\tilde{\psi}_-=\frac{(i\boldsymbol{\gamma}^{\perp}\cdot\boldsymbol{\partial}_{\perp}+m)\gamma^+\psi_+}{2i\partial_-},\label{eqn:psi_minus_tilde}
	\end{equation}
	and
	\begin{equation}
	\Upsilon=\frac{-e\boldsymbol{\gamma}^{\perp}\cdot\mathbf{A}_{\perp}\gamma^+\psi_+}{2i\partial_-}.
	\label{eqn:Upsilon}
	\end{equation}
	Then, as shown in Ref.~\cite{KS}, the light-front fermion instantaneous diagram depicted in Fig.~\ref{fig:Instantaneous_Electron_Vertex} corresponds to the interaction Hamiltonian density given by  
	$\bar{\Upsilon}(i\gamma^-\partial_-)\Upsilon$.
	This reveals that the instantaneous contribution to the fermion propagator given by 
	Eq.~(\ref{eqn:electron_LF_time_ordered_propagator}) is obtained through the 
	``interaction" part of the constraint field $\psi_-$. 
	
	We may define
	\begin{equation}
	\psi=\tilde{\psi}+\delta_{\mathbb{C} 0}\Upsilon.
	\label{eqn:tilde_and_Upsilon}
	\end{equation}
	When $ \mathbb{C}\neq0 $, $ \psi=\tilde{\psi} $ is the free fermion field. When $ \mathbb{C}=0 $, $ \psi $ can be split into $ \psi=\psi_++\psi_- $, where only $ \psi_+=\tilde{\psi}_+ $ is independent. The constraint field $ \psi_- $ can be further split into $ \psi_-=\tilde{\psi}_-+\Upsilon $, where $ \Upsilon $ is the ``interaction'' part of the field. We write $ \tilde{\psi}=\tilde{\psi}_++\tilde{\psi}_- $, then $ \tilde{\psi} $ is the free part of the field in any interpolating angle $ 0\leq\delta\leq\frac{\pi}{4} $.
	
	We discuss this unique feature of the fermion propagator in LFD further 
	illustrating the old-fashioned perturbation theory in sub-subsection~\ref{ssb:hamiltonian} 
	and presenting the physical processes such as 
	the electron-positron annihilation to the pair production of two photons ($e^+ e^- \rightarrow \gamma \gamma$)
	and the Compton scattering ($e \gamma \rightarrow e \gamma$) in Sections \ref{sec:eess} and \ref{sec:calculation}.
	
	\subsubsection{Free Fields}
	\label{ssb:free_fields}
	
	The Fourier expansion of the free fermion field $\psi(x)$ takes the form
	\begin{align}
	&\psi(x^{\wh{+}},\mathbf{x}^{\perp},x^{\wh{-}})=\int \frac{d^{2}\mathbf{p}_{\perp}dp_{\wh{-}}}{(2\pi)^{3}(2p^{\wh{+}})}\sum_{s=\pm 1/2} \left[ u^{(s)}e^{-ix^{\wh{-}}p_{\wh{-}}-i\mathbf{x}^{\perp}\cdot\mathbf{p}_{\perp}}\right.\notag\\
	&\times\left. b(\mathbf{p}_{\perp},p_{\wh{-}};s;x^{\wh{+}})
	+v^{(s)}e^{ix^{\wh{-}}p_{\wh{-}}+i\mathbf{x}^{\perp}\cdot\mathbf{p}_{\perp}}d^{\dagger}(\mathbf{p}_{\perp},p_{\wh{-}};s;x^{\wh{+}}) \right],
	\label{eqn:Free_fermion_field_in_3_momentum_space}
	\end{align}
	where the spinors of the particle $(u)$ and the antiparticle $(v)$ satisfy the Dirac equation:
	\begin{align}
	(\gamma^{\wh{\mu}}p_{\wh{\mu}}-m)u=0, \label{eqn:Dirac_eq_for_u}\\
	(\gamma^{\wh{\mu}}p_{\wh{\mu}}+m)v=0. \label{eqn:Dirac_eq_for_v}
	\end{align}
	Here, we take $u$ and $v$ to be the generalized helicity spinors $u_{H}$ and $v_{H}$ whose explicit expressions in the chiral basis have been given in Ref.~\cite{LAJ2015}.
	For simplicity, we will omit the subscript ``H'' throughout this paper.
	
	Plugging $\psi$ given by Eq.~(\ref{eqn:Free_fermion_field_in_3_momentum_space}) 
	to the free Dirac equation, 
	\begin{equation}
	(i\gamma^{\wh{\mu}}\partial_{\wh{\mu}}-m)\psi(x^{\wh{+}},\mathbf{x}^{\perp},x^{\wh{-}})=0 \label{eqn:psi_Dirac_eq},
	\end{equation}
	and using the relations in Eqs.~(\ref{eqn:Dirac_eq_for_u}) and (\ref{eqn:Dirac_eq_for_v}), we find that $b(\mathbf{p}_{\perp},p_{\wh{-}};s;x^{\wh{+}})$ and $d^{\dagger}(\mathbf{p}_{\perp},p_{\wh{-}};s;x^{\wh{+}})$ satisfy the following differential equations:
	\begin{align}
	[i\gamma^{\wh{+}}\partial_{\wh{+}}-\gamma^{\wh{+}}p_{\wh{+}}]\,b(\mathbf{p}_{\perp},p_{\wh{-}};s;x^{\wh{+}})=0,  \label{eqn:b_differential_Eq}\\
	[i\gamma^{\wh{+}}\partial_{\wh{+}}+\gamma^{\wh{+}}p_{\wh{+}}]\,d^{\dagger}(\mathbf{p}_{\perp},p_{\wh{-}};s;x^{\wh{+}})=0.  \label{eqn:d_differential_Eq}
	\end{align}
	Solving these equations, we get
	\begin{align}
	b(\mathbf{p}_{\perp},p_{\wh{-}};s;x^{\wh{+}})=e^{-ix^{\wh{+}}p_{\wh{+}}}\,b(\mathbf{p}_{\perp},p_{\wh{-}};s;0),
	\label{eqn:b_solution}\\
	d^{\dagger}(\mathbf{p}_{\perp},p_{\wh{-}};s;x^{\wh{+}})=e^{ix^{\wh{+}}p_{\wh{+}}}\,d^{\dagger}(\mathbf{p}_{\perp},p_{\wh{-}};s;0).
	\label{eqn:d_solution}
	\end{align}
	Since the time dependence decouples from the rest of the operator, we may drop the time labels and define
	\begin{align}
	b(\mathbf{p}_{\perp},p_{\wh{-}};s)\equiv b(\mathbf{p}_{\perp},p_{\wh{-}};s;0),\label{eqn:b_new_defined}\\
	d^{\dagger}(\mathbf{p}_{\perp},p_{\wh{-}};s)\equiv d^{\dagger}(\mathbf{p}_{\perp},p_{\wh{-}};s;0).\label{eqn:d_new_defined}
	\end{align}
	Then, the free fermion field can be summarized as
	\begin{align}
	\psi(x)=\int \frac{d^{2}\mathbf{p}_{\perp}dp_{\wh{-}}}{(2\pi)^{3}2p^{\wh{+}}}\sum_{s=\pm 1/2}&\left[ u^{(s)}e^{-ix^{\wh{\mu}}p_{\wh{\mu}}}\,b(\mathbf{p}_{\perp},p_{\wh{-}};s) \right. \notag \\
	&\left. +v^{(s)}e^{ix^{\wh{\mu}}p_{\wh{\mu}}}\,d^{\dagger}(\mathbf{p}_{\perp},p_{\wh{-}};s) \right]. \label{eqn:Free_fermion_field_in_4_momentum_space}
	\end{align}
	Following a similar procedure, we can also find the free photon field as 
	\begin{align}
	A^{\wh{\mu}}(x)=\int \frac{d^{2}\mathbf{p}_{\perp}dp_{\wh{-}}}{(2\pi)^{3}2p^{\wh{+}}}\sum_{\lambda=\pm}\epsilon^{\wh{\mu}}(p,\lambda) &\left[ e^{-ix^{\wh{\mu}}p_{\wh{\mu}}}\,a(\mathbf{p}_{\perp},p_{\wh{-}};s) \right. \notag \\
	&\left. +e^{ix^{\wh{\mu}}p_{\wh{\mu}}}\,a^{\dagger}(\mathbf{p}_{\perp},p_{\wh{-}};s) \right],
	\label{eqn:Free_photon_field_in_4_momentum_space}
	\end{align}
	where again the polarization vectors $\epsilon^{\wh{\mu}}(p,\pm)$ are explicitly given in Ref.~\cite{JLS2015}.
	
	\subsubsection{Energy-Momentum and Angular Momentum Tensors}
	\label{ssb:momentum_and_angular_momentum}
	Using Noether's theorem, the conserved energy-momentum tensor and angular momentum tensor can be written as
	\begin{align}
	{T^{\wh{\mu}}}_{\wh{\nu}}&=i\bar{\psi}\gamma^{\wh{\mu}}\partial_{\wh{\nu}}\psi-F^{\wh{\mu}\wh{\lambda}}\partial_{\wh{\nu}}A_{\wh{\lambda}}-{g^{\wh{\mu}}}_{\wh{\nu}}\mathcal{L}, \label{eqn:energy_momentum_tensor}\\
	{J^{\wh{\lambda}}}_{\wh{\mu}\wh{\nu}}&=x_{\wh{\mu}}{T^{\wh{\lambda}}}_{\wh{\nu}}-x_{\wh{\nu}}{T^{\wh{\lambda}}}_{\wh{\mu}}+{S^{\wh{\lambda}}}_{\wh{\mu}\wh{\nu}}, \label{eqn:angular_momentum}
	\end{align}
	where
	\begin{align}
	{S^{\wh{\lambda}}}_{\wh{\mu}\wh{\nu}}&=i\frac{1}{4}\bar{\psi}\gamma^{\wh{\lambda}}[\gamma_{\wh{\mu}},\gamma_{\wh{\nu}}]\psi+{F^{\wh{\lambda}}}_{\wh{\mu}}A_{\wh{\nu}}-{F^{\wh{\lambda}}}_{\wh{\nu}}A_{\wh{\mu}}.\label{eqn:spin_part_of_angular_momentum}
	\end{align}
	In particular, the total four-momentum and total angular momentum given by
	\begin{align}
	P_{\wh{\mu}}=\int d^{2}\mathbf{x}^{\perp}dx^{\wh{-}}{T^{\wh{+}}}_{\wh{\mu}},\label{eqn:total_momentum}\\
	M_{\wh{\mu} \wh{\nu}}=\int  d^{2}\mathbf{x}^{\perp}dx^{\wh{-}}{J^{\wh{+}}}_{\wh{\mu} \wh{\nu}}\label{eqn:total_angular_momentum}
	\end{align}
	are constants of motion. In particular, the kinematic generators which do not alter the interpolating time $x^{\wh{+}}$, such as 
	$P_{1}, P_{2}, P_{\wh{-}}, M_{12}, M_{2\wh{-}}, M_{1\wh{-}}$, are provided by their corresponding densities given by
	\begin{align}
	&{T^{\wh{+}}}_{i}=\; i\bar{\psi}\gamma^{\wh{+}}\partial_{i}\psi-\partial_{i}A_{j}(\partial^{\wh{+}}A^{j}-\partial^{j}A^{\wh{+}}),
	\\
	&{T^{\wh{+}}}_{\wh{-}}=\; i\bar{\psi}\gamma^{\wh{+}}\partial_{\wh{-}}\psi-\partial_{\wh{-}}A_{j}(\partial^{\wh{+}}A^{j}-\partial^{j}A^{\wh{+}}),\\
	&{J^{\wh{+}}}_{12}=\; x_{1}{T^{\wh{+}}}_{2}-x_{2}{T^{\wh{+}}}_{1}+\dfrac{1}{2}i\bar{\psi}\gamma^{\wh{+}}\gamma_{1}\gamma_{2}\psi\nonumber\\
	&\quad+A^{2}\partial^{\wh{+}}A^{1}-A^{1}\partial^{\wh{+}}A^{2}+A^{1}\partial^{2}A^{\wh{+}}-A^{2}\partial^{1}A^{\wh{+}},\label{eqn:J_12}\\
	&{J^{\wh{+}}}_{1\wh{-}}=\; x_{1}{T^{\wh{+}}}_{\wh{-}}-x_{\wh{-}}{T^{\wh{+}}}_{1}+\dfrac{1}{2}i\bar{\psi}\gamma^{\wh{+}}\gamma_{1}\gamma_{\wh{-}}\psi\nonumber\\
	&\quad+A_{\wh{-}}\partial^{\wh{+}}A_{1}-A_{1}\partial^{\wh{+}}A_{\wh{-}}+A_{1}\partial_{\wh{-}}A^{\wh{+}}-A_{\wh{-}}\partial_{1}A^{\wh{+}},\label{eqn:J_13}\\
	&{J^{\wh{+}}}_{2\wh{-}}=\; x_{2}{T^{\wh{+}}}_{\wh{-}}-x_{\wh{-}}{T^{\wh{+}}}_{2}+\dfrac{1}{2}i\bar{\psi}\gamma^{\wh{+}}\gamma_{2}\gamma_{\wh{-}}\psi\nonumber\\
	&\quad+A_{\wh{-}}\partial^{\wh{+}}A_{2}-A_{2}\partial^{\wh{+}}A_{\wh{-}}+A_{2}\partial_{\wh{-}}A^{\wh{+}}-A_{\wh{-}}\partial_{2}A^{\wh{+}},\label{eqn:J_23}
	\end{align}
	where $A_{\wh{-}}$ and $A^{\wh{+}}$ are given by Eqs.~(\ref{eqn:A_mT_EOM_convinient_form}) and (\ref{eqn:A^pT_EOM_convinient_form}), and thus these operators involve only independent dynamical fields $\psi$ and $A^{j} (j=1,2)$.
	
	Finally, the most important operator of the theory is of course the interpolating Hamiltonian density:
	\begin{align}
	{T^{\wh{+}}}_{\wh{+}}&=\bar{\psi}\left(-i\gamma^{j}\partial_{j}-i \gamma^{\wh{-}}\partial_{\wh{-}}+m \right)\psi+ eA_{\wh{\mu}}\bar{\psi}\gamma^{\wh{\mu}}\psi \notag\\
	& +\dfrac{1}{4}F^{\wh{\mu}\wh{\nu}}F_{\wh{\mu}\wh{\nu}}-F^{\wh{+} j}\partial_{\wh{+}}A_{j}-F^{\wh{+} \wh{-}}\partial_{\wh{+}}A_{\wh{-}},
	\label{eqn:Hamiltonian_T_+_+}
	\end{align}
	where the transverse index $ j$ is summed over according to the summation convention. 
	\subsubsection{Old-fashioned Perturbation Theory}
	\label{ssb:hamiltonian}
	With Eqs.~(\ref{eqn:A_mT_EOM_convinient_form}) - (\ref{eqn:A_pT_EOM_convinient_form}), as well as Eqs.~(\ref{eqn:psi_minus_tilde}) - (\ref{eqn:tilde_and_Upsilon}), we can rewrite ${T^{\wh{+}}}_{\wh{+}}$ in terms of the independent degrees of freedom $A^{1}$, $A^{2}$, $\tilde{\psi}$, and separate out the interaction part of the Hamiltonian density from the free part. The detailed derivation is given in Appendix \ref{app:Hamiltonian}. 
	Eq.~(\ref{eqn:Hamiltonian_T_+_+}) becomes
	\begin{equation}
	\mathcal{H}\equiv {T^{\wh{+}}}_{\wh{+}}=
	\mathcal{H}_{0}+\mathcal{V}
	\label{eqn:H=H0+V}
	\end{equation}
	with
	\begin{align}
	\mathcal{H}_{0}&=\bar{\tilde{\psi}}(-i\gamma^{j}\partial_{j}-i\gamma^{\wh{-}}\partial_{\wh{-}}+m)\tilde{\psi}\notag\\
	&+\dfrac{1}{4}\tilde{F}^{\wh{\mu}\wh{\nu}}\tilde{F}_{\wh{\mu}\wh{\nu}}-\tilde{F}^{\wh{+} j}\partial_{\wh{+}}\tilde{A}_{j}-\tilde{F}^{\wh{+} \wh{-}}\partial_{\wh{+}}\tilde{A}_{\wh{-}}\label{eqn:H0_QED},\\
	\mathcal{V}&=e\tilde{A}_{\wh{\mu}}\bar{\tilde{\psi}}\gamma^{\wh{\mu}}\tilde{\psi}+\delta_{\mathbb{C}0}\bar{\Upsilon}(i\gamma^-\partial_-)\Upsilon+\frac{1}{2}e\phi J^{\wh{+}},
	\label{eqn:V_QED}
	\end{align}
	where we have defined $\tilde{A}_{\wh{\mu}}$ 
	as
	\begin{equation}
	(\tilde{A}_{\wh{+}},\tilde{A}_{1},\tilde{A}_{2},\tilde{A}_{\wh{-}})\equiv(\mathbb{S}\dfrac{\boldsymbol\partial_{\perp}\cdot\mathbf{A}_{\perp}}{\partial_{\wh{-}}},A_{1},A_{2},A_{\wh{-}}),
	\label{eqn:tilde{A}}
	\end{equation}
	and
	\begin{align}
	\label{eqn:phi}
	&\ \ \phi(x)\equiv A_{\wh{+}}(x)-\tilde{A}_{\wh{+}}(x)=\frac{A^{\wh{+}}}{\mathbb{C}}\notag\\
	&=-\frac{eJ^{\wh{+}}(x)}{\mathbb{C}\boldsymbol{\partial}_{\perp}^2+\partial_{\wh{-}}^2}\notag\\
	&=e \int d^{2}\mathbf{X'}^{\perp}dX'^{\wh{-}} \frac{ J^{\wh{+}}(X')}{4\pi\sqrt{(\mathbf{X}^{\perp}-\mathbf{X'}^{\perp})^{2}+(X^{\wh{-}}-X'^{\wh{-}})^{2}}},
	\end{align}
	where we switched the simplified notation in the second line into the expression of integration in the third one. The capital $ X^{\wh{\mu}}\equiv(x^{\wh{+}},\frac{x^1}{\sqrt{\mathbb{C}}},\frac{x^2}{\sqrt{\mathbb{C}}},x^{\wh{-}}) $ is introduced previously above Eq.~(\ref{eqn:transformed_EOM_Laplace}).
	Eq.~(\ref{eqn:phi}) may be considered as a generalization of Eq. (4.58) in Ref. \cite{KS} for the quantization interpolating
	between IFD and LFD.
	
	We can then calculate the scattering matrix element $S_{fi} = <f| S | i>$ between the initial and final states $|i>$ and $|f>$ 
	with the ``old-fashioned'' perturbation theory expansion
	\begin{align}
	S_{fi}&=\delta_{fi} -i2\pi\delta(P_{\wh{+} i}-P_{\wh{+} f})
	\notag \\
	&\times <f|\left[V+V(P_{\wh{+}}-P_{\wh{+} 0}+i\epsilon)^{-1}V+\cdots\right]|i>,
	\label{eqn:S_matrix_QED}
	\end{align}
	where $P_{\wh{+} 0}=\int d^{2}\mathbf{x}^{\perp}dx^{\wh{-}} \mathcal{H}_{0}$ and \
	$V=\int d^{2}\mathbf{x}^{\perp}dx^{\wh{-}} \mathcal{V}$.
	This leads to the same rules for $x^{\wh{+}}$-ordered diagrams which we obtained in subsection~(\ref{ssb:Rules_for_x^(pt)-ordered_diagrams}) by directly decomposing the covariant Feynman diagrams.
	This can be seen by calculating a few matrix elements of the interaction Hamiltonian $V$.
	
	The first term in Eq.~(\ref{eqn:V_QED}) after volume integration gives the interaction at equal interpolating time $x^{\wh{+}}=0$:
	\begin{align}
	-iV_{1}=-i e \int d^{2}\mathbf{x}^{\perp}dx^{\wh{-}}  \tilde{A}_{\wh{\mu}}(0,\mathbf{x}^{\perp},x^{\wh{-}})\bar{\tilde{\psi}}(0,\mathbf{x}^{\perp},x^{\wh{-}})\gamma^{\wh{\mu}}\notag\\
	\times\tilde{\psi}(0,\mathbf{x}^{\perp},x^{\wh{-}}),
	\label{eqn:V1}
	\end{align}
	which is the ``ordinary'' vertex interaction as demonstrated in Fig.~\ref{fig:Normal_Vertex}.
	
	With Eq.~(\ref{eqn:Upsilon}), the second term in Eq.~(\ref{eqn:V_QED}) can be shown to provide the fermion instantaneous interaction
	\begin{align}
	-iV_{2}&=
	-\dfrac{1}{2}e^{2}\delta_{\mathbb{C}0}\int d^{2}\mathbf{x}^{\perp}dx^{-}  \bar{\tilde{\psi}}(0,\mathbf{x}^{\perp},x^{-})\gamma^{i}\tilde{A}_{i}(0,\mathbf{x}^{\perp},x^{-})\notag\\
	&\times
	\frac{\gamma^+}{\partial_-} \tilde{A}_{j}(0,\mathbf{x}^{\perp},x^{-})\gamma^j\tilde{\psi}(0,\mathbf{x}^{\perp},x^{-})\notag\\
	&=
	-\dfrac{1}{4}e^{2}\delta_{\mathbb{C}0}\int d^{2}\mathbf{x}^{\perp}dx^{-}  \bar{\tilde{\psi}}(0,\mathbf{x}^{\perp},x^{-})\gamma^{i}\tilde{A}_{i}(0,\mathbf{x}^{\perp},x^{-})
	\gamma^+\notag\\
	&\times\int dx'^-\epsilon(x^--x'^-) \tilde{A}_{j}(0,\mathbf{x}^{\perp},x'^{-})\gamma^j\tilde{\psi}(0,\mathbf{x}^{\perp},x'^{-}).
	\label{eqn:V2}
	\end{align}
	Using
	\begin{equation}
	\frac{1}{2}\int dx'^-\epsilon(x^--x'^-)\,e^{ -iq^+(x^--x'^-)}=\frac{i}{q^+} ,
	\end{equation}
	Eq.~(\ref{eqn:V2}) can be shown to yield the vertex of Fig.~\ref{fig:Instantaneous_Electron_Vertex}, as discussed in Ref. \cite{KS}.
	
	The third term in Eq.~(\ref{eqn:V_QED}) written out in full is
	\begin{align}
	-iV_{3}
	&=\frac{1}{2}ie^{2}
	\int d^{2}\mathbf{x}^{\perp}dx^{\wh{-}} \bar{\psi}(0,\mathbf{x}^{\perp},x^{\wh{-}})\gamma^{\wh{+}}\psi(0,\mathbf{x}^{\perp},x^{\wh{-}})\notag\\
	&\times\frac{1}{\boldsymbol{\partial}_{\perp}^{2}\mathbb{C}+\partial_{\wh{-}}^{2}} \bar{\psi}(0,\mathbf{x}^{\perp},x^{\wh{-}})\gamma^{\wh{+}}\psi(0,\mathbf{x}^{\perp},x^{\wh{-}})\notag\\
	&=-\frac{1}{2}ie^{2}
	\int d^{2}\mathbf{x}^{\perp}dx^{\wh{-}} \bar{\psi}(0,\mathbf{x}^{\perp},x^{\wh{-}})\gamma^{\wh{+}}\psi(0,\mathbf{x}^{\perp},x^{\wh{-}})\notag\\
	&\times\int d^{2}\mathbf{X}'^{\perp}dX'^{\wh{-}} \frac{\bar{\psi}(0,\mathbf{X}'^{\perp},X'^{\wh{-}})\gamma^{\wh{+}}\psi(0,\mathbf{X}'^{\perp},X'^{\wh{-}})}{4\pi\sqrt{(\mathbf{X}^{\perp}-\mathbf{X}'^{\perp})^{2}+(X^{\wh{-}}-X'^{\wh{-}})^{2}}}.
	\label{eqn:V3}
	\end{align}
	In the scaled transverse space with the variable of $ \mathbf{X}^{\perp} $, one should note that the corresponding transverse momentum becomes $ \sqrt{\mathbb{C}}\mathbf{q}_{\perp} $ due to the equality given by $\mathbf{q}_{\perp}\cdot\mathbf{x}^{\perp}=\sqrt{\mathbb{C}}\mathbf{q}_{\perp}\cdot \mathbf{X}^{\perp}$. Using
	\begin{align}
	&\int d^{2}\mathbf{X}'^{\perp}dX'^{\wh{-}} \frac{ e^{-i\left[\sqrt{\mathbb{C}} \mathbf{q}_{\perp}\cdot(\mathbf{X}^{\perp}-\mathbf{X}'^{\perp})+q_{\wh{-}}(X^{\wh{-}}-X'^{\wh{-}})\right]}}{4\pi\sqrt{(\mathbf{X}^{\perp}-\mathbf{X}'^{\perp})^{2}+(X^{\wh{-}}-X'^{\wh{-}})^{2}}} \notag\\
	=&\frac{1}{\mathbb{C}\mathbf{q}_{\perp}^2+q_{\wh{-}}^2},\label{eqn:3d_spatial_integration_formular}
	\end{align}
	where $ q $ is the momentum transfer at the vertex, as depicted in Fig.~\ref{fig:Instantaneous_Photon_Vertex}. We find that 
	the interaction $-iV_3$ yields the ``Coulomb" vertices of Fig.  \ref{fig:Instantaneous_Photon_Vertex}. We note that this interpolation result coincides with the IFD result for $ \mathbb{C}=1 $, where $ \psi=\tilde{\psi} $ is the free fermion field, while for 
	$ \mathbb{C}=0 $, the $ \psi $ field in Eq.~(\ref{eqn:V3}) changes naturally to $ \tilde{\psi} $ due to the $ \gamma^{+2}=0 $ property of the LF, so that the $ \Upsilon $ field does not contribute to the $ + $ component of the current. The transverse components of the momentum in Eq.~(\ref{eqn:3d_spatial_integration_formular}) also drop off naturally due to the $ \mathbb{C} $ factor in  front, 
	reproducing smoothly Kogut and Soper's result in Ref. \cite{KS}.

	Thus, when we calculate the scattering matrix formally in the interpolating QED, we get the same rules as we summarized in sub-subsection~\ref{ssb:Rules_for_x^(pt)-ordered_diagrams} when we decompose the covariant Feynman diagrams directly~\cite{thesis}.

	
	

	\section{Toy calculation of $ e^+ e^- $ annihilation producing two scalar particles}\label{sec:eess}
	\begin{figure}
		\centering
		\includegraphics[width=0.7\linewidth]{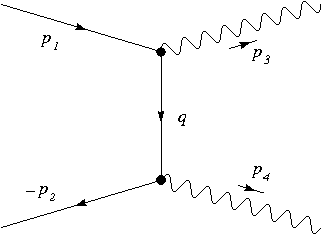}
		\caption{Feynman diagram for $ e^+ e^- \to \gamma \gamma $ process. While this figure is drawn for the t-channel Feynman diagram, the crossed channel (or u-channel) can be drawn by crossing
                 the two final state particles.} 
		\label{fig:PairAnnihilationto2s_tchannel}
	\end{figure}


	Having laid out the foundation of interpolating QED, we can now make some calculations. The first simple heuristic example we consider is $ e^+ e^- $ annihilation producing two scalar particles. In the next section, we will consider the typical QED process of $ e^+e^-\to \gamma \gamma $, as well as $ e\gamma\to e\gamma $, but for now we don't consider the photon polarization to make things simpler. While the Feynman diagram of $ e^+e^-\to \gamma \gamma $ is shown in Fig.~\ref{fig:PairAnnihilationto2s_tchannel},
the photon ($\gamma$) line should be understood as the scalar particle line for the production process of two scalar particles.
	
\begin{figure}
		\centering
		\subfloat[]{\includegraphics[width=0.49\columnwidth]{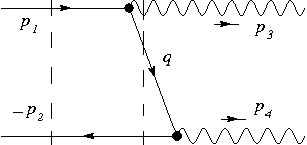}\label{fig:PairAnnihilationto2s_tchannel_a}}
		\vspace{0pt}
		\centering
		\subfloat[]{\includegraphics[width=0.49\columnwidth]{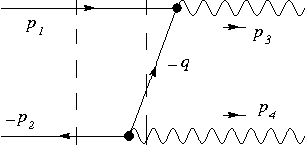}\label{fig:PairAnnihilationto2s_tchannel_b}}
		\caption{\label{fig:PairAnnihilationto2s_tchannel_TO}Time-ordered diagrams (a) and (b) for $ e^+ e^- \to \gamma \gamma $ annihilation process. The u-channel amplitudes can be obtained by crossing the two final state particles. }
	\end{figure}

	As mentioned in the Introduction, the covariant propagator of the intermediate virtual fermion is given by
	\begin{equation}\label{eqn:total_propagator}
	\Sigma=\frac{\slashed{q}+m}{q^2-m^2}.
	\end{equation}
	In the instant form where the system evolves with ordinary time $ t $, this covariant Feynman amplitude can be decomposed into two time-ordered ones, as shown in Figs. \ref{fig:PairAnnihilationto2s_tchannel_a} and \ref{fig:PairAnnihilationto2s_tchannel_b},
	where again the photon ($\gamma$) line should be understood as the scalar particle line for the production process of two scalar particles. Figs. \ref{fig:PairAnnihilationto2s_tchannel_a} and \ref{fig:PairAnnihilationto2s_tchannel_b} correspond to the following time-ordered amplitudes
	\begin{align}
	\Sigma_a^{\mathsf{IFD}}&=\frac{1}{2q_{on}^0}\;\frac{\slashed{q}_a+m}{q^0-q_{on}^0},\\
	\Sigma_b^{\mathsf{IFD}}&=\frac{1}{2q_{on}^0}\;\frac{-\slashed{q}_b+m}{-q^0-q_{on}^0}.
	\end{align}
	Here, $ q_{on} $ is the momentum 4-vector with its spacial part equal to that of $ q\,(=p_1-p_3=q_a) $ but satisfies the Einstein energy-momentum relationship, and $ q_b $ corresponds to the negative energy (anti-particle) contribution with $ q_b=-q_a=-q $.

	The sum of the two propagators can easily be verified to be equal to the covariant one, Eq.~(\ref{eqn:total_propagator}):
	\begin{align}
	\Sigma_a^{\mathsf{IFD}}+\Sigma_b^{\mathsf{IFD}}&=\frac{1}{2q_{on}^0}\left( \frac{\slashed{q}+m}{q^0-q_{on}^0}-\frac{\slashed{q}+m}{q^0+q_{on}^0}\right) \notag\\
	&=\frac{1}{2q_{on}^0}\frac{2q_{on}^0(\slashed{q}+m)}{(q^0)^2-(q_{on}^0)^2}\notag\\
	&=\frac{\slashed{q}+m}{q^2-m^2},
	\end{align}
	where the on-shell condition $ q_{on}^0=\sqrt{\vec{q}^2+m^2} $ is used.
	
	Such time-ordering also exists in the interpolating dynamics, whose ``time" means the interpolating time $ x^{\wh{+}} $.
	The interpolating time-ordered diagrams are also Figs. \ref{fig:PairAnnihilationto2s_tchannel_a} and \ref{fig:PairAnnihilationto2s_tchannel_b}, and the propagators  of the intermediate virtual fermion for each time-ordering are given by
	\begin{align}\label{eqn:propagator_TOa}
	\Sigma_a&=\frac{1}{2Q^{\wh{+}}}\;\frac{\slashed{Q}_a+m}{q_{\wh{+}}-Q_{a\wh{+}}}\\\label{eqn:propagator_TOb}
	\Sigma_b&=\frac{1}{2Q^{\wh{+}}}\;\frac{-\slashed{Q}_b+m}{-q_{\wh{+}}-Q_{b\wh{+}}},
	\end{align}
	where $Q_{a\wh{+}}$ and $Q_{b\wh{+}}$ are the interpolating on-mass-shell energy of the intermediate propagating fermion as mentioned in the introduction and
	again their expressions are explicitly given by 
		\begin{align} \label{eqn:qanew+}
	Q_{a\wh{+}}&=
	\frac{-\mathbb{S}q_{a\wh{-}}+Q^{\wh{+}}}{\mathbb{C}},\\ \label{eqn:qbnew+}
	Q_{b\wh{+}}&=\frac{-\mathbb{S}q_{b\wh{-}}+Q^{\wh{+}}}{\mathbb{C}},
	\end{align}
with $Q^{\wh{+}}$ denoting the on-mass-shell value of $q^{\wh{+}}$ as given by Eq.~(\ref{eqn:Q^+}).
If the interpolating longitudinal momentum $q_{\wh{-}}$ is positive, i.e. $q_{\wh{-}}>0$, then 
the intermediate propagating fermion in the time-ordered amplitude in Fig.~\ref{fig:PairAnnihilationto2s_tchannel_a}
is ``forward" moving and the corresponding time-ordered amplitude $\Sigma_a$ is equivalent to $\Sigma_F$ given by Eq.~(\ref{eqn:Sigma}),
while the time-ordered amplitude in Fig.~\ref{fig:PairAnnihilationto2s_tchannel_b} with the ``backward" moving ($-q_{\wh{-}}<0$) intermediate
fermion corresponds to $\Sigma_B$ in the same equation, Eq.~(\ref{eqn:Sigma}).
	 Using Eqs.~(\ref{eqn:qanew+}) - (\ref{eqn:qbnew+}), the sum of these two interpolating propagators can also be verified to be equal to Eq.~(\ref{eqn:total_propagator})
as shown in Appendix \ref{app:Propagator}.

When we take the limit to the LFD, i.e., $ \delta\to\frac{\pi}{4} $ or $ \mathbb{C}\to0 $, the expressions in Eq.~(\ref{eqn:propagator_TOa}) and (\ref{eqn:propagator_TOb}) change
to the so-called ``on-mass-shell propagating contribution" and the ``instantaneous fermion contribution", respectively, if and only if $q^+>0$. For $q^+>0$, the time-ordered diagram shown in Fig.~\ref{fig:PairAnnihilationto2s_tchannel_b} has the ``backward" moving intermediate fermion (for $\mathbb{C} = 0$, $-q_{\wh{-}} = -q^+ < 0$), and the LF energy for the intermediate virtual fermion, $ Q_{b\wh{+}} $, goes to infinity, however the existence of the spin sum on the numerator makes it altogether a finite result. The finite result turns out to be the instantaneous fermion contribution unique in the LFD as formally discussed in Sec.~\ref{sec:formal}. This can be now shown explicitly as follows:
	\begin{align}
	&\Sigma_{b,\delta\to\frac{\pi}{4}}=\lim\limits_{\mathbb{C}\to 0}\left( \frac{1}{2Q^{\wh{+}}}\;\frac{\slashed{Q}_b-m}{q_{\wh{+}}+\frac{\mathbb{S}q_{\wh{-}}+Q^{\wh{+}}}{\mathbb{C}}}\right) \notag\\
	&=\lim\limits_{\mathbb{C}\to 0} \frac{1}{2Q^{\wh{+}}}\;\frac{\mathbb{C}\left( \gamma^{\wh{+}}\frac{\mathbb{S}q_{\wh{-}}+Q^{\wh{+}}}{\mathbb{C}}-\gamma^{\wh{-}}q_{\wh{-}}-\gamma^{\perp}.q_{\perp}-m\right) }{\mathbb{C}q_{\wh{+}}+\mathbb{S}q_{\wh{-}}+Q^{\wh{+}}} \notag\\
	&=\frac{\gamma^+\left( q_-+Q^+\right) }{2q^+\left( q_-+Q^+\right)}\notag\\
	&=\frac{\gamma^+}{2q^+}.
	\label{eqn:LFinst}
	\end{align}
At the same time, the first diagram shown in Fig.~\ref{fig:PairAnnihilationto2s_tchannel_a} turns out to be the  on-mass-shell contribution as shown explicitly in the following:
	\begin{align}
	&\Sigma_{a,\delta\to\frac{\pi}{4}}=\lim\limits_{\mathbb{C}\to 0} \left( \frac{1}{2Q^{\wh{+}}}\;\frac{\slashed{Q}_a+m}{q_{\wh{+}}-Q_{a\wh{+}}}\right) \notag\\
	&=\frac{1}{2q_-}\;\frac{\slashed{Q}_a+m}{q^--Q_a^-}\notag\\
	&=\frac{\slashed{q}_{on}+m}{2q^+\left( q^--q_{on}^-\right) }\notag\\
	&=\frac{\slashed{q}_{on}+m}{q^2-m^2}.
	\label{eqn:LFon}
	\end{align}
This proves the decomposition of the covariant fermion propagator in LFD~\cite{SJChang} given by
Eq.~(\ref{eqn:LF-fermion-propagator}) as discussed in the introduction (Sec.~\ref{sec:intro}) 
as well as in the formal derivation (Sec.~\ref{sec:formal}).
	
	
	Let's now compute the time-ordered amplitudes for the $ e^+ e^- $ annihilation into two scalar particles using the interpolating formulation, which
	 are given by
	\begin{equation}\label{eqn:Ma}
	\mathcal{M}_a^{\lambda_1, \lambda_2} =\bar{v}_{\lambda_2}(p_2)\cdot\Sigma_a\cdot u_{\lambda_1}(p_1)
	\end{equation}
	and
	\begin{equation}\label{eqn:Mb}
	\mathcal{M}_b^{\lambda_1, \lambda_2} =\bar{v}_{\lambda_2}(p_2)\cdot\Sigma_b\cdot u_{\lambda_1}(p_1),
	\end{equation}
	where  $\lambda_1$ and $\lambda_2$ represent
	the helicities of the initial $e^-$ and $e^+$ spinors, respectively, and the overall factor such as the coupling constant $ e $, etc., is taken to be 1. Here, $ \Sigma_a $ and $ \Sigma_b $ are given by Eqs.~(\ref{eqn:propagator_TOa}) and (\ref{eqn:propagator_TOb}). If $q=p_1 - p_3$, then these amplitudes are the t-channel amplitudes
	which we may denote as $\mathcal{M}_{a,t}^{\lambda_1, \lambda_2}$ and $\mathcal{M}_{a,t}^{\lambda_1, \lambda_2}$. Similarly, if $q=p_1 - p_4$, then we may denote them as the u-channel amplitudes 	$\mathcal{M}_{a,u}^{\lambda_1, \lambda_2}$ and $\mathcal{M}_{b,u}^{\lambda_1, \lambda_2}$, respectively.
	
	The spinors in the interpolation form were studied in Ref.~\cite{LAJ2015} and the results were given by
	\begin{align*}
	u_H^{(+1/2)}(P)&=\left( \begin{array}{c}
	\sqrt{\frac{P_{\wh{-}}+\mathbb{P}}{2\mathbb{P}}}\sqrt{\frac{P^{\wh{+}}+\mathbb{P}}{\sin\delta+\cos\delta}}\\
	P^R\sqrt{\frac{\sin\delta+\cos\delta}{2\mathbb{P}(\mathbb{P}+P_{\wh{-}})}}\sqrt{P^{\wh{+}}+\mathbb{P}}\\
	\sqrt{\frac{P_{\wh{-}}+\mathbb{P}}{2\mathbb{P}}}\sqrt{\frac{P^{\wh{+}}-\mathbb{P}}{\cos\delta-\sin\delta}}\\
	P^R\sqrt{\frac{\cos\delta-\sin\delta}{2\mathbb{P}(\mathbb{P}+P_{\wh{-}})}}\sqrt{P^{\wh{+}}-\mathbb{P}}
	\end{array}\right) ,\\	u_H^{(-1/2)}(P)&=\left( \begin{array}{c}
	-P^L\sqrt{\frac{\cos\delta-\sin\delta}{2\mathbb{P}(\mathbb{P}+P_{\wh{-}})}}\sqrt{P^{\wh{+}}-\mathbb{P}}\\
	\sqrt{\frac{P_{\wh{-}}+\mathbb{P}}{2\mathbb{P}}}\sqrt{\frac{P^{\wh{+}}-\mathbb{P}}{\cos\delta-\sin\delta}}\\
	-P^L\sqrt{\frac{\sin\delta+\cos\delta}{2\mathbb{P}(\mathbb{P}+P_{\wh{-}})}}\sqrt{P^{\wh{+}}+\mathbb{P}}\\
	\sqrt{\frac{P_{\wh{-}}+\mathbb{P}}{2\mathbb{P}}}\sqrt{\frac{P^{\wh{+}}+\mathbb{P}}{\sin\delta+\cos\delta}}		
	\end{array}\right),
	\end{align*}
	where $ P^R=P^1+iP^2 $ and $ P^L=P^1-iP^2 $, and the antiparticle spinors are obtained by charge conjugation. %
	
\begin{figure}
		\centering
		\includegraphics[width=0.7\linewidth]{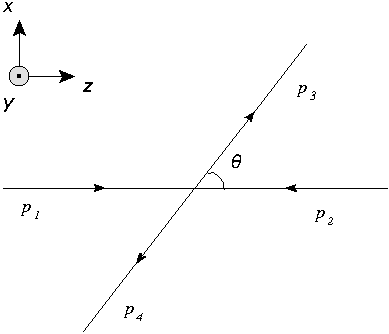}
		\caption{$ e^+e^- $ pair annihilation process at angle $ \theta $ in center of mass frame}
		\label{fig:Annihilation_kinematics}
	\end{figure}

	To make the numerical calculations, we need to specify the kinematics for the process, as shown in Fig.~\ref{fig:Annihilation_kinematics}. We choose the initial reference frame to be the $ e^+e^- $ center of mass frame (CMF), and study the whole landscape of the amplitude change under the boost operation in the $\hat{z}$-direction as well as the change of the interpolation angle $ \delta $. The moving direction of the incoming electron is chosen as the $ +\hat{z} $-direction. Then, the 4-momenta of the initial and final particles can be written as

\begin{align}
\label{eqn:p1234_Annihilation}
		p_1&=(E_0 ,0 ,0 ,P_e) \notag \\
		p_2&=(E_0 ,0 ,0 , -P_e) \notag \\
		p_3&=(E_0 ,E_0\sin\theta ,0 ,E_0\cos\theta) \notag\\
		p_4&=(E_0 ,-E_0\sin\theta ,0 ,-E_0\cos\theta).
		\end{align}

The angular distribution of each helicity amplitude in IFD which depends on the reference frame will be contrasted with the corresponding angular distribution 
of LFD helicity amplitude. 
In the kinematics given by Eq.~(\ref{eqn:p1234_Annihilation}), we note that the intermediate propagating fermion momentum in the time-ordered process depicted 
in Fig.~\ref{fig:PairAnnihilationto2s_tchannel_a} is given by $q=p_1-p_3=(0,-E_0\sin\theta ,0,0,P_e -E_0\cos\theta)$ and thus its light-front plus component 
$q^+=P_e -E_0\cos\theta$ can be negative as well as positive depending on the scattering angle $\theta$. 
Thus, for the kinematic region of $q^+<0$ in LFD, the t-channel process in Fig.~\ref{fig:PairAnnihilationto2s_tchannel_a}  corresponds
to the ``backward" process $\Sigma_B$ although it corresponds to the ``forward" process $\Sigma_F$ for the kinematic region of $q^+>0$. 
The critical scattering angle which separates the kinematic region between $q^+>0$ and $q^+<0$ can of course be obtained by
$q^+=0$ in the corresponding process. In the present kinematics given by Eq.~(\ref{eqn:p1234_Annihilation}), 
the critical scattering angles for t-channel with $q=p_1 - p_3$ and u-channel with $q=p_1 - p_4$ are respectively given by 
	\begin{equation}\label{critical_angle_t}
	\theta_{c,t}=\arccos\left( \frac{P_e}{E_0}\right) ,
	\end{equation}
	\begin{equation}\label{critical_angle_u}
	\theta_{c,u}=\arccos\left( -\frac{P_e}{E_0}\right) . 
	\end{equation}
One should realize that the same amplitude, e.g. $\mathcal{M}_{a,t}^{\lambda_1,\lambda_2}$ given by Eq.~(\ref{eqn:Ma}) with $q=p_1 -p_3$, can correspond to either the 
``on-mass-shell propagating contribution" or the ``instantaneous fermion contribution" in LFD depending on the scattering angle,
e.g. $\theta > \theta_{c,t}$ or $\theta < \theta_{c,t}$, respectively.

\subsection{Collinear Scattering/Annihilation, $\theta = \pi$}
\label{subsec:collinear} 

Before we discuss the angular dependence of the interpolating helicity amplitudes, 
we first consider the collinear amplitude taking the the center of mass angle $\theta$ between the moving direction of incoming electron (particle 1) and outgoing photon (particle 3) as $\pi$, i.e. the collinear back-to-back scattering/annihilation process, in order to exhibit the essential landscape of the helicity amplitudes depending on the reference frame, i.e. the center-of-mass momentum in the $\hat{z}$-direction $ P^z $, and the interpolation angle $\delta$. In this collinear kinematics, the two time-ordered t-channel processes depicted in Figs. \ref{fig:PairAnnihilationto2s_tchannel_a} and \ref{fig:PairAnnihilationto2s_tchannel_b} correspond to the ``forward" moving and ``backward" moving processes without any complication.
Thus, the amplitudes  $\mathcal{M}_{a,t}^{\lambda_1,\lambda_2}$ and $\mathcal{M}_{b,t}^{\lambda_1,\lambda_2}$ correspond to the 
``on-mass-shell propagating contribution" and ``instantaneous fermion contribution", respectively. 
In order not to concern ourselves with the absolute values, we also scale all the energy and momentum values by the electron mass $ m_e $, and take the scalar particles as massless. 
For the simple illustration, we take the initial energy of each particle as $ 2 m_e $, i.e. $ E_0 = 2 m_e$ and $P_e = \sqrt{3} m_e$. 
		
\begin{figure}
		\centering
		\subfloat[]{
			\includegraphics[width=0.48\columnwidth]{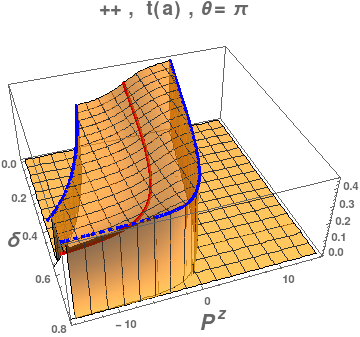}
			\label{fig:eessRRtapi}}
		\centering
		\subfloat[]{
			\includegraphics[width=0.48\columnwidth]{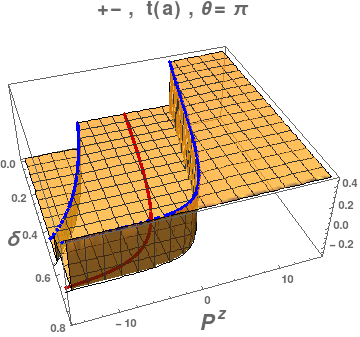}
			\label{fig:eessRLtapi}}
		\\
		\centering
		\subfloat[]{
			\includegraphics[width=0.48\columnwidth]{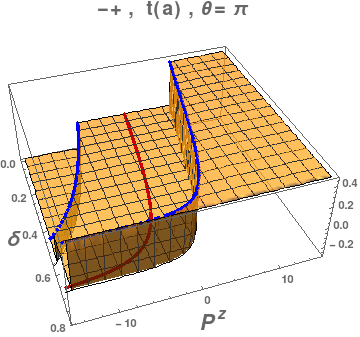}
			\label{fig:eessLRtapi}}
		\centering
		\subfloat[]{
			\includegraphics[width=0.48\columnwidth]{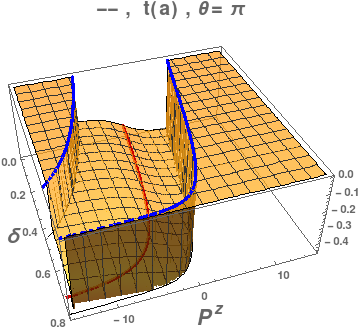}
			\label{fig:eessLLtapi}}
		\caption{\label{fig:Scalar_theta=pi_ta}Annihilation amplitudes for $ e^+ e^- $ to two scalars t channel time-ordering 
		process-a : for (a) helicity $ ++ $ , (b) helicity $ +- $ , (c) helicity $ -+ $ and (d) helicity $ -- $ .}
	\end{figure}

\begin{figure}
		\centering
		\subfloat[]{
			\includegraphics[width=0.48\columnwidth]{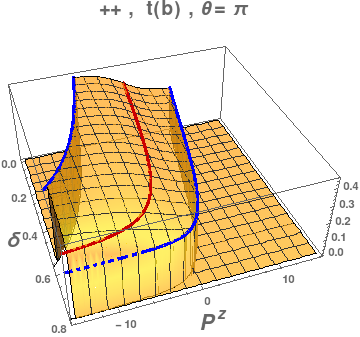}
			\label{fig:eessRRtbpi}}
		\centering
		\subfloat[]{
			\includegraphics[width=0.48\columnwidth]{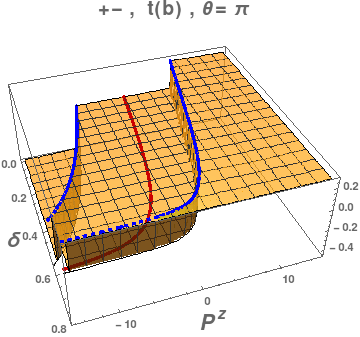}
			\label{fig:eessRLtbpi}}
		\\
		\centering
		\subfloat[]{
			\includegraphics[width=0.48\columnwidth]{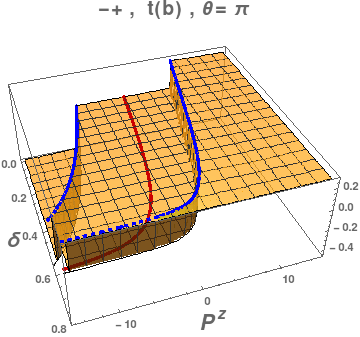}
			\label{fig:eessLRtbpi}}
		\centering
		\subfloat[]{
			\includegraphics[width=0.48\columnwidth]{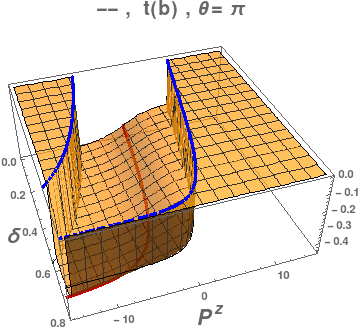}
			\label{fig:eessLLtbpi}}
		\caption{\label{fig:Scalar_theta=pi_tb}Annihilation amplitudes for $ e^+ e^- $ to two scalars t channel time-ordering 
		process-b: for (a) helicity $ ++ $ , (b) helicity $ +- $ , (c) helicity $ -+ $ and (d) helicity $ -- $ .}
	\end{figure}

\begin{figure}
		\centering
		\subfloat[]{
			\includegraphics[width=0.48\columnwidth]{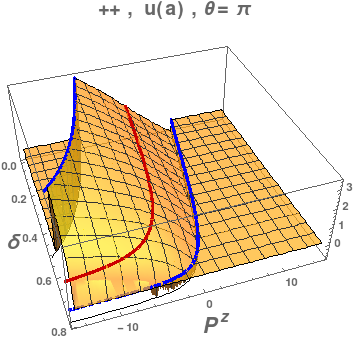}
			\label{fig:eessRRuapi}}
		\centering
		\subfloat[]{
			\includegraphics[width=0.48\columnwidth]{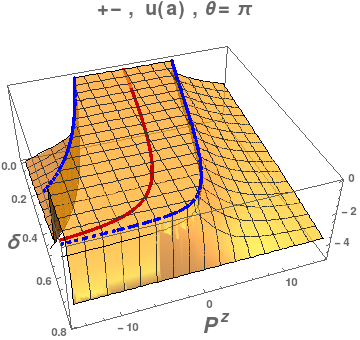}
			\label{fig:eessRLuapi}}
		\\
		\centering
		\subfloat[]{
			\includegraphics[width=0.48\columnwidth]{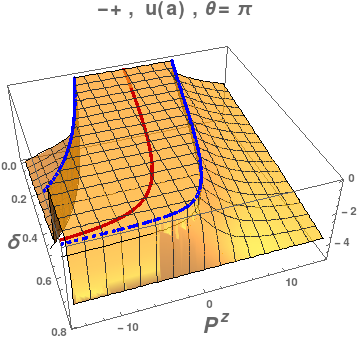}
			\label{fig:eessLRuapi}}
		\centering
		\subfloat[]{
			\includegraphics[width=0.48\columnwidth]{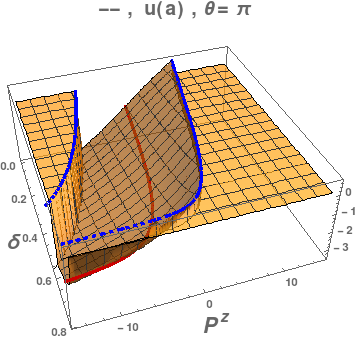}
			\label{fig:eessLLuapi}}
		\caption{\label{fig:Scalar_theta=pi_ua}Annihilation amplitudes for $ e^+ e^- $ to two scalars u channel time-ordering process-a: for (a) helicity $ ++ $ , (b) helicity $ +- $ , (c) helicity $ -+ $ and (d) helicity $ -- $ .}
	\end{figure}

\begin{figure}
		\centering
		\subfloat[]{
			\includegraphics[width=0.48\columnwidth]{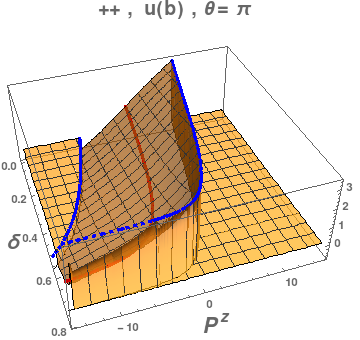}
			\label{fig:eessRRubpi}}
		\centering
		\subfloat[]{
			\includegraphics[width=0.48\columnwidth]{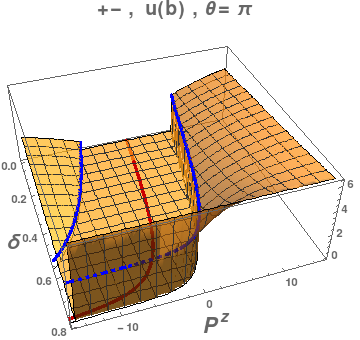}
			\label{fig:eessRLubpi}}
		\\
		\centering
		\subfloat[]{
			\includegraphics[width=0.48\columnwidth]{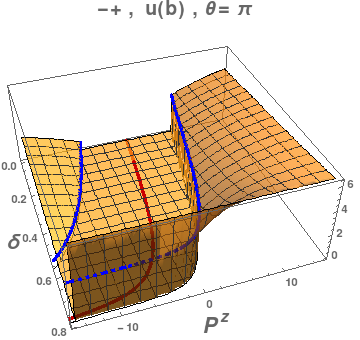}
			\label{fig:eessLRubpi}}
		\centering
		\subfloat[]{
			\includegraphics[width=0.48\columnwidth]{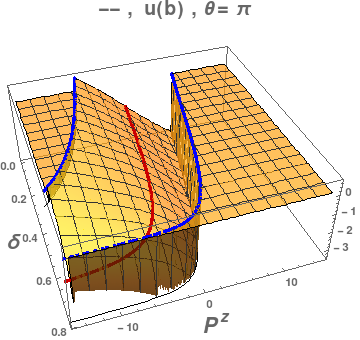}
			\label{fig:eessLLubpi}}
		\caption{\label{fig:Scalar_theta=pi_ub}Annihilation amplitudes for $ e^+ e^- $ to two scalars u channel time-ordering process-b: for (a) helicity $ ++ $ , (b) helicity $ +- $ , (c) helicity $ -+ $ and (d) helicity $ -- $ .}
\end{figure}

\begin{figure}
		\centering
		\includegraphics[width=0.7\linewidth]{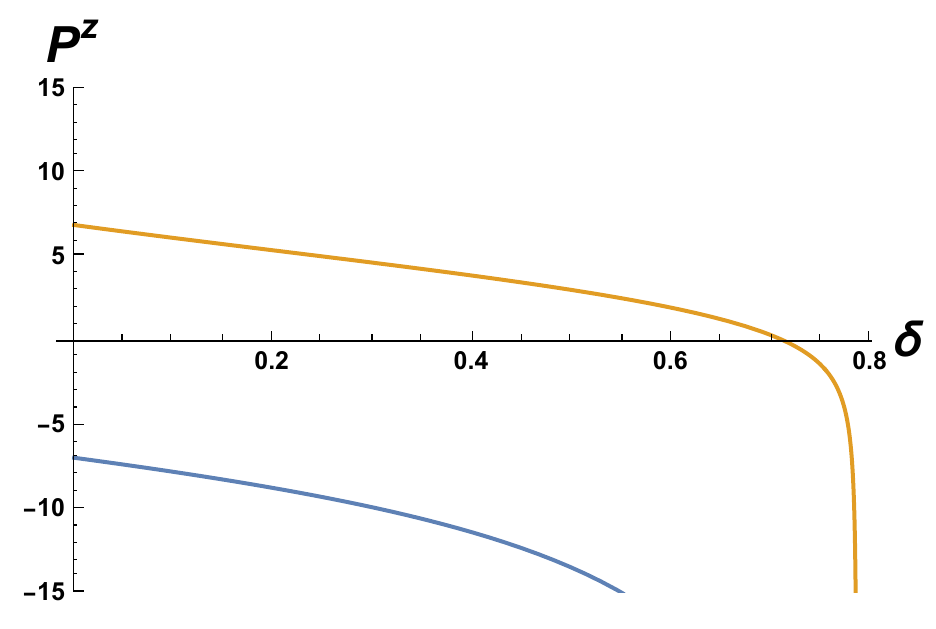}
		\caption{Two Boundaries.}
		\label{fig:TwoBoundaries}
	\end{figure}

The results of the collinear back-to-back scattering/annihilation, i.e. $\theta = \pi$, are shown in Figs. \ref{fig:Scalar_theta=pi_ta} and \ref{fig:Scalar_theta=pi_tb}, where we use ``$ + $'' and ``$ - $'' to denote the helicity of the initial fermions. For example, ``$ +- $'' means a right-handed electron and a left-handed positron annihilation. As the final state particles are scalars, they don't have any designation of helicities. Here, t(a) means the first time-ordering of t channel, corresponding to the diagram Fig.~\ref{fig:PairAnnihilationto2s_tchannel_a}, t(b) means Fig.~\ref{fig:PairAnnihilationto2s_tchannel_b}, etc. There is also the u channel, which can be obtained by swapping the two outgoing particles, and the two time-ordering of u channel can be drawn in a similar way. The results of the u channel are shown in Figs. \ref{fig:Scalar_theta=pi_ua} and \ref{fig:Scalar_theta=pi_ub}. The amplitudes are plotted as a function of $ P^z $ and $ \delta $. When $ \delta\to0 $, i.e. the back ends of the figures, the IFD results are obtained, while $ \delta\to\pi/4 $, i.e. the front ends of the figures, the LFD results are obtained. The red solid line in the middle of all the figures is given by
	\begin{equation}
	\label{J-curve}
	P^z=-\sqrt{\frac{s(1-\mathbb{C})}{2\mathbb{C}}},
	\end{equation}
	where $\sqrt{s} $ is the center of mass energy.
This characteristic curve called ``J-curve" has been discussed extensively in our previous works \cite{Alfredo}~\cite{JLS2015}~\cite{LAJ2015} in conjunction with
the zero-mode in $P^z \to - \infty$ limit where the plus component of the light-front momentum for all the particles involved
in the scattering/annihilation process vanishes, i.e. $p_i^+ \to 0$ ($i = 1,2,3,4$). We note here that this characteristic ``J-curve"
corresponds to the zero of the interpolating total longitudinal momentum, $P_{\wh{-}} = 0$. 	
As discussed in Ref.~\cite{LAJ2015}, ``J-curve" sits in between the two boundaries indicated by blue dashed lines in all of the figures, Figs.~\ref{fig:Scalar_theta=pi_ta}, \ref{fig:Scalar_theta=pi_tb}, \ref{fig:Scalar_theta=pi_ua} and \ref{fig:Scalar_theta=pi_ub}, across which the amplitude changes abruptly. The reason for this abrupt change, as we also discussed in our previous spinor work \cite{LAJ2015}, is because the electron and positron moving along z direction have the speed less than the speed of light $c$ so that the direction of the particle motion can be swapped to the opposite direction in the frame which moves faster than the particle. Namely, the helicity defined in IFD is not invariant but dependent on the reference frame. For a given helicity amplitude in IFD, the particle's spin must flip when its moving direction flips to maintain the given helicity. This results in a sudden abrupt change in each helicity amplitude. In other words, a different spin configuration appears going across the boundary. For example, the left and right boundaries drawn in all the panels of Fig.~\ref{fig:Scalar_theta=pi_ta} correspond respectively to $ p_{1 \wh{-}}=0 $ (zero longitudinal interpolating momentum for the electron) and $ p_{2 \wh{-}}=0 $ (zero longitudinal interpolating momentum for the positron). The change of the helicity depending on the reference frame has been extensively discussed in our previous spinor work \cite{LAJ2015}. 
In particular, the LF helicity of the particle moving in the $-\hat z$ direction is opposite to the Jacob-Wick helicity defined in the IFD. Such swap of the helicity between the IFD and LFD
for the particle moving in the $-\hat z$ direction has been extensively discussed in Ref.\cite{LAJ2015} and the application in the deeply virtual Compton scattering has been
reviewed in Ref.\cite{JB2013}. We find indeed that the behavior of the angle between the momentum direction and the spin direction bifurcates at a critical interpolation angle and the IFD and the LFD separately belong to the two different branches bifurcated at this critical interpolation angle.  
The details of the discussion on the boundaries in the helicity amplitudes, similar to the left and right boundaries in Fig.~\ref{fig:Scalar_theta=pi_ta}, can be found in Ref. \cite{LAJ2015} with the examples of $ e\mu\to e\mu $ and $ e^+e^-\to\mu^+\mu^- $ processes. Solving the equation $ p_{1\wh{-}}=0 $, we get
	\begin{equation}
	\tan\delta=-\frac{E_0 P^z+P_e\sqrt{(2E_0)^2+(P^z)^2}}{P_e P^z+E_0\sqrt{(2E_0)^2+(P^z)^2}}
	\label{eq:electron_boundary}
	\end{equation}
	for the electron and similarly from $ p_{2\wh{-}}=0 $ we get
	\begin{equation}
	\tan\delta=-\frac{E_0 P^z-P_e\sqrt{(2E_0)^2+(P^z)^2}}{E_0\sqrt{(2E_0)^2+(P^z)^2}-P_e P^z}
	\label{eq:positron_boundary}
	\end{equation}
	for the positron.
These two boundaries are depicted in Fig.~\ref{fig:TwoBoundaries}.
At $P^z = 0$, the critical interpolating angle $\delta_c$ corresponding to the boundary due to the positron's helicity swap is given by
$\delta_c = \tan^{-1}(P_e /E_0)$. For $E_0 = 2 m_e$ and $P_e = \sqrt{3} m_e$, this critical value is given by
$\delta_c  = \tan^{-1}(\sqrt{3}/2) \approx 0.713724$ as one can see from Fig.~\ref{fig:TwoBoundaries}.
The bifurcation of the two helicity branches, one belongs to the IFD side and the other belongs to the LFD side, occurs 
exactly at $\delta = \delta_c$ in the CMF ($P^z = 0$) and the abrupt change of the helicity amplitudes crossing from one branch to another branch, e.g.
$0 \leq \delta < \delta_c \approx 0.713724$ and $\delta_c \approx 0.713724 < \delta \leq \pi/4$, can be understood in the example 
presented in Figs.~\ref{fig:Scalar_theta=pi_ta}, \ref{fig:Scalar_theta=pi_tb}, \ref{fig:Scalar_theta=pi_ua} and \ref{fig:Scalar_theta=pi_ub}
as well as in our previous works~\cite{LAJ2015}. One should note that this bifurcation of the two helicity branches is independent of
the scattering angle $\theta$ and thus persists even in the non-collinear helicity amplitudes that we discuss later in this section.  

However, one should note that the LFD result is completely outside of these boundaries, as it appears as a straight line on the LF end. This is due to the boost invariance of the helicity in LFD as we emphasize in the present work as well as in our previous works~\cite{Alfredo}~\cite{JLS2015}~\cite{LAJ2015}. In LFD, we note that the results depicted 
in Figs. \ref{fig:Scalar_theta=pi_tb} and \ref{fig:Scalar_theta=pi_ub} correspond to the instantaneous fermion contribution as shown in Eq.~(\ref{eqn:LFinst}). 
One may note\cite{BL} that the amplitude $\bar{v}\gamma^+ u$ vanishes for the helicity non-flip case, i.e. $++$ and $--$, while it survives for the helicity flip case, i.e. $+-$ and $-+$.
This demonstrates that the LFD ($\delta=\pi/4$) results of $++$ and $--$ helicity amplitudes, $\mathcal{M}_{b,t}^{+,+}$, $\mathcal{M}_{b,t}^{-,-}$, $\mathcal{M}_{b,u}^{+,+}$ and $\mathcal{M}_{b,u}^{-,-}$, respectively, are zero 
while the LFD results of $+-$ and $-+$ helicity amplitudes,
$\mathcal{M}_{b,t}^{+,-}$, $\mathcal{M}_{b,t}^{-,+}$, $\mathcal{M}_{b,u}^{+,-}$ and $\mathcal{M}_{b,u}^{-,+}$, respectively, are non-zero as shown in Figs. \ref{fig:Scalar_theta=pi_tb} and \ref{fig:Scalar_theta=pi_ub}.
	
For this collinear back-to-back scattering/annihilation process, the apparent angular momentum conservation can be rather easily seen in all of Figs.~\ref{fig:Scalar_theta=pi_ta}, \ref{fig:Scalar_theta=pi_tb}, \ref{fig:Scalar_theta=pi_ua} and \ref{fig:Scalar_theta=pi_ub}.
Because the initial electron and positron are spin $ \frac{1}{2} $ particles and the final state particles are spin-less, only the spin singlet system of the two spin-half particles
can annihilate and produce two scalar particles in the center-of-mass frame (i.e. $P^z=0$) due to the angular momentum conservation.
Thus, only when the initial particles have their spins in opposite direction, the amplitude can be non-zero. In Figs.~\ref{fig:Scalar_theta=pi_ta}, 
\ref{fig:Scalar_theta=pi_tb}, \ref{fig:Scalar_theta=pi_ua} and \ref{fig:Scalar_theta=pi_ub}, we note that the $ +- $ and $ -+ $ helicity amplitudes between the two blue line boundaries vanish as 
they correspond to the spin triplet configuration not satisfying the angular momentum conservation. Also, the relative sign between the non-vanishing
$++$ and $--$ helicity amplitudes in the same kinematic region is opposite revealing the nature of spin singlet configuration. Moreover, these results are consistent with the 
the well-known symmetry based on parity conservation that the amplitudes in helicity basis must satisfy\cite{JB2013}
		\begin{equation}\label{eqn:amplitude_relation_scalar_case}
		\mathcal{M}(-\lambda_1,-\lambda_2)=(-1)^{-\lambda_1-\lambda_2}\mathcal{M}(\lambda_1,\lambda_2)
		\end{equation} 
	where $ \lambda_1 $ and $ \lambda_2 $ are the helicities of the incoming electron and positron. 
 
\begin{figure}
		\centering
		\subfloat[]{
			\includegraphics[width=0.48\columnwidth]{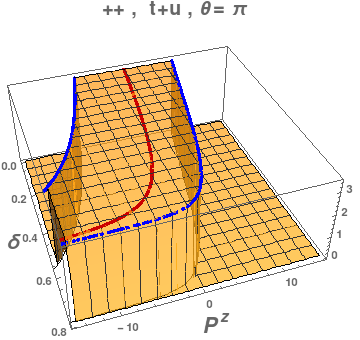}
			\label{fig:eessRRtupi}}
		\centering
		\subfloat[]{
			\includegraphics[width=0.48\columnwidth]{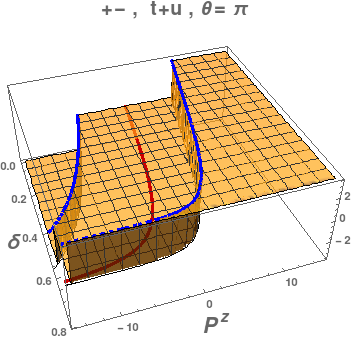}
			\label{fig:eessRLtupi}}
		\\
		\centering
		\subfloat[]{
			\includegraphics[width=0.48\columnwidth]{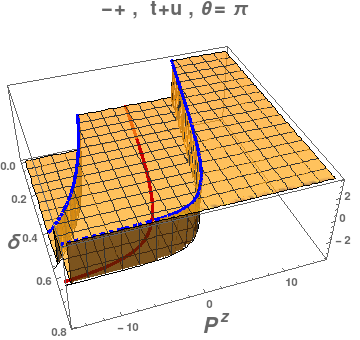}
			\label{fig:eessLRtupi}}
		\centering
		\subfloat[]{
			\includegraphics[width=0.48\columnwidth]{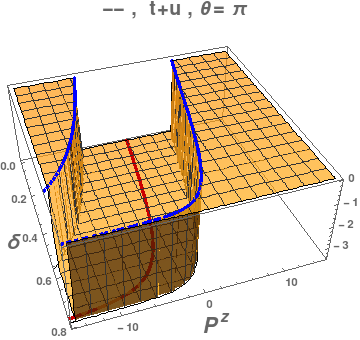}
			\label{fig:eessLLtupi}}
		\caption{\label{fig:Scalar_theta=pi_total}Total annihilation amplitudes for $ e^+ e^- $ to two scalars: for (a) helicity $ ++ $ , (b) helicity $ +- $ , (c) helicity $ -+ $ and (d) helicity $ -- $ .}
		\label{fig:eesstotal}
	\end{figure}

The sum of t-channel and u-channel amplitude of each initial helicity state is shown in Fig.~\ref{fig:Scalar_theta=pi_total}.
All the symmetry that each channel and time-ordered amplitude individually satisfy of course work in the sum of the individual amplitude as well. 
Thus, again in Fig.~\ref{fig:Scalar_theta=pi_total}, the angular momentum conservation and the spin singlet nature of the system are also manifest, i.e.  $ +- $ and $ -+ $ helicity amplitudes between the two blue line boundaries vanish and the non-vanishing
$++$ and $--$ helicity amplitudes in the same kinematic region have opposite sign to each other.
They are again consistent with Eq.~(\ref{eqn:amplitude_relation_scalar_case}).

In Figs.~\ref{fig:Scalar_theta=pi_ta}, \ref{fig:Scalar_theta=pi_tb}, \ref{fig:Scalar_theta=pi_ua}, \ref{fig:Scalar_theta=pi_ub} and \ref{fig:Scalar_theta=pi_total},
we note that the IFD results in $P^z \to +\infty$ appear to yield the corresponding LFD results as one can see the smooth connection of each and every amplitude in the 
right region outside the right boundary. This may suggest that the IFD result in the infinite momentum frame (IMF) yields the LFD result. However, one should note that
the IFD results in $P^z \to -\infty$ are not only different from the corresponding LFD results but also incapable of achieving the LFD results as they are apart by the two
blue boundaries in between. Thus, the IMF in the left region outside the left boundary in IFD cannot yield the desired LFD result although the IMF in the right region
outside the right boundary may do the job. One should be cautious in the prevailing notion of the equivalence between the IFD at the IMF and the LFD. 

Of course, if each helicity amplitude shown in Fig.~\ref{fig:Scalar_theta=pi_total} is squared and summed over all four helicity states, then the result  
is completely independent of $ P^z $ and $ \delta $ as a flat constant in the entire region of $P^z$ and $\delta$ space.

\subsection{Non-collinear Scattering/Annihilation, $0 < \theta < \pi$}
\label{subsec:noncollnear}

\begin{figure}
		\centering
		\subfloat[]{
			\includegraphics[width=0.48\columnwidth]{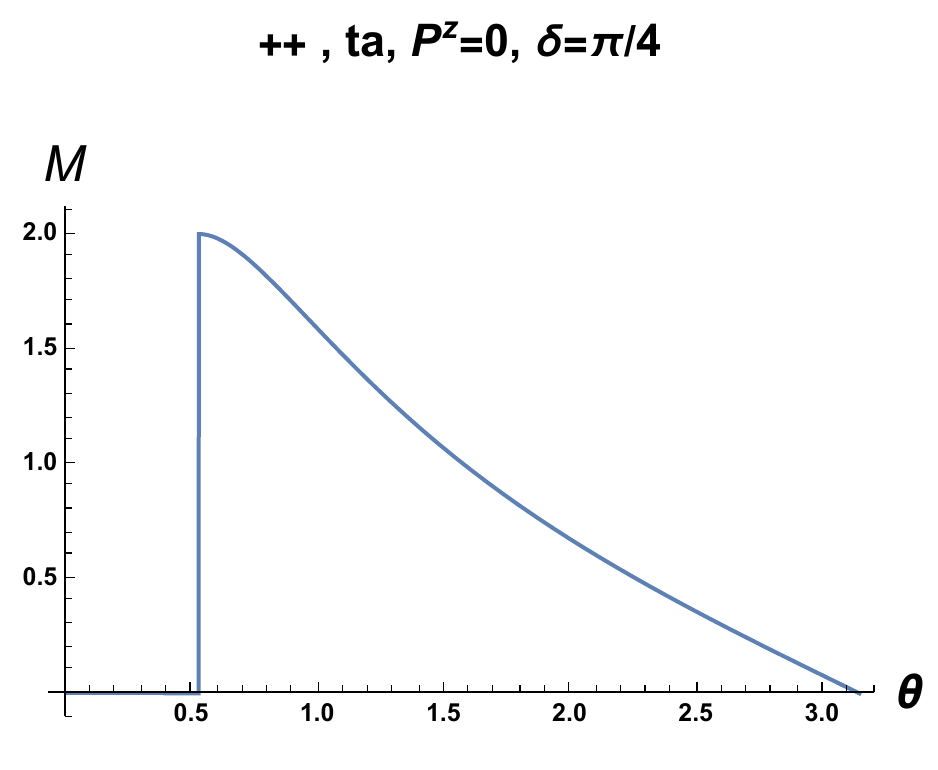}
			\label{fig:RRtaLFDangdist}}
		\centering
		\subfloat[]{
			\includegraphics[width=0.48\columnwidth]{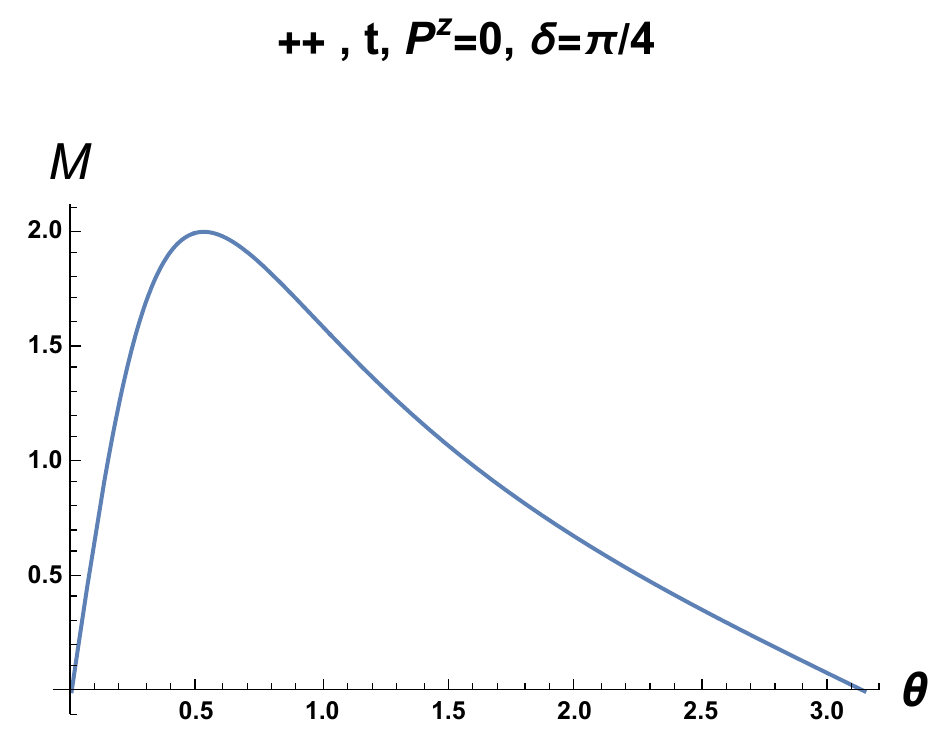}
			\label{fig:RRtabLFDangdist}}	\\			
		\centering
		\subfloat[]{
			\includegraphics[width=0.48\columnwidth]{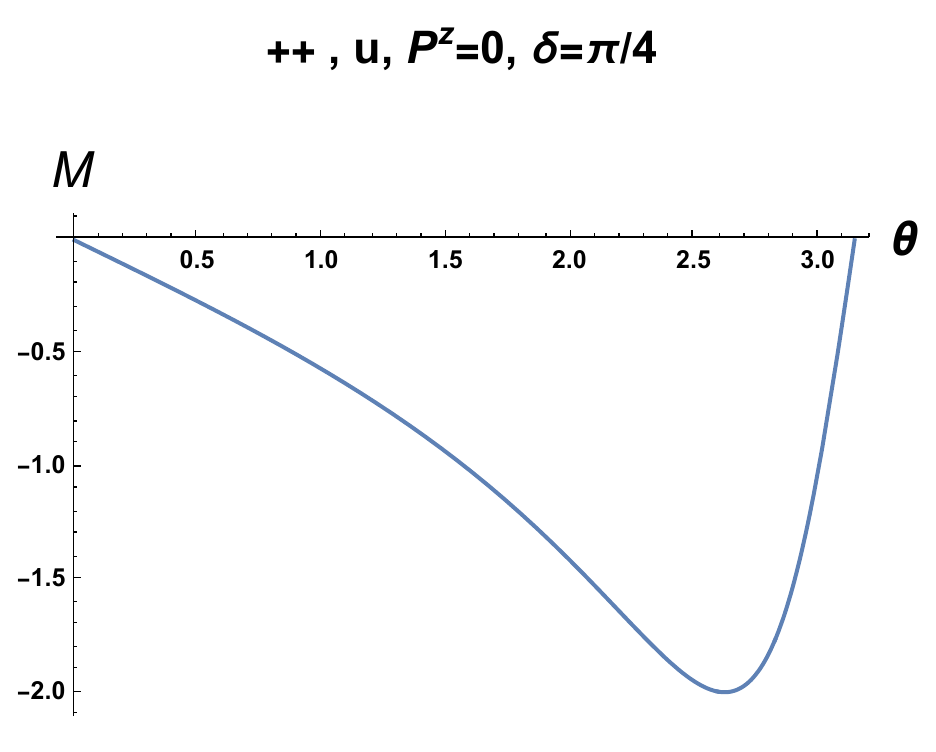}
			\label{fig:RRuabLFDangdist}}
		\caption{\label{fig:ta_t_pp} $ ++ $ annihilation helicity amplitudes for: (a) $ \mathcal{M}^{+,+}_{a,t} $ ,
			(b) $ \mathcal{M}^{+,+}_{a,t} + \mathcal{M}^{+,+}_{b,t} $, and (c)  $ \mathcal{M}^{+,+}_{a,u} + \mathcal{M}^{+,+}_{b,u} $ .}
			\label{fig:RRtuLFDangdist}
	\end{figure}

Now, the non-collinear helicity amplitudes can be computed by varying the center of mass angle $\theta$ in the scattering/annihilation process.
As discussed earlier, for the non-collinear kinematics, the same amplitude can correspond to either the 
``on-mass-shell propagating contribution" or the ``instantaneous fermion contribution" in LFD depending on the region of the scattering angle.
For example, the amplitude $\mathcal{M}_{a,t}^{\lambda_1,\lambda_2}$ corresponds to 
the ``instantaneous fermion contribution" in LFD for the region $\theta < \theta_{c,t}$ while it corresponds to the 
``on-mass-shell propagating contribution" for the region $\theta > \theta_{c,t}$, where $\theta_{c,t}$ is given by Eq.~(\ref{critical_angle_t}).
For $E_0 = 2 m_e$ and $P_e = \sqrt{3} m_e$, $\theta_{c,t} = \pi/6$ and  $\theta_{c,u} = 5\pi/6$ from Eq.~(\ref{critical_angle_t}) and Eq.~(\ref{critical_angle_u}), respectively.
To demonstrate the existence of this critical angle only at LFD, we may take a look closely at each light-front helicity amplitude and contrast
its behavior with the ones off the value of $\delta=\pi/4$. As an example, in Fig.~\ref{fig:RRtuLFDangdist}, we show the result of the angular
distribution for the $++$ helicity amplitudes: (a) $\mathcal{M}_{a,t}^{+,+}$ (b) $\mathcal{M}_{a,t}^{+,+}$+$\mathcal{M}_{b,t}^{+,+}$ and (c) $\mathcal{M}_{a,u}^{+,+}$+$\mathcal{M}_{b,u}^{+,+}$ at the exact light-front, i.e. $\delta=\pi/4$. For $\mathcal{M}_{a,t}^{+,+}$ shown in Fig.~\ref{fig:RRtaLFDangdist}, the left side of $\theta_{c,t} =\frac{\pi}{6}\approx 0.523599$ (i.e. $\theta < \theta_{c,t}$) is the ``instantaneous fermion contribution" and thus the amplitude is zero as expected from the light-front instantaneous propagator, $ \frac{\gamma^+}{2q^+} $, due to $\bar{v}^{\uparrow}\gamma^+ u^{\uparrow}=0 $\cite{BL}. 
On the other hand, the right side of the critical angle (i.e. $\theta > \theta_{c,t}$) is the ``on-mass-shell propagating contribution" for $\mathcal{M}_{a,t}^{+,+}$.
These two distinguished contributions for $\theta < \theta_{c,t}$ and $\theta > \theta_{c,t}$ yield a dramatic ``cliff" feature for $\mathcal{M}_{a,t}^{+,+}$ as shown in Fig.~\ref{fig:RRtaLFDangdist}.
Due to the sign change of the intermediate fermion momentum $ q_b=-q_a=-q $ for the other time-ordered amplitude $\mathcal{M}_{b,t}^{+,+}$,
the angle regions for the ``instantaneous fermion contribution" and the ``on-mass-shell propagating contribution" swap in $\mathcal{M}_{b,t}^{+,+}$ with respect to 
$\mathcal{M}_{a,t}^{+,+}$,i.e. the right side  ($\theta > \theta_{c,t}$) becomes the ``instantaneous fermion contribution" and the left side ($\theta < \theta_{c,t}$) becomes the ``on-mass-shell propagating contribution" for $\mathcal{M}_{b,t}^{+,+}$, while it was the other way around for $\mathcal{M}_{a,t}^{+,+}$ as discussed above.
The addition of the two time-ordered amplitudes, $\mathcal{M}_{a,t}^{+,+}$+$\mathcal{M}_{b,t}^{+,+}$, is shown in Fig.~\ref{fig:RRtabLFDangdist}.
Since the ``instantaneous fermion contribution" for the $++$ helicity amplitudes in LFD is always zero due to $\bar{v}^{\uparrow}\gamma^+ u^{\uparrow}=0$,
the ``on-mass-shell propagating contribution" for $\mathcal{M}_{b,t}^{+,+}$ is rather easily figured out by subtracting the curve depicted in Fig.~\ref{fig:RRtaLFDangdist} 
from the curve depicted in Fig.~\ref{fig:RRtabLFDangdist}. Essentially the same procedure of obtaining the t-channel amplitude can be applied to the u-channel
amplitude by exchanging the two final state scalar particles, i.e. $p_3 \leftrightarrow p_4$. Thus, $q$ becomes $p_1 - p_4$ in the u-channel while it was $p_1 - p_3$ in
the t-channel and the result of $\mathcal{M}_{a,u}^{+,+}$+$\mathcal{M}_{b,u}^{+,+}$ is obtained as shown in Fig.~\ref{fig:RRuabLFDangdist}.

\begin{figure}
		\centering
		\subfloat[]{
			\includegraphics[width=0.48\columnwidth]{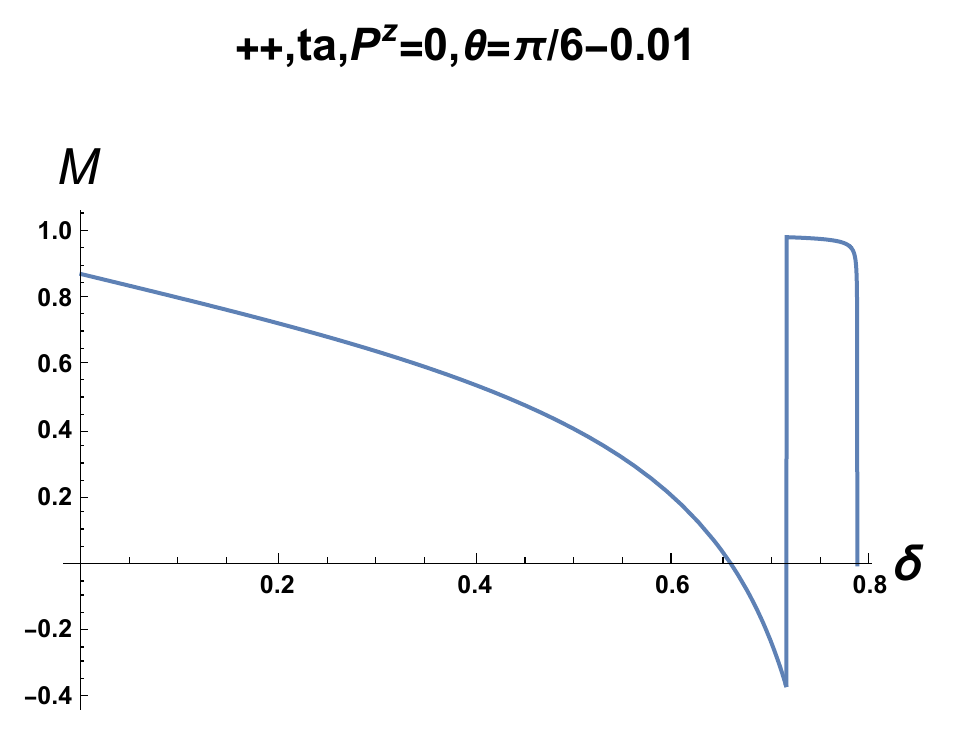}
			\label{fig:RRtathcmInterpol}}
			\centering
		\subfloat[]{
			\includegraphics[width=0.48\columnwidth]{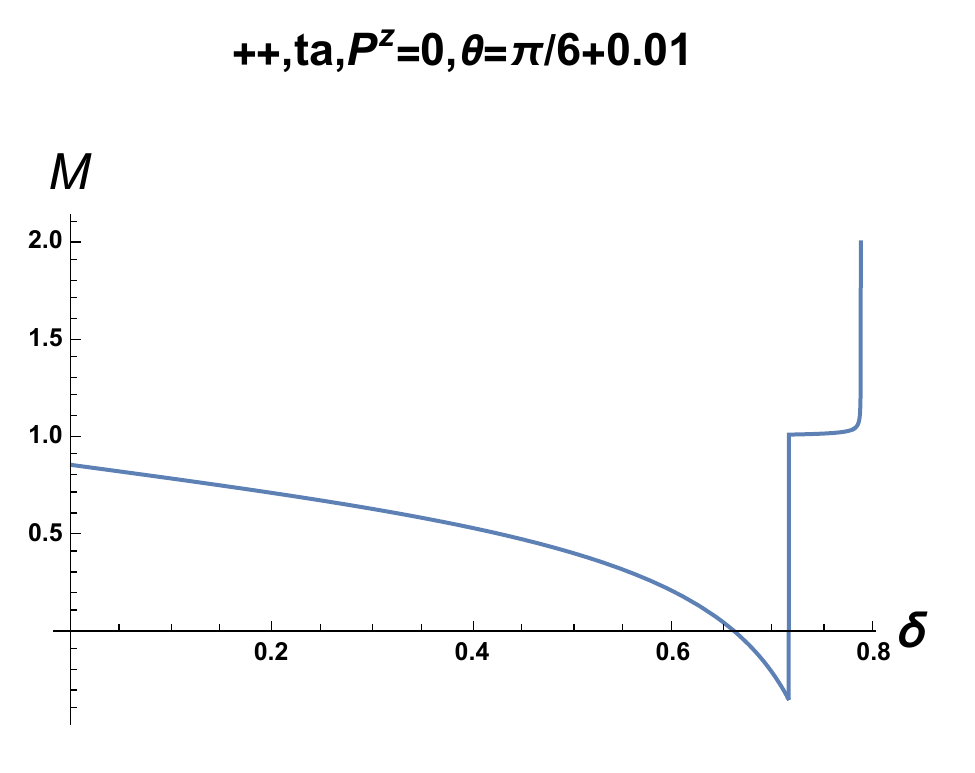}
			\label{fig:RRtathcpInterpol}}\\
			\centering
		\subfloat[]{
			\includegraphics[width=0.48\columnwidth]{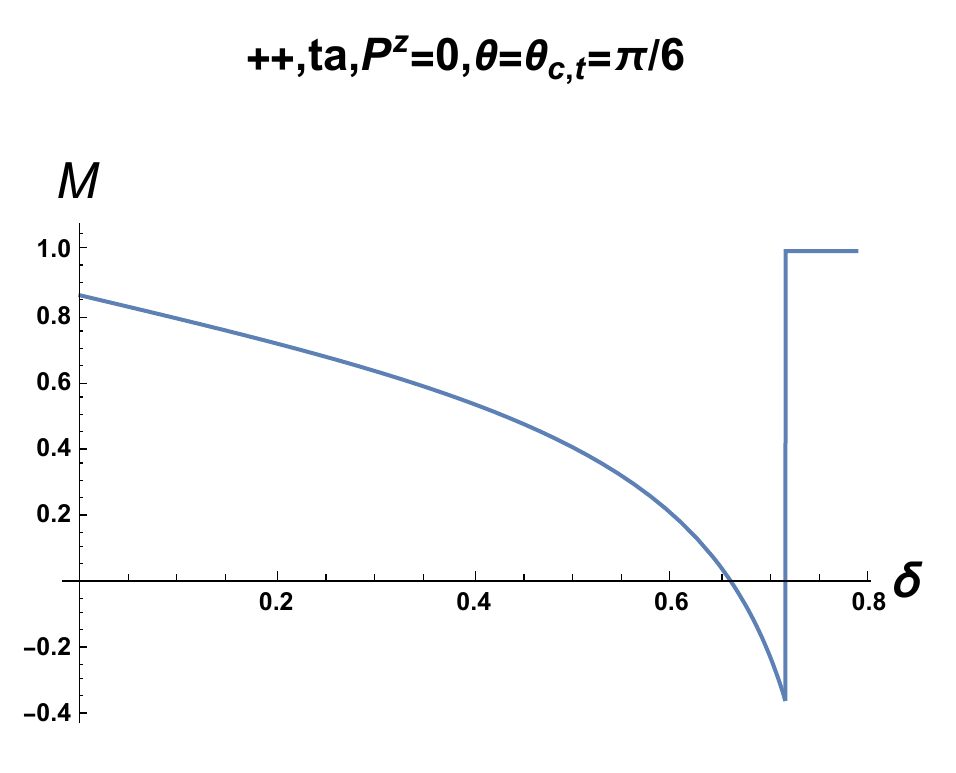}
			\label{fig:RRtathcInterpol}}
		\caption{\label{fig:ta_Interpol_pp} Interpolation angle dependence of $ ++ $ annihilation helicity amplitudes $ \mathcal{M}_{a,t}^{+,+} $  for: (a) $\theta < \theta_{c,t}$ ,
		(b) $\theta > \theta_{c,t}$, and
		(c) $\theta = \theta_{c,t}$.}
			\label{fig:RRthcaInterpol}
	\end{figure}

To exhibit that the ``instantaneous fermion contribution" is the unique feature only in LFD ($\delta=\pi/4$), we take a look at the interpolation angle $\delta$ dependence of
the amplitude $\mathcal{M}_{a,t}^{+,+}$ by slightly varying the scattering angle $\theta$ around the critical angle $\theta_{c,t}$. In Fig.~\ref{fig:ta_Interpol_pp}, we show the $ \delta $-dependence of $\mathcal{M}_{a,t}^{+,+}$ at (a) $ \theta = \theta_{c,t} - 0.01 \approx 0.513599$ (b) $ \theta = \theta_{c,t} + 0.01 \approx 0.533599 $ and (c) $ \theta = \theta_{c,t} \approx 0.523599$. These three values of the angle $\theta$ chosen for Fig.~\ref{fig:ta_Interpol_pp} correspond to slightly left of the ``cliff", at the ``cliff", and slightly right of the ``cliff" in Fig.~\ref{fig:RRtaLFDangdist}, respectively. 
Since the values of the amplitude $\mathcal{M}_{a,t}^{+,+}$ dramatically change around the critical angle $\theta_{c,t}$ from 0.0 on the left ($\theta < \theta_{c,t}$) to around 2.0 on the right immediately passing the critical angle $\theta_{c,t}$ as depicted in Fig.~\ref{fig:RRtaLFDangdist}, we should be able to see the corresponding dramatic change also in Fig.~\ref{fig:ta_Interpol_pp}. We see indeed this dramatic change in Fig.~\ref{fig:ta_Interpol_pp}\footnote{Note that the scale of Fig.~\ref{fig:RRtathcpInterpol} is doubled from Figs.~\ref{fig:RRtathcmInterpol} and \ref{fig:RRtathcInterpol} to fit them 
all in one collective figure of Fig.~\ref{fig:ta_Interpol_pp}.} on top of the abrupt change of the helicity amplitude due to the bifurcation of two helicity branches discussed above in
the collinear ($\theta = \pi$) helicity amplitudes as well as in our previous work~\cite{LAJ2015} extensively, one in the side of IFD and the other in the side of LFD, divided by the critical interpolating angle $\delta_c \approx 0.713724$ discussed below Eq.~(\ref{eq:positron_boundary}) and depicted in Fig.~\ref{fig:TwoBoundaries}. 
In Fig.~\ref{fig:RRtathcmInterpol}, the value of the amplitude $\mathcal{M}_{a,t}^{+,+}$ 
at the right end ($ \delta=\frac{\pi}{4} $) is 0.00 while the value for $\delta_c < \delta < \frac{\pi}{4}$ (not including $\delta = \frac{\pi}{4}$) is around 1.0 and falls off to get linked to the smoothly behaving curve for the region 
$\delta < \delta_c$ that belongs to the helicity branch on the IFD side.
In Fig.~\ref{fig:RRtathcpInterpol}, however,  the value of the amplitude $\mathcal{M}_{a,t}^{+,+}$ 
at the right end ($ \delta=\frac{\pi}{4} $) is around 2.00 while the value for $\delta_c < \delta < \frac{\pi}{4}$ (again not including $\delta = \frac{\pi}{4}$) is still around 1.0 and again falls off to get linked to the smoothly behaving curve for the region $\delta < \delta_c$ that belongs to the helicity branch on the IFD side.
Thus, the helicity amplitude $\mathcal{M}_{a,t}^{+,+}$ doesn't change much except its value at $\delta=\pi/4$ or at LFD.
In the region $\delta_c < \delta \leq \frac{\pi}{4}$ that belongs to the helicity branch on the LFD side, one can see the dramatic change of the helicity amplitude at the right end point
$\delta = \pi/4$, i.e. only at LFD but not anywhere else. This clearly demonstrates that the ``instantaneous fermion contribution" exists only in the LFD.
 
\begin{figure}
		\centering
		\subfloat[]{
			\includegraphics[width=0.48\columnwidth]{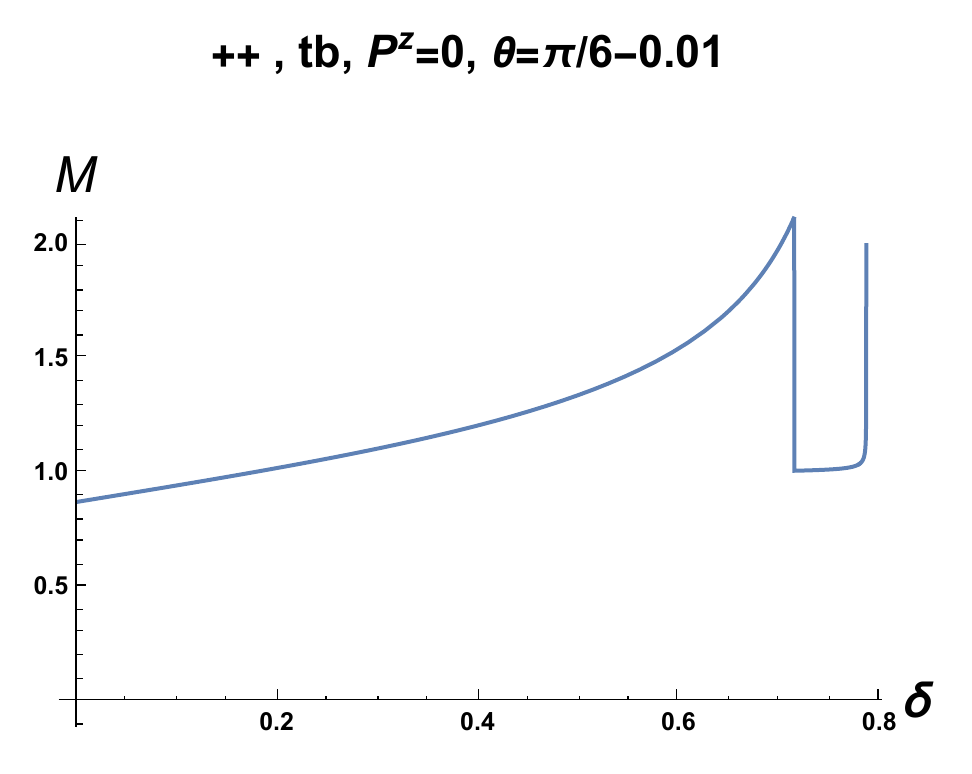}
			\label{fig:RRtbthcmInterpol}}
			\centering
		\subfloat[]{
			\includegraphics[width=0.48\columnwidth]{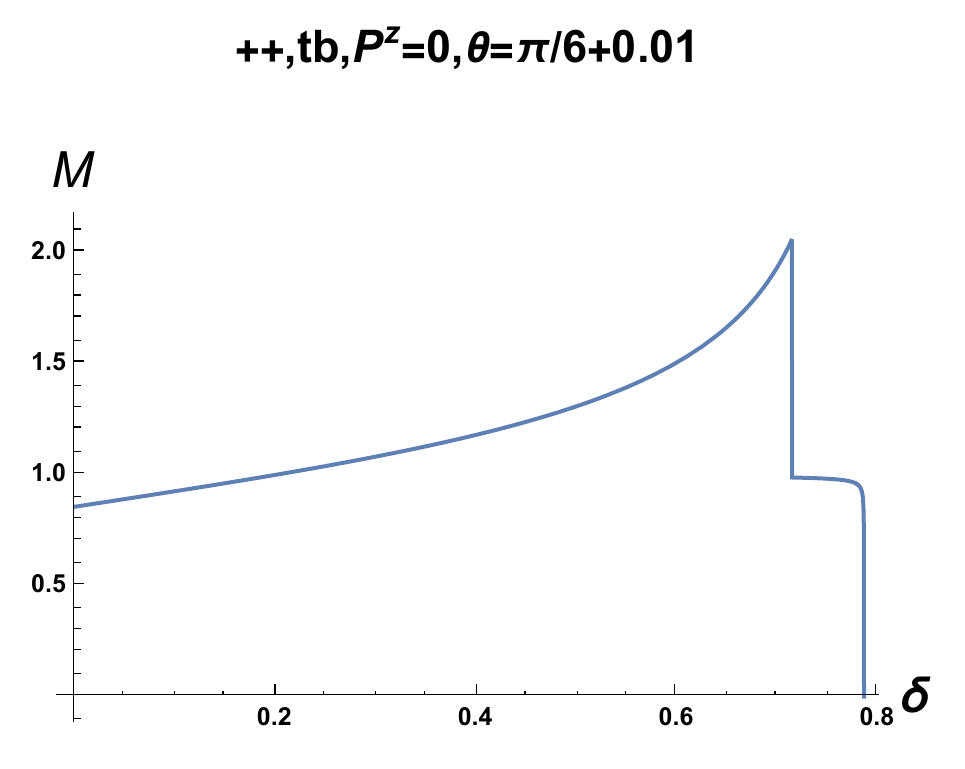}
			\label{fig:RRtbthcpInterpol}}
			\\
	\centering
		\subfloat[]{
			\includegraphics[width=0.48\columnwidth]{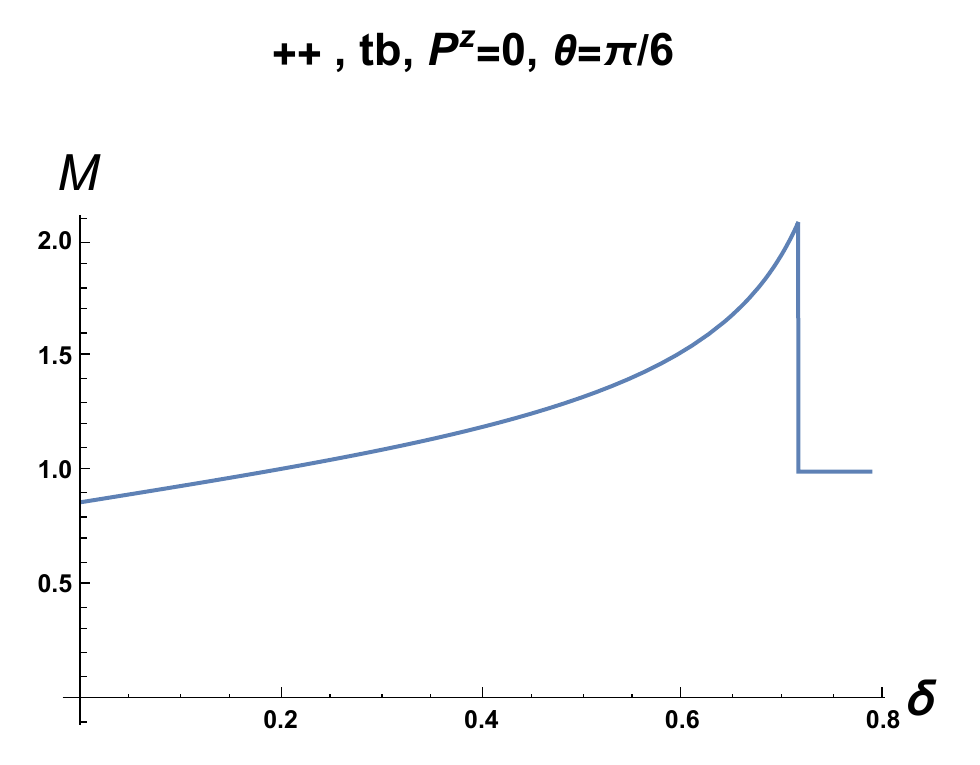}
			\label{fig:RRtbthcInterpol}}
	\caption{\label{fig:tb_Interpol_pp} Interpolation angle dependence of $ ++ $ annihilation helicity amplitudes $ \mathcal{M}_{b,t}^{+,+} $ for: (a) $\theta < \theta_{c,t}$ ,
		(b) $\theta > \theta_{c,t}$, and
		(c) $\theta = \theta_{c,t}$.}
			\label{fig:RRthcbInterpol}
	\end{figure}

Similarly, in Fig.~\ref{fig:RRthcbInterpol}, we show the $ \delta $-dependence of the other time-ordered $ ++ $ helicity amplitude $\mathcal{M}_{b,t}^{+,+}$ at (a) $ \theta = \theta_{c,t} - 0.01 \approx 0.513599$ (b) $ \theta = \theta_{c,t} + 0.01 \approx 0.533599 $ and (c) $ \theta = \theta_{c,t} \approx 0.523599$. 
As discussed earlier, the angle regions for the ``instantaneous fermion contribution" and the ``on-mass-shell propagating contribution" swap in $\mathcal{M}_{b,t}^{+,+}$ with respect to 
$\mathcal{M}_{a,t}^{+,+}$ due to the sign change of the intermediate fermion momentum $ q_b=-q_a=-q $ for 
$\mathcal{M}_{b,t}^{+,+}$, i.e. the right side ($\theta > \theta_{c,t}$) and the left side ($\theta < \theta_{c,t}$) become the ``instantaneous fermion contribution" and 
the ``on-mass-shell propagating contribution" for $\mathcal{M}_{b,t}^{+,+}$.
Since the ``instantaneous fermion contribution" for the $++$ helicity amplitudes in LFD is always zero (again due to $\bar{v}^{\uparrow}\gamma^+ u^{\uparrow}=0$),
we now should be able to see $\mathcal{M}_{b,t}^{+,+} = 0$ for $\theta > \theta_{c,t}$ while $\mathcal{M}_{b,t}^{+,+} \neq 0$ for $\theta < \theta_{c,t}$ in LFD.
We indeed see this expected LFD result in Fig.~\ref{fig:RRthcbInterpol} as the value of  $\mathcal{M}_{b,t}^{+,+}$ at $\delta = \pi/4$ turns out to be exactly 0.0 for 
$\theta = \theta_{c,t} + 0.01$ in Fig.~\ref{fig:RRtbthcpInterpol} while it is around 2.0 for $\theta = \theta_{c,t} - 0.01$ in Fig.~\ref{fig:RRtbthcmInterpol}.
This dramatic change at LFD again clearly demonstrates that the ``instantaneous fermion contribution" exists only in the LFD.
For $\delta_c < \delta < \frac{\pi}{4}$ (not including $\delta = \frac{\pi}{4}$), however, the value of $\mathcal{M}_{b,t}^{+,+}$  is around 1.0 and rises up to get linked to the smoothly behaving curve for the region $\delta < \delta_c$ that belongs to the helicity branch on the IFD side. As shown in Fig.~\ref{fig:RRthcbInterpol},
the helicity amplitude $\mathcal{M}_{b,t}^{+,+}$ doesn't change much except its value at $\delta=\pi/4$ or at LFD.

		\begin{figure}
		\centering
		\subfloat[]{
			\includegraphics[width=0.48\columnwidth]{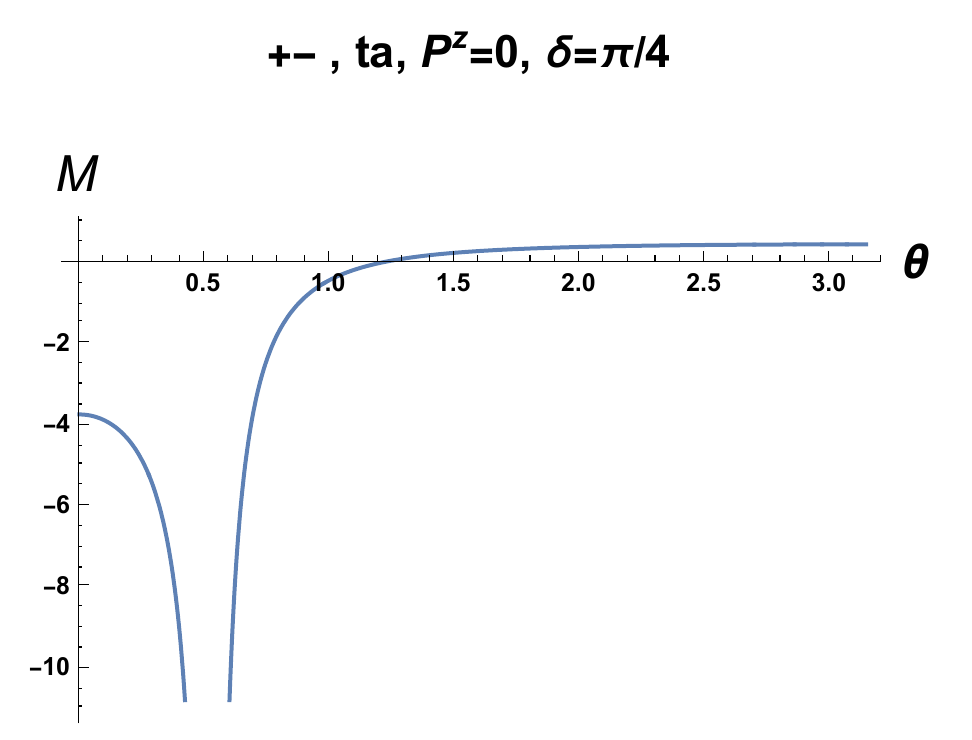}
			\label{fig:RLtaLFDangdist}}
		\centering
		\subfloat[]{
			\includegraphics[width=0.48\columnwidth]{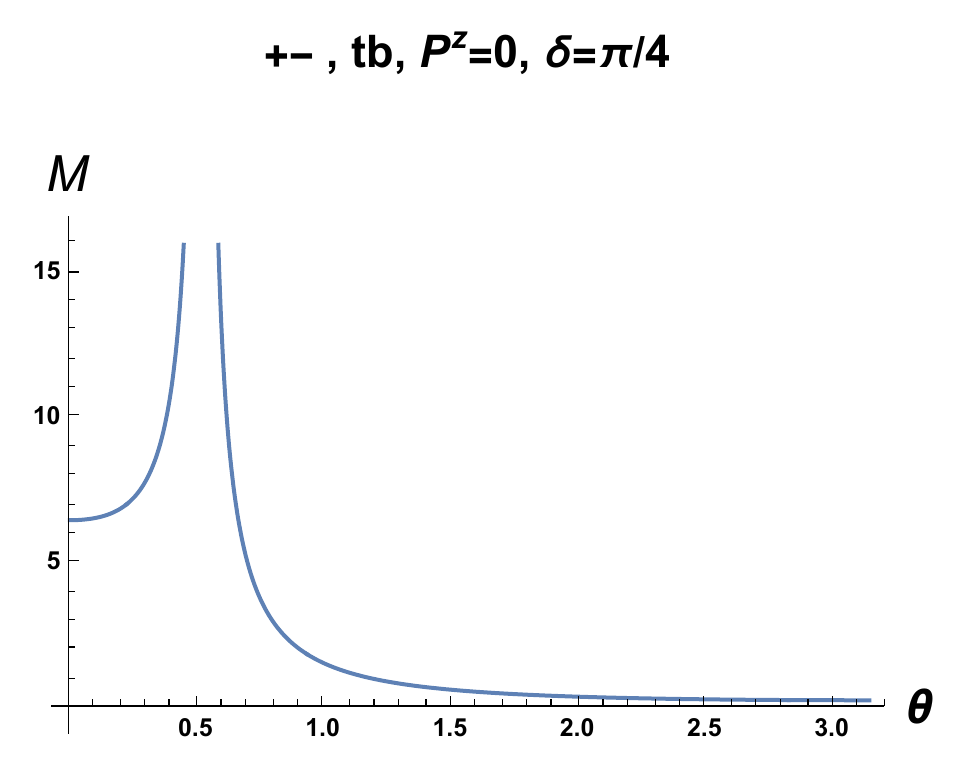}
			\label{fig:RLtbLFDangdist}}
		\\
		\centering
		\subfloat[]{
			\includegraphics[width=0.48\columnwidth]{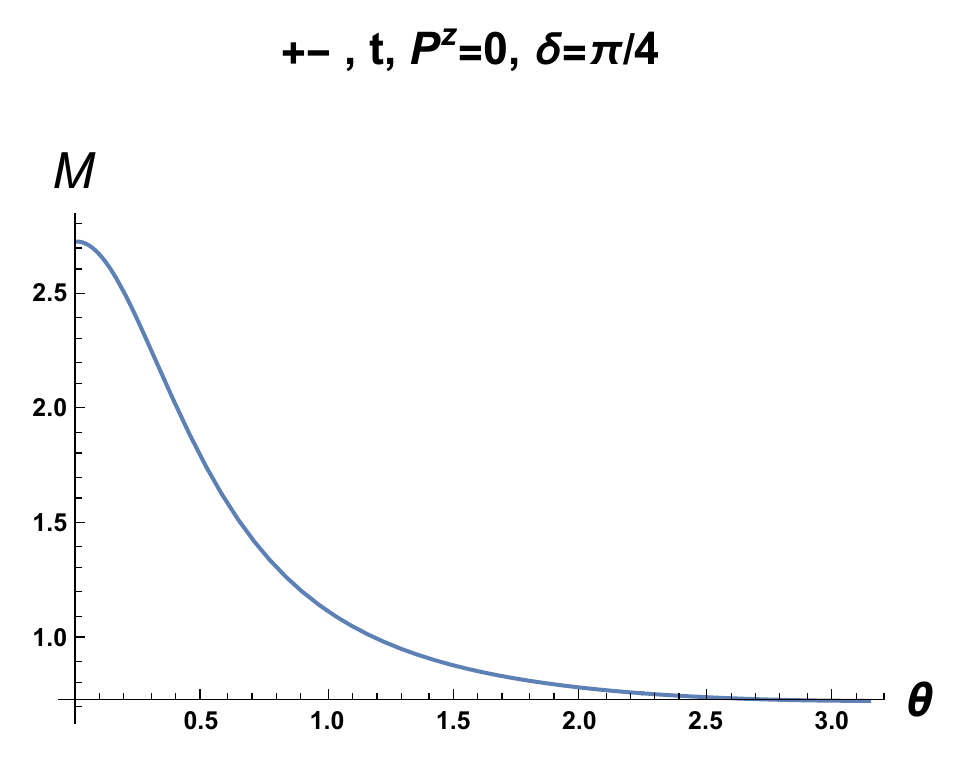}
			\label{fig:RLtabLFDangdist}}
		\centering
		\subfloat[]{
			\includegraphics[width=0.48\columnwidth]{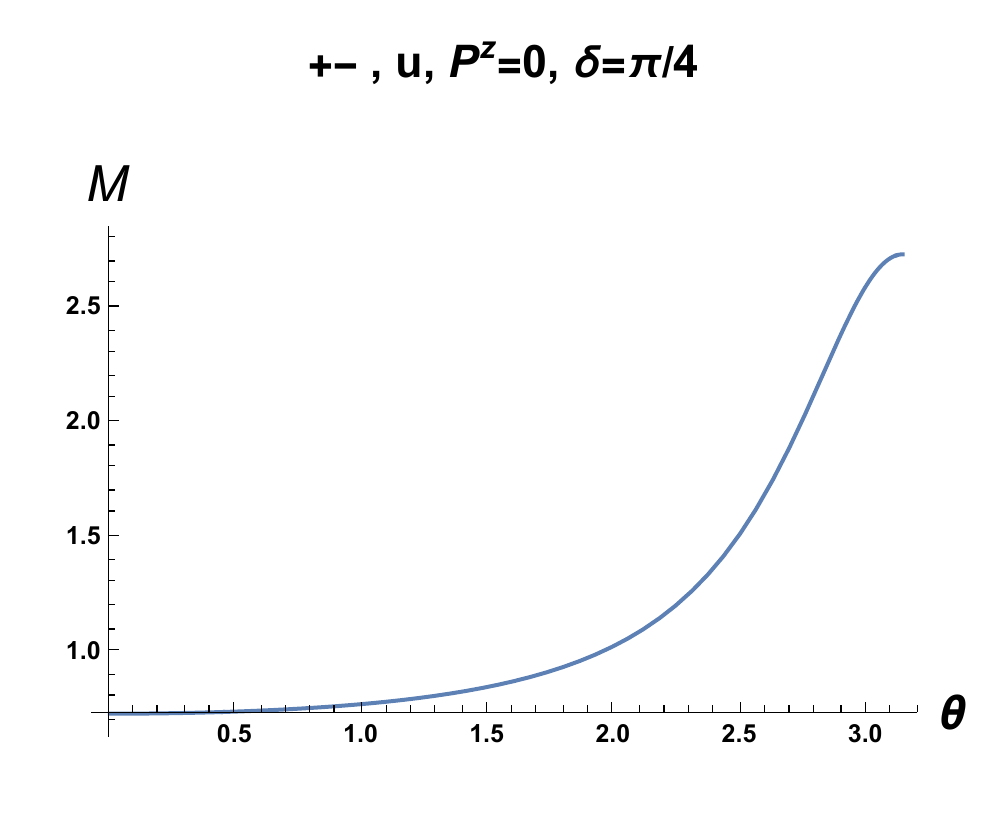}
			\label{fig:RLuabLFDangdist}}
		\caption{\label{fig:tb_Interpol_pm} $ +- $ annihilation helicity amplitudes for: (a) $ \mathcal{M}^{+,-}_{a,t} $ ,
			(b) $ \mathcal{M}^{+,-}_{b,t} $ 
			(c) $ \mathcal{M}^{+,-}_{a,t} + \mathcal{M}^{+,-}_{b,t} $, and (d)  $ \mathcal{M}^{+,-}_{a,u} + \mathcal{M}^{+,-}_{b,u} $ .}
		\label{fig:RLthcInterpol}
	\end{figure}

				\begin{figure}
		\centering
		\subfloat[]{
			\includegraphics[width=0.48\columnwidth]{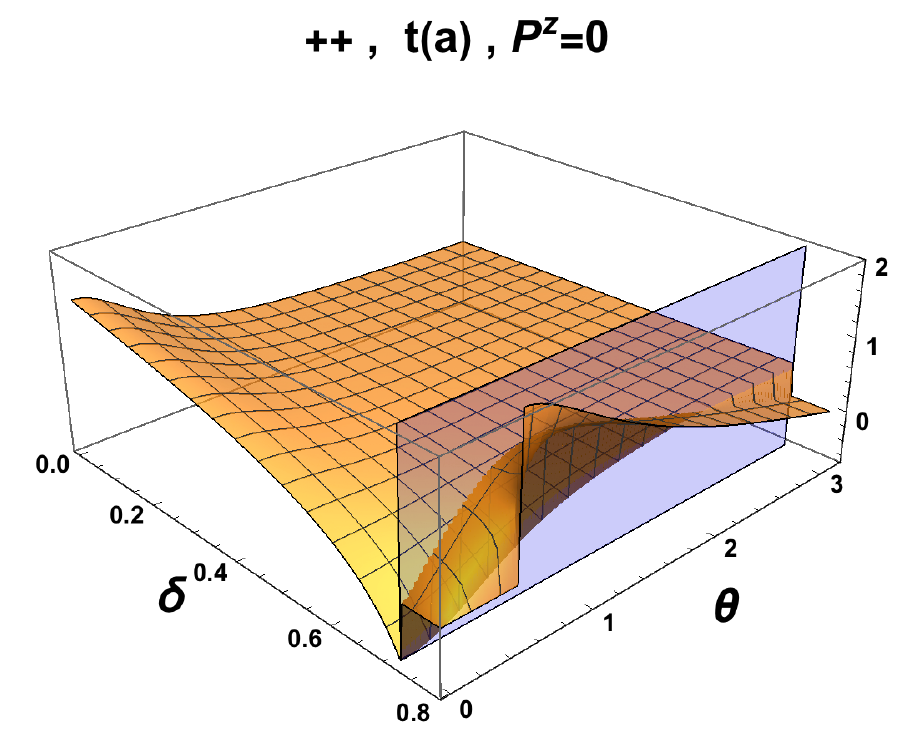}
			\label{fig:eessAngDistRRta}}
		\centering
		\subfloat[]{
			\includegraphics[width=0.48\columnwidth]{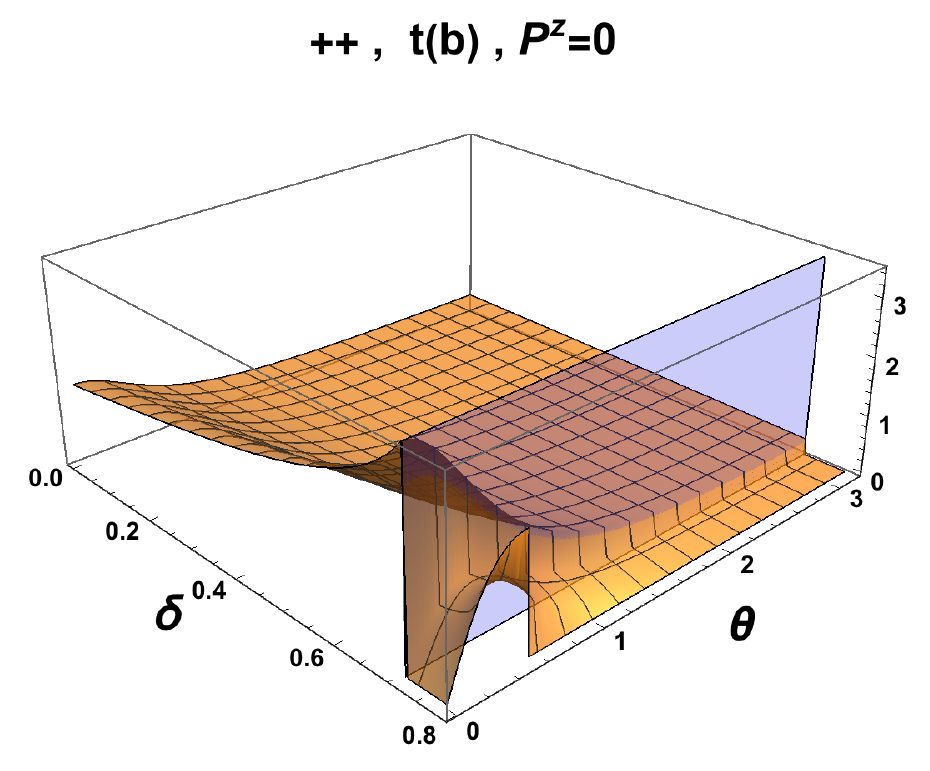}
			\label{fig:eessAngDistRRtb}}
		\\
		\centering
		\subfloat[]{
			\includegraphics[width=0.48\columnwidth]{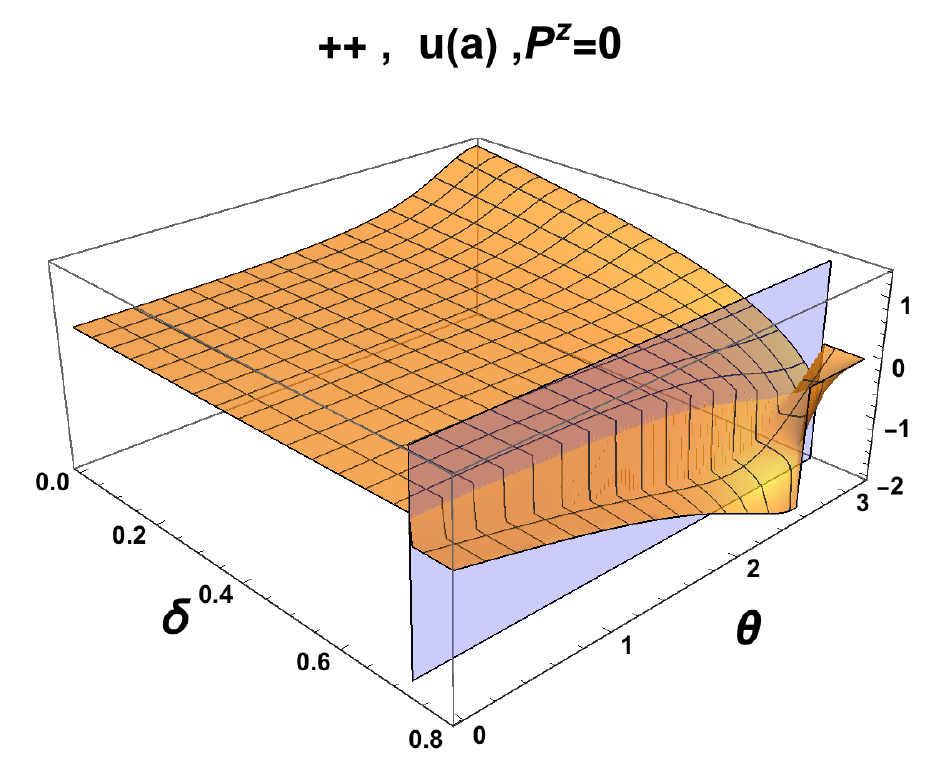}
			\label{fig:eessAngDistRRua}}
		\centering
		\subfloat[]{
			\includegraphics[width=0.48\columnwidth]{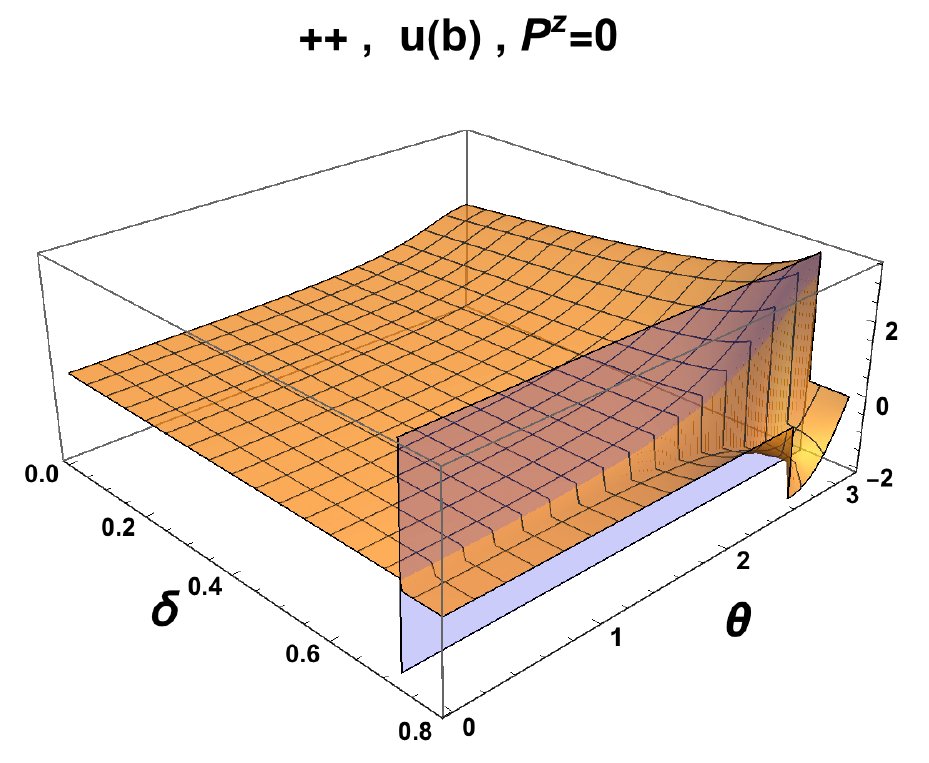}
			\label{fig:eessAngDistRRub}}
		\caption{\label{fig:eessAngDistRR}Angular distribution of the helicity amplitude $ ++ $ 
		for (a) t-channel time-ordering process-a, $\mathcal{M}_{a,t}^{+,+}$  (b) t-channel time-ordering process-b, $\mathcal{M}_{b,t}^{+,+}$ 
		(c) u-channel time-ordering process-a, $\mathcal{M}_{a,u}^{+,+}$ (d) u-channel time-ordering process-b, $\mathcal{M}_{b,u}^{+,+}$.}
	\end{figure}

				\begin{figure}
		\centering
		\subfloat[]{
			\includegraphics[width=0.48\columnwidth]{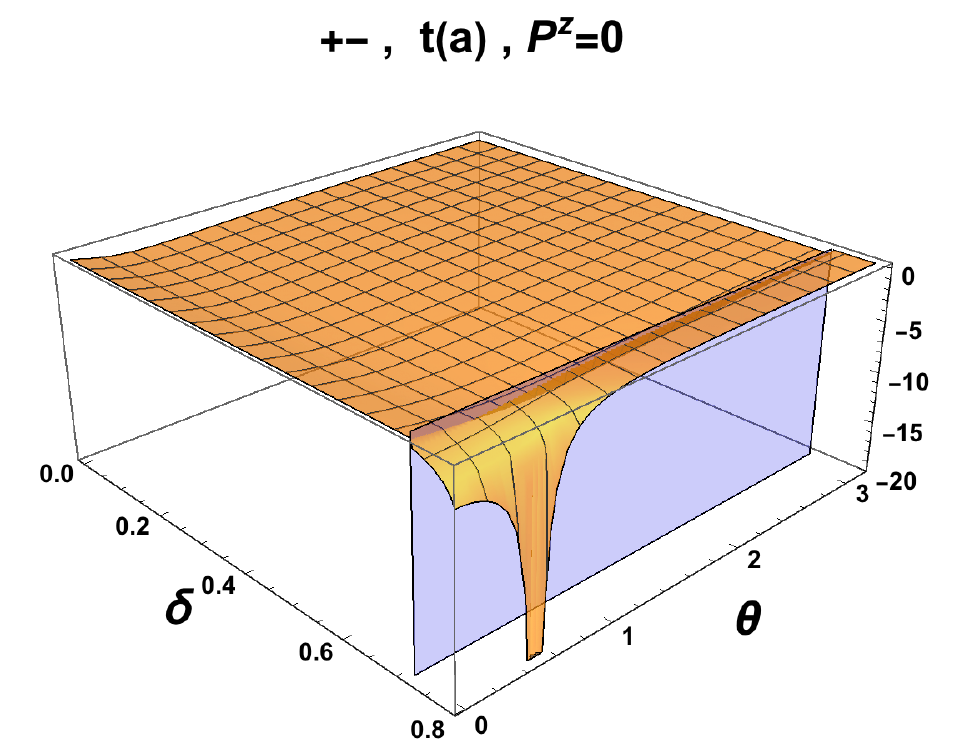}
			\label{fig:eessAngDistRLta}}
		\centering
		\subfloat[]{
			\includegraphics[width=0.48\columnwidth]{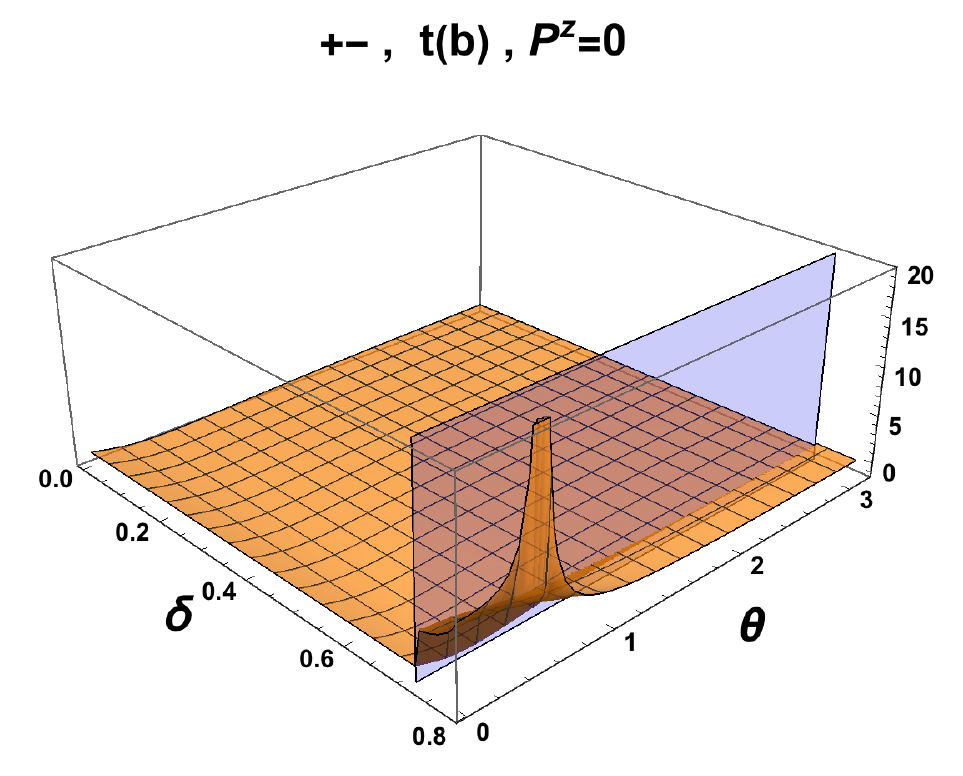}
			\label{fig:eessAngDistRLtb}}
		\\
		\centering
		\subfloat[]{
			\includegraphics[width=0.48\columnwidth]{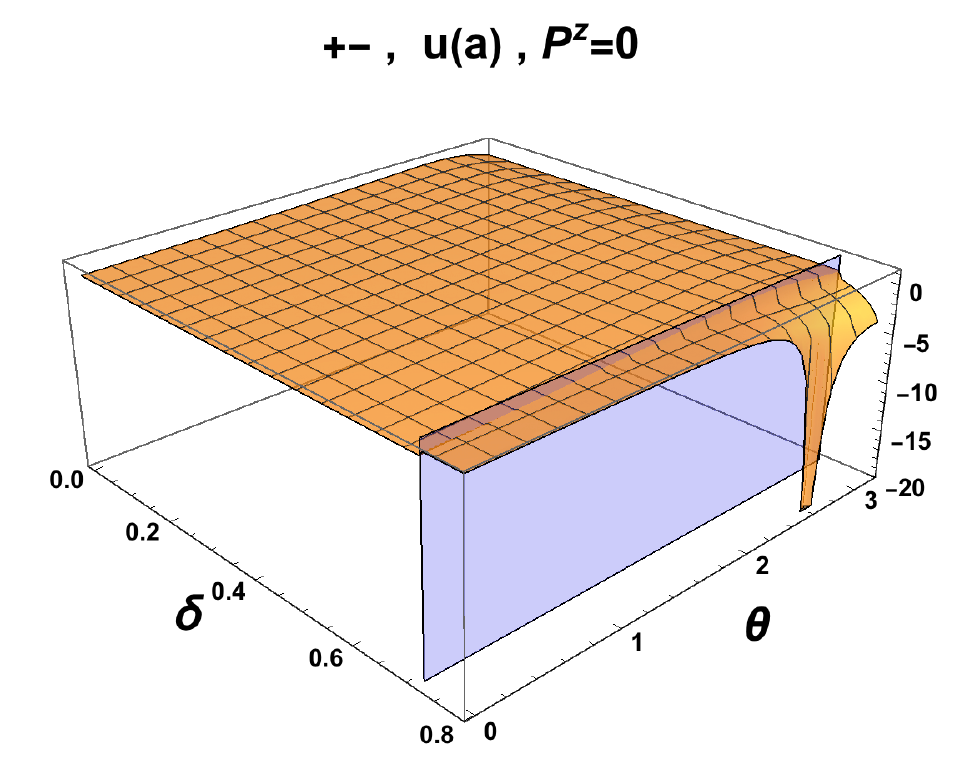}
			\label{fig:eessAngDistRLua}}
		\centering
		\subfloat[]{
			\includegraphics[width=0.48\columnwidth]{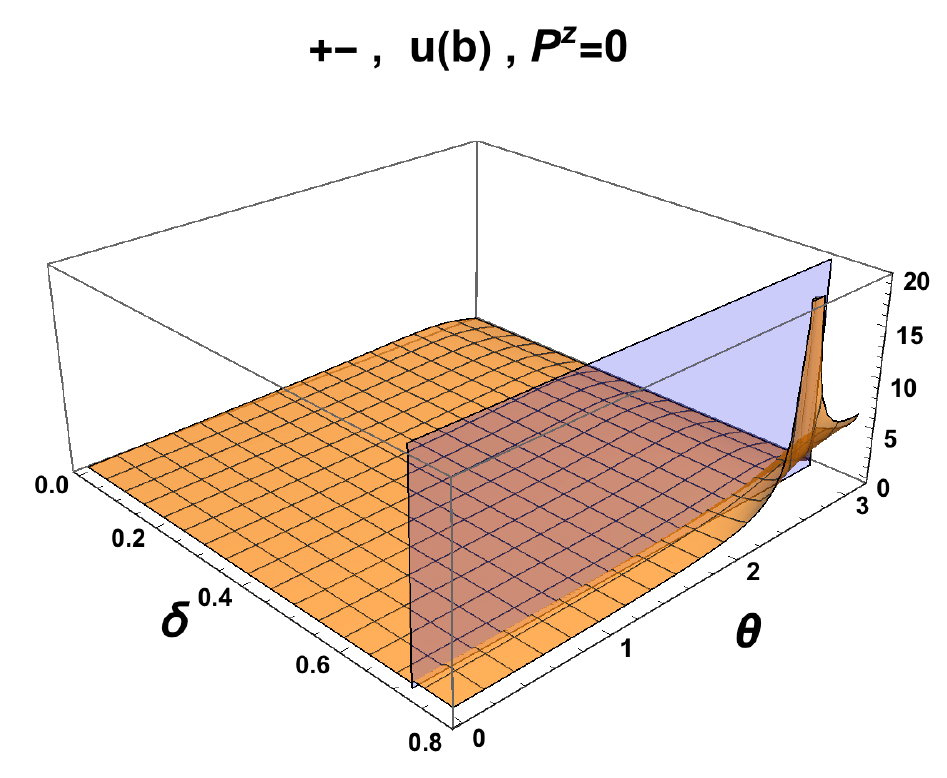}
			\label{fig:eessAngDistRLub}}
		\caption{\label{fig:eessAngDistRL}Angular distribution of the helicity amplitude $ +- $ 
		for (a) t-channel time-ordering process-a, $\mathcal{M}_{a,t}^{+,-}$  (b) t-channel time-ordering process-b, $\mathcal{M}_{b,t}^{+,-}$ 
		(c) u-channel time-ordering process-a, $\mathcal{M}_{a,u}^{+,-}$ (d) u-channel time-ordering process-b, $\mathcal{M}_{b,u}^{+,-}$.}
	\end{figure}
	
				\begin{figure}
		\centering
		\subfloat[]{
			\includegraphics[width=0.48\columnwidth]{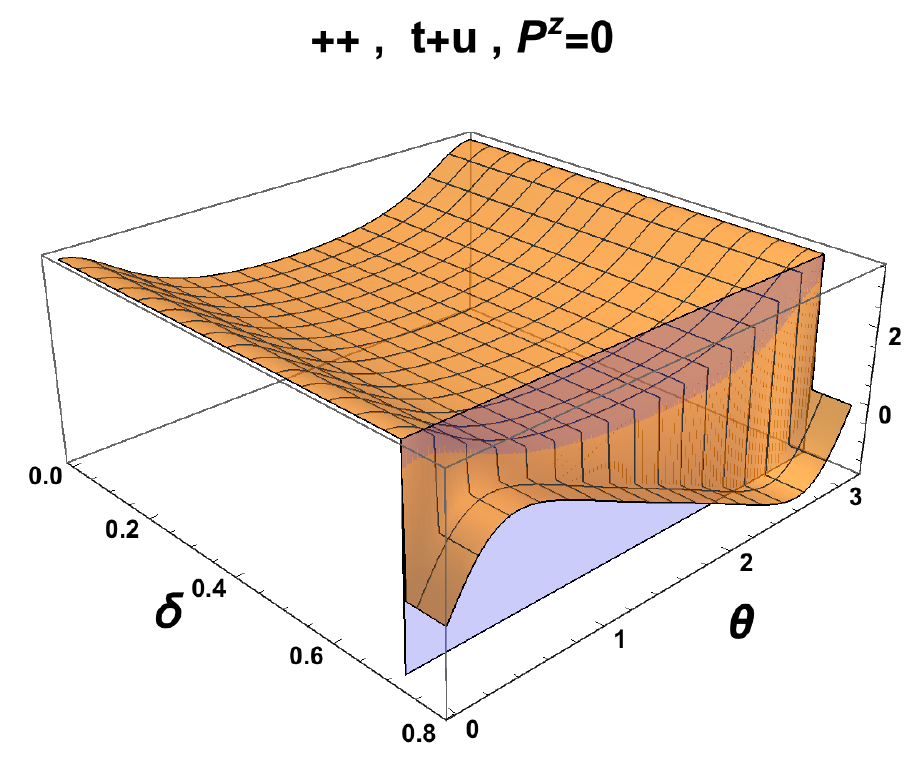}
			\label{fig:eessAngDistRRtu}}
		\centering
		\subfloat[]{
			\includegraphics[width=0.48\columnwidth]{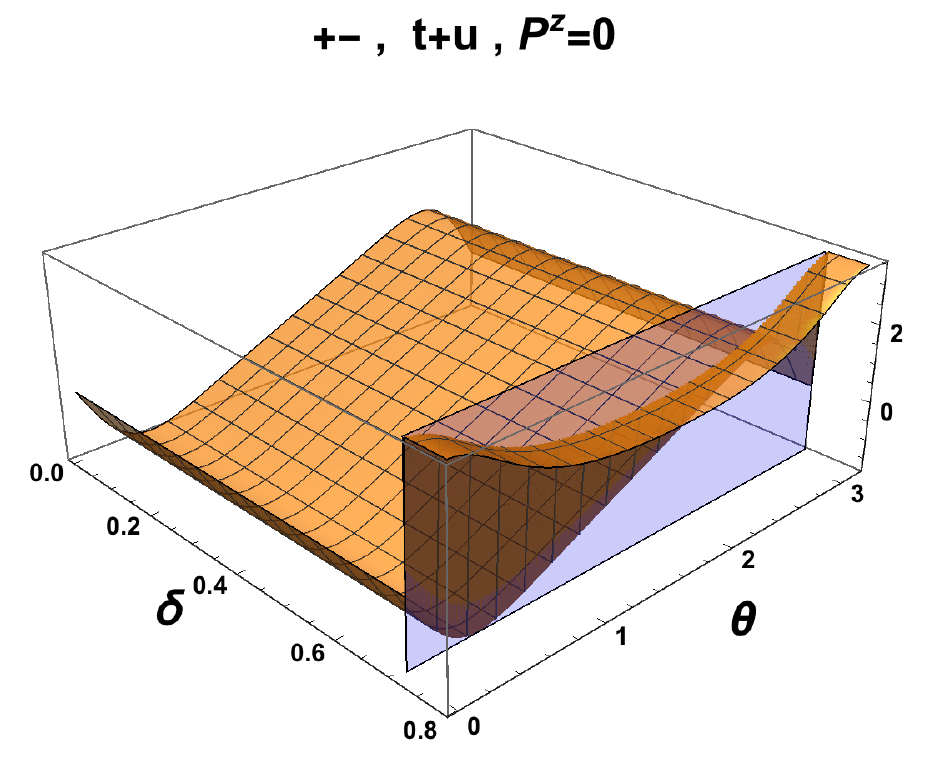}
			\label{fig:eessAngDistRLtu}}
		\\
		\centering
		\subfloat[]{
			\includegraphics[width=0.48\columnwidth]{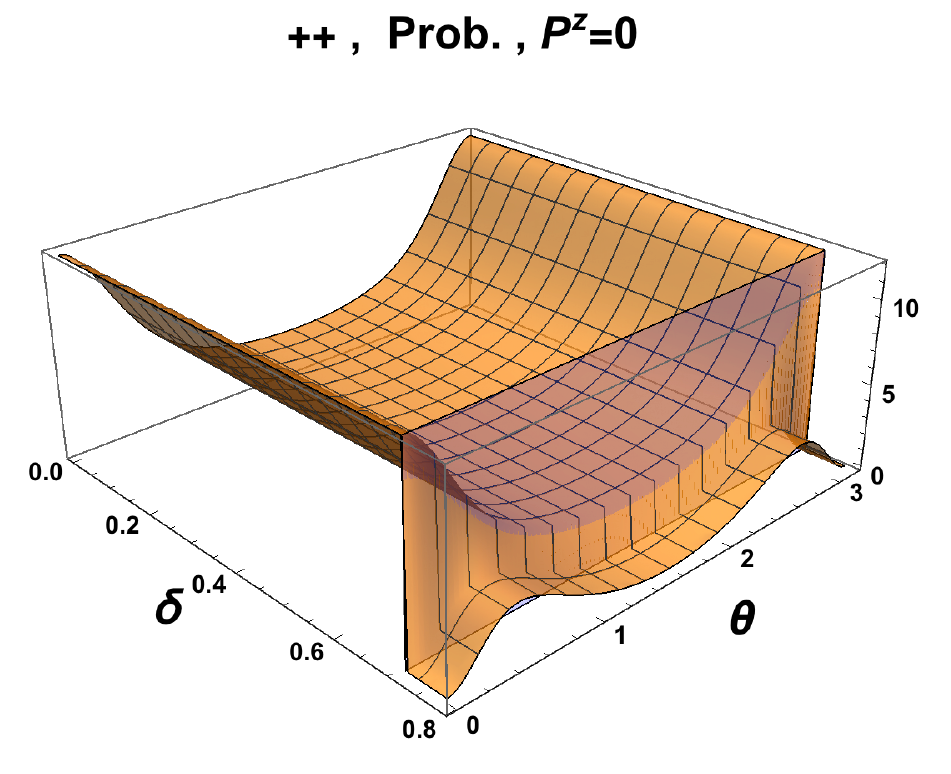}
			\label{fig:eessAngDistRRprob}}
		\centering
		\subfloat[]{
			\includegraphics[width=0.48\columnwidth]{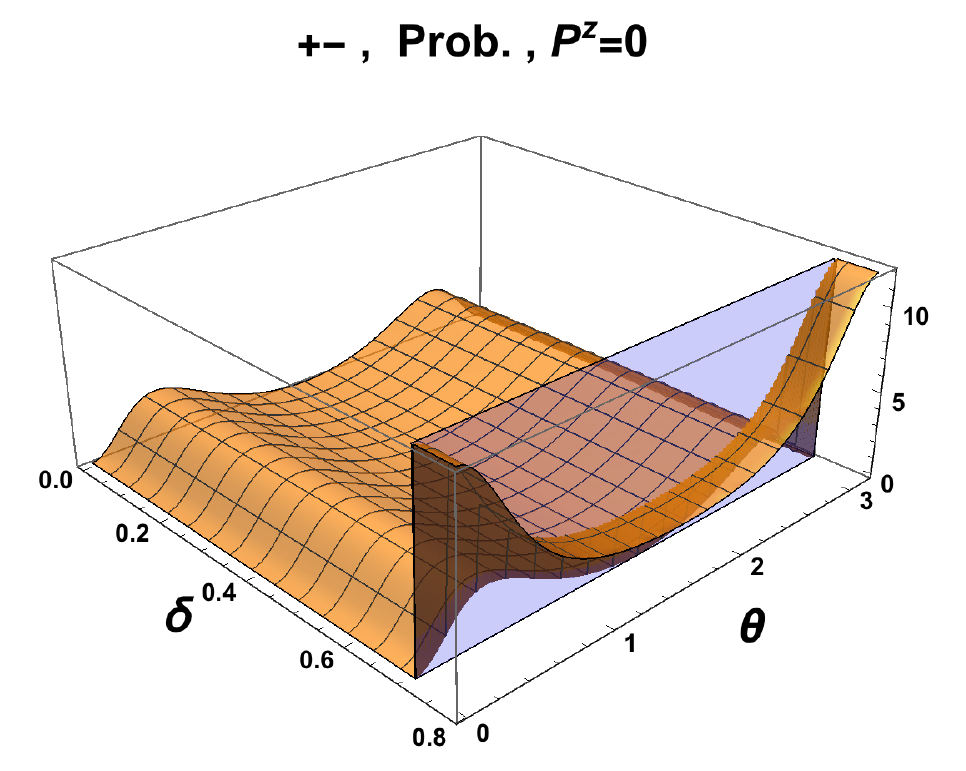}
			\label{fig:eessAngDistRLprob}}
		\caption{\label{fig:eessAngDistAmpAndProb} (a) $\mathcal{M}_{a,t}^{+,+}+\mathcal{M}_{b,t}^{+,+}+\mathcal{M}_{a,u}^{+,+}+\mathcal{M}_{a,u}^{+,+}$ 
		(b) $\mathcal{M}_{a,t}^{+,-}+\mathcal{M}_{b,t}^{+,-}+\mathcal{M}_{a,u}^{+,-}+\mathcal{M}_{a,u}^{+,-}$ 
		(c) $|\mathcal{M}_{a,t}^{+,+}+\mathcal{M}_{b,t}^{+,+}+\mathcal{M}_{a,u}^{+,+}+\mathcal{M}_{a,u}^{+,+}|^2$  
		(d) $|\mathcal{M}_{a,t}^{+,-}+\mathcal{M}_{b,t}^{+,-}+\mathcal{M}_{a,u}^{+,-}+\mathcal{M}_{a,u}^{+,-}|^2$ }
	\end{figure}
	
				\begin{figure}
		\centering
		\subfloat[]{
			\includegraphics[width=0.48\columnwidth]{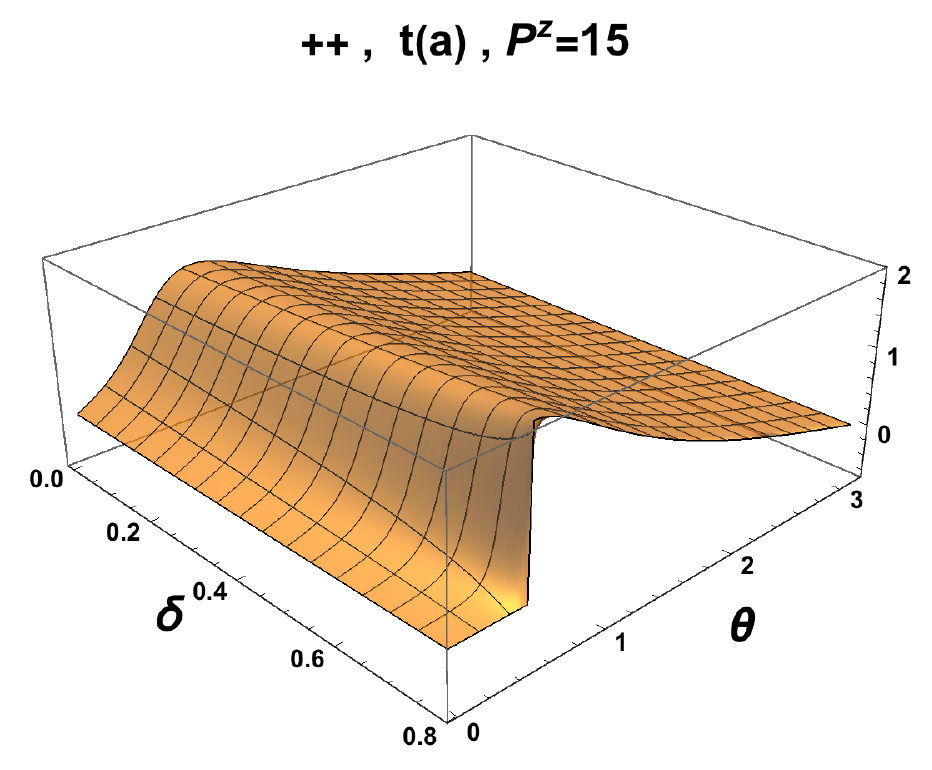}
			\label{fig:eessAngDistRRtaPzp}}
		\centering
		\subfloat[]{
			\includegraphics[width=0.48\columnwidth]{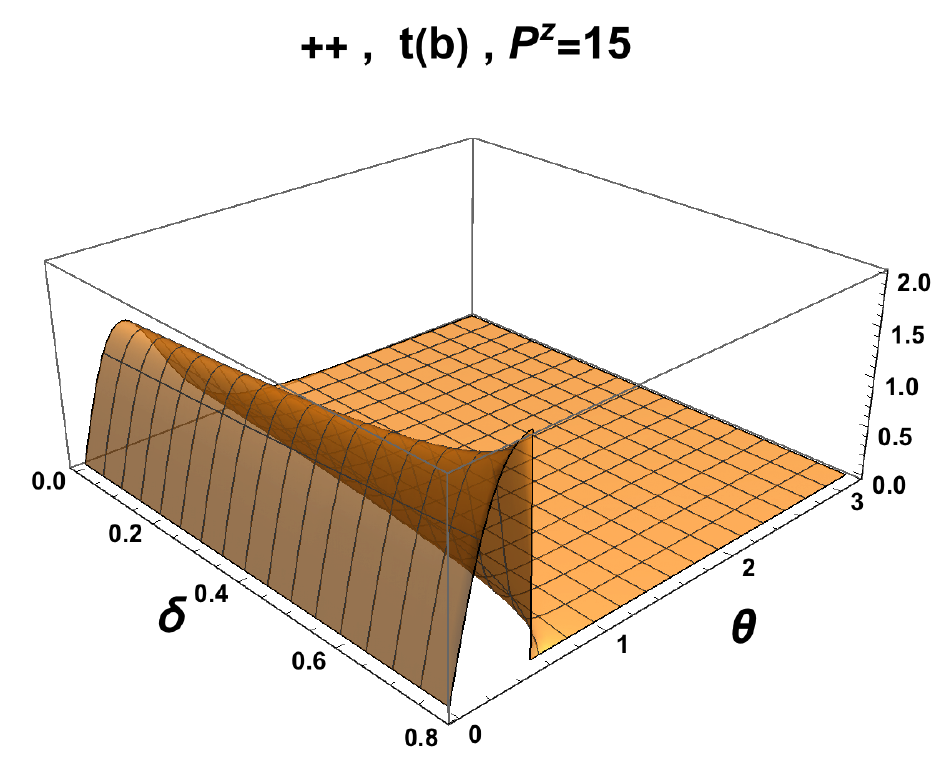}
			\label{fig:eessAngDistRRtbPzp}}
		\\
		\centering
		\subfloat[]{
			\includegraphics[width=0.48\columnwidth]{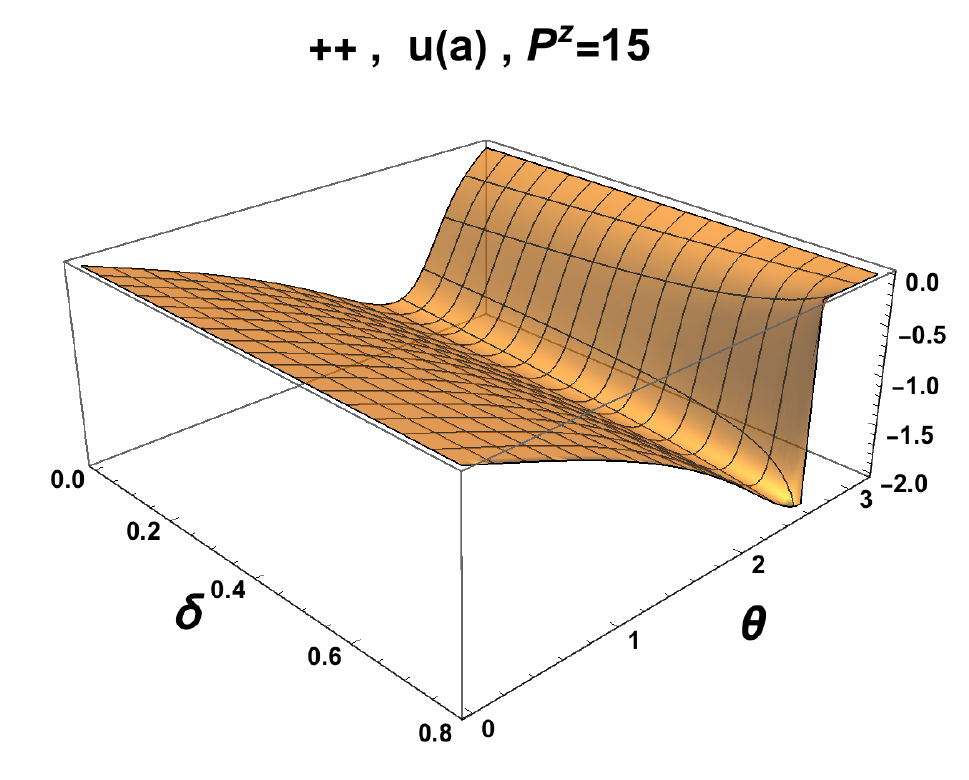}
			\label{fig:eessAngDistRRuaPzp}}
		\centering
		\subfloat[]{
			\includegraphics[width=0.48\columnwidth]{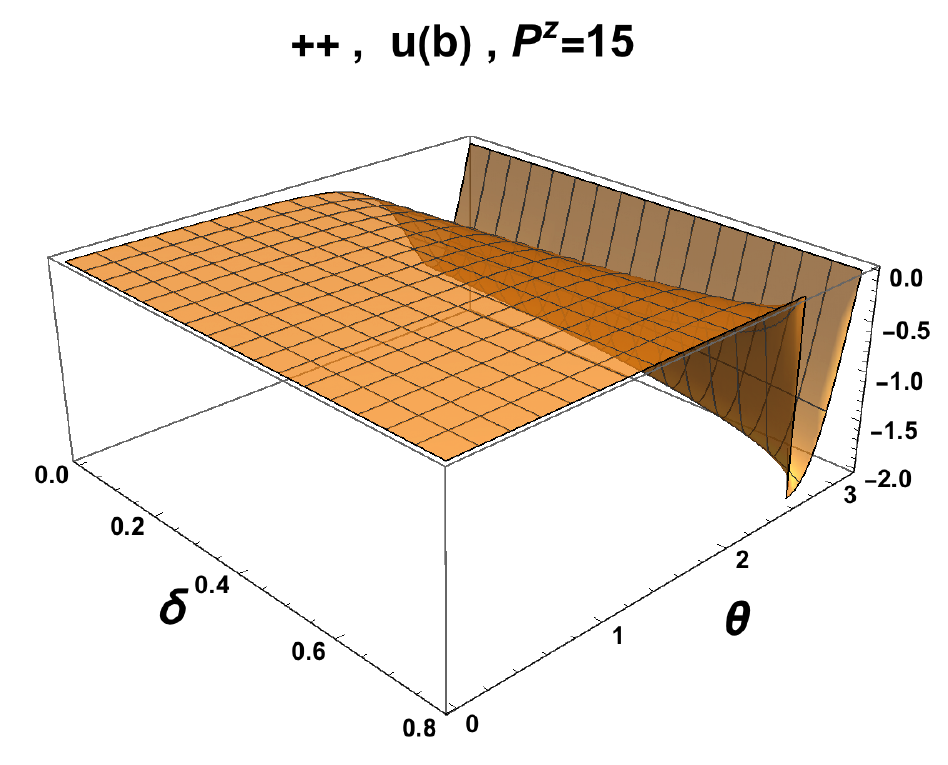}
			\label{fig:eessAngDistRRubPzp}}
		\caption{\label{fig:eessAngDistRRPzp}Angular distribution of the helicity amplitude $ ++ $ 
		for (a) t-channel time-ordering process-a, $\mathcal{M}_{a,t}^{+,+}$  (b) t-channel time-ordering process-b, $\mathcal{M}_{b,t}^{+,+}$ 
		(c) u-channel time-ordering process-a, $\mathcal{M}_{a,u}^{+,+}$ (d) u-channel time-ordering process-b, $\mathcal{M}_{b,u}^{+,+}$.}
	\end{figure}

				\begin{figure}
		\centering
		\subfloat[]{
			\includegraphics[width=0.48\columnwidth]{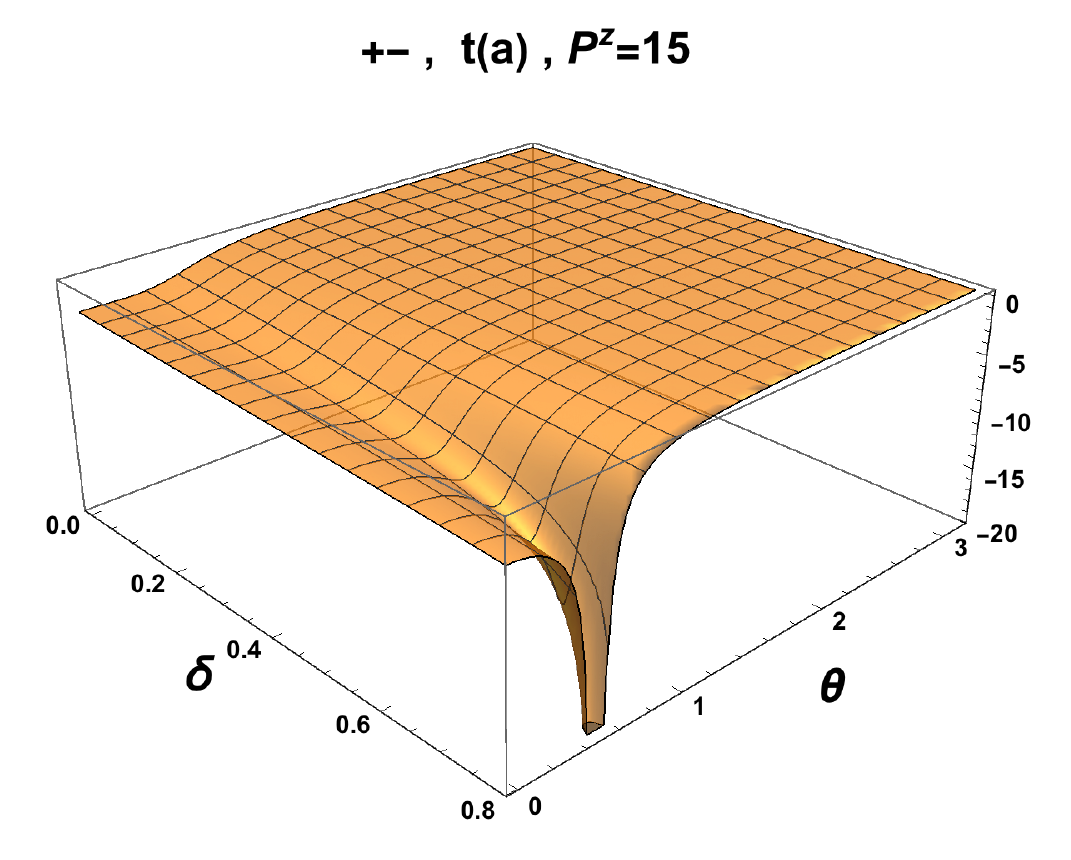}
			\label{fig:eessAngDistRLtaPzp}}
		\centering
		\subfloat[]{
			\includegraphics[width=0.48\columnwidth]{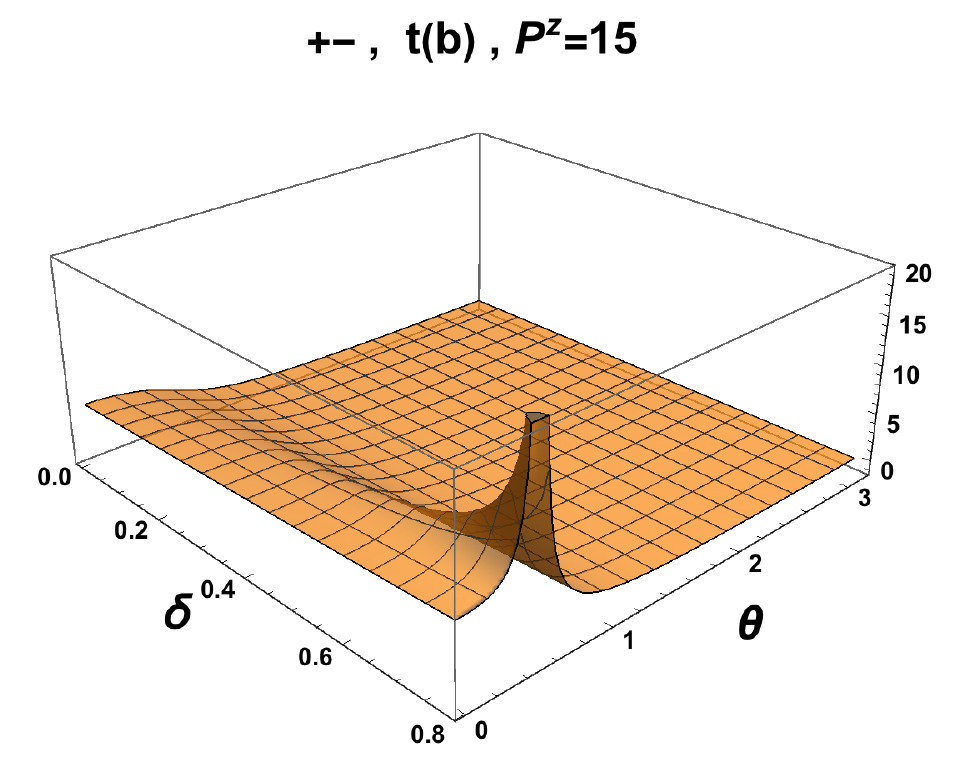}
			\label{fig:eessAngDistRLtbPzp}}
		\\
		\centering
		\subfloat[]{
			\includegraphics[width=0.48\columnwidth]{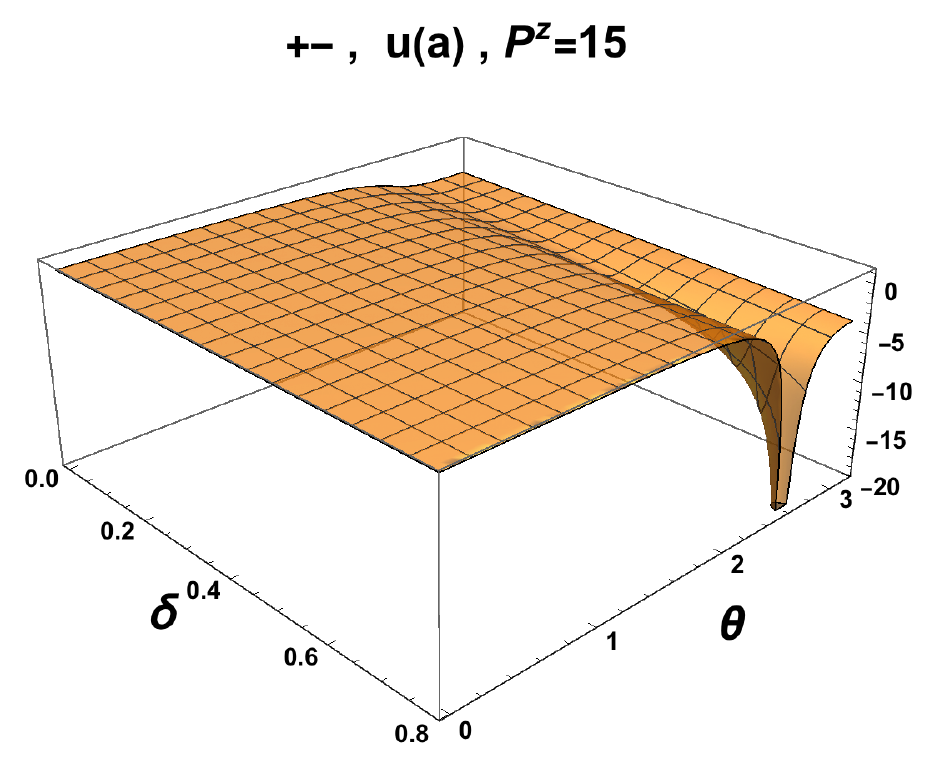}
			\label{fig:eessAngDistRLuaPzp}}
		\centering
		\subfloat[]{
			\includegraphics[width=0.48\columnwidth]{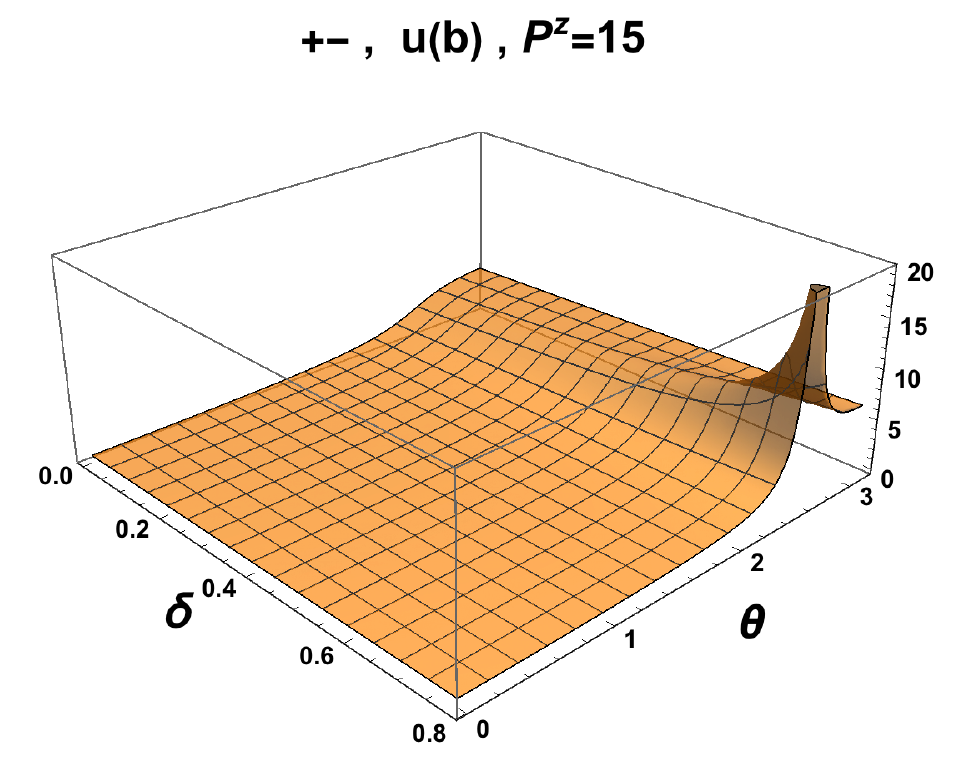}
			\label{fig:eessAngDistRLubPzp}}
		\caption{\label{fig:eessAngDistRLPzp}Angular distribution of the helicity amplitude $ +- $ 
		for (a) t-channel time-ordering process-a, $\mathcal{M}_{a,t}^{+,-}$  (b) t-channel time-ordering process-b, $\mathcal{M}_{b,t}^{+,-}$ 
		(c) u-channel time-ordering process-a, $\mathcal{M}_{a,u}^{+,-}$ (d) u-channel time-ordering process-b, $\mathcal{M}_{b,u}^{+,-}$.}
	\end{figure}
	
				\begin{figure}
		\centering
		\subfloat[]{
			\includegraphics[width=0.48\columnwidth]{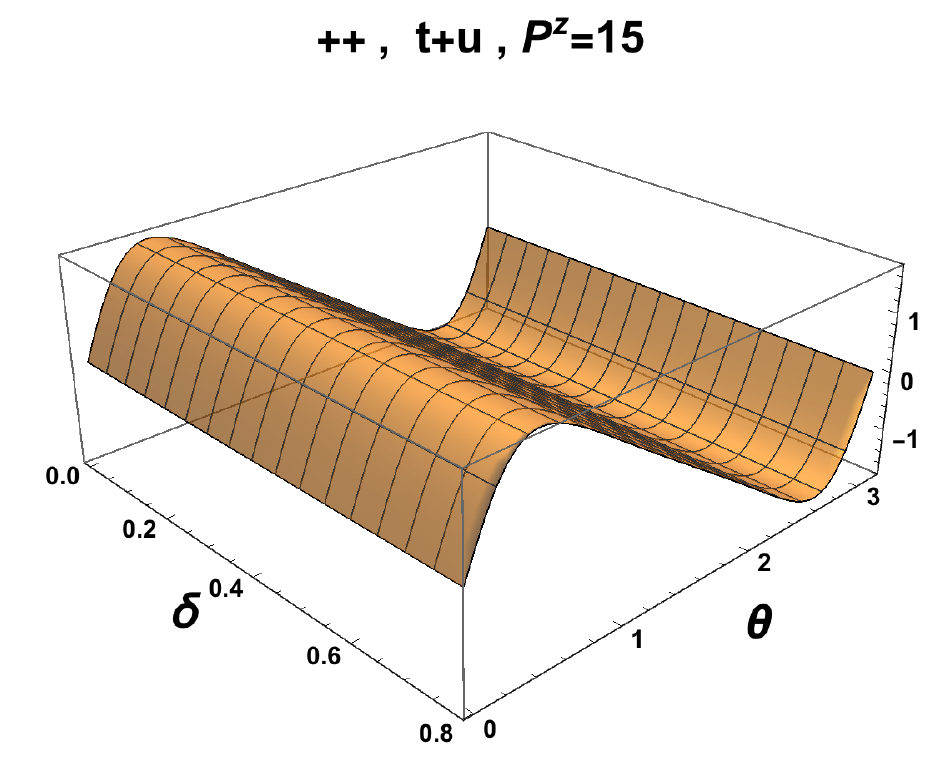}
			\label{fig:eessAngDistRRtuPzp}}
		\centering
		\subfloat[]{
			\includegraphics[width=0.48\columnwidth]{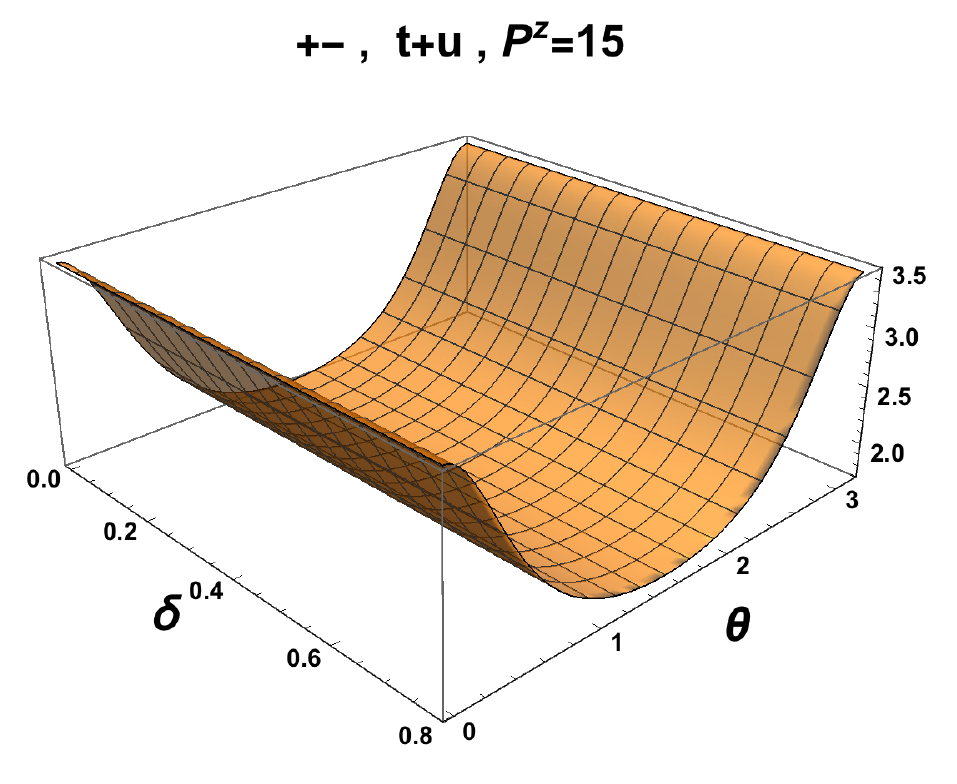}
			\label{fig:eessAngDistRLtuPzp}}
		\\
		\centering
		\subfloat[]{
			\includegraphics[width=0.48\columnwidth]{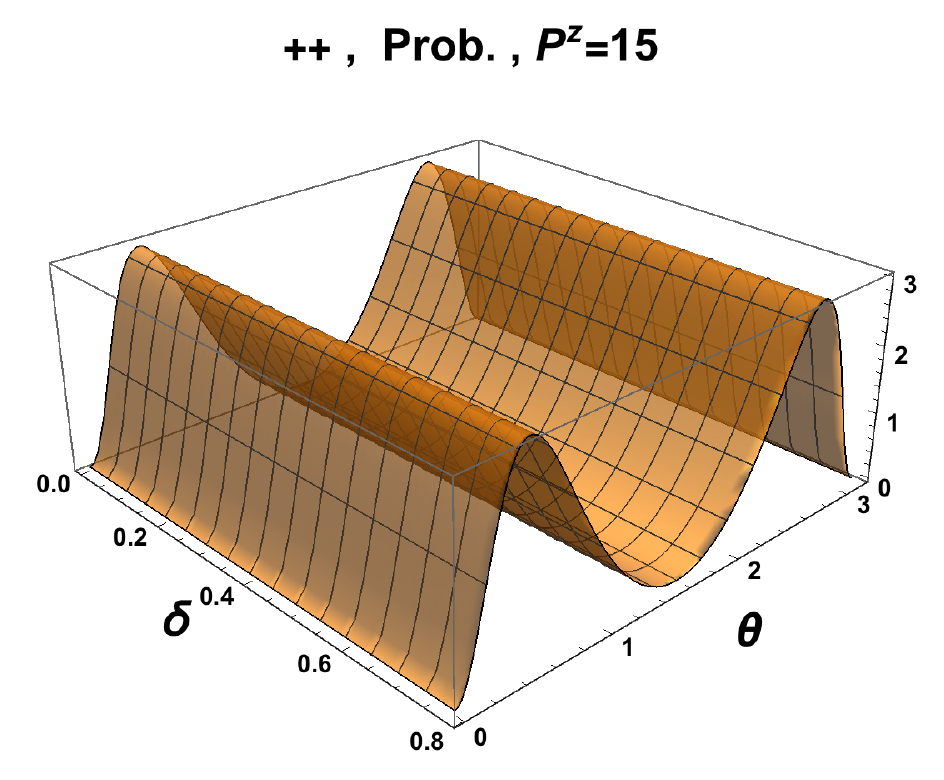}
			\label{fig:eessAngDistRRprobPzp}}
		\centering
		\subfloat[]{
			\includegraphics[width=0.48\columnwidth]{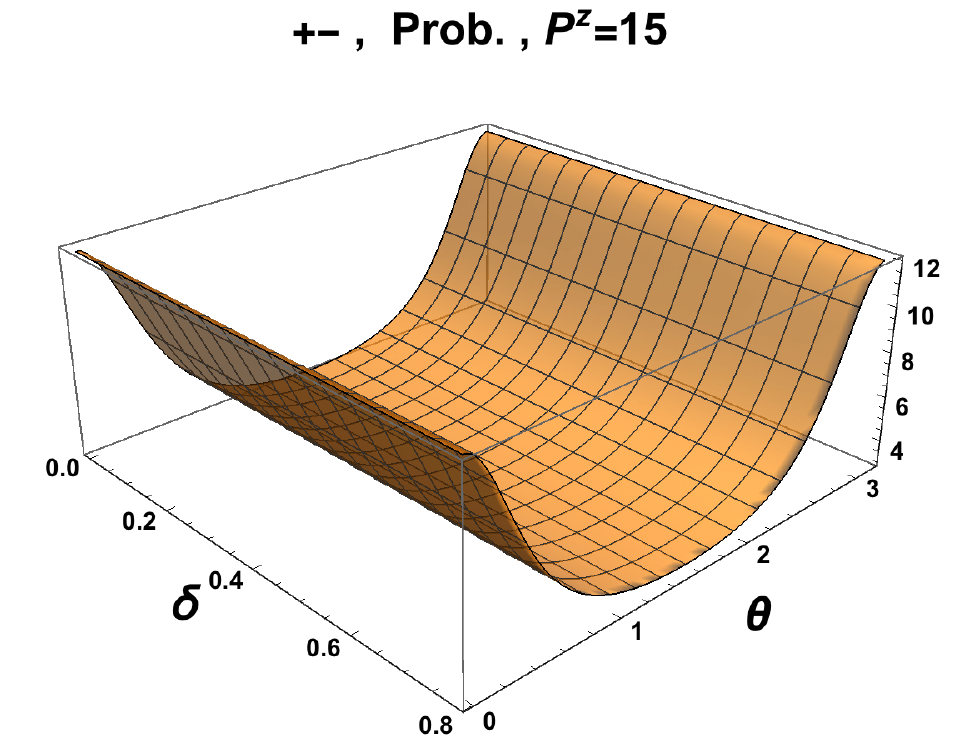}
			\label{fig:eessAngDistRLprobPzp}}
		\caption{\label{fig:eessAngDistAmpAndProbPzp} (a) $\mathcal{M}_{a,t}^{+,+}+\mathcal{M}_{b,t}^{+,+}+\mathcal{M}_{a,u}^{+,+}+\mathcal{M}_{a,u}^{+,+}$ 
		(b) $\mathcal{M}_{a,t}^{+,-}+\mathcal{M}_{b,t}^{+,-}+\mathcal{M}_{a,u}^{+,-}+\mathcal{M}_{a,u}^{+,-}$ 
		(c) $|\mathcal{M}_{a,t}^{+,+}+\mathcal{M}_{b,t}^{+,+}+\mathcal{M}_{a,u}^{+,+}+\mathcal{M}_{a,u}^{+,+}|^2$  
		(d) $|\mathcal{M}_{a,t}^{+,-}+\mathcal{M}_{b,t}^{+,-}+\mathcal{M}_{a,u}^{+,-}+\mathcal{M}_{a,u}^{+,-}|^2$ }
	\end{figure}
				\begin{figure}
		\centering
		\subfloat[]{
			\includegraphics[width=0.48\columnwidth]{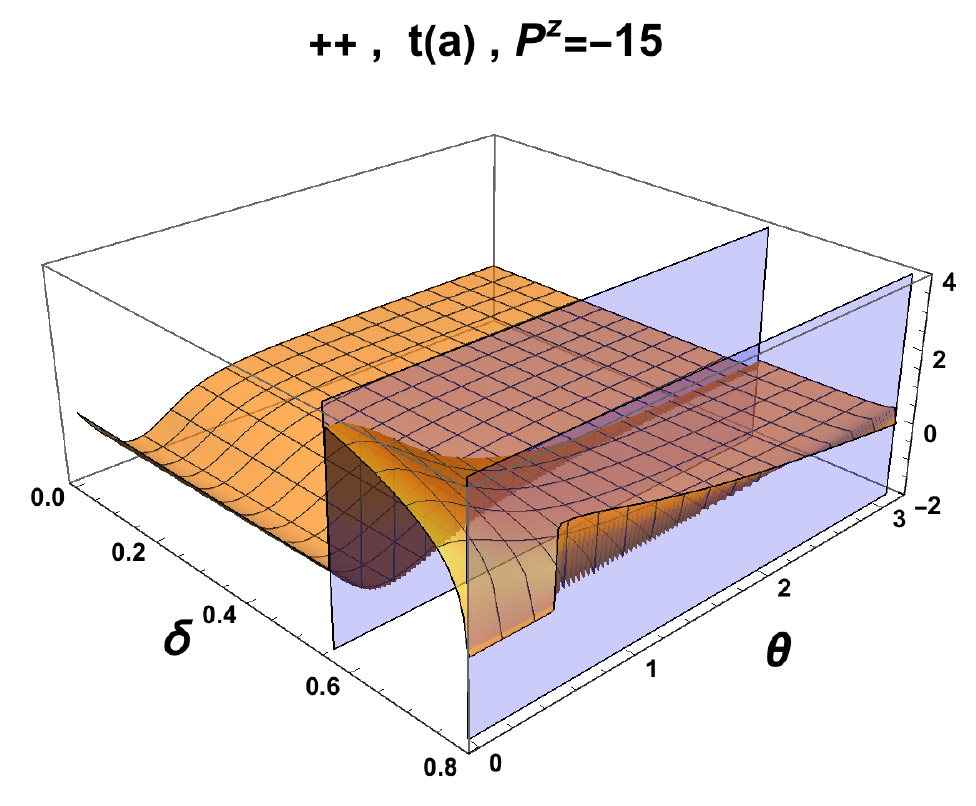}
			\label{fig:eessAngDistRRtaPzm}}
		\centering
		\subfloat[]{
			\includegraphics[width=0.48\columnwidth]{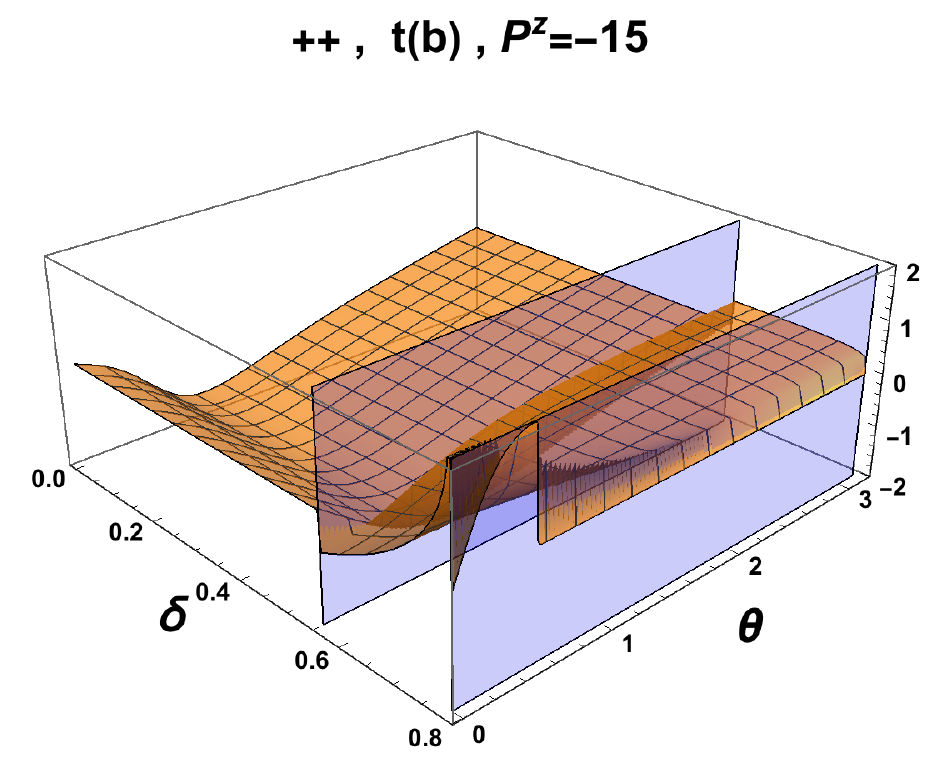}
			\label{fig:eessAngDistRRtbPzm}}
		\\
		\centering
		\subfloat[]{
			\includegraphics[width=0.48\columnwidth]{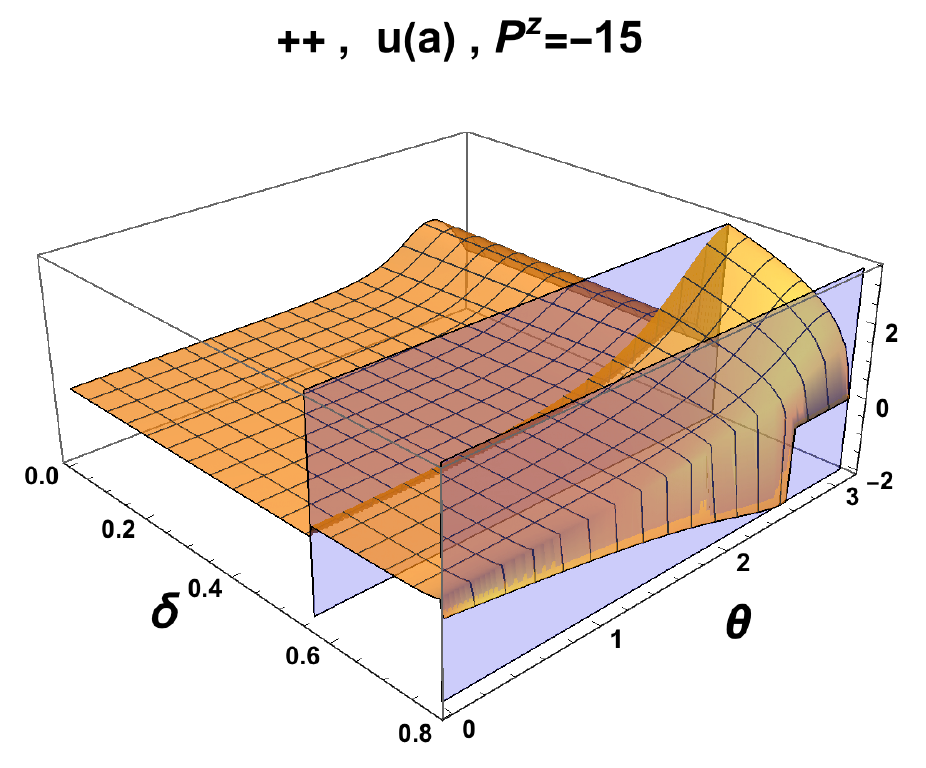}
			\label{fig:eessAngDistRRuaPzm}}
		\centering
		\subfloat[]{
			\includegraphics[width=0.48\columnwidth]{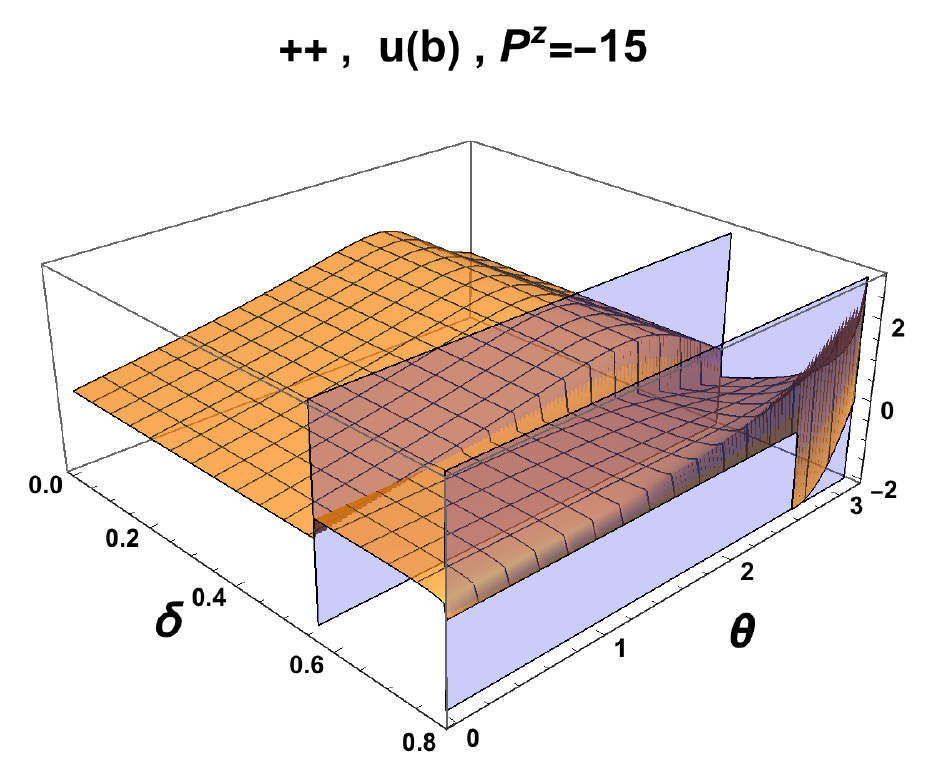}
			\label{fig:eessAngDistRRubPzm}}
		\caption{\label{fig:eessAngDistRRPzm}Angular distribution of the helicity amplitude $ ++ $ 
		for (a) t-channel time-ordering process-a, $\mathcal{M}_{a,t}^{+,+}$  (b) t-channel time-ordering process-b, $\mathcal{M}_{b,t}^{+,+}$ 
		(c) u-channel time-ordering process-a, $\mathcal{M}_{a,u}^{+,+}$ (d) u-channel time-ordering process-b, $\mathcal{M}_{b,u}^{+,+}$.}
	\end{figure}

				\begin{figure}
		\centering
		\subfloat[]{
			\includegraphics[width=0.48\columnwidth]{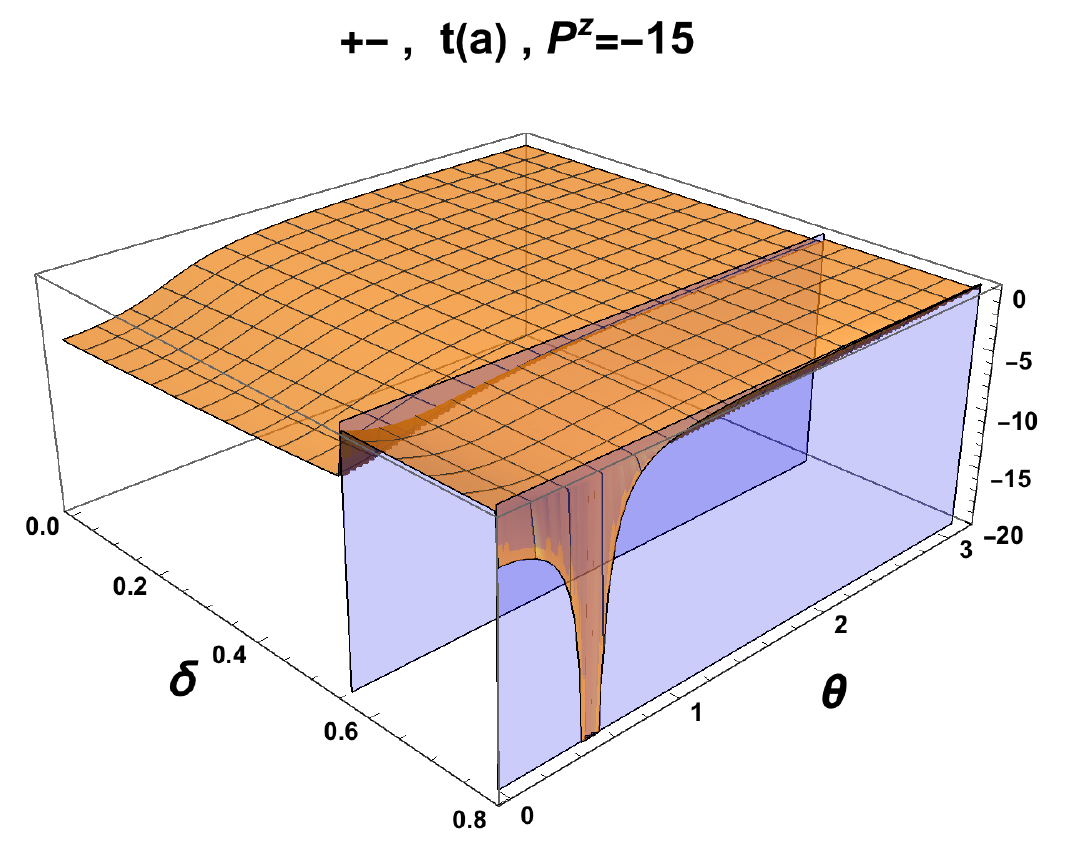}
			\label{fig:eessAngDistRLtaPzm}}
		\centering
		\subfloat[]{
			\includegraphics[width=0.48\columnwidth]{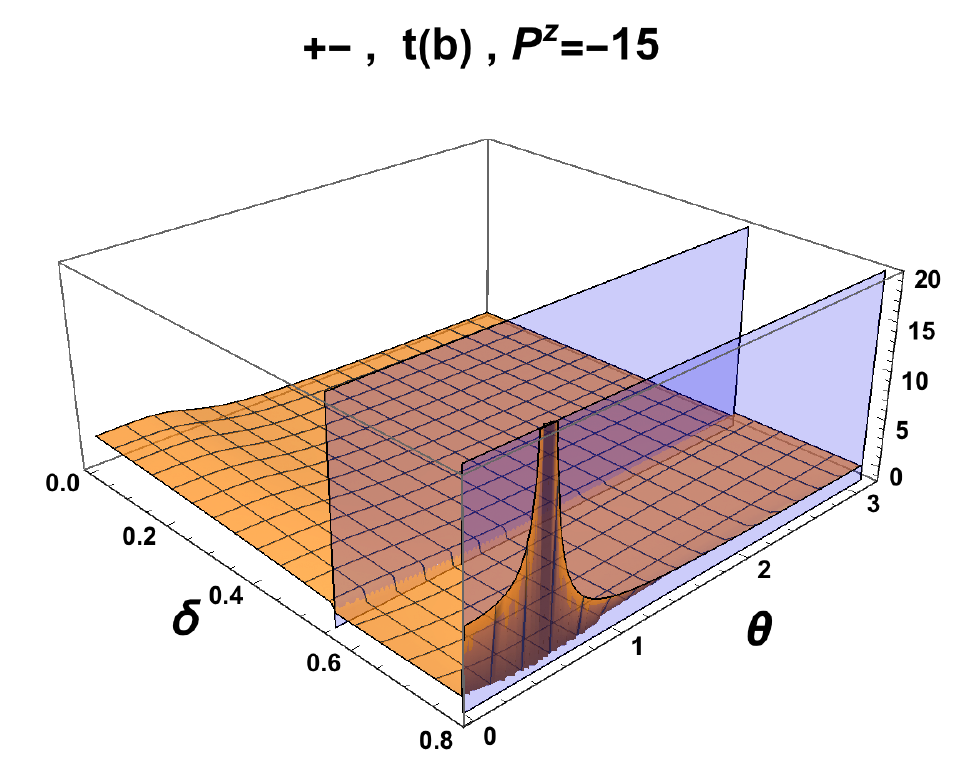}
			\label{fig:eessAngDistRLtbPzm}}
		\\
		\centering
		\subfloat[]{
			\includegraphics[width=0.48\columnwidth]{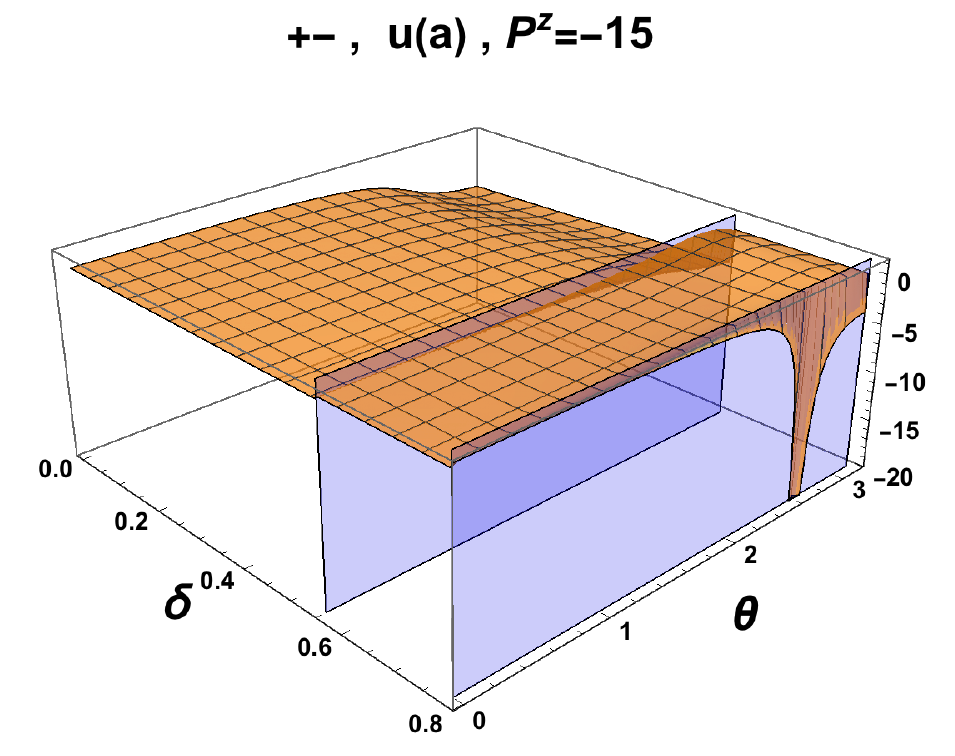}
			\label{fig:eessAngDistRLuaPzm}}
		\centering
		\subfloat[]{
			\includegraphics[width=0.48\columnwidth]{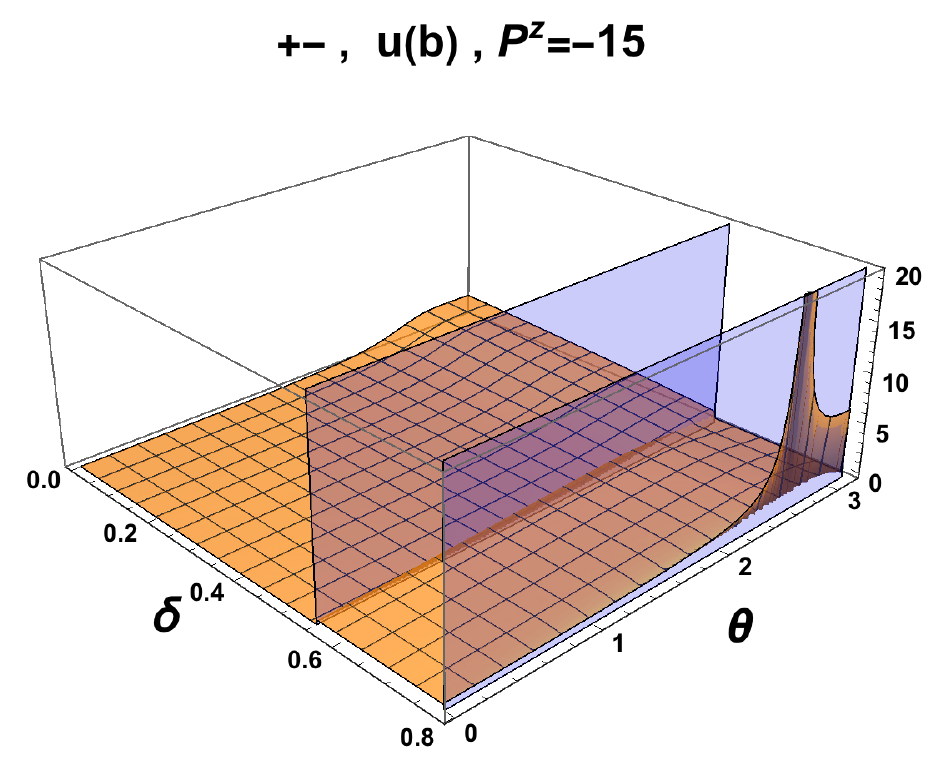}
			\label{fig:eessAngDistRLubPzm}}
		\caption{\label{fig:eessAngDistRLPzm}Angular distribution of the helicity amplitude $ +- $ 
		for (a) t-channel time-ordering process-a, $\mathcal{M}_{a,t}^{+,-}$  (b) t-channel time-ordering process-b, $\mathcal{M}_{b,t}^{+,-}$ 
		(c) u-channel time-ordering process-a, $\mathcal{M}_{a,u}^{+,-}$ (d) u-channel time-ordering process-b, $\mathcal{M}_{b,u}^{+,-}$.}
	\end{figure}
	
				\begin{figure}
		\centering
		\subfloat[]{
			\includegraphics[width=0.48\columnwidth]{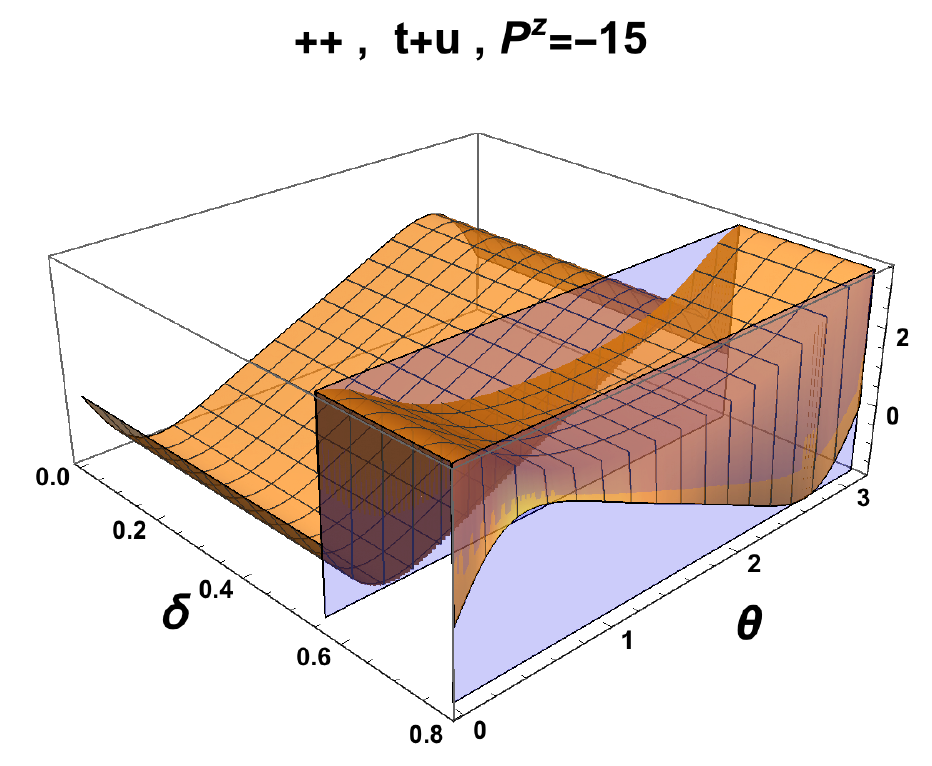}
			\label{fig:eessAngDistRRtuPzm}}
		\centering
		\subfloat[]{
			\includegraphics[width=0.48\columnwidth]{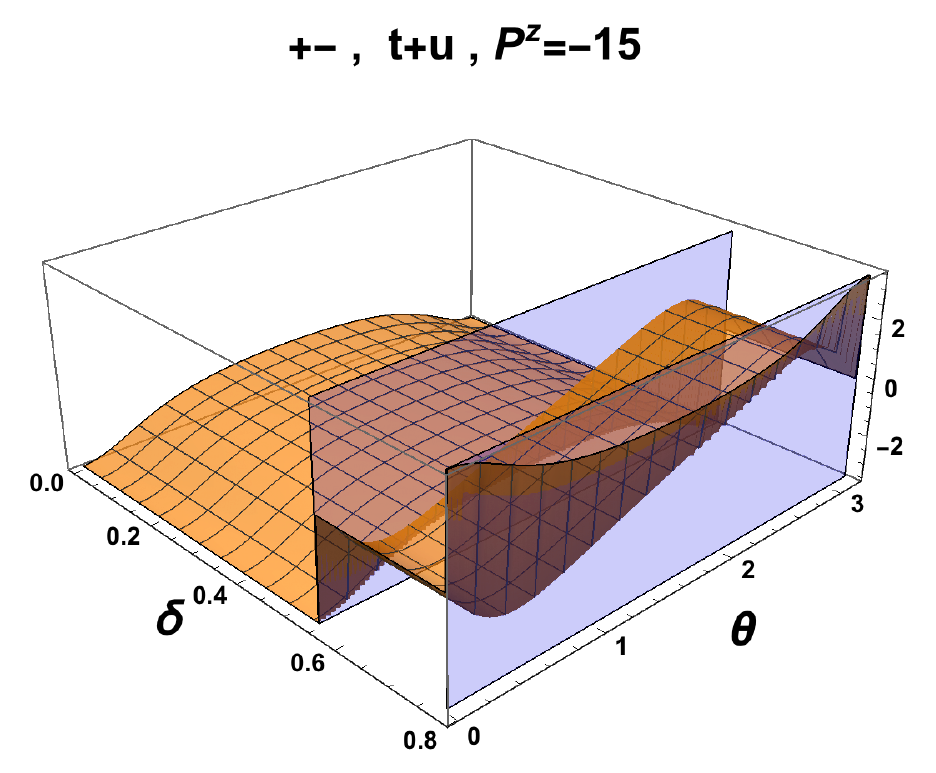}
			\label{fig:eessAngDistRLtuPzm}}
		\\
		\centering
		\subfloat[]{
			\includegraphics[width=0.48\columnwidth]{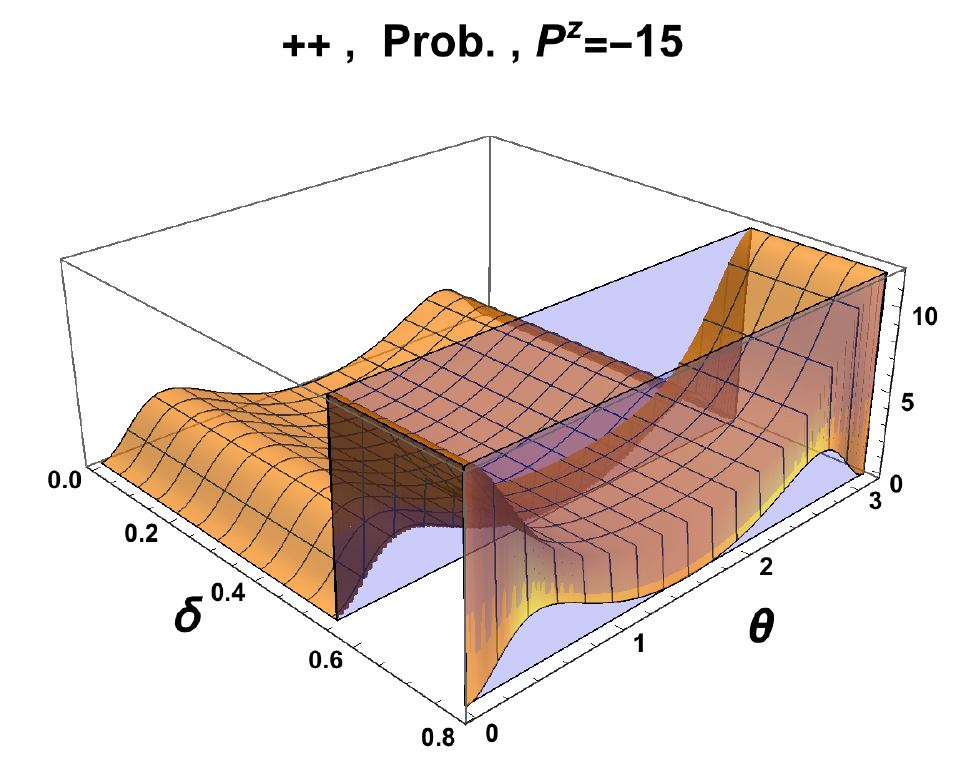}
			\label{fig:eessAngDistRRprobPzm}}
		\centering
		\subfloat[]{
			\includegraphics[width=0.48\columnwidth]{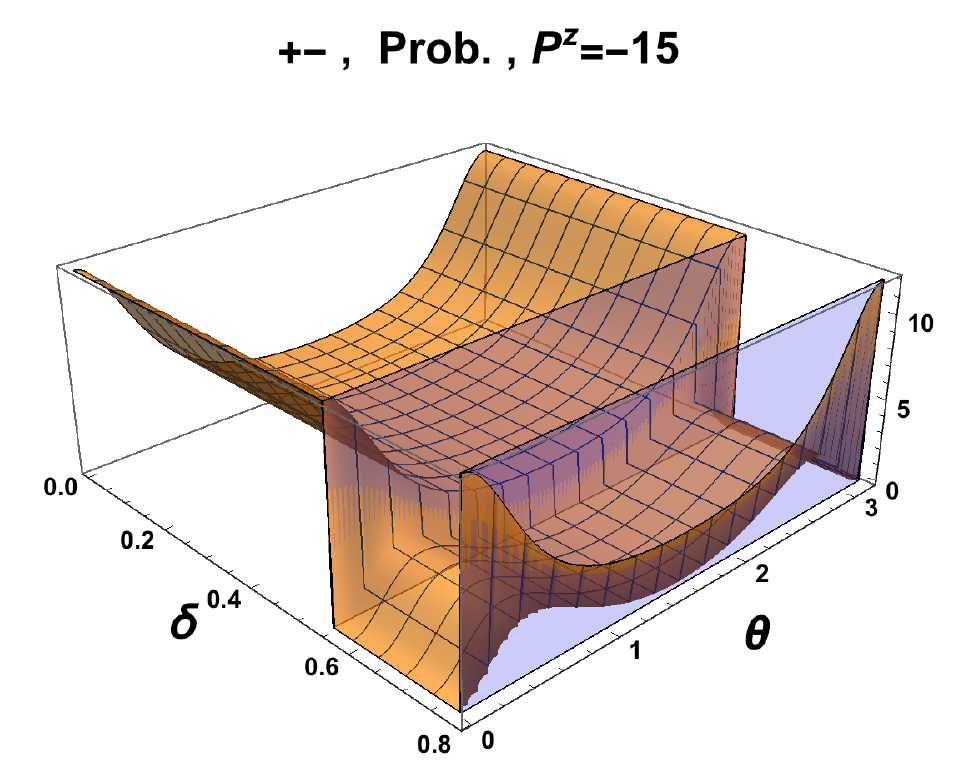}
			\label{fig:eessAngDistRLprobPzm}}
		\caption{\label{fig:eessAngDistAmpAndProbPzm} (a) $\mathcal{M}_{a,t}^{+,+}+\mathcal{M}_{b,t}^{+,+}+\mathcal{M}_{a,u}^{+,+}+\mathcal{M}_{a,u}^{+,+}$ 
		(b) $\mathcal{M}_{a,t}^{+,-}+\mathcal{M}_{b,t}^{+,-}+\mathcal{M}_{a,u}^{+,-}+\mathcal{M}_{a,u}^{+,-}$ 
		(c) $|\mathcal{M}_{a,t}^{+,+}+\mathcal{M}_{b,t}^{+,+}+\mathcal{M}_{a,u}^{+,+}+\mathcal{M}_{a,u}^{+,+}|^2$  
		(d) $|\mathcal{M}_{a,t}^{+,-}+\mathcal{M}_{b,t}^{+,-}+\mathcal{M}_{a,u}^{+,-}+\mathcal{M}_{a,u}^{+,-}|^2$ }
	\end{figure}

\begin{figure}
		\centering
		\subfloat[]{
			\includegraphics[width=0.48\columnwidth]{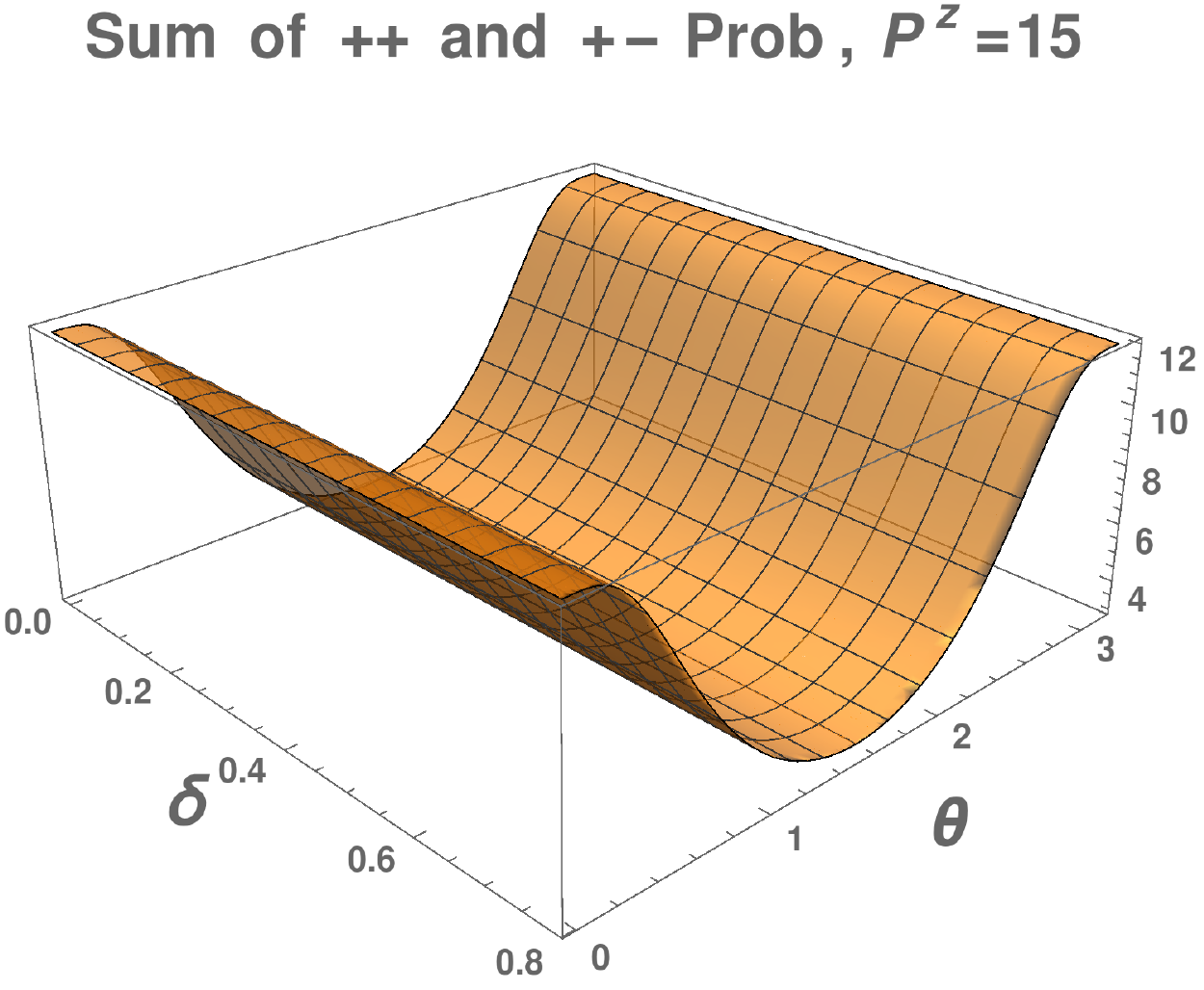}
			\label{fig:HelicityProbabilitySumPzp}}
			\centering
		\subfloat[]{
			\includegraphics[width=0.48\columnwidth]{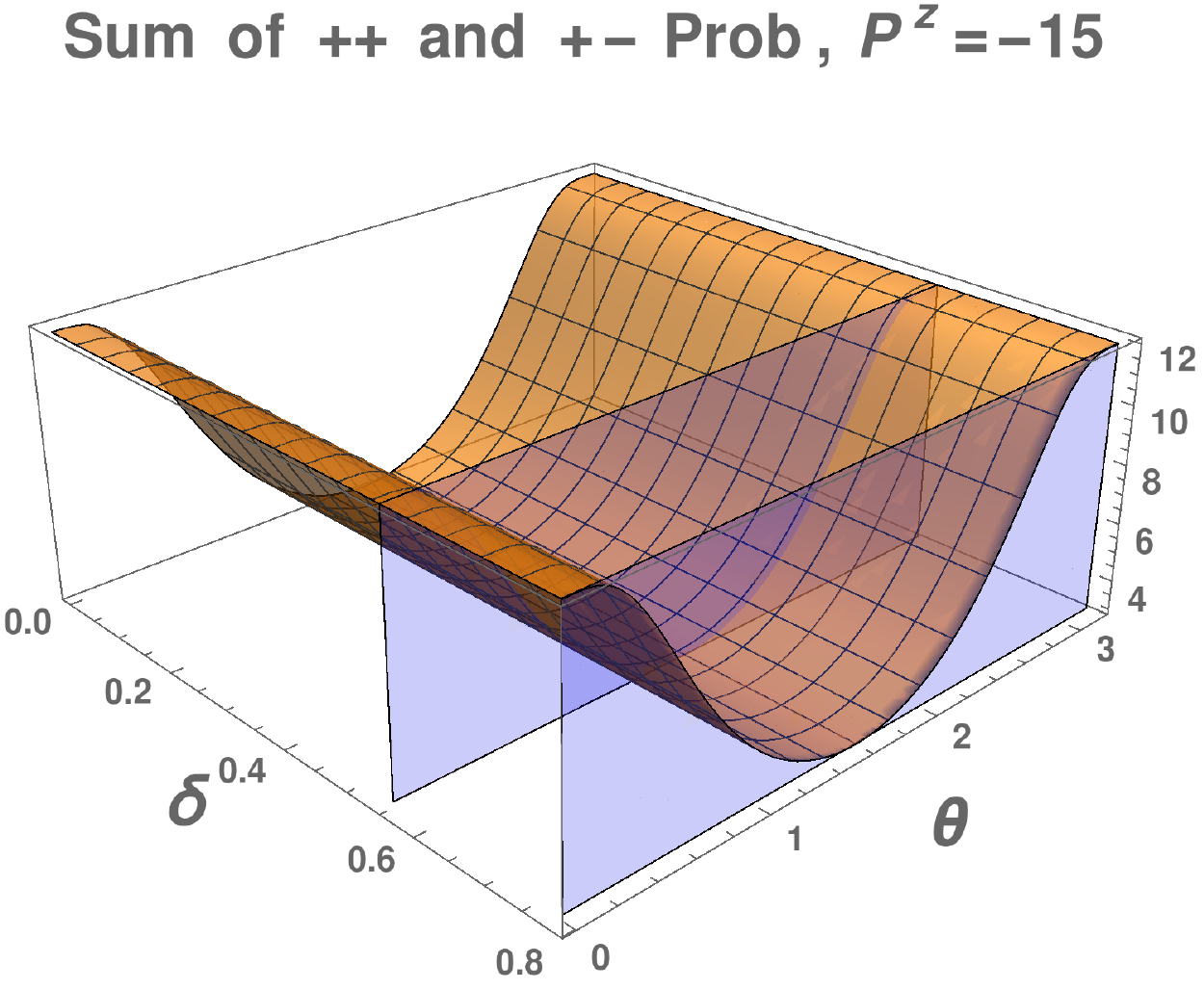}
			\label{fig:HelicityProbabilitySumPzm}}
			\\
	\centering
		\subfloat[]{
			\includegraphics[width=0.48\columnwidth]{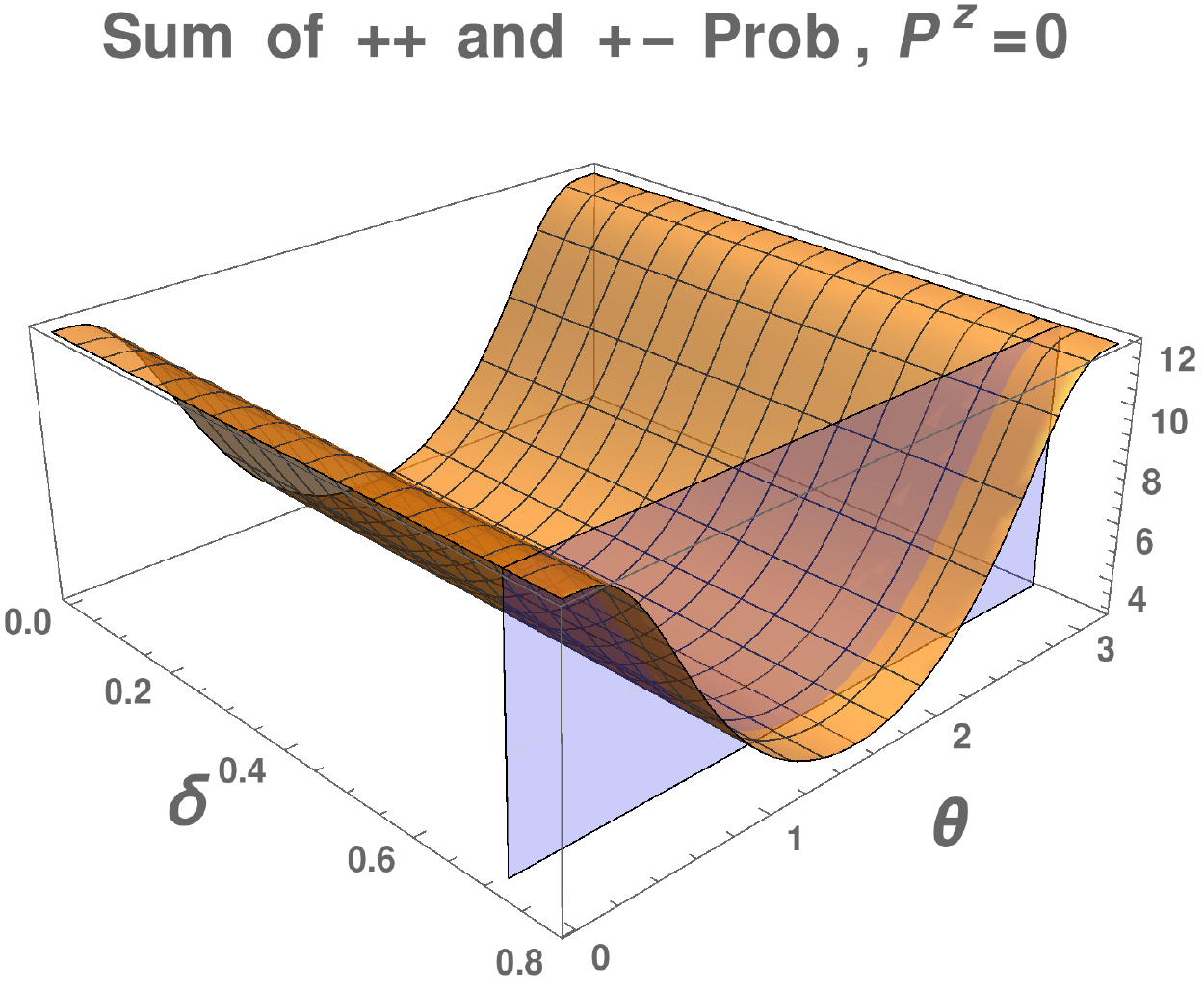}
			\label{fig:HelicityProbSumPz0}}
	\caption{\label{fig:HelicityProbabilitySum}Sum of $++$ and $+-$ Helicity Probabilities for: (a) $P^z = +15 m_e$ ,
		(b) $P^z = -15 m_e$, and
		(c) $P^z = 0$ (CMF).}
	\end{figure}

Even more distinct feature of LFD can be noticed in Fig.~\ref{fig:RLthcInterpol} where we present the $+-$ helicity amplitudes in LFD ($\delta = \pi/4$)
(a) $\mathcal{M}_{a,t}^{+,-}$, (b) $\mathcal{M}_{b,t}^{+,-}$, (c) $\mathcal{M}_{a,t}^{+,-} + \mathcal{M}_{b,t}^{+,-}$
and (d) $\mathcal{M}_{a,u}^{+,-} + \mathcal{M}_{b,u}^{+,-}$. In contrast to $\mathcal{M}_{a,t}^{+,+}$ discussed above,  the ``instantaneous fermion contribution" 
to $\mathcal{M}_{a,t}^{+,-}$ doesn't vanish due to $\bar{v}^{\downarrow}\gamma^+ u^{\uparrow} \neq 0 $\cite{BL}
and thus the amplitude shown in Fig.~\ref{fig:RLtaLFDangdist} gets the singularity from the light-front instantaneous propagator, $ \frac{\gamma^+}{2q^+} $, as $q^+ = 0$ occurs at $\theta = \theta_{c,t}$. 
The singularity from the same origin but with the opposite sign due to $q_b=-q_a=-q$ for $\mathcal{M}_{b,t}^{+,-}$ shown in Fig.~\ref{fig:RLtbLFDangdist}
cancels the singularity shown in Fig.~\ref{fig:RLtaLFDangdist} and the net result of $\mathcal{M}_{a,t}^{+,-} + \mathcal{M}_{b,t}^{+,-}$ is finite and well behaved
as shown in Fig.~\ref{fig:RLtabLFDangdist}. It is interesting to note that the singularities in different light-front time-ordered processes corroborate each other to cancel
themselves and make the Lorentz invariant amplitude finite and well behaved. The crossed channel total amplitude $\mathcal{M}_{a,u}^{+,-} + \mathcal{M}_{b,u}^{+,-}$
is of course also finite and well behaved with the apparent symmetry $\theta \rightarrow \pi-\theta$ between the t-channel and the u-channel as shown in 
Figs.~\ref{fig:RLtabLFDangdist} and \ref{fig:RLuabLFDangdist}.

The helicity $-+$ and $--$ amplitudes, $\mathcal{M}_{a,t}^{-,+}$, $\mathcal{M}_{b,t}^{-,+}$, $\mathcal{M}_{a,u}^{-,+}$, $\mathcal{M}_{b,u}^{-,+}$, $
\mathcal{M}_{a,t}^{-,-}$, $\mathcal{M}_{b,t}^{-,-}$, $\mathcal{M}_{a,u}^{-,-}$, $\mathcal{M}_{b,u}^{-,-}$, have all been computed as well and 
Eq.~(\ref{eqn:amplitude_relation_scalar_case}) based on the parity conservation has been verified explicitly among all the helicity amplitudes for the present 
$e^+ e^-$ scattering/annihilation process. Thus, the helicity $-+$ and $--$ amplitudes can be rather easily figured out once the helicity $++$ and $+-$ amplitudes 
are given. In Figs.~\ref{fig:eessAngDistRR}, \ref{fig:eessAngDistRL} and \ref{fig:eessAngDistAmpAndProb},
we provide the whole landscape of the interpolation angle ($\delta$) dependence for the angular distributions of the helicity $++$ and $+-$ amplitudes at CMF (i.e. $P^z=0$). 
In each and every figure, the critical interpolation angle $\delta_c$ which separates the IFD side and the LFD side of helicity branches is denoted by
a thin boundary sheet at $\delta = \delta_c \approx 0.713724$ in CMF ($P^z=0$).
In Fig.~\ref{fig:eessAngDistRR}, we show the angular distribution of the helicity $ ++ $ amplitudes (a) $\mathcal{M}_{a,t}^{+,+}$  (b) $\mathcal{M}_{b,t}^{+,+}$ 
(c) $\mathcal{M}_{a,u}^{+,+}$ (d) $\mathcal{M}_{b,u}^{+,+}$.
Similarly, in Fig.~\ref{fig:eessAngDistRL}, we show the angular distribution of the helicity $ +- $ amplitudes 
(a) $\mathcal{M}_{a,t}^{+,-}$  (b) $\mathcal{M}_{b,t}^{+,-}$ 
(c) $\mathcal{M}_{a,u}^{+,-}$ (d) $\mathcal{M}_{b,u}^{+,-}$.
At $\delta = \pi/4$ (LFD), the profiles of the ``instantaneous fermion contribution" and the ``on-mass-shell propagating contribution" depicted 
in Figs.~\ref{fig:RRtaLFDangdist} and \ref{fig:RLtaLFDangdist} are visible in Figs.~\ref{fig:eessAngDistRRta} and \ref{fig:eessAngDistRLta}, respectively.
Adding both t-channel and u-channel time-ordered amplitudes all together, we get the results shown in Fig.~\ref{fig:eessAngDistAmpAndProb}.
In Figs.~\ref{fig:eessAngDistRRtu} and \ref{fig:eessAngDistRLtu}, the sum of $++$ helicity amplitude 
$\mathcal{M}_{a,t}^{+,+}+\mathcal{M}_{b,t}^{+,+}+\mathcal{M}_{a,u}^{+,+}+\mathcal{M}_{a,u}^{+,+}$ and 
the sum of $+-$ helicity amplitude 
$\mathcal{M}_{a,t}^{+,-}+\mathcal{M}_{b,t}^{+,-}+\mathcal{M}_{a,u}^{+,-}+\mathcal{M}_{a,u}^{+,-}$
are respectively shown. The corresponding amplitude squares (or probabilities) are also shown 
in Figs.~\ref{fig:eessAngDistRRprob} and \ref{fig:eessAngDistRLprob}, respectively. 
Here, we note a remarkable correspondence between the IFD and LFD profiles of the $++$ amplitude in Fig.~\ref{fig:eessAngDistRRtu} and 
the LFD and IFD profiles of the $+-$ amplitude in Fig.~\ref{fig:eessAngDistRLtu} modulo the overall signs, respectively. 
This remarkable correspondence between the IFD and LFD profiles is further self-evident in 
Figs.~\ref{fig:eessAngDistRRprob} and ~\ref{fig:eessAngDistRLprob} as the overall sign doesn't matter in the amplitude square
or the probability. As discussed earlier, the LF helicity of the particle moving in the $-\hat z$ direction is opposite to the Jacob-Wick helicity defined in the IFD. 
Since the incident $e^- e^+$ annihilation takes place along the $z$-axis and the positron ($e^+$) is moving in the $-\hat z$ direction, 
the swap of the helicity between the IFD and LFD for the positron can be understood as we see the IFD/LFD profile correspondence in Fig.~\ref{fig:eessAngDistAmpAndProb}.

To examine the frame dependence of the whole landscape, we have computed all the helicity amplitudes discussed above with the non-zero center of momentum ($P^z \neq 0$) 
as well. In particular, we took a large enough center of momentum to pass the helicity boundaries given by Eqs.~(\ref{eq:electron_boundary}) and (\ref{eq:positron_boundary}) that we have discussed extensively in the previous subsection \ref{subsec:collinear}.  
In Figs.~\ref{fig:eessAngDistRRPzp}, \ref{fig:eessAngDistRLPzp} and \ref{fig:eessAngDistAmpAndProbPzp}, 
we show the results for $P^z = +15 m_e$ while we do for $P^z = -15 m_e$ in Figs.~\ref{fig:eessAngDistRRPzm}, \ref{fig:eessAngDistRLPzm} and \ref{fig:eessAngDistAmpAndProbPzm}. In these figures, the whole landscapes of the interpolation angle ($\delta$) dependence for the angular distributions of the helicity $++$ and $+-$ amplitudes 
are presented for the boosted frames with $P^z = +15 m_e$ and $P^z = -15 m_e$. As we have shown in the collinear case presented in the last subsection \ref{subsec:collinear},
no helicity boundaries exist between IFD and LFD in the frame with $P^z = +15 m_e$ while there are two distinct helicity boundaries, one from electron and the other from positron
(see Eqs.~(\ref{eq:electron_boundary}) and (\ref{eq:positron_boundary}), respectively), between IFD and LFD for $ P^z=-15m_e $. 
While all of these figures (Figs.~\ref{fig:eessAngDistRRPzp}, \ref{fig:eessAngDistRLPzp}, and \ref{fig:eessAngDistAmpAndProbPzp} for $P^z = +15 m_e$ and Figs.~\ref{fig:eessAngDistRRPzm}, \ref{fig:eessAngDistRLPzm} and \ref{fig:eessAngDistAmpAndProbPzm} for $P^z = -15 m_e$) were depicted in terms of the CMF angle $\theta$, 
they all can be also shown in terms of the apparent angle $\theta_{\mathrm{app}}$ in the boosted frame using the relationship between $\theta_{\mathrm{app}}$ and $\theta$, i.e.
\begin{equation}
	\tan\theta_{\mathrm{app}}=\frac{\sin\theta}{\gamma \left( \beta+\cos\theta\right) },
	\label{eqn:thetaapprelation}
	\end{equation}
where the $\gamma$ factor in the boosted frame is given by $\gamma = 1/\sqrt{1-\beta^2} = \sqrt{1+\left( \frac{P^z}{2E_0}\right)^2 }$ in terms of 
the total momentum $P^z$ in the boosted frame and the total energy $2 E_0$ in the CMF.  
All of those figures in terms of  $\theta_{\mathrm{app}}$ as well as the derivation of Eq.(\ref{eqn:thetaapprelation}) are shown in Appendix \ref{app:ApprantAngle}.

For $P^z = + 15 m_e$, the angular distribution of the helicity $ ++ $ amplitudes are shown in Fig.~\ref{fig:eessAngDistRRPzp} (a) $\mathcal{M}_{a,t}^{+,+}$  
(b) $\mathcal{M}_{b,t}^{+,+}$ (c) $\mathcal{M}_{a,u}^{+,+}$ (d) $\mathcal{M}_{b,u}^{+,+}$, while the angular distribution of the helicity $ +- $ amplitudes are shown in 
Fig.~\ref{fig:eessAngDistRLPzp} (a) $\mathcal{M}_{a,t}^{+,-}$  (b) $\mathcal{M}_{b,t}^{+,-}$ (c) $\mathcal{M}_{a,u}^{+,-}$ (d) $\mathcal{M}_{b,u}^{+,-}$.
The profiles of the ``instantaneous fermion contribution" and the ``on-mass-shell propagating contribution"  at $\delta = \pi/4$ (LFD) discussed at CMF ($P^z=0$)
survive invariantly although significant changes for the region $0 \leq \delta < \pi/4$ are apparent in the landscape without any helicity boundaries as expected in 
this boosted frame with $P^z = + 15 m_e$.
The net results adding both t-channel and u-channel time-ordered amplitudes all together are shown in Fig.~\ref{fig:eessAngDistAmpAndProbPzp}.
In Figs.~\ref{fig:eessAngDistRRtuPzp} and \ref{fig:eessAngDistRLtuPzp}, the sum of $++$ helicity amplitude 
$\mathcal{M}_{a,t}^{+,+}+\mathcal{M}_{b,t}^{+,+}+\mathcal{M}_{a,u}^{+,+}+\mathcal{M}_{a,u}^{+,+}$ and 
the sum of $+-$ helicity amplitude 
$\mathcal{M}_{a,t}^{+,-}+\mathcal{M}_{b,t}^{+,-}+\mathcal{M}_{a,u}^{+,-}+\mathcal{M}_{a,u}^{+,-}$
are respectively shown. The corresponding amplitude squares (or probabilities) are also shown 
in Figs.~\ref{fig:eessAngDistRRprobPzp} and \ref{fig:eessAngDistRLprobPzp}, respectively. 

For $P^z = - 15 m_e$, the angular distribution of the helicity $ ++ $ amplitudes are shown in Fig.~\ref{fig:eessAngDistRRPzm} (a) $\mathcal{M}_{a,t}^{+,+}$  
(b) $\mathcal{M}_{b,t}^{+,+}$ (c) $\mathcal{M}_{a,u}^{+,+}$ (d) $\mathcal{M}_{b,u}^{+,+}$, while the angular distribution of the helicity $ +- $ amplitudes are shown in 
Fig.~\ref{fig:eessAngDistRLPzm} (a) $\mathcal{M}_{a,t}^{+,-}$  (b) $\mathcal{M}_{b,t}^{+,-}$ (c) $\mathcal{M}_{a,u}^{+,-}$ (d) $\mathcal{M}_{b,u}^{+,-}$.
The LFD profiles of the ``instantaneous fermion contribution" and the ``on-mass-shell propagating contribution"  are again invariant regardless of 
$P^z$ values ($P^z = +15 m_e, 0, -15 m_e$) exhibiting the boost invariance of the helicity amplitudes in LFD. For the region $0 \leq \delta < \pi/4$, however, there appear two critical 
interpolating angles at  $\delta = \delta_{c,e^-} \approx 0.55062$ and $\delta = \delta_{c, e^+} \approx 0.784165$, which can be estimated from 
Eqs.~(\ref{eq:electron_boundary}) and (\ref{eq:positron_boundary}), respectively. Except the LFD profiles, the whole landscapes of angular distributions
are dynamically varied both for $0 \le \delta < \delta_{c,e^-} \approx 0.55062$ and $\delta_{c,e^-} \approx 0.55062 < \delta < \delta_{c, e^+} \approx 0.784165$ 
depending on the reference frames ($P^z = +15 m_e, 0, -15 m_e$).
The net results adding both t-channel and u-channel time-ordered amplitudes all together are shown in Fig.~\ref{fig:eessAngDistAmpAndProbPzm}.
In Figs.~\ref{fig:eessAngDistRRtuPzm} and \ref{fig:eessAngDistRLtuPzm}, the sum of $++$ helicity amplitude 
$\mathcal{M}_{a,t}^{+,+}+\mathcal{M}_{b,t}^{+,+}+\mathcal{M}_{a,u}^{+,+}+\mathcal{M}_{a,u}^{+,+}$ and 
the sum of $+-$ helicity amplitude 
$\mathcal{M}_{a,t}^{+,-}+\mathcal{M}_{b,t}^{+,-}+\mathcal{M}_{a,u}^{+,-}+\mathcal{M}_{a,u}^{+,-}$
are respectively shown. The corresponding amplitude squares (or probabilities) are also shown 
in Figs.~\ref{fig:eessAngDistRRprobPzm} and \ref{fig:eessAngDistRLprobPzm}, respectively. 

Finally, Fig.~\ref{fig:HelicityProbabilitySum} shows the sum of the $++$ and $+-$ helicity amplitude squares which is the half of the total probability sum including $-+$ and $--$ helicity amplitude squares in all three reference frames ($P^z = +15 m_e, 0, -15 m_e$) discussed above. 
Although the individual helicity amplitude squares in LFD ($\delta = \pi/4$) are independent of the reference frames, the individual helicity amplitude squares for $0 \leq \delta < \pi/4$ varied depending on the reference frames as we have seen in Figs.~\ref{fig:eessAngDistRRprob}, \ref{fig:eessAngDistRRprobPzp} and \ref{fig:eessAngDistRRprobPzm} for $|\mathcal{M}_{a,t}^{+,+}+\mathcal{M}_{b,t}^{+,+}+\mathcal{M}_{a,u}^{+,+}+\mathcal{M}_{a,u}^{+,+}|^2$ as well as in Figs. \ref{fig:eessAngDistRLprob}, \ref{fig:eessAngDistRLprobPzp} and \ref{fig:eessAngDistRLprobPzm} for $|\mathcal{M}_{a,t}^{+,-}+\mathcal{M}_{b,t}^{+,-}+\mathcal{M}_{a,u}^{+,-}+\mathcal{M}_{a,u}^{+,-}|^2$.  
For the $P^z = +15 m_e$ frame, there were no helicity boundaries and the individual helicity amplitude squares were same regardless of the $\delta$ values as shown
in Figs. \ref{fig:eessAngDistRRprobPzp}  and \ref{fig:eessAngDistRLprobPzp}. However, for the other reference frames with $P^z = -15 m_e$ and $P^z = 0$ (CMF), where
there were two ($\delta = \delta_{c,e^-} \approx 0.55062$ and $\delta = \delta_{c, e^+} \approx 0.784165$) boundaries and 
one ($\delta_c \approx 0.713724$) bounary, respectively, each individual helicity amplitude squares varied significantly across the corresponding helicity boundaries.
However, the sum of helicity amplitude squares is completely independent of not only the interpolating angle $\delta$ but also the reference frames as it should be.  
The boost-invariant physical quantity must be of course completely independent of the interpolation angle, regardless of IFD, LFD or any other dynamics in between.

\subsection{Summary of $e^+ e^- \rightarrow$ two scalar particles}
\label{subsec:Toy_summary}

As we have shown in all of these results, the LFD results are completely independent of the reference frame due to the boost invariance 
while the IFD results are dependent on the reference frame. As discussed in the collinear case (see Figs.~\ref{fig:Scalar_theta=pi_ta}, \ref{fig:Scalar_theta=pi_tb}, \ref{fig:Scalar_theta=pi_ua}, \ref{fig:Scalar_theta=pi_ub} and \ref{fig:Scalar_theta=pi_total}), the LFD results are outside the spin-flip boundary and 
the LF helicity of the particle moving in the $-\hat z$ direction is opposite to the Jacob-Wick helicity defined in the IFD. With this helicity swap between the IFD and the LFD for the particle moving in the $-\hat z$ direction, we can see again the angular momentum
conservation and the spin-singlet nature in the LFD results. Namely, the $++$ and $--$ LF helicity amplitudes vanish at $\theta = 0$ and $\theta = \pi$ 
(see e.g. Fig.~\ref{fig:eessAngDistRRtu}) and they are equal in the angular dependence while the relative sign between the $+-$ and $-+$ LF helicity amplitudes is opposite to each other in accordance with Eq.~(\ref{eqn:amplitude_relation_scalar_case}). As discussed in the collinear case, 
the IFD results in $P^z \to -\infty$ does not yield the LFD results. 
Likewise, in the non-collinear case, we also note that the angular distribution of the IFD amplitudes in $P^z \to -\infty$ is opposite in sign with respect
to the corresponding angular distribution of the LFD amplitudes (see Figs. \ref{fig:eessAngDistRRtuPzm} and \ref{fig:eessAngDistRLtuPzm}), let alone
that each time-ordered amplitudes of IFD in $P^z \to -\infty$ yields far different angular distribution from the corresponding LFD results (see e.g. Figs. \ref{fig:eessAngDistRRPzm} and \ref{fig:eessAngDistRLPzm}). Although the angular distribution of each IFD helicity amplitude in $P^z \to +\infty$ is supposed to yield the identical corresponding
angular distribution of the LFD helicity amplitude, one would need to boost the $ P^z $ value much higher than $+15m_e$ 
(see e.g. Figs. \ref{fig:eessAngDistRRtuPzp} and \ref{fig:eessAngDistRLtuPzp}) in order 
to get indeed the very similar profile of ``on-mass-shell propagating contribution" and the ``instantaneous fermion contribution" in LFD. 

In the helicity amplitude square (or probability) level, we see the built-in $t-u$ symmetry in the $e^+ e^-$ annihilation process regardless of IFD or LFD as manifested 
in the $\theta \to \pi-\theta$ symmetry of the angular distributions presented in Figs.~\ref{fig:eessAngDistRRprob} and \ref{fig:eessAngDistRLprob}
as well as in Fig.~\ref{fig:HelicityProbabilitySum}. We have verified that the result shown in Fig.~\ref{fig:HelicityProbabilitySum} is in exact agreement
with the analytic result of the total amplitude square for the scalar particle pair production in $e^+ e^-$ annihilation given by
\begin{align}
&\left| \mathcal{M} \right|^2_{\rm{scalar}} \equiv \sum_{\lambda_1,\lambda_2}|\mathcal{M}_{a,t}^{\lambda_1,\lambda_2}+\mathcal{M}_{b,t}^{\lambda_1,\lambda_2}
+\mathcal{M}_{a,u}^{\lambda_1,\lambda_2}+\mathcal{M}_{b,u}^{\lambda_1,\lambda_2}|^2 \notag\\
&=\frac{2\left( ut+m^2(4s-5t+3u)-15m^4\right) }{(t-m^2)^2}\notag\\
&+\frac{2\left( tu+m^2(4s-5u+3t)-15m^4\right) }{(u-m^2)^2}\notag\\
&+\frac{2\left( (s+u)u+(s+t)t+2m^2(3s-t-u)-30m^4\right) }{(t-m^2)(u-m^2)},
\label{eqn:scalar_cross_section}
\end{align}
where $m=m_e$ and the Mandelstam variables $s=(p_1+p_2)^2$, $t=(p_1 - p_3)^2$ and $u=(p_1 - p_4)^2$ are give by
$s= 16 m_e^2$, $t= (-7 + 4 \sqrt{3} \cos \theta)m_e^2$ and $u= -(7 + 4 \sqrt{3} \cos \theta)m_e^2$ in CMF given by 
Eq.~(\ref{eqn:p1234_Annihilation}) with $E_0 = 2 m_e$ and $P_e = \sqrt{3} m_e$ for our numerical calculation.

In fact, the built-in $t-u$ symmetry in each and every helicity amplitude square is completely independent of the interpolation angle $\delta$
as shown in Figs.~\ref{fig:eessAngDistRRprob}, \ref{fig:eessAngDistRLprob}, \ref{fig:eessAngDistRRprobPzp}, \ref{fig:eessAngDistRLprobPzp}, \ref{fig:eessAngDistRRprobPzm} and \ref{fig:eessAngDistRLprobPzm}.
Essentially the same kind of $t-u$ symmetry can be found in the $e^+ e^- \to \gamma \gamma$ QED process which we now discuss
in the next section, Sec.~\ref{sec:calculation}.\\
%
	\section{Interpolating Helicity Scattering Probabilities}\label{sec:calculation}
	\subsection{\label{sub:annihilation}$ e^+ e^- $ Pair Annihilation into two photons}
	Having discussed all the helicity amplitudes of the pair production of scalar particles in $e^+ e^-$ annihilation, 
	we now look into the two photon production process in the same initial state of $e^+ e^-$ annihilation. 
	While there must be some similarity inherited from the same initial state, there must be also some difference in the helicity amplitudes
	due to the change of the final state from the spinless pair of scalar particles to the two real photons in QED.
	The identification of the real photon helicity would require a particular attention as it doesn't carry any rest mass and moves invariantly with the speed of light. 
		The lowest order t-channel QED Feynman diagram is already in place as Fig.~\ref{fig:PairAnnihilationto2s_tchannel} and the corresponding u-channel diagram can be attained by swapping the two final photons in Fig.~\ref{fig:PairAnnihilationto2s_tchannel}. 
	The two time-ordered diagrams in the t-channel are also displayed in Figs. \ref{fig:PairAnnihilationto2s_tchannel_a} and \ref{fig:PairAnnihilationto2s_tchannel_b} and the kinematic is the same with the previous calculation illustrated in Fig.~\ref{fig:Annihilation_kinematics}, and written in the previous section in Eq.~(\ref{eqn:p1234_Annihilation}).
	 
	The QED helicity amplitudes $\mathcal{M}_t^{\lambda_1,\lambda_2,\lambda_3,\lambda_4}$ and $\mathcal{M}_u^{\lambda_1,\lambda_2,\lambda_3,\lambda_4}$ 
	with the two initial lepton helicities $\lambda_1$ and $\lambda_2$
	and the final two photon helicities $\lambda_3$ and $\lambda_4$ in t and u channels, respectively, are now expressed with the interpolating Lorentz indices $\wh{\mu}$ and $\wh{\nu}$ as 
	\begin{align}
	\mathcal{M}_t^{\lambda_1,\lambda_2,\lambda_3,\lambda_4} &=\bar{v}^{\lambda_2}(p_2)\epsilon_{\wh{\nu}}^{\lambda_4}(p_4)^*\gamma^{\wh{\nu}}\Sigma_t \gamma^{\wh{\mu}}  \epsilon_{\wh{\mu}}^{\lambda_3}(p_3)^* u^{\lambda_1}(p_1), \label{eqn:TwoPhotonAmpsT}\\ 
	\mathcal{M}_u^{\lambda_1,\lambda_2,\lambda_3,\lambda_4} &=\bar{v}^{\lambda_2}(p_2)\epsilon_{\wh{\mu}}^{\lambda_3}(p_3)^*\gamma^{\wh{\mu}}\Sigma_u \gamma^{\wh{\nu}}  \epsilon_{\wh{\nu}}^{\lambda_4}(p_4)^* u^{\lambda_1}(p_1), \label{eqn:TwoPhotonAmpsU}
	\end{align}
	where $\Sigma_t$ and $\Sigma_u$ are 
	\begin{equation}
	\Sigma_t = \frac{\slashed{p}_1 - \slashed{p}_3 + m}{t - m^2}, \,\,\,\,\, \Sigma_u = \frac{\slashed{p}_1 - \slashed{p}_4 + m}{u - m^2}
         \label{eqn:t&uOperator}
         \end{equation}	
	with $t=(p_1 - p_3)^2$ and $u=(p_1 - p_4)^2$, and the polarization vector $\epsilon_{\wh{\mu}}^\lambda(P)$ is given by~\cite{JLS2015}
	\begin{widetext}
		\begin{align}
		\epsilon_{\wh{\mu}}^+(P)&=-\frac{1}{\sqrt{2}\mathbb{P}}\left( \mathbb{S}|\mathbf{P}_{\perp}|, \frac{P_1P_{\wh{-}}-iP_2\mathbb{P}}{|\mathbf{P}_{\perp}|},\frac{P_2P_{\wh{-}}+iP_1\mathbb{P}}{|\mathbf{P}_{\perp}|}, -\mathbb{C}|\mathbf{P}_{\perp}|\right) ,\notag \\
		\epsilon_{\wh{\mu}}^-(P)&=\frac{1}{\sqrt{2}\mathbb{P}}\left( \mathbb{S}|\mathbf{P}_{\perp}|, \frac{P_1P_{\wh{-}}+iP_2\mathbb{P}}{|\mathbf{P}_{\perp}|},\frac{P_2P_{\wh{-}}-iP_1\mathbb{P}}{|\mathbf{P}_{\perp}|}, -\mathbb{C}|\mathbf{P}_{\perp}|\right) ,\notag \\
		\epsilon_{\wh{\mu}}^0(P)&=\frac{P^{\wh{+}}}{M\mathbb{P}}\left( P_{\wh{+}}-\frac{M^2}{P^{\wh{+}}}, P_1, P_2, P_{\wh{-}}\right),\label{photon_polarization}
		\end{align}
	\end{widetext}
	with $ \mathbb{P}=\sqrt{P_{\wh{-}}^2+\mathbf{P}_{\perp}^2\mathbb{C}}=\sqrt{(P^{\wh{+}})^2-M^2\mathbb{C}} $.
Note that this interpolating polarization vector $\epsilon_{\wh{\mu}}^\lambda(P)$ respects the gauge condition	
$A^{\pT}=0$ and $\partial_{\mT}A_{\mT}+\boldsymbol\partial_{\perp}\cdot\mathbf{A}_{\perp}\Cc=0$, which 
links the light-front gauge $A^{+}=0$ in the LFD and the Coulomb gauge $\boldsymbol\nabla \cdot \mathbf{A}=0$ in IFD as 
discussed in Ref.\cite{JLS2015}.		
For the sake of generality, we kept here the generic femion and gauge boson mass as $m$ and $M$, respectively.
The real photon helicity $\lambda$ takes only $+$ or $-$ but not $0$ as $M \to 0$ limit and thus there is no issue involved in taking the massless limit. 
One should note that not only the final state momenta $p_3$ and $p_4$ are swapped but also the QED vertices
with the $\gamma$ matrices are exchanged between the t-channel amplitude and the u-channel amplitude given by 
Eqs.~(\ref{eqn:TwoPhotonAmpsT}) and (\ref{eqn:TwoPhotonAmpsU}), respectively.
 
The symmetry of the helicity amplitudes based on the parity conservation\cite{JB2013} given by Eq.~(\ref{eqn:amplitude_relation_scalar_case}) for the pair production of
the scalar particles is also now extended for the two-photon production as  
	\begin{widetext}
		\begin{equation}\label{eqn:amplitude_relation_Annihilation}
		\mathcal{M}^{-\lambda_1,-\lambda_2,-\lambda_3,-\lambda_4} = (-1)^{\lambda_3+\lambda_4-\lambda_1-\lambda_2}\mathcal{M}^{\lambda_1,\lambda_2,\lambda_3,\lambda_4} ,
		\end{equation} 
	\end{widetext}
	where $ \lambda_3 $ and $ \lambda_4 $ are the helicities of the outgoing photons while 
 $ \lambda_1 $ and $ \lambda_2 $ are the incoming electron and positron helicities, respectively.
 As this symmetry works identically both for t and u channels, the subscripts $t$ and $u$ in the helicity amplitude above are suppressed
 in Eq.~(\ref{eqn:amplitude_relation_Annihilation}). 
	
	Now, recalling Eqs.~(\ref{eqn:propagator_TOa}) and (\ref{eqn:propagator_TOb}), the time-ordered amplitudes in t-channel can be written in short-hand notations 
	without specifying the helicities as
	\begin{equation}
	\mathcal{M}_{a,t}=\bar{v}(p_2) \slashed{\epsilon}(p_4)^* \left( \frac{1}{2Q_t^{\wh{+}}}\frac{\slashed{Q}_{a,t}+m}{q_{t\wh{+}}-Q_{a,t\wh{+}}}\right)  \slashed{\epsilon}(p_3)^* u(p_1),
	\end{equation}
	and
	\begin{equation}
	\mathcal{M}_{b,t}=\bar{v}(p_2) \slashed{\epsilon}(p_4)^* \left( \frac{1}{2Q_t^{\wh{+}}}\frac{-\slashed{Q}_{b,t}+m}{-q_{t\wh{+}}-Q_{b,t\wh{+}}}\right) \slashed{\epsilon}(p_3)^* u(p_1),
	\end{equation}
where
$q_{a,t} = q_t \equiv p_1 - p_3$, $q_{b,t} = -q_{a,t} = -q_t$, and $Q_{a,t\wh{+}}$ and $Q_{b,t\wh{+}}$ are the interpolating on-mass-shell energy of the intermediate propagating fermion given by 		
\begin{align} \label{eqn:qatnew+}
	Q_{a,t\wh{+}}&=\frac{-\mathbb{S}q_{a,t\wh{-}}+Q_t^{\wh{+}}}{\mathbb{C}},\\ \label{eqn:qbtnew+}
	Q_{b,t\wh{+}}&=\frac{-\mathbb{S}q_{b,t\wh{-}}+Q_t^{\wh{+}}}{\mathbb{C}},
	\end{align}
with $Q_t^{\wh{+}}$ denoting the on-mass-shell value of $q_t^{\wh{+}}$ as 
	\begin{equation}
	Q_t^{\wh{+}}\equiv\sqrt{q_{t\wh{-}}^{2}+\mathbb{C} (\mathbf{q}_{t\perp}^{2}+m^{2})}.
	\label{eqn:Q_t^+}
	\end{equation}
The kinematics here is of course identical to the ones given in the last section, Sec.~\ref{sec:eess}, despite the explicit notations to specify $t$ and $u$ channels
which now involve the swap of not only the final state particle momenta but also the QED photon and fermion vertices. Thus, we elaborate the notations 
to designate the t and u channels more explicitly for this section. 

As the final state photons with momentum $p_3$ and $p_4$ must be swapped for the u-channel amplitudes, we denote 
the intermediate fermion momentum between the two photon vertices as $q_u = p_1 - p_4$ and correspondingly 
designate all other time-ordered variables replacing $q_{a,t}$ and $q_{b,t}$ in the t channel time-ordered amplitudes by $q_{a,u} = q_u = p_1 - p_4$ 
and $q_{b,u} = -q_{a,u} = -q_u$, respectively. Consequently, the interpolating on-mass-shell energy of the 
intermediate propagagting fermion $Q_{a,u\wh{+}}$ and $Q_{b,u\wh{+}}$ for the two time-ordered amplitudes are also given by 
replacing $q_{a,t}$ and $q_{b,t}$ by $q_{a,u}$ and $q_{b,u}$, respectively, in Eqs.~(\ref{eqn:qatnew+}) and (\ref{eqn:qbtnew+})
together with the replacement of $Q_t^{\wh{+}}$ in Eq.~(\ref{eqn:Q_t^+}) by $Q_u^{\wh+}$ as 
	\begin{equation}
	Q_u^{\wh{+}}\equiv\sqrt{q_{u\wh{-}}^{2}+\mathbb{C} (\mathbf{q}_{u\perp}^{2}+m^{2})}.
	\label{eqn:Q_u^+}
	\end{equation}
While the notations are more elaborated in this section as described here, there's no change in the kinematics from the ones provided in the last section, Sec.~\ref{sec:eess}. 

	To make the numerical calculations, we take the same initial energy of each particle (i.e. $E_e = 2 m_e$ and $P_e = \sqrt{3} m_e$) and and the same three different reference frames (i.e. CMF given by Eq.~(\ref{eqn:p1234_Annihilation}) and boosted frames with $P^z = 15 m_e$ and $P^z = -15 m_e$) used in Sec.~\ref{sec:eess} for the angular distribution analysis of the interpolating helicity amplitudes. While we focus on the CMF result in this section, the results in the boosted frames ($P^z = 15 m_e$ and 
$P^z = -15 m_e$) are summarized in the Appendix \ref{app:BoostedAnnihilation} and   
the $P^z$ dependence of the interpolating helicity amplitudes for 
a particular scattering, e.g. $\theta = \pi/3$ case, is shown in the Appendix \ref{app:PzDepThetaPiOver3}.

	\begin{figure*}
		\centering
		\subfloat[]{\includegraphics[width=0.49\textwidth]{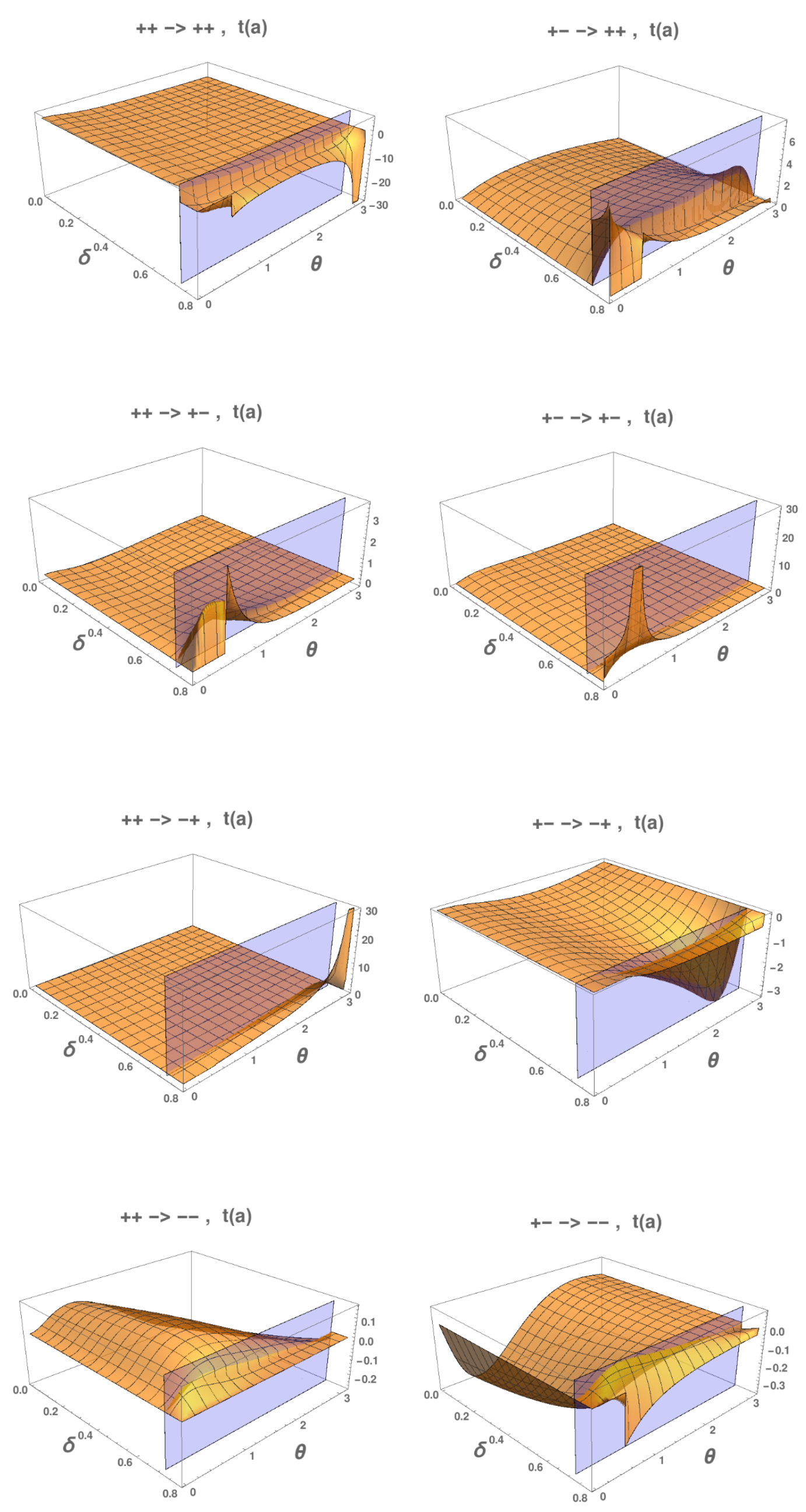}\label{fig:JoinedFigmta}}
		\centering
		\subfloat[]{\includegraphics[width=0.49\textwidth]{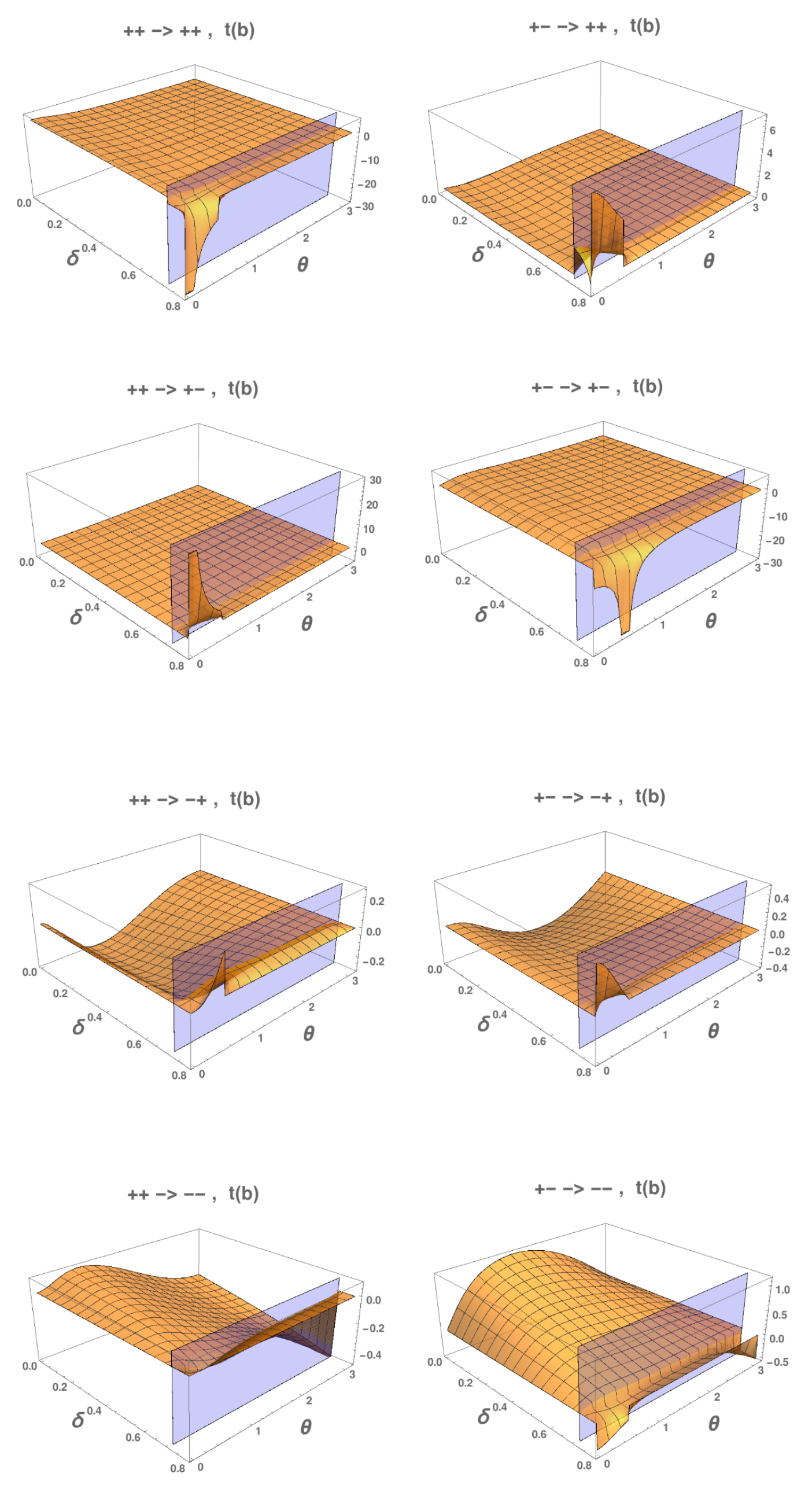}\label{fig:JoinedFigmtb}}
		\caption{\label{fig:JoinedFigmtaandtb}Angular distribution of the helicity amplitudes for (a) t-channel time-ordering 
		process-a and (b) t-channel time-ordering process-b}
	\end{figure*}
	
	\begin{figure*}
		\centering
		\includegraphics[width=1.0\textwidth]{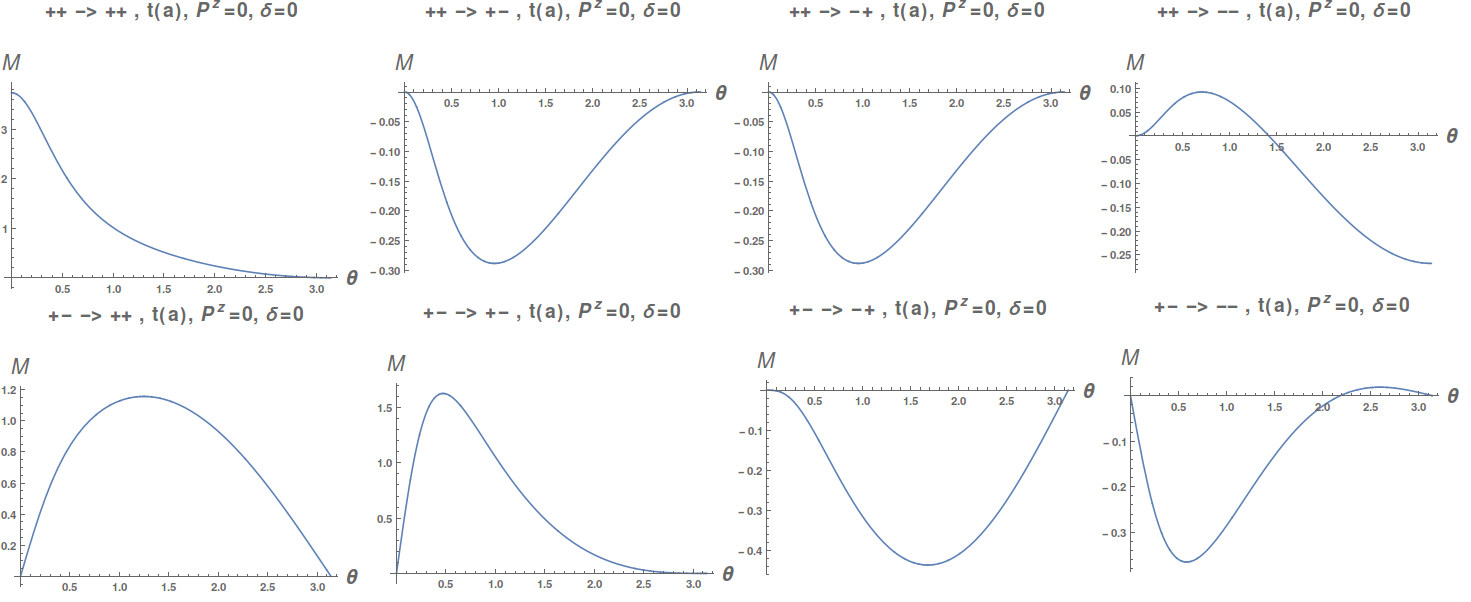}
		\caption{Helicity amplitudes for the non-zero electron mass. As the chirality is not conserved, the upper four amplitudes are not zero.}
		\label{fig:chirality2}
	\end{figure*}

			\begin{figure*}
		\centering
		\includegraphics[width=1.0\textwidth]{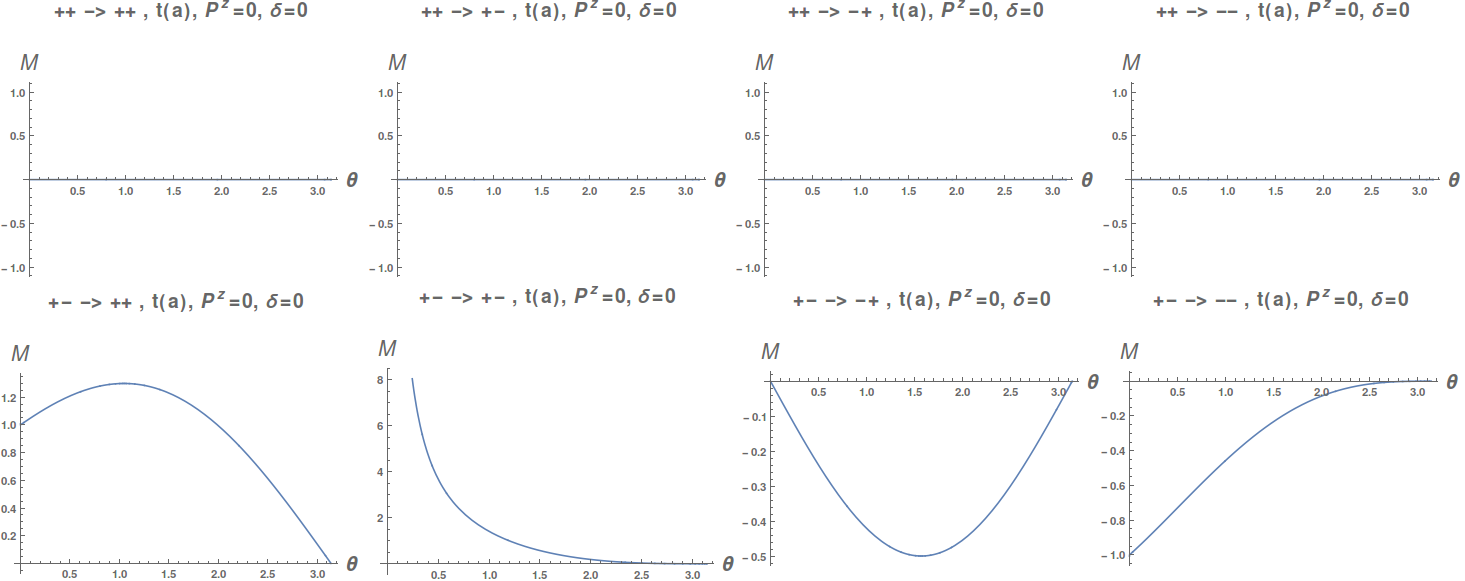}
		\caption{Helicity amplitudes for the massless electron. When setting electron mass equal to zero, chirality is conserved. The upper four amplitudes are zero as the initial helicity of the electron and positron are the same.}
		\label{fig:chirality1}
	\end{figure*}

In Fig.~\ref{fig:JoinedFigmtaandtb}, we show the whole landscape of the interpolation angle ($\delta$) dependence for the angular distributions of the helicity amplitudes 
with the notation $\lambda_1 \lambda_2 \to \lambda_3 \lambda_4$ for $++\to++$, $++\to+-$, $++\to-+$, $++\to--$, as well as $+-\to++$, $+-\to+-$, $+-\to-+$, $+-\to--$ at CMF (i.e. $P^z=0$) in (a) $ \mathcal{M}^{\lambda_1,\lambda_2,\lambda_3,\lambda_4}_{a,t} $ and (b) $ \mathcal{M}^{\lambda_1,\lambda_2,\lambda_3,\lambda_4}_{b,t} $. 
The far most left two columns of Fig.~\ref{fig:JoinedFigmtaandtb} show the helicity amplitudes $\mathcal{M}^{+,+,\lambda_3,\lambda_4}_{a,t}$ and $\mathcal{M}^{+,-,\lambda_3,\lambda_4}_{a,t}$ with the final four helicity configurations
of the photon pairs $\{\lambda_3,\lambda_4\} = \{+,+\}, \{+,-\}, \{-,+\}, \{-,-\}$ but with the initial $++$ and $+-$ helicity configurations of $e^+ e^-$ pair annihilation. While the results here are shown for the nonzero fermion mass $m = m_e$, one may 
check first the consistency with chiral symmetry by taking the massless limit $m\to0$. We did this check for 
the IFD ($\delta=0$) amplitudes which might be more accessible for intuitive understanding due to the more familiar
Jacob-Wick helicity used in the IFD. In the massless limit, the chirality coincides with the helicity.  Consistency check 
with the chiral symmetry here is thus equivalent to the helicity conservation of the massless fermion fields 
in the electromagnetic vector coupling.  For the illustration, the IFD ($\delta=0$) profiles of the left two columns in Fig.~\ref{fig:JoinedFigmtaandtb} are shown in Fig.~\ref{fig:chirality2} and the corresponding profiles for the massless limit of the fermion, $ m\to0 $ limit, in Fig.~\ref{fig:chirality1}. While the helicity amplitudes are nonzero both for $\mathcal{M}^{+,+,\lambda_3,\lambda_4}_{a,t}$ and $\mathcal{M}^{+,-,\lambda_3,\lambda_4}_{a,t}$ as shown in Fig.~\ref{fig:chirality2} for $m = m_e$, 
the helicity amplitudes $\mathcal{M}^{+,+,\lambda_3,\lambda_4}_{a,t}$ all vanish for $m = 0$ as shown in Fig.~\ref{fig:chirality1}.
One may understand this result as a consequence of chiral symmetry and the helicity conservation in the $m = 0$ limit.
As the electromagnetic interaction preserves the chirality/helicity in the massless limit, one may understand why all the helicity amplitudes $\mathcal{M}^{+,+,\lambda_3,\lambda_4}_{a,t}$ vanish for $m = 0$ as shown in Fig.~\ref{fig:chirality1}. 

	\begin{figure*}
			\centering
			\subfloat[]{\includegraphics[width=0.49\textwidth]{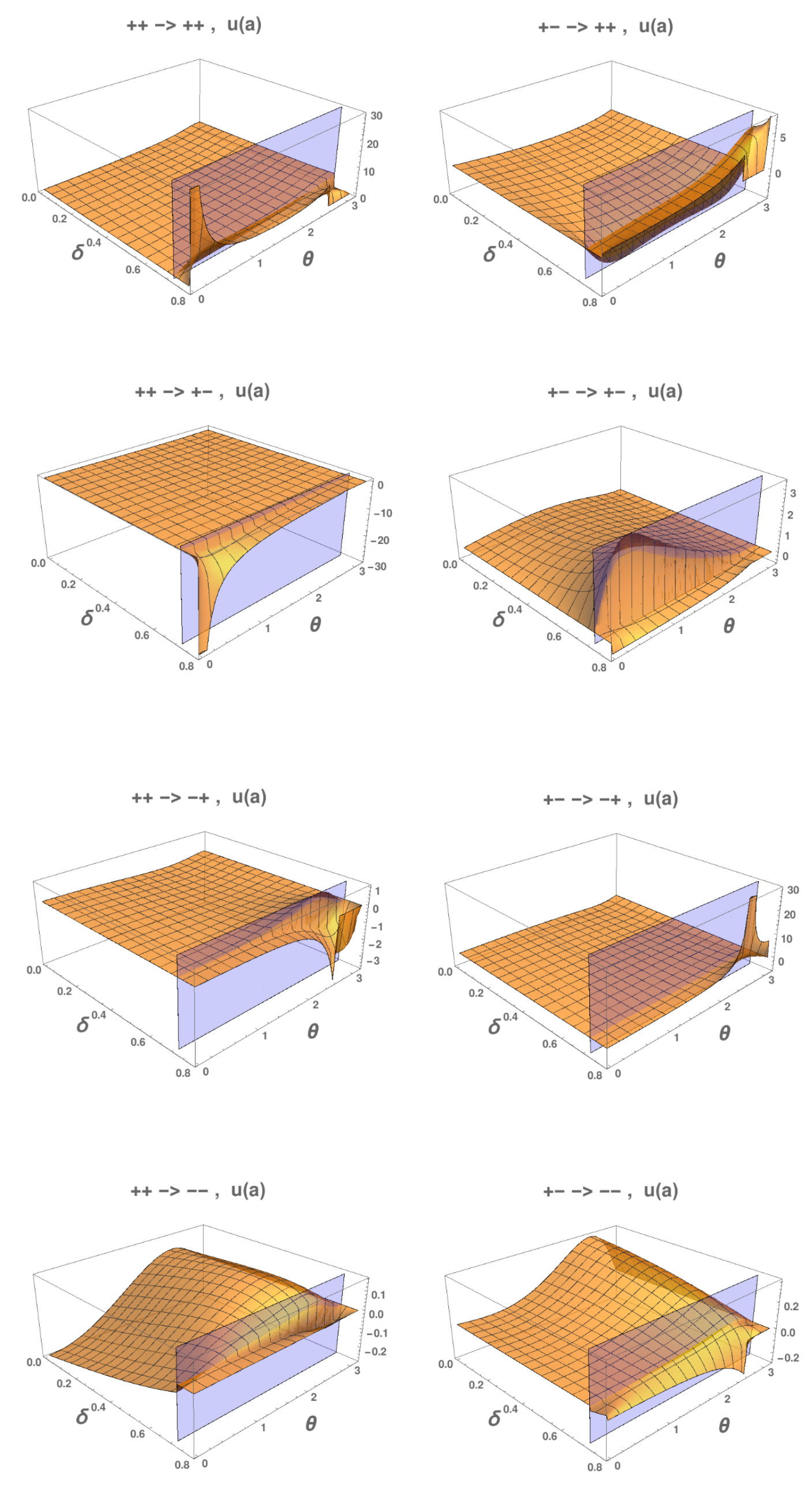}\label{fig:JoinedFigmua}}
			\centering
			\subfloat[]{\includegraphics[width=0.49\textwidth]{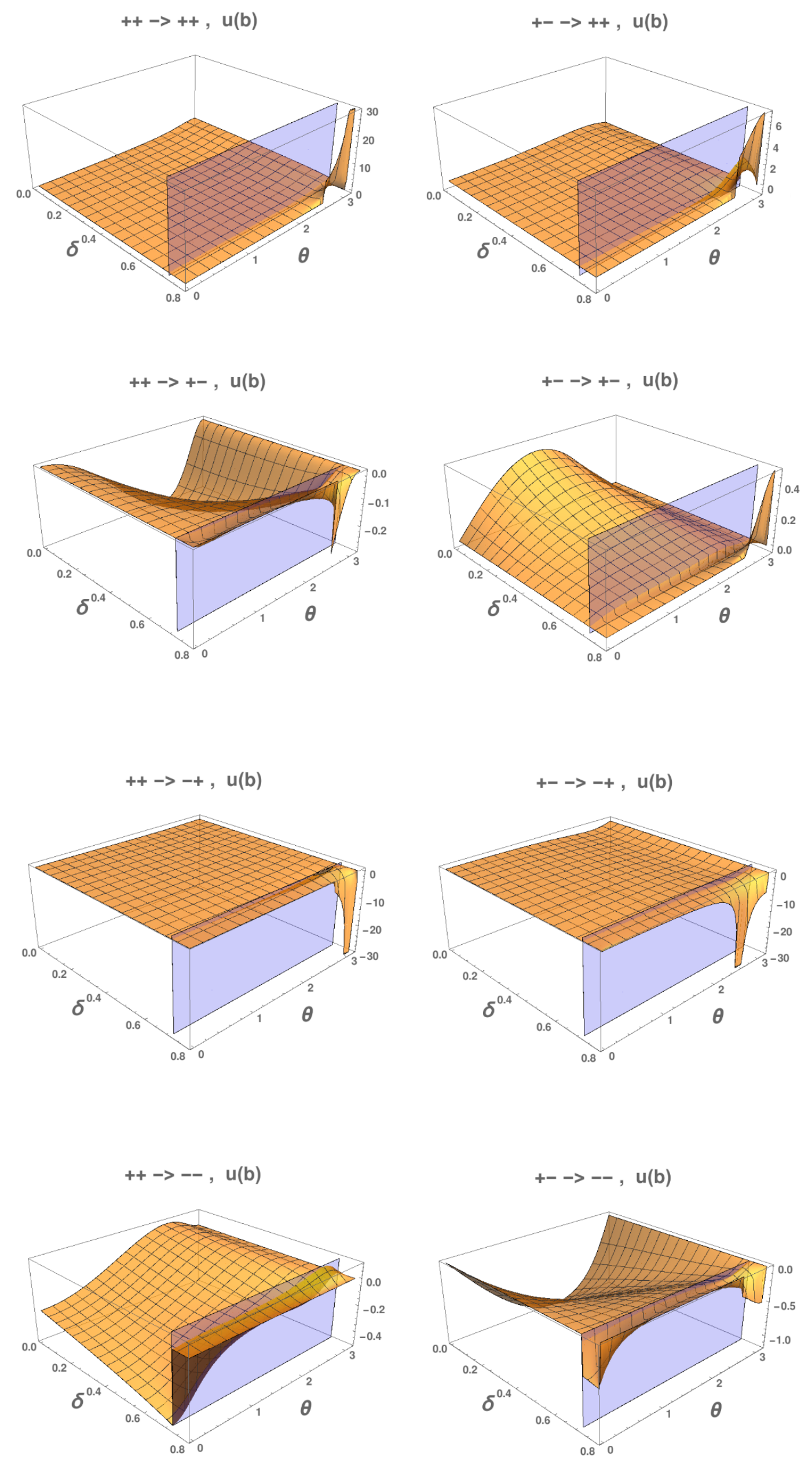}\label{fig:JoinedFigmub}}
			\caption{\label{fig:JoinedFigmuaandub}Angular distribution of the helicity amplitudes for (a) u-channel time-ordering process-a and (b) u-channel time-ordering process-b
			}
		\end{figure*}
	\begin{figure}
		\centering
	 \includegraphics[width=0.49\textwidth]{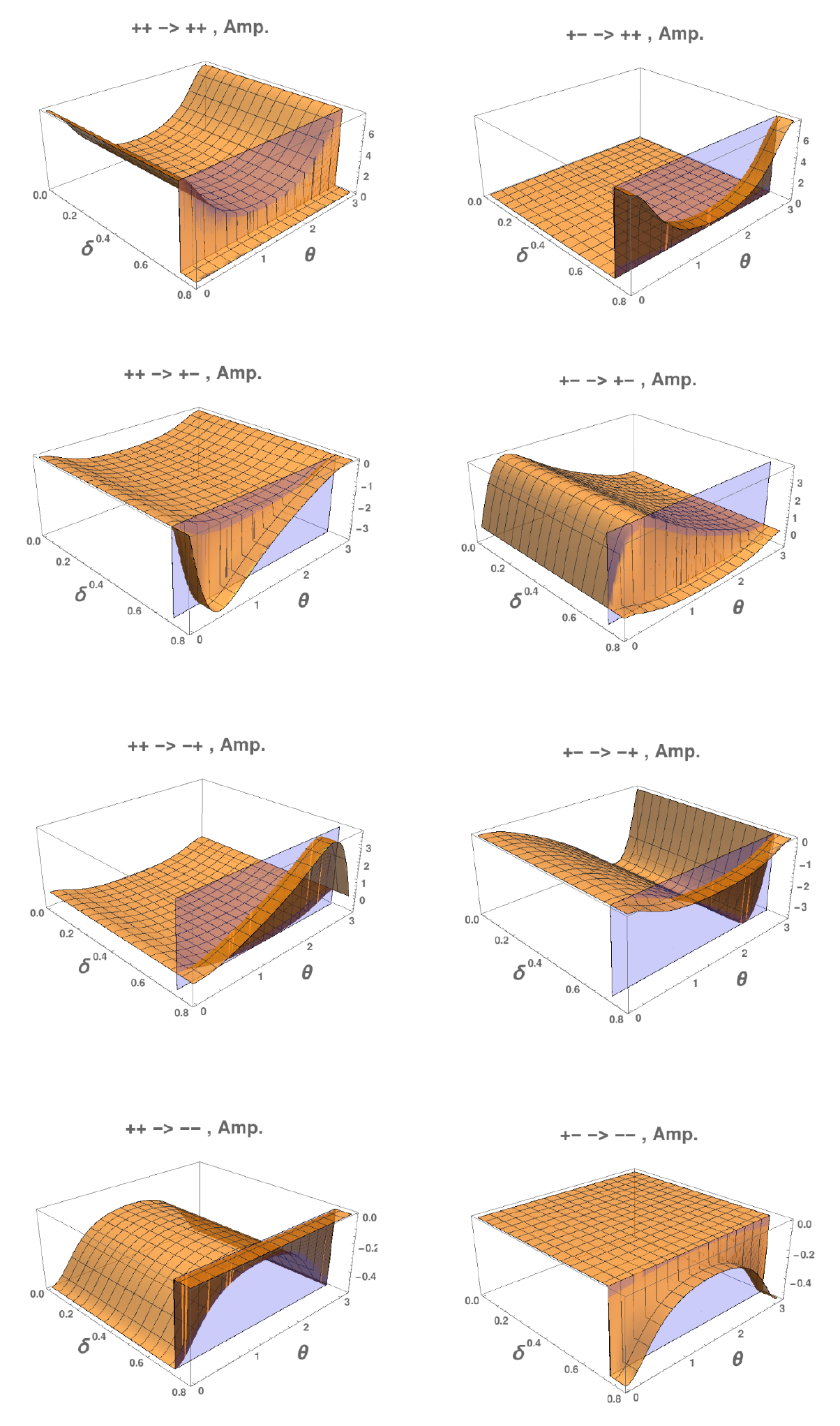}
		\caption{\label{fig:JoinedFigamp}Angular distribution of the t+u helicity amplitudes}
		\end{figure}		
	\begin{figure}		
		\centering
		\includegraphics[width=0.49\textwidth]{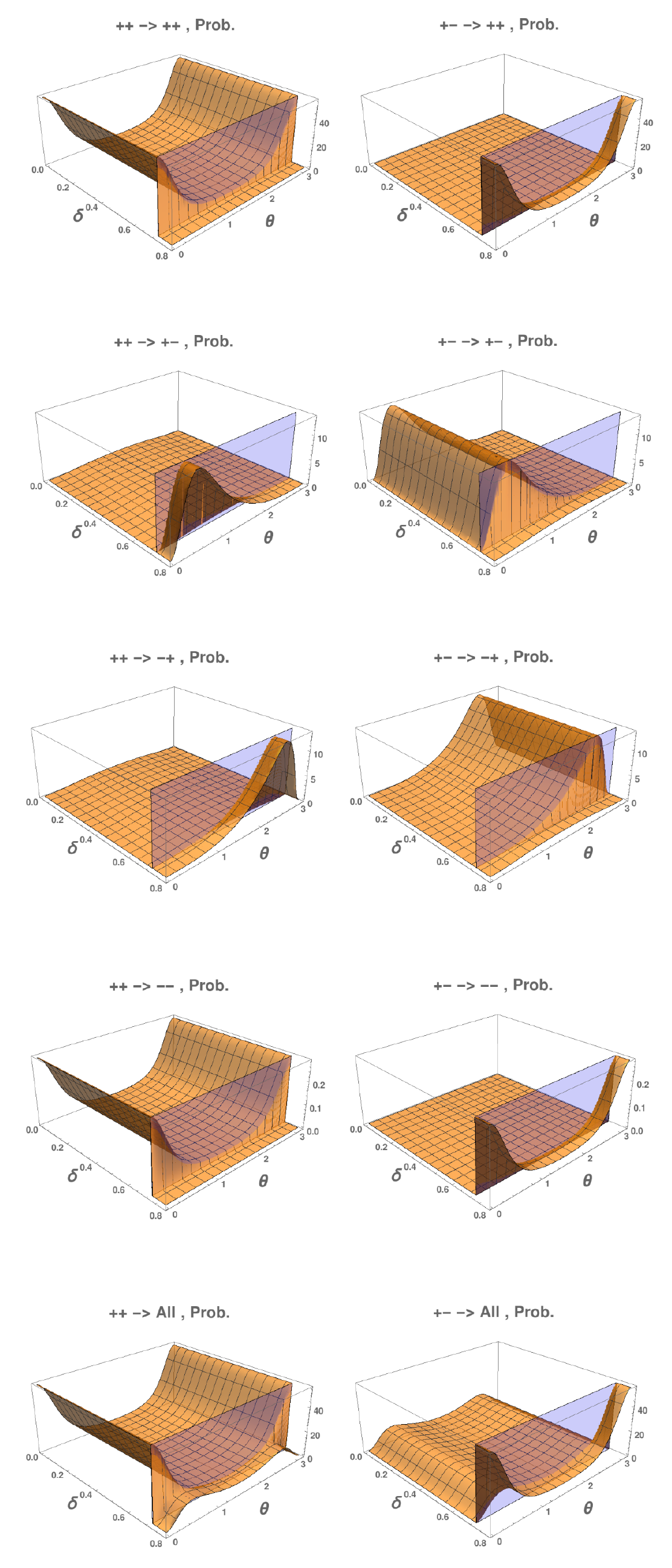}
		\caption{\label{fig:JoinedFigprob}Angular distribution of the t+u helicity probabilities. The figures in the last row are the results of summing over all the figures above each of them.
		}
	\end{figure}

Having checked the consistency of our results with respect to the chiral symmetry and helicity conservation in the massless limit,
we now turn to each individual helicity amplitudes and examine their individual characteristics of the angular distribution
in the whole landscape of the interpolating helicity amplitudes. 
The far most left column of Fig.~\ref{fig:JoinedFigmtaandtb} shows the helicity amplitudes $\mathcal{M}^{+,+,\lambda_3,\lambda_4}_{a,t}$ with the final four helicity configurations
of the photon pairs $\{\lambda_3,\lambda_4\} = \{+,+\}, \{+,-\}, \{-,+\}, \{-,-\}$ but with the same initial $++$ helicity configuration of $e^+ e^-$ pair annihilation. 
This column may be compared with Fig.~\ref{fig:eessAngDistRRta} which shows the helicity amplitude of the pair production of scalar particles with the same initial $++$ helicity configuration of $e^+ e^-$ pair annihilation. As in Fig.~\ref{fig:eessAngDistRRta}, a thin boundary sheet at $\delta = \delta_c \approx 0.713724$ in CMF ($P^z=0$) is shown 
in each and every figure of Fig.~\ref{fig:JoinedFigmtaandtb} to denote the critical interpolation angle $\delta_c$ which separates the IFD side and the LFD side of helicity branches.
Although there are four final helicity configurations in the photon pair production, the basic structure of the initial $++$ helicity configuration of $e^+ e^-$ pair annihilation is inherited as one can see the clear separation of the ``instantaneous fermion contribution" from the ``on-mass-shell propagating 
contribution" in LFD with the critical angle $\theta_{c,t}$ given by Eq.~(\ref{critical_angle_t}). As discussed in Sec.~\ref{sec:eess}, the critical angle $\theta_{c,t}$
turns out to be $\theta_{c,t} = \pi/6 \approx 0.523599$ for $E_0 = 2 m_e$ and $P_e = \sqrt{3} m_e$ and it's apparent that the unique feature of LFD with respect to
separation of the ``instantaneous fermion contribution" from the ``on-mass-shell propagating contribution" is persistent whether the final states are the pair of the scalar particles
or the pair of two photons. For the photon pair production, however, a dramatic new feature appears due to the photon polarization given by Eq.~(\ref{photon_polarization}).
In particular, $\epsilon^\lambda_{\wh{\mu}}$ in Eq.~(\ref{photon_polarization}) reveals a singular feature as $\delta \to \pi/4$. For $\lambda = +$ as an example, 
at $\delta = \pi/4$, i.e. in LFD, the polarization component $\epsilon^+_+ =  -\frac{|\mathbf{P}_{\perp}|}{\sqrt{2}P^+}$ behaves as 
$\epsilon^+_+ = -\frac{\sin \theta}{\sqrt{2}(1+\cos\theta)}$ so that $\epsilon^+_+ \approx -\frac{\sqrt{2}}{\varepsilon}$ for $\theta = \pi-\varepsilon$ with small $\varepsilon$.
This explains the singular behavior near $\theta = \pi$ for $\mathcal{M}^{+,+,+,+}_{a,t}$ in LFD shown in the top far left figure of Fig.~\ref{fig:JoinedFigmtaandtb}.
For $\theta \approx \pi$, one should note that $p_3^+\approx 0$ and the corresponding photon's polarization component $\epsilon^+_+$ yields the singular behavior 
exhibited in the LFD result of $\mathcal{M}^{+,+,+,+}_{a,t}$. This light-front singularity in $\mathcal{M}^{+,+,+,+}_{a,t}$ turns out to be cancelled by the 
same with the opposite sign in $\mathcal{M}^{+,+,+,+}_{b,u}$ as one may see in Fig.~\ref{fig:JoinedFigmuaandub}. Similarly, the light-front singularity appearing
in $\mathcal{M}^{+,+,+,+}_{b,t}$ for $\theta \approx 0$ due to $p_4^+ \approx 0$ in the (b) time-ordered process is cancelled by the same with the opposite sign in 
$\mathcal{M}^{+,+,+,+}_{a,u}$. Thus, the total helicity amplitude summing all
the $t$ and $u$ channel time-ordered amplitudes, i.e. $\mathcal{M}^{+,+,+,+}_{a,t}+\mathcal{M}^{+,+,+,+}_{b,t}+\mathcal{M}^{+,+,+,+}_{a,u}+\mathcal{M}^{+,+,+,+}_{b,u}$,
is free from any singular behavior as shown in Fig.~\ref{fig:JoinedFigamp}. 
One may also notice that the effect of the overall sign change between $\epsilon^+_{\wh{\mu}}$ and $\epsilon^-_{\wh{\mu}}$, i.e. $(\epsilon^+_{\wh{\mu}})^* = - 
\epsilon^-_{\wh{\mu}}$, in Eq.~(\ref{photon_polarization}) is reflected in the negative vs. positive sign difference of the helicity amplitudes and 
ultimately the light-front singularity between $\mathcal{M}^{+,+,+,+}_{a,t}$ and $\mathcal{M}^{+,+,-,+}_{a,t}$ as shown in Fig.~\ref{fig:JoinedFigmtaandtb}.
Similar to the cancellation of the light-front singularity between $\mathcal{M}^{+,+,+,+}_{a,t}$ and $\mathcal{M}^{+,+,+,+}_{a,t}$,
the light-front singularity in $\mathcal{M}^{+,+,-,+}_{a,t}$ turns out to be cancelled by the same with the opposite sign in $\mathcal{M}^{+,+,-,+}_{b,u}$ as shown in Fig.~\ref{fig:JoinedFigmuaandub}.  Again, the total helicity amplitude summing all
the t and u channel time-ordered amplitudes, i.e. $\mathcal{M}^{+,+,-,+}_{a,t}+\mathcal{M}^{+,+,-,+}_{b,t}+\mathcal{M}^{+,+,-,+}_{a,u}+\mathcal{M}^{+,+,-,+}_{b,u}$,
is completely free from any singular behavior as shown in Fig.~\ref{fig:JoinedFigamp}. 
However, one should also note that the survival of this singular behavior depends on the time-ordering of the process as well as the helicities of the particles in the process 
as not only the longitudinal component but also the transverse component of the polarization vector also matters in affecting the removal or survival of the singular behavior in the helicity amplitude. As an example, one can see that the singular behavior from the zero-mode $p_3^+ \approx 0$ for $\theta \approx \pi$ is removed in $\mathcal{M}^{+,+,+,-}_{a,t}$
while it shows up in $\mathcal{M}^{+,+,+,+}_{a,t}$ and $\mathcal{M}^{+,+,-,+}_{a,t}$. 
As we have already discussed, the reason why all of the four t-channel (a) time-ordered helicity amplitudes ($\mathcal{M}^{+,+,+,+}_{a,t}$, $\mathcal{M}^{+,+,+,-}_{a,t}$, $\mathcal{M}^{+,+,-,+}_{a,t}$, $\mathcal{M}^{+,+,-,-}_{a,t}$) with the initial $++$ helicity configuration of $e^+ e^-$ pair annihilation vanish for $\theta < \theta_{c,t} = \pi/6 \approx 0.523599$
is because the region $\theta < \theta_{c,t} = \pi/6 \approx 0.523599$ belongs to the ``instantaneous fermion contribution" and 
$\bar{v}^{\uparrow}\gamma^+ u^{\uparrow}=0$, i.e. the $\gamma^+$ operator of the instantaneous contribution in LFD cannot 
link between the initial electron and positron pair with the same helicity. Conversely, in the t-channel (b) time-ordered process,
the region of ``instantaneous fermion contribution" is for $\theta > \theta_{c,t} = \pi/6 \approx 0.523599$ and 
all of the four t-channel (b) time-ordered helicity amplitudes ($\mathcal{M}^{+,+,+,+}_{b,t}$, $\mathcal{M}^{+,+,+,-}_{b,t}$, $\mathcal{M}^{+,+,-,+}_{b,t}$, $\mathcal{M}^{+,+,-,-}_{b,t}$) with the initial $++$ helicity configuration of $e^+ e^-$ pair annihilation vanish in the region $\theta > \theta_{c,t} = \pi/6 \approx 0.523599$ as shown in
Fig.~\ref{fig:JoinedFigmtb}. In the region $\theta < \theta_{c,t} = \pi/6 \approx 0.523599$, however, these amplitudes are non-vanishing and 
$\mathcal{M}^{+,+,+,+}_{b,t}$ and $\mathcal{M}^{+,+,+,-}_{b,t}$ exhibit even the singular behavior near $\theta \approx 0$ due to the $p_4^+ \approx 0$ zero-mode as depicted
in Fig.~\ref{fig:JoinedFigmtb}. 

Having discussed the helicity amplitudes $\mathcal{M}^{+,+,\lambda_3,\lambda_4}_{a,t}+\mathcal{M}^{+,+,\lambda_3,\lambda_4}_{b,t}+\mathcal{M}^{+,+,\lambda_3,\lambda_4}_{a,u}+\mathcal{M}^{+,+,\lambda_3,\lambda_4}_{b,u}$ in Fig.~\ref{fig:JoinedFigamp}, we note here the IFD/LFD profile correspondence similar to what we have noted 
in Fig.~\ref{fig:eessAngDistAmpAndProb}. Namely, for the outgoing photon helicities $\lambda_3$ and $\lambda_4$, the IFD profile of the incident $+ +$ helicity amplitude
corresponds to the LFD profile of the incident $+ -$ helicity amplitude modulo overall signs of the helicity amplitudes, and vice versa. 
While the reason for this correspondence is again partly due to the swap of the helicity between the IFD and LFD for the incident positron moving in the $-\hat z$ direction
as we have already discussed for the results of Fig.~\ref{fig:eessAngDistAmpAndProb} in Section~\ref{sec:eess}, we should also note the interesting characteristic of the outgoing
real photon helicities $\lambda_3$ and $\lambda_4$. 
The relationship between the LF helicity and the Jacob-Wick helicity defined in the IFD is generally given by a Wigner rotation~\cite{Carlson-Ji}. For the massless particle such as the real photon, the relationship gets particularly simplified as unity unless the massless particle is  
moving in the $-\hat z$ direction. Thus, for the region $0< \theta < \pi$ without involving exact boundary values of $\theta =0$ and $\theta = \pi$, the LF helicity and 
the Jacob-Wick helicity coincide so that there is no difference between the LF helicity and the Jacob-Wick helicity for the real photons. 
For this reason, the helicity amplitude $\mathcal{M}^{+,+,\lambda_3,\lambda_4}_{a,t}+\mathcal{M}^{+,+,\lambda_3,\lambda_4}_{b,t}+\mathcal{M}^{+,+,\lambda_3,\lambda_4}_{a,u}+\mathcal{M}^{+,+,\lambda_3,\lambda_4}_{b,u}$ in IFD/LFD corresponds to $\mathcal{M}^{+,-,\lambda_3,\lambda_4}_{a,t}+\mathcal{M}^{+,-,\lambda_3,\lambda_4}_{b,t}+\mathcal{M}^{+,-,\lambda_3,\lambda_4}_{a,u}+\mathcal{M}^{+,-,\lambda_3,\lambda_4}_{b,u}$ in LFD/IFD, respectively, for the region $0< \theta < \pi$.
As an example, in Fig.~\ref{fig:JoinedFigamp}, the correspondence between the profile of the total amplitude 
$\mathcal{M}^{+,-,+,-}_{a,t} + \mathcal{M}^{+,-,+,-}_{b,t} + \mathcal{M}^{+,-,+,-}_{a,u} + \mathcal{M}^{+,-,+,-}_{b,u}$ in LFD and 
the profile of the total amplitude $\mathcal{M}^{+,+,+,-}_{a,t} + \mathcal{M}^{+,+,+,-}_{b,t} + \mathcal{M}^{+,+,+,-}_{a,u} + \mathcal{M}^{+,+,+,-}_{b,u}$ in IFD is manifest.
Likewise, the IFD/LFD profile correspondence of the probability for each and every $\{\lambda_3,\lambda_4\}$ pair of photon helicities is self-evident as shown
in Fig.~\ref{fig:JoinedFigprob}. 

For the exact boundary values $\theta =0$ and $\theta = \pi$, one of the outgoing real photons moves in the $-\hat z$ direction
and thus the only care that one has to take is to swap the values of the LF helicity amplitudes according to the correspondence between the LF helicity and 
the Jacob-Wick helicity defined in the IFD as discussed above for the particle moving in the $-\hat z$ direction. For $\theta=0$, $p_3 = E_0 (1,0,0,1)$ and $p_4 = E_0 (1,0,0,-1)$
in the CMF kinematics given by Eq.~(\ref{eqn:p1234_Annihilation}). Thus, the Jacob-Wick helicity pair $\{\lambda_3,\lambda_4\}$ in IFD corresponds to 
the LF helicity pair $\{\lambda_3,-\lambda_4\}$ in LFD at exact $\theta=0$. Likewise, the Jacob-Wick helicity pair $\{\lambda_3,\lambda_4\}$ in IFD corresponds to 
the LF helicity pair $\{-\lambda_3,\lambda_4\}$ in LFD at exact $\theta=\pi$. This treacherous point of the LF helicity identification at the exact boundary values
of $\theta =0$ and $\pi$ can be analyzed with the care of procedure in taking massless limit ($M \to 0$) for the gauge boson polarization vector given by Eq.~(\ref{photon_polarization}) and the details of analysis will be presented elsewhere. In this work, although we keep in mind of the treacherous LF helicity identification
at the exact boundary values, we present our work focusing on the region $0< \theta < \pi$ without involving the exact boundary values of $\theta =0$ and $\theta = \pi$.

	\begin{figure}
		\centering
		\subfloat[]{
			\includegraphics[width=0.48\columnwidth]{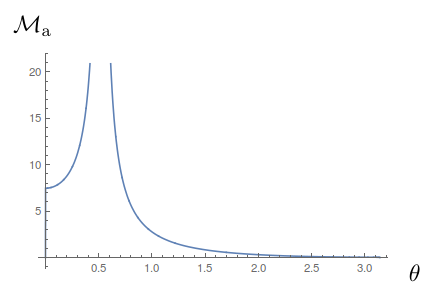}
			\label{fig:ta_plot}}
		\centering
		\subfloat[]{
			\includegraphics[width=0.48\columnwidth]{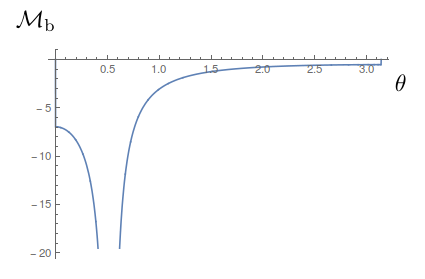}
			\label{fig:tb_plot}}
		\caption{	\label{fig:tatbLF}(a) Profile of the t-channel (a) time-ordered annihilation amplitude in LFD for ``$ +-\to+- $",
		(b) Profile of the t-channel (b) time-ordered annihilation amplitude in LFD for ``$ +-\to+- $''}
	\end{figure}
	\begin{figure}
		\centering
		\subfloat[]{
			\includegraphics[width=0.48\columnwidth]{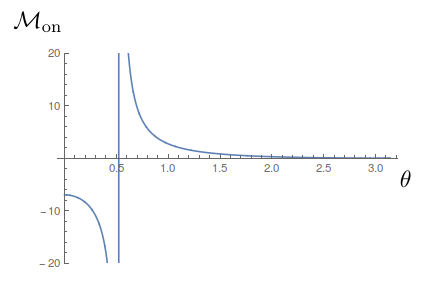}
			\label{fig:onshell_plot}}
		\centering
		\subfloat[]{
			\includegraphics[width=0.48\columnwidth]{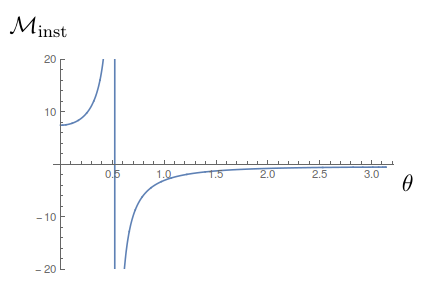}
			\label{fig:inst_plot}}
		\caption{	\label{fig:oninstLF}Time-ordered annihilation amplitudes (a) on-shell and (b) instantaneous contributions at LF for ``$ +-\to+- $''}
	\end{figure}
	\begin{figure}
		\centering
			\includegraphics[width=0.48\columnwidth]{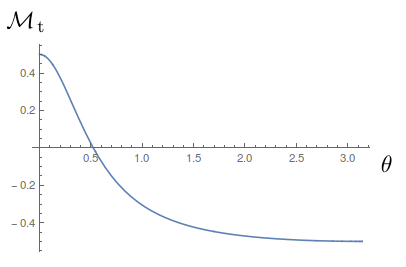}			
		\caption{\label{fig:Mot+-+-}Sum of time-ordered annihilation amplitudes  
		in LFD for ``$ +-\to+- $''}
	\end{figure}
The result of the total amplitude $\mathcal{M}^{+,-,\lambda_3,\lambda_4}_{a,t} + \mathcal{M}^{+,-,\lambda_3,\lambda_4}_{b,t} + \mathcal{M}^{+,-,\lambda_3,\lambda_4}_{a,u} + \mathcal{M}^{+,-,\lambda_3,\lambda_4}_{b,u}$ may be further analyzed by taking a look at each channel and time-ordered process separately shown in Figs.~\ref{fig:JoinedFigmtaandtb} and ~\ref{fig:JoinedFigmuaandub} for the region $0< \theta <\pi$. 
For  $\mathcal{M}^{+,-,\lambda_3,\lambda_4}_{a,t}$ with  the initial $e^+ e^-$ helicity pair $\{+ -\}$ and the final four helicity configurations of the photon pairs $\{\lambda_3,\lambda_4\} = \{+,+\}, \{+,-\}, \{-,+\}, \{-,-\}$ depicted in the second column of Fig.~\ref{fig:JoinedFigmtaandtb}, 
one may compare the result with Fig.~\ref{fig:eessAngDistRLta} which shows the helicity amplitude of the pair production of scalar particles with 
the same initial $+-$ helicity configuration of $e^+ e^-$ pair annihilation. Not all of the four t-channel (a) time-ordered helicity amplitudes with the initial $+-$ helicity configuration of $e^+ e^-$ pair annihilation vanish for $\theta < \theta_{c,t} = \pi/6 \approx 0.523599$ in LFD although the region $\theta < \theta_{c,t} = \pi/6 \approx 0.523599$ belongs to 
the light-front ``instantaneous fermion contribution", because $\bar{v}^{\downarrow}\gamma^+ u^{\uparrow}\neq0$, i.e. the $\gamma^+$ operator of the instantaneous contribution in LFD can link between the initial electron and positron pair with the opposite helicity. For example, $\mathcal{M}^{+,-,+,-}_{a,t}$ is clearly nonzero as shown in the second figure from the top of the second column of Fig.~\ref{fig:JoinedFigmtaandtb}. Depending on the final photon helicities, however, the amplitude can still vanish as in the case of $\mathcal{M}^{+,-,-,+}_{a,t}$ and $\mathcal{M}^{+,-,-,-}_{a,t}$. Moreover, it is interesting to note the dramatic rise of the amplitude $\mathcal{M}^{+,-,+,+}_{a,t}$ as the scattering/annihilation process becomes collinear ($\theta \approx 0$) due to the light-front zero-mode $p_4^+ \approx 0$ yielding nonzero finite amplitude although the amplitude $\mathcal{M}^{+,-,+,+}_{a,t}$ appears to vanish for the region $0< \theta < \theta_{c,t} = \pi/6 \approx 0.523599$ as depicted in the top figure of the second column of Fig.~\ref{fig:JoinedFigmtaandtb}. 
In particular, the profile of $\mathcal{M}^{+,-,+,-}_{a,t}$ and $\mathcal{M}^{+,-,+,-}_{b,t}$ in LFD appears as shown in Fig.~\ref{fig:tatbLF}.
In LFD, as discussed previously, the regions $0 <\theta< \theta_{c,t} = \pi/6 \approx 0.523599$ 
and $\theta_{c,t} = \pi/6 \approx 0.523599 <\theta< \pi$ provide the ``instantaneous fermion contribution" and the 
``on-mass-shell propagating contribution" for the light-front (a) time-ordered amplitude, while the regions are swapped for the light-front (b) time-ordered amplitude.
As shown in Fig.~\ref{fig:oninstLF}, one may collect the ``instantaneous fermion contribution" and the ``on-mass-shell propagating contribution"
by themselves separately to show the combined (a)+(b) time-ordered amplitude. Whichever way we present the result, both Figs.~\ref{fig:tatbLF} and \ref{fig:oninstLF} manifest the cancellation of the light-front singular features and yield the finite total t-channel amplitude as shown in Fig.~\ref{fig:Mot+-+-}. We note that the total t-channel amplitude at $ \theta=\theta_{c,t}=\pi/6 \approx 0.523599$ is zero. As $\theta \to \theta_{c,t}$, $ q_t^+ = p_1^+ - p_3^+ \to 0 $ and the interaction behaves as if a contact interaction  while the propagator shrinks to a point. For the case of contact interaction, squaring the diagram can yield either a fermion loop or a boson loop. Due to the $ (-1) $ factor difference between the fermion loop and the boson loop, the only consistent value of the 
amplitude square must be zero for the contact interaction. This reasoning may offer the understanding of zero amplitude at 
$ \theta=\theta_{c,t}=\pi/6 \approx 0.523599$ in Fig.~\ref{fig:Mot+-+-}.

Likewise, the u-channel helicity amplitudes $\mathcal{M}^{+,-,\lambda_3,\lambda_4}_{a,u}$ and $\mathcal{M}^{+,-,\lambda_3,\lambda_4}_{b,u}$ shown in Fig.~\ref{fig:JoinedFigmuaandub} can be understood by realizing the symmetry under the exchange of the outgoing pair of the photons as well as the forward-backward correspondence $\theta \leftrightarrow \pi-\theta$. It may not be too difficult to see the $\theta \leftrightarrow \pi-\theta$ correspondence between 
$\mathcal{M}^{+,-,\pm,\pm}_{a,u}$ and  $\mathcal{M}^{+,-,\pm,\pm}_{a,t}$ as well as $\mathcal{M}^{+,-,\pm,\pm}_{b,u}$ and  $\mathcal{M}^{+,-,\pm,\pm}_{b,t}$ modulo overall sign change of the amplitudes in the IFD side ($0<\delta<\delta_c \approx 0.713724$). The similar correspondence between 
$\mathcal{M}^{+,-,\pm,\mp}_{a,u}$ and  $\mathcal{M}^{+,-,\mp,\pm}_{a,t}$ as well as $\mathcal{M}^{+,-,\pm,\mp}_{b,u}$ and  $\mathcal{M}^{+,-,\mp,\pm}_{b,t}$ can be observed
without much difficulty comparing Figs.~\ref{fig:JoinedFigmtaandtb} and ~\ref{fig:JoinedFigmuaandub}. 
It is evident that the same symmetry is inherited in the sum of the amplitudes presented in Fig.~\ref{fig:JoinedFigamp}
as one can see the $\theta \leftrightarrow \pi-\theta$ symmetry in the $\theta \leftrightarrow \pi-\theta$ correspondence between 
$\mathcal{M}^{+,\pm,\lambda_3,\lambda_4}_{a,t} + \mathcal{M}^{+,\pm,\lambda_3,\lambda_4}_{b,t}+\mathcal{M}^{+,\pm,\lambda_3,\lambda_4}_{a,u} + \mathcal{M}^{+,\pm,\lambda_3,\lambda_4}_{b,u}$
and $\mathcal{M}^{+,\pm,\lambda_4,\lambda_3}_{a,t} + \mathcal{M}^{+,\pm,\lambda_4,\lambda_3}_{b,t}+\mathcal{M}^{+,\pm,\lambda_4,\lambda_3}_{a,u} + 
\mathcal{M}^{+,\pm,\lambda_4,\lambda_3}_{b,u}$ for any photon helicity $\lambda_3$ and $\lambda_4$. 
Due to Eq.~(\ref{eqn:amplitude_relation_Annihilation}), the same correspondence applies to 
$\mathcal{M}^{-,\pm,\lambda_3,\lambda_4}_{a,t} + \mathcal{M}^{-,\pm,\lambda_3,\lambda_4}_{b,t}+\mathcal{M}^{-,\pm,\lambda_3,\lambda_4}_{a,u} + \mathcal{M}^{-,\pm,\lambda_3,\lambda_4}_{b,u}$
and $\mathcal{M}^{-,\pm,\lambda_4,\lambda_3}_{a,t} + \mathcal{M}^{-,\pm,\lambda_4,\lambda_3}_{b,t}+\mathcal{M}^{-,\pm,\lambda_4,\lambda_3}_{a,u} + 
\mathcal{M}^{-,\pm,\lambda_4,\lambda_3}_{b,u}$ as well.

\begin{figure}
	\centering
		\subfloat[]{
			\includegraphics[width=0.48\columnwidth]{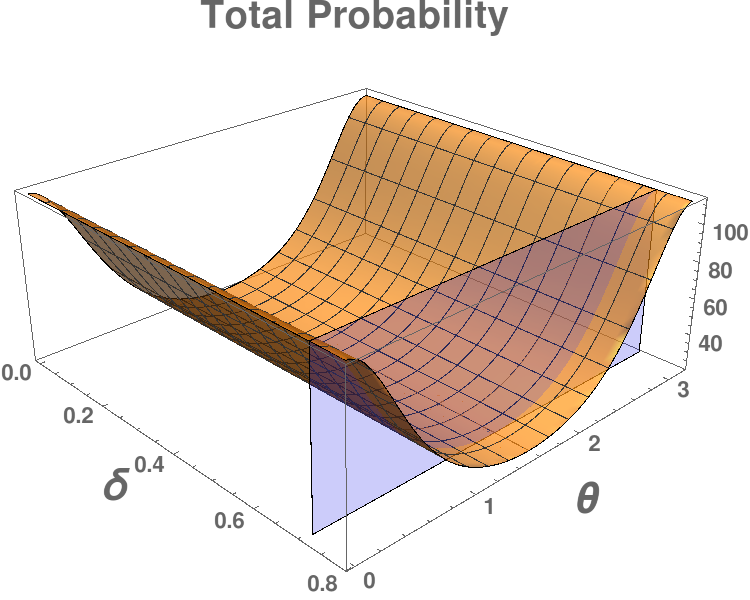}
			\label{fig:Prob}}
	\centering
		\subfloat[]{\includegraphics[width=0.48\columnwidth]{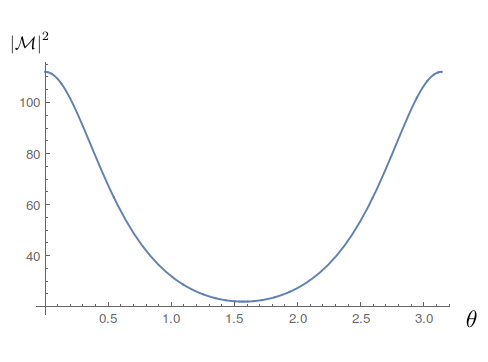}\label{fig:Annihilation_totalprob_theo}}
	\caption{\label{fig:TwoGammaHelicityProbabilitySumCMF}$e^+ e^- \to \gamma \gamma$ (a) Sum of helicity probabilities for $P^z = 0$ (CMF), 
	(b) Result from the manifestly Lorentz invariant formula given by Eq.~(\ref{eqn:spin_averaged_amplitude_Annihilation})}
	\end{figure}
	

The corresponding probabilities $|\mathcal{M}^{+,\pm,\lambda_3,\lambda_4}_{a,t} + \mathcal{M}^{+,\pm,\lambda_3,\lambda_4}_{b,t}+\mathcal{M}^{+,\pm,\lambda_3,\lambda_4}_{a,u} + \mathcal{M}^{+,\pm,\lambda_3,\lambda_4}_{b,u}|^2$
and $|\mathcal{M}^{+,\pm,\lambda_4,\lambda_3}_{a,t} + \mathcal{M}^{+,\pm,\lambda_4,\lambda_3}_{b,t}+\mathcal{M}^{+,\pm,\lambda_4,\lambda_3}_{a,u} + 
\mathcal{M}^{+,\pm,\lambda_4,\lambda_3}_{b,u}|^2$ shown in Fig.~\ref{fig:JoinedFigprob} of course exhibits the same symmetry with the definite positive sign everywhere.
The bottom two figures in Fig.~\ref{fig:JoinedFigprob} summing the final helicities, $\sum\limits_{\lambda_3,\lambda_4}|\mathcal{M}^{+,\pm,\lambda_3,\lambda_4}_{a,t} + \mathcal{M}^{+,\pm,\lambda_3,\lambda_4}_{b,t}+\mathcal{M}^{+,\pm,\lambda_3,\lambda_4}_{a,u} + \mathcal{M}^{+,\pm,\lambda_3,\lambda_4}_{b,u}|^2$, exhibit the 
swap of the helicity between the IFD and LFD for the particle moving in the $-\hat z$ direction which we have discussed previously. Namely, the IFD result 
of $\sum\limits_{\lambda_3,\lambda_4}|\mathcal{M}^{+,\pm,\lambda_3,\lambda_4}_{a,t} + \mathcal{M}^{+,\pm,\lambda_3,\lambda_4}_{b,t}+\mathcal{M}^{+,\pm,\lambda_3,\lambda_4}_{a,u} + \mathcal{M}^{+,\pm,\lambda_3,\lambda_4}_{b,u}|^2$ is identical to the LFD result of $\sum\limits_{\lambda_3,\lambda_4}|\mathcal{M}^{+,\mp,\lambda_3,\lambda_4}_{a,t} + \mathcal{M}^{+,\mp,\lambda_3,\lambda_4}_{b,t}+\mathcal{M}^{+,\mp,\lambda_3,\lambda_4}_{a,u} + \mathcal{M}^{+,\mp,\lambda_3,\lambda_4}_{b,u}|^2$ and vice versa. By adding the two initial helicity states as well, we may now compare our total result
with the well-known manifestly Lorentz invariant result given by 
	\begin{widetext}
		\begin{align}\label{eqn:spin_averaged_amplitude_Annihilation}
		\left| \mathcal{M} (e^+e^- \to \gamma \gamma) \right|^2 
		&\equiv\sum_{\lambda_1,\lambda_2,\lambda_3,\lambda_4} 
		|\mathcal{M}^{\lambda_1,\lambda_2,\lambda_3,\lambda_4}_{a,t} + \mathcal{M}^{\lambda_1,\lambda_2,\lambda_3,\lambda_4}_{b,t}+\mathcal{M}^{\lambda_1,\lambda_2,\lambda_3,\lambda_4}_{a,u} + \mathcal{M}^{\lambda_1,\lambda_2,\lambda_3,\lambda_4}_{b,u}|^2 \notag \\
		&= 2 \sum_{\lambda_2,\lambda_3,\lambda_4}|\mathcal{M}^{+,\lambda_2,\lambda_3,\lambda_4}_{a,t} + \mathcal{M}^{+,\lambda_2,\lambda_3,\lambda_4}_{b,t}+\mathcal{M}^{+,\lambda_2,\lambda_3,\lambda_4}_{a,u} + \mathcal{M}^{+,\lambda_2,\lambda_3,\lambda_4}_{b,u}|^2 \notag \\
		&= 8 \left[\frac{u_m}{t_m}+\frac{t_m}{u_m}+2m^2\left( \frac{s_m}{t_mu_m}-\frac{1}{t_m}-\frac{1}{u_m}\right)-4m^4\left( \frac{1}{t_m^2}+\frac{1}{u_m^2}\right)   \right] ,
		\end{align}
	\end{widetext}
where $s_m=s-4m^2$, $t_m=t-m^2$, $u_m=u-m^2$ and the electric charge factor is taken to be one.
Taking the specific values, $m=m_e$, $s= 16 m_e^2$, $t= (-7 + 4 \sqrt{3} \cos \theta)m_e^2$ and $u= -(7 + 4 \sqrt{3} \cos \theta)m_e^2$,
given just below Eq.~(\ref{eqn:scalar_cross_section}) for our numerical calculation in CMF, we find that 
the two results, (a) the twice of summing the bottom two figures in Fig.~\ref{fig:JoinedFigprob} and (b)
the analytic result given by Eq.~(\ref{eqn:spin_averaged_amplitude_Annihilation}) coincide each other  
as shown in Fig.~\ref{fig:TwoGammaHelicityProbabilitySumCMF}. The result shown in the left panel of Fig.~\ref{fig:TwoGammaHelicityProbabilitySumCMF}
is of course completely independent of the interpolation angle $\delta$ as it should be.
The analytic result in Eq.~(\ref{eqn:spin_averaged_amplitude_Annihilation}) is apparently symmetric under $t \leftrightarrow u$ exchange as it must be and 
gets reduced to the well-known textbook result~\cite{Halzen-Martin} in the massless limit ($m \to 0$) given by 
		\begin{equation}\label{eqn:spin_averaged_amplitude_neg_Annihilation}
	\left| \mathcal{M} (e^+e^- \to \gamma \gamma) \right|^2 =8\left(\frac{u}{t}+\frac{t}{u} \right) .
	\end{equation}
It may be interesting to compare this result with the massless limit of Eq.~(\ref{eqn:scalar_cross_section}) for the pair production of spinless particles (or ``scalar photons")
given by  
		\begin{equation}\label{eqn:spin_averaged_amplitude_neg_scalarAnnihilation}
		\left| \mathcal{M} \right|^2_{\rm{scalar}} =2\left(\frac{u}{t}+\frac{t}{u}-2 \right) ,
		\end{equation}
where the normalization is reduced by the factor 4 due to the lack of final spin (or helicity) degrees of freedom. 
When $ t=u $, i.e. $ \theta=\pi/2 $ in the massless limit of the initial fermions, we may note that the probability of producing two ``scalar photons" is zero
while the probability of producing two real photons is non-zero.
This may be understood from the fact that the two final ``scalar photons" do not carry enough number of degrees of freedom
while the real photon carries the transverse spin-1 polarization to offer the matching of the number of degrees of freedom
between the initial and final states involving both spin singlet and triplet configurations in the annihilation/production process. 

As we have now shown that the square of the sum of all the individual channel and time-ordered helicity amplitudes in CMF ($P^z=0$) is identical to the completely Lorentz-invariant expression in terms of the Mandelstam variables ($s,t,u$),
we are assured that our CMF result in Eq.~(\ref{eqn:spin_averaged_amplitude_Annihilation}) 
must be reproduced even if each individual channel and time-ordered helicity amplitudes are computed in other boosted
frames, e.g. $P^z=15m_e$ or $P^z=-15m_e$. Nevertheless, each individual amplitudes are not boost invariant except 
the LFD ($\delta=\pi/4$) profiles. The IFD ($\delta=0$) profiles in the $P^z=-15m_e$ are vastly different not only from the corresponding IFD ($\delta=0$) profiles in the $P^z=15m_e$ but also from the corresponding LFD ($\delta=\pi/4$) profiles. 
As we have already discussed in Sec.~\ref{sec:eess}, it requires a great caution in the prevailing notion of the equivalence
between the IFD in IMF and the LFD. The results in the boosted frames ($P^z = 15 m_e$ and $P^z = -15 m_e$) are summarized 
in the Appendix \ref{app:BoostedAnnihilation}. We have also shown the $P^z$ dependence of the interpolating helicity amplitudes for 
a particular scattering, e.g. $\theta = \pi/3$ case in the Appendix \ref{app:PzDepThetaPiOver3}. 
	
	\subsection{\label{sub:compton}Compton Scattering}

	\begin{figure}
		\centering
		\subfloat[]{\includegraphics[width=0.45\columnwidth]{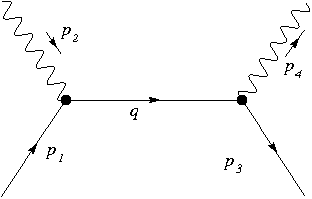}\label{fig:RealComptonScattering_schannel}}
			\hspace{10pt}
		\centering
		\subfloat[]{\includegraphics[width=0.45\columnwidth]{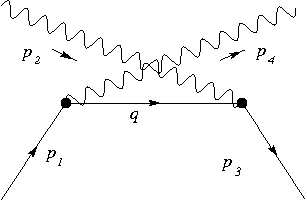}\label{fig:RealComptonScattering_uchannel}}
		\caption{\label{fig:RealComptonScattering_Feynman} s-channel and u-channel Feynman diagrams for Compton scattering}
	\end{figure}

	\begin{figure}
		\centering
		\subfloat[]{\includegraphics[width=0.49\columnwidth]{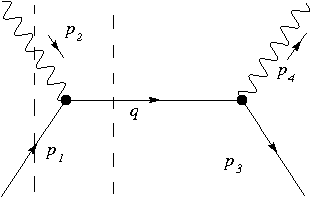}\label{fig:RealComptonScattering_schannel_a}}
		\centering
		\subfloat[]{\includegraphics[width=0.49\columnwidth]{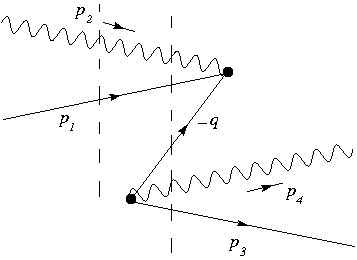}\label{fig:RealComptonScattering_schannel_b}}		
\\
		\centering
		\subfloat[]{\includegraphics[width=0.49\columnwidth]{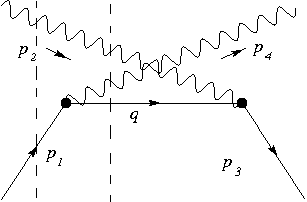}\label{fig:RealComptonScattering_uchannel_a}}
		\vspace{0pt}
		\centering
		\subfloat[]{\includegraphics[width=0.49\columnwidth]{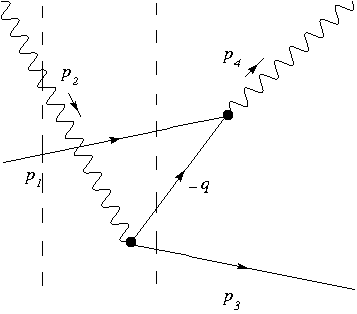}\label{fig:RealComptonScattering_uchannel_b}}
		\caption{\label{fig:RealComptonScattering_s&u_channel_TO}Time-ordered diagrams for 
		s and u channel Compton scattering}
	\end{figure}

Another important physical scattering processes in QED which involves the fermion propagator in the lowest order
is the Compton scattering $e \gamma \to e \gamma$. Similar to the $e^+ e^- \to \gamma \gamma$ process shown in 
Fig.~\ref{fig:PairAnnihilationto2s_tchannel} which we have extensively discussed in the previous subsection, 
the lowest Feynman diagrams for the Compton scattering process is shown in Fig.~\ref{fig:RealComptonScattering_Feynman}.
For the obvious reason from the Compton kinematics and the corresponding Mandelstam variables given by 
	\begin{align}
	s&=(p_1+p_2)^2=2p_1\cdot p_2+m^2\\
	t&=(p_1-p_3)^2=-2p_1\cdot p_3+2m^2\\
	u&=(p_1-p_4)^2=-2p_1\cdot p_4+m^2,
	\end{align}
 we call the diagram shown in Fig.~\ref{fig:RealComptonScattering_schannel} as the s-channel diagram and 
 the crossed diagram in Fig.~\ref{fig:RealComptonScattering_uchannel} as the u-channel diagram. 

 The s-channel Feynman diagram is then equivalent to 
 the sum of the top two time-ordered diagrams (a) and (b) 
 shown in Fig.~\ref{fig:RealComptonScattering_s&u_channel_TO}.
 Similarly, the two time-ordered diagrams for the u-channel (c) and (d) 
 are shown in the bottom of Fig.~\ref{fig:RealComptonScattering_s&u_channel_TO}.
 For clarity and simplicity, we call these two u-channel time-ordered diagrams as 
 the u-channel (a) and (b) time-ordered diagrams for the rest of presentation.

Now, the s-channel and u-channel Compton amplitudes are given by  
	\begin{align}
		\mathcal{M}_s^{\lambda_1,\lambda_2,\lambda_3,\lambda_4} &=\bar{u}^{\lambda_3}(p_3)\epsilon_{\wh{\nu}}^{\lambda_4}(p_4)^*\gamma^{\wh{\nu}}\Sigma_s \gamma^{\wh{\mu}}  \epsilon_{\wh{\mu}}^{\lambda_2}(p_2) u^{\lambda_1}(p_1), \notag \\
		\mathcal{M}_u^{\lambda_1,\lambda_2,\lambda_3,\lambda_4} &=\bar{u}^{\lambda_3}(p_3)
		 \epsilon_{\wh{\mu}}^{\lambda_2}(p_2) \gamma^{\wh{\mu}} \Sigma_u 
		\epsilon_{\wh{\nu}}^{\lambda_4}(p_4)^*\gamma^{\wh{\nu}}  u^{\lambda_1}(p_1), 
		\label{eqn:sChannelCompton}
	\end{align}
where $\Sigma_s$ and $\Sigma_u$ are 
	\begin{equation}
	\Sigma_s = \frac{\slashed{q}_s + m}{s - m^2} \quad \mathrm{and}  \quad \Sigma_u= \frac{\slashed{q}_u + m}{u - m^2}
	 \label{eqn:s_Operator}
         \end{equation}	
	with $q_s = p_1 + p_2$ and $s=q_s^2$, while $q_u = p_1 - p_4$ and $u=q_u^2$. 
Then, the time-ordered amplitudes of the s-channel Compton scattering can be written in short-hand notations 
	without specifying the helicities as
	\begin{equation}
	\mathcal{M}_{a,s}=\bar{u}(p_3) \slashed{\epsilon}(p_4)^* \left( \frac{1}{2Q_s^{\wh{+}}}\frac{\slashed{Q}_{a,s}+m}{q_{s\wh{+}}-Q_{a,s\wh{+}}}\right)  \slashed{\epsilon}(p_2) u(p_1),
	\end{equation}
	and
	\begin{equation}
	\mathcal{M}_{b,s}=\bar{u}(p_3) \slashed{\epsilon}(p_4)^* \left( \frac{1}{2Q_s^{\wh{+}}}\frac{-\slashed{Q}_{b,s}+m}{-q_{s\wh{+}}-Q_{b,s\wh{+}}}\right) \slashed{\epsilon}(p_s) u(p_1),
	\end{equation}
where
$Q_{a,s\wh{+}}$ and $Q_{b,s\wh{+}}$ are the interpolating on-mass-shell energy of the intermediate propagating fermion given by 	
\begin{align} \label{eqn:qasnew+}
	Q_{a,s\wh{+}}&=\frac{-\mathbb{S}q_{a,s\wh{-}}+Q_s^{\wh{+}}}{\mathbb{C}},\\ \label{eqn:qbsnew+}
	Q_{b,s\wh{+}}&=\frac{-\mathbb{S}q_{b,s\wh{-}}+Q_s^{\wh{+}}}{\mathbb{C}},
	\end{align}
with $q_{a,s}=q_s$, $q_{b,s}=-q_s$ and $Q_s^{\wh{+}}$ denoting the on-mass-shell value of $q_s^{\wh{+}}$ as 
	\begin{equation}
	Q_s^{\wh{+}}\equiv\sqrt{q_{s\wh{-}}^{2}+\mathbb{C} (\mathbf{q}_{s\perp}^{2}+m^{2})}.
	\label{eqn:Q_s^+}
	\end{equation}
Similarly, the time-ordered amplitudes of the u-channel Compton scattering can be written by replacing the 
s-channel variables by the corresponding u-channel variables.  

In contrast to the time-ordered processes in $e^+ e^- \to \gamma \gamma$ (see Fig.~\ref{fig:PairAnnihilationto2s_tchannel_TO}), 
the s-channel time-ordered processes (a) and (b) in Compton scattering, $e \gamma \to e \gamma$, involve one-particle and five-particle Fock states, respectively, while both of the u-channel time-ordered processes (c) and (d) in Compton scattering involve three-particle Fock states as one can see in Fig.~\ref{fig:RealComptonScattering_s&u_channel_TO}.
In particular, the one-particle intermediate state in the s-channel time-ordered process (a) in Compton scattering
provides immediately the positivity of $q_s^+ = (s + \mathbf{q}_{s\perp}^{2})/(\frac{\mathbf{p}_{1\perp}^{2}+m^2}{p_1^+} +\frac{\mathbf{p}_{2\perp}^{2}}{p_2^+}) > 0$ no matter what the kinematics are chosen. There is no need to figure out the critical scattering angles as we have obtained in the case of the $e^+ e^- \to \gamma \gamma$ process such as Eqs.~(\ref{critical_angle_t}) and (\ref{critical_angle_u}). Regardless of kinematics the Compton scattering, the positivity of $q_s^+ >0$ allows the use of Eqs.~(\ref{eqn:LFinst}) and (\ref{eqn:LFon}) to identify immediately the ``on-mass-shell propagating contribution" and the ``instantaneous contribution" in LFD as corresponding to the s-channel time-ordered processes (a) and (b) in Fig.~\ref{fig:RealComptonScattering_s&u_channel_TO}, respectively. 
For the u-channel Compton scattering, however, the identification of the ``on-mass-shell propagating contribution" and the ``instantaneous contribution" in LFD depends on the kinematics similar to the $e^+ e^- \to \gamma \gamma$ case. 
Nevertheless, we note that the CMF kinematics in the Compton scattering allows the identification of the ``on-mass-shell propagating contribution" and the ``instantaneous contribution" in LFD as corresponding to the u-channel time-ordered processes (c) and (d) in Fig.~\ref{fig:RealComptonScattering_s&u_channel_TO}, respectively, regardless of the scattering angle.
For the immediate identification of the ``on-mass-shell propagating contribution" and the ``instantaneous contribution" in LFD
both for the s and u channels with the correspondence to the time-ordered processes shown in Fig.~\ref{fig:RealComptonScattering_s&u_channel_TO}, we choose the CMF in this work for the rest of the discussion on the Compton scattering. The well-known Klein-Nishina formular~\cite{KN} in the target rest frame and the Thomson limit in the low energy Compton scattering, etc. will be discussed separately elsewhere.   

The kinematics pictured in Fig.~\ref{fig:Annihilation_kinematics} can be applied in the Compton scattering and written as the following:
		\begin{align}\label{eqn:p1234_Compton_CMframe}
		p_1&=(E_0 ,0 ,0 ,P_e) \notag \\
		p_2&=(P_e ,0 ,0 , -P_e) \notag \\
		p_3&=(E_0 ,P_e\sin\theta ,0 ,P_e\cos\theta) \notag \\
		p_4&=(P_e ,-P_e\sin\theta ,0 ,-P_e\cos\theta),
		\end{align}
	where $ P_e=\sqrt{E_0^2-m_e^2} $. 
In this work, we discuss the whole landscape of Compton scattering with respect to the interpolation angle $\delta$
and the C.M. momentum $P^z$ to show the frame dependence of each and every 
time-ordered scattering amplitudes in both s and u channels. 
For the numerical calculation of the interpolating helicity amplitudes, we 
scale all the energy and momentum values by the electron mass as done previously
and take $ m=m_e $, $ E_0=2 m_e $ and $ \theta=\pi/3$. Any further discussion such as the angular distribution, 
the energy ($E_0$) dependence, etc. in CMF will be presented together with the discussion of the target rest frame 
elsewhere as mentioned earlier.
	\begin{figure*}
		\centering
		\includegraphics[width=1.0\linewidth]{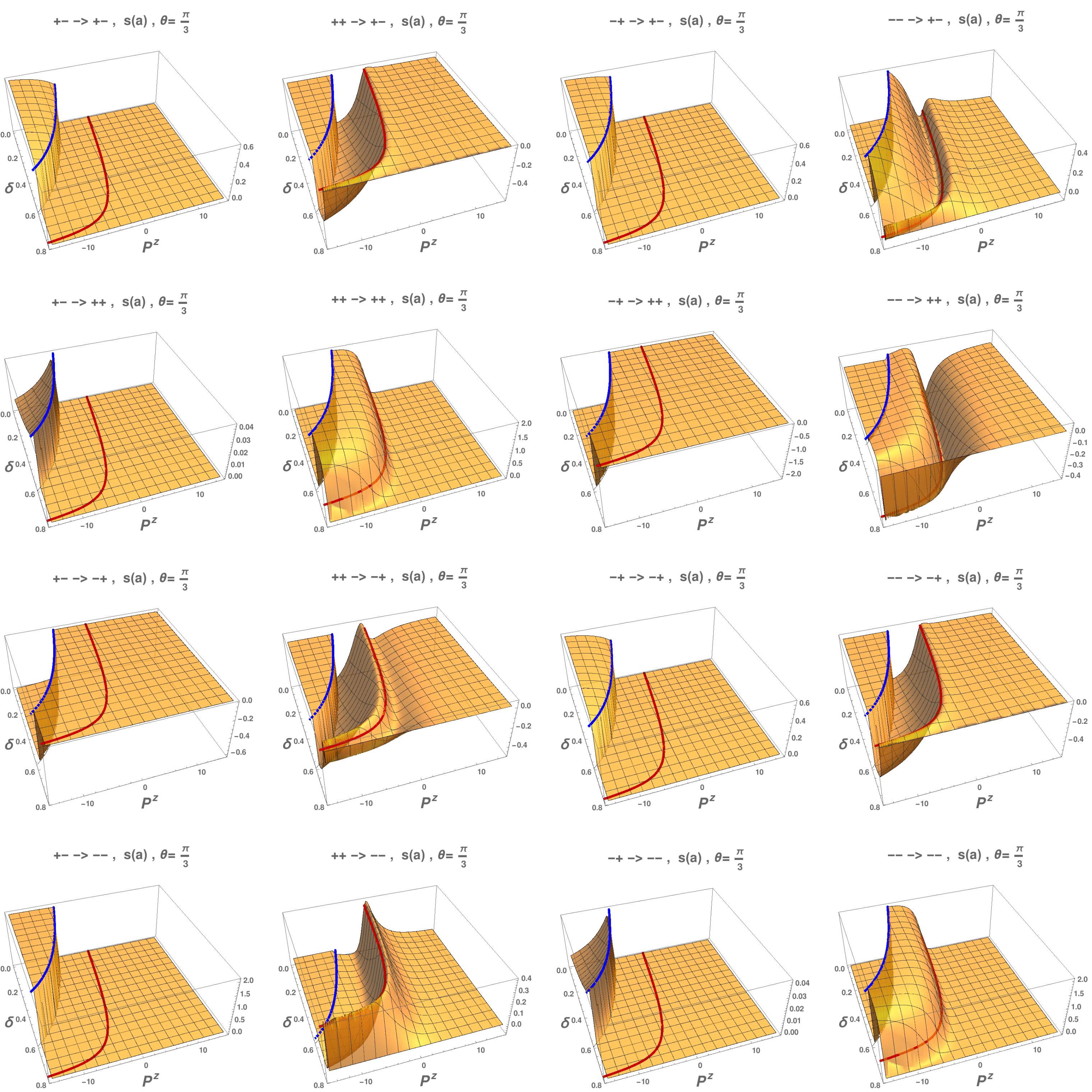}
		\caption{Compton Scattering Amplitudes --- s channel, time-ordering (a)}
		\label{fig:Compton_JoinedFigmsa}
	\end{figure*}
	\begin{figure*}
		\centering
		\includegraphics[width=1.0\linewidth]{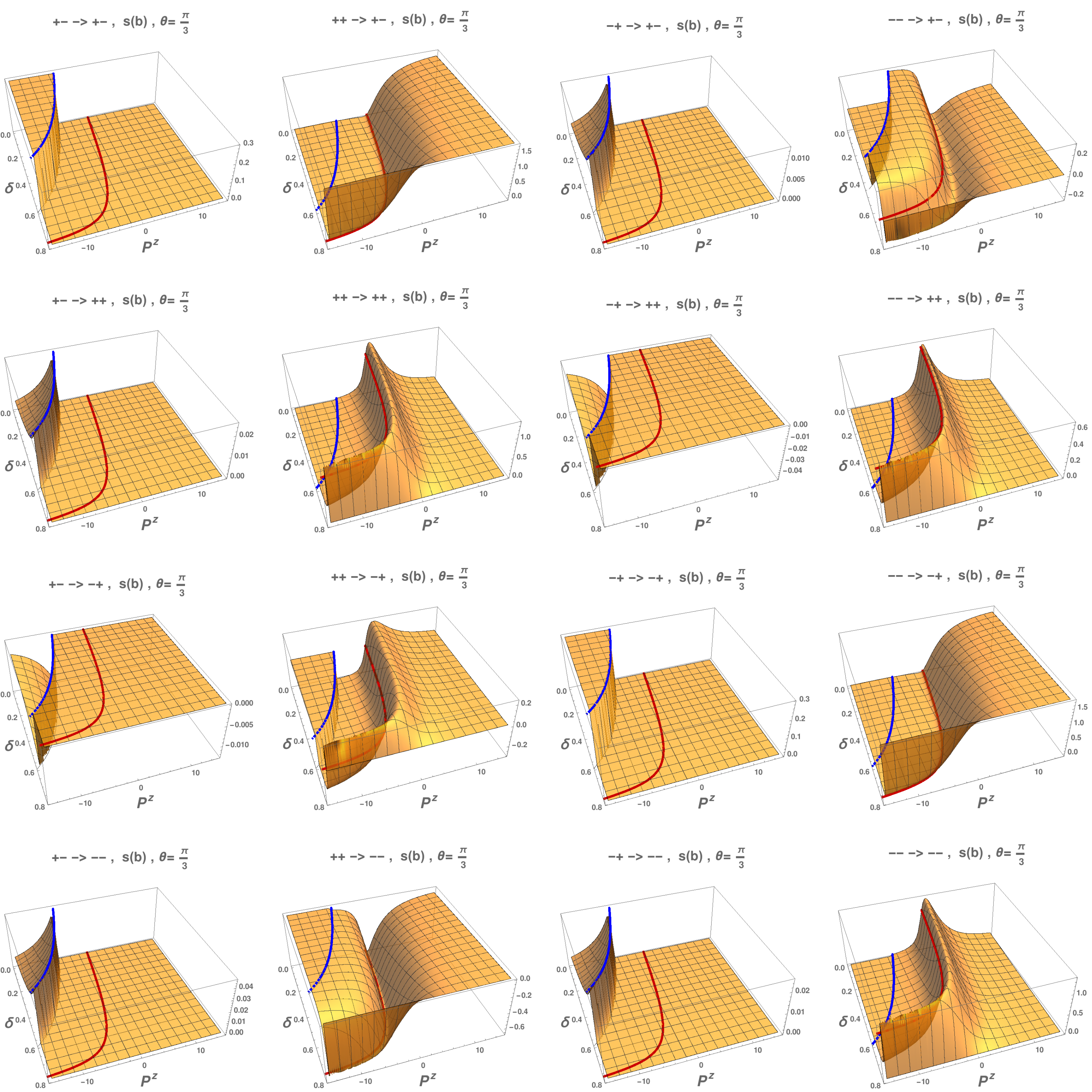}
		\caption{Compton Scattering Amplitudes --- s channel, time-ordering (b)}
		\label{fig:Compton_JoinedFigmsb}
	\end{figure*}
	\begin{figure*}
		\centering
		\includegraphics[width=1.0\linewidth]{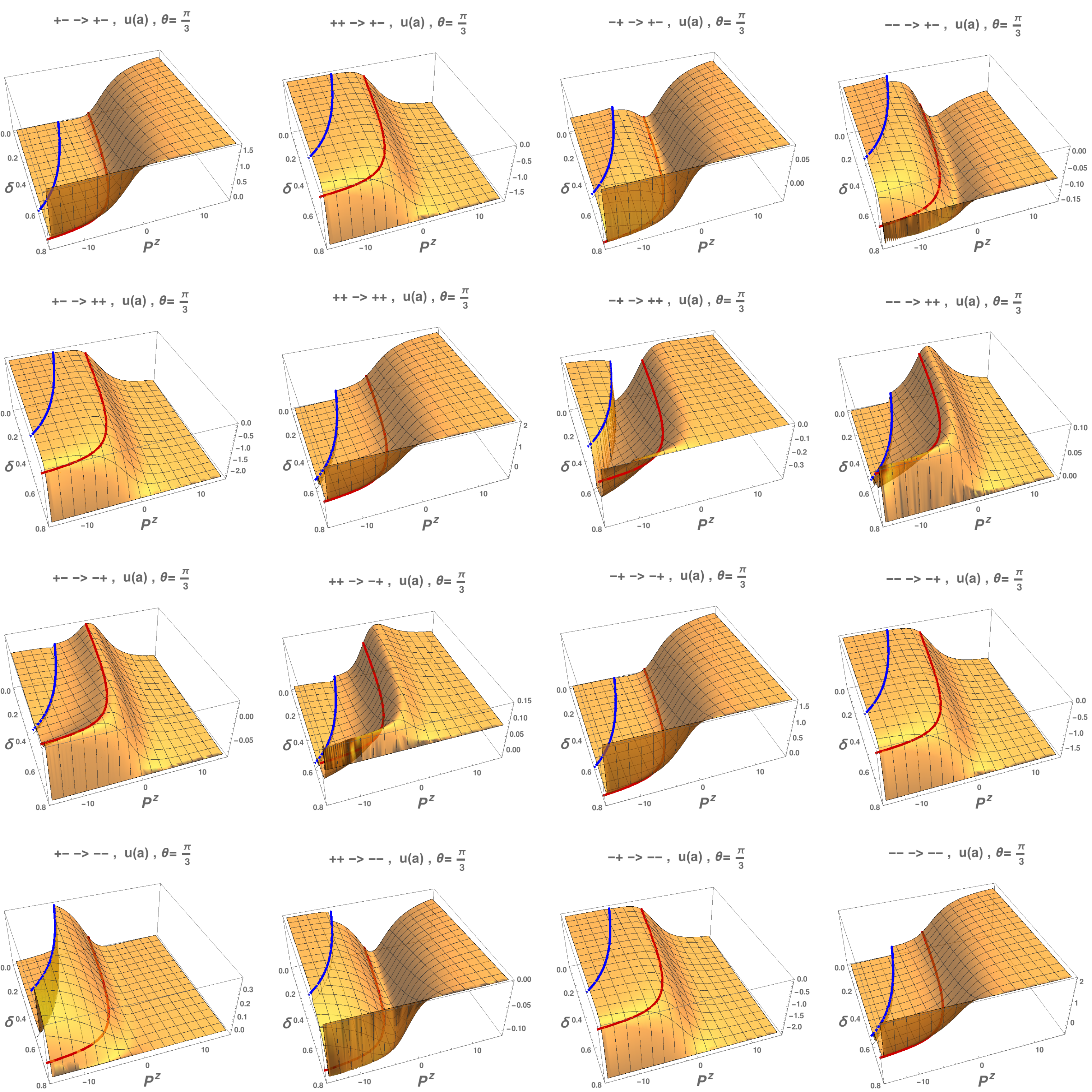}
		\caption{Compton Scattering Amplitudes --- u channel, time-ordering (a)}
		\label{fig:Compton_JoinedFigmua}
	\end{figure*}
	\begin{figure*}
		\centering
		\includegraphics[width=1.0\linewidth]{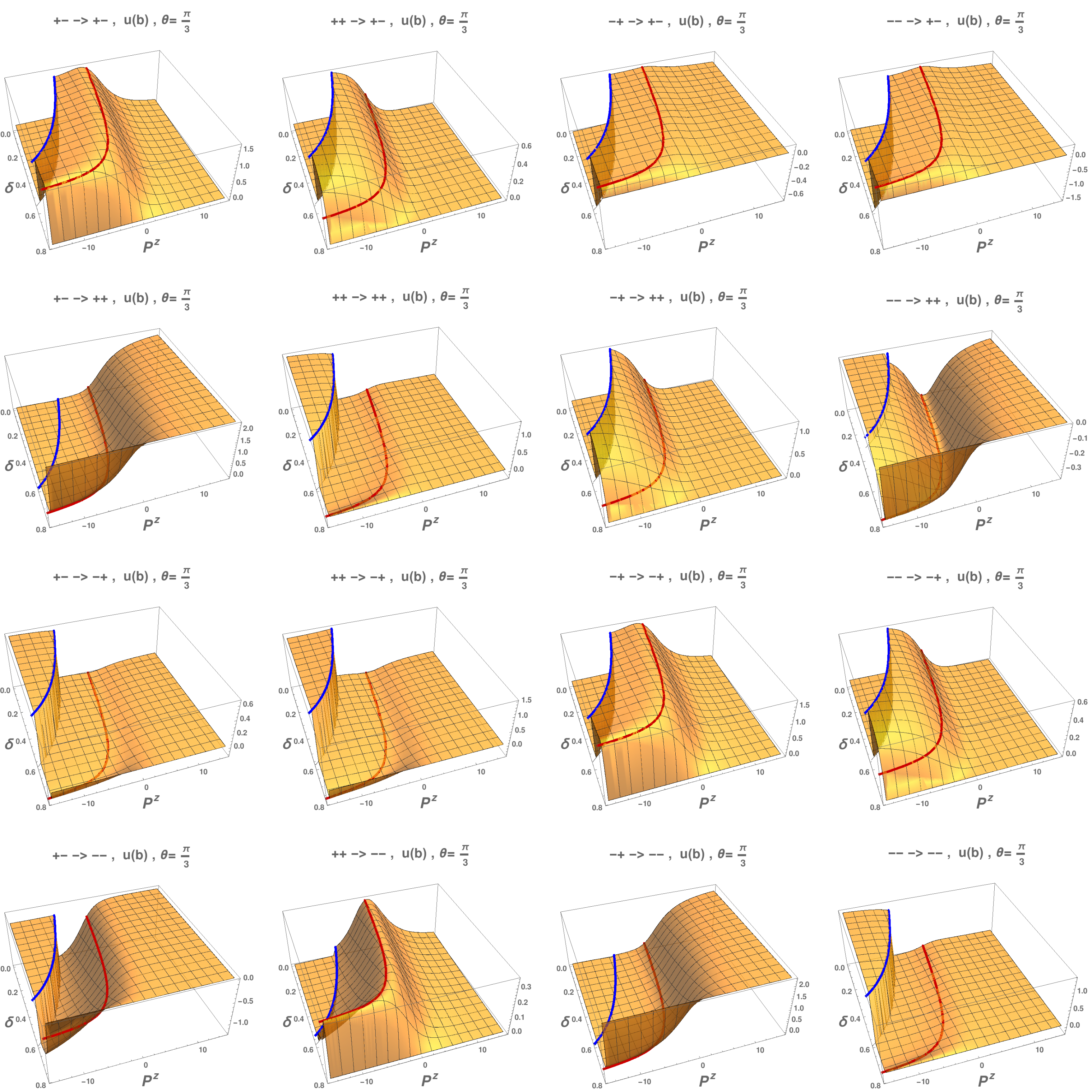}
		\caption{Compton Scattering Amplitudes --- u channel, time-ordering (b)}
		\label{fig:Compton_JoinedFigmub}
	\end{figure*}
	
\begin{figure*}
		\centering
		\includegraphics[width=1.0\linewidth]{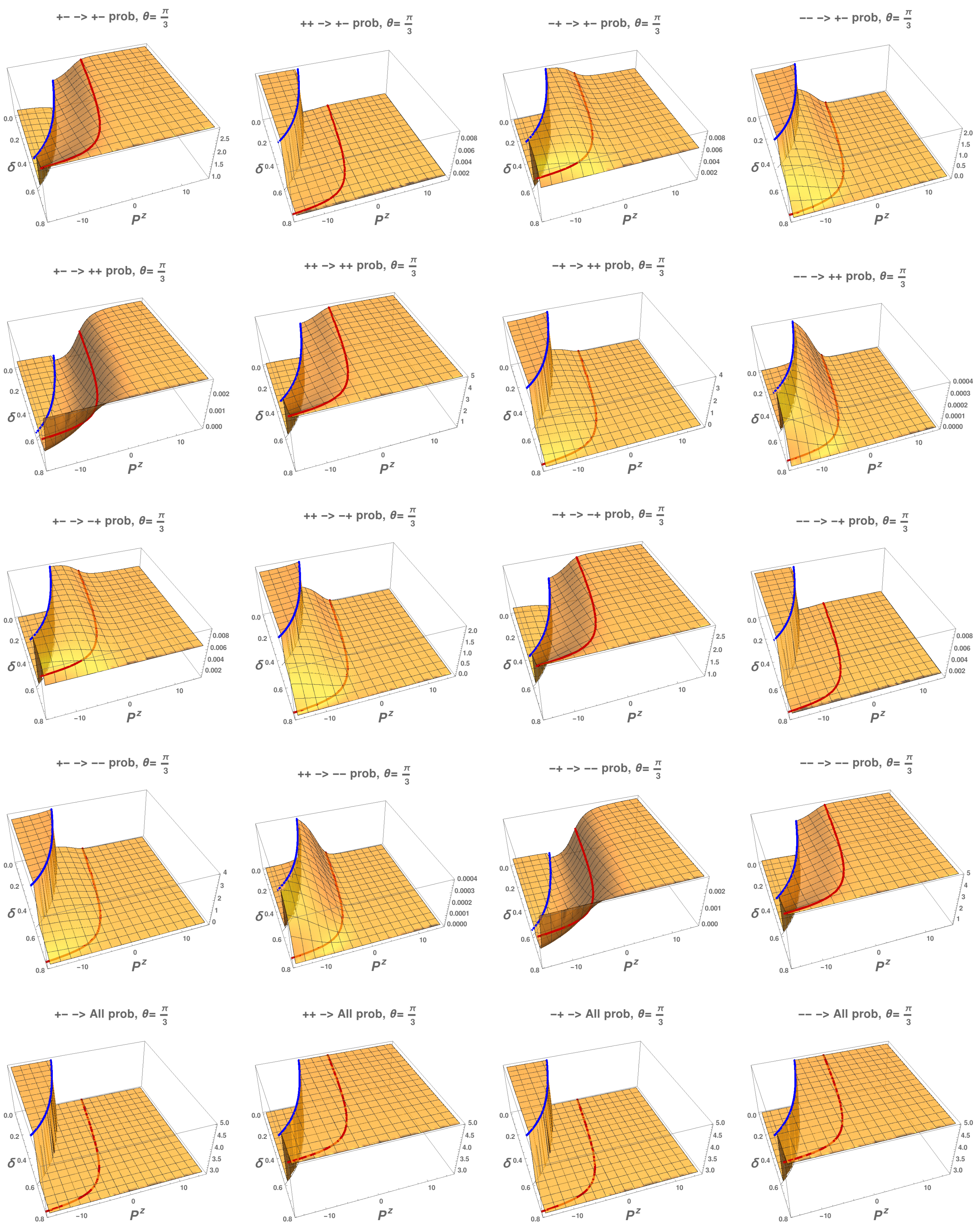}
		\caption{Compton Scattering probabilities in center of mass frame. The last row is sum over all final states for each initial state.}
		\label{fig:Compton_JoinedFigprob20}
	\end{figure*}

	The results of s-channel (a) and (b) as well as u-channel (a) and (b)time-ordered helicity amplitudes are shown in 
Figs.~\ref{fig:Compton_JoinedFigmsa}-\ref{fig:Compton_JoinedFigmub}, respectively. The probabilities, or the square of the sum of each and every helicity amplitudes, are also shown in Fig.~\ref{fig:Compton_JoinedFigprob20}.
In all of these figures, the boundary of bifurcated helicity branches between IFD and LFD due to the initial electron moving in
$\hat{z}$ direction given by Eq.~(\ref{eq:electron_boundary}) (i.e. $p_{1\wh{-}}=0$) is denoted by the blue curve while the 
characteristic ``J-curve" given by Eq.~(\ref{J-curve}) (i.e. $P_{\wh{-}} = 0$) existing in the frames boosted in -$\hat{z}$ direction  is depicted as the red curve. It is also apparent that the relationship between different helicity amplitudes given by 
Eq.~(\ref{eqn:amplitude_relation_Annihilation}) is satisfied by noting that $\lambda_2$ and $ \lambda_4 $ are now the helicities of the incoming and outgoing photons while $ \lambda_1 $ and $ \lambda_3 $ are the incoming and outgoing electrons' helicities, respectively, in Eq.~(\ref{eqn:amplitude_relation_Annihilation}).
This relationship holds as one can see that the upper left block of 2 by 2 figures are identical to the lower right block of 2 by 2 figures
while the upper right block of 2 by 2 figures and the lower left bock of 2 by 2 figures are same but with the opposite sign to each other.  For the square of amplitudes shown in Fig.~\ref{fig:Compton_JoinedFigprob20}, the same correspondence holds 
without any sign difference as it should be.

Computing the s-channel (a) time-ordered diagram shown in Fig.~\ref{fig:RealComptonScattering_schannel_a},
we obtain the results presented in Fig.~\ref{fig:Compton_JoinedFigmsa} for all 16 helicity amplitudes 
$\mathcal{M}_{a,s}^{\lambda_1,\lambda_2,\lambda_3,\lambda_4}$ for $\lambda_i = \pm$ ($i = 1,2,3,4$).
All of the LFD profiles ($\delta = \pi/4$) appear as straight lines indicating the $P^z$ independence or the frame independence
of the light-front helicity amplitudes as they should be, while the results for all other interpolation angles $0 \le \delta < \pi/4$
depend on $P^z$, i.e. frame dependent. As discussed earlier, the s-channel (a) time-ordered diagram shown in Fig.~\ref{fig:RealComptonScattering_schannel_a} corresponds to the ``on-mass-shell propagating contribution" 
in LFD. However, it is remarkable that the ``on-mass-shell propagating contribution" in LFD turned out to be absent
as the values of the LFD profiles shown in Fig.~\ref{fig:RealComptonScattering_schannel_a} are identically zero 
regardless of the initial and final helicities. We note that this triviality of the LFD results here is due to the fact that 
the initial photon is incident in the $-\hat{z}$ direction in the kinematics chosen for this calculation 
(see Eq.~(\ref{eqn:p1234_Compton_CMframe})) and thus gets only the zero-mode $p_2^+ = 0$ and $\mathbf{p}_{2\perp} = 0$. 
The zero-mode contributions are apparently absent in the ``on-mass-shell propagating contribution" in LFD.
The question is then where the nontrivial LFD result can be realized. It turns out that the nontrivial LFD result is realized 
in the ``instantaneous contribution" corresponding to the process depicted in Fig.~\ref{fig:RealComptonScattering_schannel_b}
for the kinematics given by Eq.~(\ref{eqn:p1234_Compton_CMframe}) which we use in the present calculation.
  
Fig.~\ref{fig:Compton_JoinedFigmsb} shows the results of all 16 helicity amplitudes 
$\mathcal{M}_{b,s}^{\lambda_1,\lambda_2,\lambda_3,\lambda_4}$ for $\lambda_i = \pm$ ($i = 1,2,3,4$)
which were obtained by computing the s-channel (b) time-ordered diagram shown in Fig.~\ref{fig:RealComptonScattering_schannel_b}. The ``instantaneous contribution" in LFD corresponds
the process shown in Fig.~\ref{fig:RealComptonScattering_schannel_b} with the understanding of the correspondence given by  
	\begin{equation}
	\lim\limits_{\mathbb{C}\to 0}\left( \frac{1}{2Q_s^{\wh{+}}}\;\frac{-\slashed{Q}_{b,s}+m}{-q_{s\wh{+}}-Q_{b,s\wh{+}}}\right) 
	=\frac{\gamma^+}{2q_s^+},
	\label{eqn:sLFinst}
	\end{equation}
where $q_s^+ = (E_0 + P_e)/ \sqrt{2}$ in the kinematics provided by Eq.~(\ref{eqn:p1234_Compton_CMframe}).
As mentioned above, due to the absence of the ``on-mass-shell propagating contribution" in LFD for the s-channel in the present kinematics, the entire s-channel contribution in LFD should be obtained from the ``instantaneous contribution". 
Due to $\{\gamma^{+},\gamma^{+}\}= {\gamma^+}^2 = 0$, the only non-vanishing ``instantaneous contribution"
to the s-channel helicity amplitudes in the light-front gauge $A^+ = 0$ are provided by only the transverse components of the photon polarization vectors for the helicity non-flip matrix elements between the initial and final electron spinors generically given by     
	\begin{align}
	&\bar{u}^{\lambda_3}(p_3) \slashed{\epsilon}(p_4)^*\frac{\gamma^+}{2q_s^+} \slashed{\epsilon}(p_s) u^{\lambda_1}(p_1) 
	\notag \\
	&\sim
	\bar{u}^{\lambda_3}(p_3) \mathbf{\gamma}_{\perp}^i \gamma^+ \mathbf{\gamma}_{\perp}^j u^{\lambda_1}(p_1) \notag \\
	&= 4 \delta_{\lambda_1,\lambda_3} \sqrt{p_1^+ p_3^+} (\delta^{ij}+i\epsilon^{ij}), 
	\end{align}
where $\delta^{ij}$ and $\epsilon^{ij}$ are the two-dimensional (${i,j} = {1,2}$) Kronecker delta and Levi-Civita symbol, respectively. 
As one can see in Fig.~\ref{fig:RealComptonScattering_schannel_b}, only the two helicity amplitudes $\mathcal{M}_{b,s}^{+,+,+,-}$ and $\mathcal{M}_{b,s}^{-,-,-,+}$ which are equal to each other appear to be non-zero for $\delta \approx \pi/4$. 
Besides the caveat in assigning the light-front helicity for the real photon moving in the $-\hat z$ direction, which was discussed 
earlier, it is remarkable that the ``instantaneous contribution" of effectively only one helicity amplitude in LFD provides the entire 
s-channel Compton amplitude.  

Likewise, the u-channel (a) and (b) time-ordered helicity amplitudes, 
$\mathcal{M}_{a,u}^{\lambda_1,\lambda_2,\lambda_3,\lambda_4}$ and 
$\mathcal{M}_{b,u}^{\lambda_1,\lambda_2,\lambda_3,\lambda_4}$, for $\lambda_i = \pm$ ($i = 1,2,3,4$) are shown in 
Figs.~\ref{fig:Compton_JoinedFigmua} and ~\ref{fig:Compton_JoinedFigmub}, respectively.
As mentioned earlier, the u-channel time-ordered processes (c) and (d) in Fig.~\ref{fig:RealComptonScattering_s&u_channel_TO}
correspond to the ``on-mass-shell propagating contribution" and the ``instantaneous contribution" in LFD for the CMF kinematics
as we take in this work. From Eq.~(\ref{eqn:p1234_Compton_CMframe}), $p_4^+ = P_e (1- \cos\theta) = \sqrt{3}m_e/2 \neq 0$ for $P_e = \sqrt{E_0^2-m_e^2} = \sqrt{3} m_e$ with $E_0 = 2 m_e$ and $\theta=\pi/3$ and 
the ``on-mass-shell propagating contribution" in LFD corresponding to the u-channel (a) time-ordered process shown in Fig.~\ref{fig:RealComptonScattering_uchannel_a} is nontrivial in contrast to the trivial s-channel (a) time-ordered result. 
However, the ``instantaneous contribution" in LFD corresponding to the u-channel (b) time-ordered process shown in Fig.~\ref{fig:RealComptonScattering_uchannel_b} gets again effectively only one helicity amplitude in LFD due to $\{\gamma^{+},\gamma^{+}\}= {\gamma^+}^2 = 0$ and the light-front gauge $A^+ = 0$ as discussed in the s-channel ``instantaneous contribution". 
As one can see in Fig.~\ref{fig:Compton_JoinedFigmub}, only non-zero helicity amplitudes  for $\delta \approx \pi/4$ are $\mathcal{M}_{b,u}^{+,-,+,+}$ and $\mathcal{M}_{b,u}^{-,+,-,-}$ which are equal to each other.

Now, summing all the s-channel and u-channel time-ordered amplitudes shown in Fig.~\ref{fig:RealComptonScattering_s&u_channel_TO} and squaring the total amplitude,
we obtain the Compton scattering probabilities for each and every helicities shown in Fig.~\ref{fig:Compton_JoinedFigprob20}.
As these results are the helicity amplitude squares, one may regard them as the polarization observables exhibiting the change 
of the predicted magnitudes depending on the reference frames in the range of total center of momentum $-15m_e <P^z< 15m_e$
from the lowest order interpolating QED computation for the Compton scattering process in the range of interpolation angle between IFD ($\delta=0$) and LFD ($\delta=\pi/4$). These results again alert the caution in the prevailing notion of the equivalence between the IMF formulated in IFD and the LFD as the IFD results in $P^z \to -\infty$ appear incapable of achieving the LFD results 
although the IFD results in large $P^z > 0$ seem to yield the corresponding LFD results. As the clear differences between the $P^z \to -\infty$ IFD and the LFD show up in level of physical observables, one should be cautious in the prevailing notion of the equivalence between the IFD at the IMF and the LFD. 

\begin{figure}[h!]
	\centering
	\includegraphics[width=0.85\linewidth]{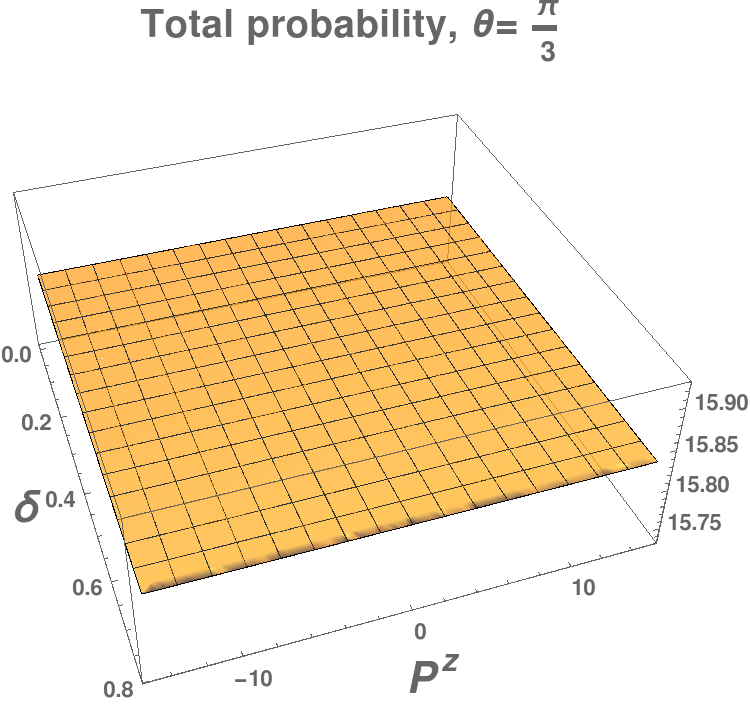}
	\caption{Total probability of Compton Scattering in center of mass frame}
	\label{fig:Compton_FigProb}
\end{figure}

The sum of the probabilities over all final helicity states for each initial helicities are shown in the last row of Fig.~\ref{fig:Compton_JoinedFigprob20} and the sum over initial helicity states (i.e. the sum over all sixteen 
total helicity amplitude squares) turns out to be completely independent of $\delta$ and $P^z$ as it must be (see Fig.~\ref{fig:Compton_FigProb}). Indeed, this result is in complete agreement with the well-known manifestly Lorentz invariant result given by 
	\begin{widetext}
		\begin{align}\label{eqn:spin_averaged_Compton_amplitude}
		\left| \mathcal{M} (e\gamma \to e\gamma) \right|^2 
		&\equiv\sum_{\lambda_1,\lambda_2,\lambda_3,\lambda_4} 
		|\mathcal{M}^{\lambda_1,\lambda_2,\lambda_3,\lambda_4}_{a,s} + \mathcal{M}^{\lambda_1,\lambda_2,\lambda_3,\lambda_4}_{b,s}+\mathcal{M}^{\lambda_1,\lambda_2,\lambda_3,\lambda_4}_{a,u} + \mathcal{M}^{\lambda_1,\lambda_2,\lambda_3,\lambda_4}_{b,u}|^2 \notag \\		
		&= - 8 \left[\frac{u_m}{s_m}+\frac{s_m}{u_m}+2m^2\left( \frac{t_m}{s_mu_m}-\frac{1}{s_m}-\frac{1}{u_m}\right)-4m^4\left( \frac{1}{s_m^2}+\frac{1}{u_m^2}\right)   \right] ,
		\end{align}
	\end{widetext}
where $s_m=s-m^2$, $t_m=t-4m^2$, $u_m=u-m^2$ and the electric charge factor is taken to be one.
For the kinematics given by Eq.~(\ref{eqn:p1234_Compton_CMframe}) with $E_0 = 2 m_e, \theta = \pi/3$ and $m = m_e$
used in our numerical computation, 
the value from the analytic result given by Eq.~(\ref{eqn:spin_averaged_Compton_amplitude}) yields 
$\left| \mathcal{M} (e\gamma \to e\gamma) \right|^2 = (4/169) (991 - 186 \sqrt{3}) \approx 15.8305$ which is in precise agreement
with the total probability obtained in Fig.~\ref{fig:Compton_FigProb}.
In the high energy limit, Eq.~(\ref{eqn:spin_averaged_Compton_amplitude}) in the massless limit ($m \to 0$) reduces to 
the well-know textbook~\cite{Halzen-Martin} Compton result given by
	\begin{equation}\label{eqn:spin_averaged_amplitude_neg_Compton}
	\left| \mathcal{M} (e\gamma \to e\gamma) \right|^2 =-8 \left(\frac{u}{s}+\frac{s}{u} \right).
	\end{equation}
The crossing symmetry between the $e \gamma \to e \gamma$ process and the $e^+ e^- \to \gamma \gamma$ process 
is reflected by the $s \leftrightarrow t$ symmetry between Eqs.~(\ref{eqn:spin_averaged_Compton_amplitude}) and 
(\ref{eqn:spin_averaged_amplitude_Annihilation}) as well as Eqs.~(\ref{eqn:spin_averaged_amplitude_neg_Compton}) and 
(\ref{eqn:spin_averaged_amplitude_neg_Annihilation}) with the overall sign consistent to each other for the positivity of the amplitude square.

	\section{Summary and Conclusion}\label{sec:conclusion}
In this work, we have completed the interpolation of Quantum Electrodynamics between the instant form and the front form proposed by Dirac~\cite{Dirac} in 1949. We started from the QED Lagrangian and presented the interpolating Hamiltonian formulation 
introducing a parameter $\delta$ which corresponds between the instant form dynamics (IFD) at $\delta=0$ and the front-form dynamics which we call the light-front dynamics (LFD) at $\delta=\pi/4$. Not only have we summarized
the interpolating time-ordered diagram rules for the computation of QED processes in terms of the 
interpolation angle parameter $0 \le \delta \le \pi/4$ as presented in Sec.~\ref{sec:formal}, but also we have applied
these rules to the typical QED processes such as $e^+ e^- \to \gamma \gamma$ and $e \gamma \to e \gamma$ which 
involve the fermion propagator beyond what we have already presented in our previous works~\cite{JLS2015,LAJ2015}.
Entwining the fermion propagator interpolation with our previous works of the interpolating helicity spinors and the electromagnetic gauge field interpolation, we have now fastened the bolts and nuts necessary in launching the interpolating QED.

Our interpolating formulation reveals that there exists the constraint fermion degree of freedom in LFD ($\delta=\pi/4$) 
distinguished from the ordinary equal-time fermion degrees of freedom.
The constraint component of the fermion degrees of freedom in LFD results in the instantaneous contribution to the fermion propagator distinguished from the ordinary equal-time forward and backward propagation of relativistic fermion degrees of freedom. It is interesting to note that the manifestly covariant fermion propagator decouples to the ``on-mass-shell propagating contribution" and the ``instantaneous fermion contribution" only in LFD but not in any
other interpolating dynamics ($0 \le \delta < \pi/4$).  The helicity of the on-mass-shell fermion spinors in LFD is also distinguished from the ordinary Jacob-Wick helicity in the IFD with respect to whether the helicity depends on the reference frame or not~\cite{LAJ2015}.  

To exemplify these distinguished features of the fermion degrees of freedom in LFD, we have computed the annihilation process of the fermion and anti-fermion pair 
interpolating the fermion degrees of freedom between the IFD and the LFD.
We presented the leading order QED processes ($ e^+e^-\to \gamma \gamma $ and $ e\gamma\to e\gamma $), providing the whole landscape of helicity amplitudes from the IFD to the LFD. In the cross-section level, we showed the precise agreement of our result with the textbook formula. The helicity conservation in the chiral limit was discussed, and the angular momentum conservation was checked in each case. 

Our analysis clarifies any conceivable confusion in the prevailing
notion of the equivalence between the IMF approach in the IFD and the LFD. By investigating the dependence of the helicity amplitudes on the reference frame, i.e. $ P^z $-dependence, we find that in IFD, $ P^z\to+\infty $ and $ P^z\to-\infty $ yield very different results from each other, and that one has to be very cautious about the direction of boost in approaching to the IMF when one tries to obtain the equivalent LFD result. We have shown that although in some cases one can indeed reproduce the LFD result by boosting the system to the correct direction, in some other cases a finite, large momentum boost yields only qualitatively similar result. On the other hand, all the helicity amplitudes in LFD are independent of the reference frame, and  certain simplifications to the theory (e.g. suppression of vacuum fluctuations, vanishing of a number of diagrams, etc. ) can be realized even in the rest frame of the system. Since the helicity definition in LFD is frame-independent, no boundaries exist for the light-front helicity amplitude. 
One should also note that for the massless particle moving in the $ -\hat{z} $-direction 
the helicity defined in the LFD is opposite to the Jacob-Wick helicity defined in the IFD.
Further treacherous correspondence between IFD and LFD will be studied in our future work, extending the interpolation to the the loop-level computation and ultimately to the QCD. 

	\section*{Acknowledgements}
This work was supported by the U.S. Department of Energy Grant No. DE-FG02-03ER41260.
	\appendix
		
		\section{Fermion propagator in the position space}
		\label{app:position_space}
		
		The Feynman propagator in the position space is given by
		\begin{align}
		\label{Feynman}
		\Delta_{\rm F}(x) & =  i\int \frac{d^4 q}{(2\pi)^4}\,\frac{{\rm e}^{-iq_{\mu}x^{\mu}}}{\left(q^2-m^2+i\varepsilon\right)}.
		\end{align}
		
		In the interpolation form, it can be written as
		\begin{align}
		\label{Feynman general}
		\Delta_{\rm F}(x)
		&=i\int \frac{d^2{\bf q}_{\perp}dq_{\wh-}dq_{\wh+}}{(2\pi)^4}\notag\\
		&\times\frac{{\rm e}^{-i(q_{\wh+}x^{\wh+}+q_{\wh-}x^{\wh-}+{\bf q}_{\perp}{\bf x}^{\perp})}}{\left({\mathbb C}q_{\wh+}^{2}+2{\mathbb{S}}q_{\wh+}q_{\wh-}-{\mathbb{C}}q_{\wh-}^{2}-{\bf q}_{\perp}^{2}-m^2+i\varepsilon\right)}.
		\end{align}
		
		Solving for the quadratic expression in the denominator of the Feynman propagator in order to separate the two distinct poles, we have the two poles of $q_{\wh +}$ 
		\begin{align}
		q_{\wh +}^{(\rm a)} & =  {\cal A}_{\wh +}-i\varepsilon^\prime, \\
		q_{\wh +}^{(\rm b)} & =  -{\cal B}_{\wh +}+i\varepsilon^\prime,
		\end{align}
		where the real part of the two poles are defined as
		\begin{align}
		{\cal A}_{\wh+} & \equiv  -\frac{\mathbb S q_{\wh-}}{\mathbb C} +\frac{\sqrt{q_{\wh-}^2+\mathbb C\left({\bf q}_{\perp}^2+m^2\right)}}{\mathbb C}, \label{eqn:poleA}\\
		{\cal B}_{\wh+} & \equiv  \frac{\mathbb S q_{\wh-}}{\mathbb C} +\frac{\sqrt{q_{\wh-}^2+\mathbb C\left({\bf q}_{\perp}^2+m^2\right)}}{\mathbb C}, \label{eqn:poleB}
		\end{align}
		and the imaginary part of the poles is given by
		\be
		\label{eqn:epsilon_prime}
		\varepsilon^{\prime} \equiv \frac{\varepsilon}{2\sqrt{q_{\wh-}^2+\mathbb C\left({\bf q}_{\perp}^2+m^2\right)}}.
		\ee
		
		Define the square root part as
		\be
		\label{eqn:PP}
		Q^{\wh+} \equiv \sqrt{q_{\wh-}^2+\mathbb C\left({\bf q}_{\perp}^2+m^2\right)}.
		\ee
		
		From the expressions in Eqs.~(\ref{eqn:poleA}) and (\ref{eqn:poleB}), we 
		%
		%
		%
		%
		see that for any sign 
		of $q_{\wh-}$, ${\mathcal A}_{\wh+}$ is always positive and corresponds to the positive energy solution, while $-{\mathcal B}_{\wh+}$ is always negative and corresponds to the negative energy solution. Therefore, we see the pole structure in the $q_{\wh +}$ complex plane is that ${\mathcal A}_{\wh+}-i\varepsilon^\prime$, located in the fourth quadrant, and $-{\mathcal B}_{\wh+}+i\varepsilon^\prime$, located in the second quadrant.
		

		In order to perform the integration in the ``energy'' variable $q_{\wh +}$ in Eq.~(\ref{Feynman general}), we use the Cauchy residue theorem. 
		We may consider three possibilities: $x^{\wh+}>0$, $x^{\wh+}<0$, and $x^{\wh+}=0$. We now analyze these three situations case by case.
		
		For $x^{\wh+}>0$, this implies that in order to have a converging exponential factor in the integrand, we must have $\mathfrak {Im}\  q_{\wh+}<0$. This means that the semi-circle ${C}_{ R}$ that closes the contour must be located in the lower half of the complex $q_{\wh+}$ plane, in a clockwise direction. A closed contour in this sense encloses the pole $q_{\wh+}={\mathcal A}_{\wh+}-i\varepsilon^{\prime}$. We thus have for this case:	
		\begin{align}
		\label{CauchyA}
		& \oint \frac{dq_{\wh+}}{(2\pi)}\frac{{\rm e}^{-iq_{\wh+}x^{\wh+}}}{\mathbb C\left(q_{\wh+} - {\cal A}_{\wh+} + i\varepsilon^{\prime}\right)\left( q_{\wh+} + {\cal B}_{\wh+} - i\varepsilon^{\prime}\right)}  \nonumber \\
		=& \lim_{R\rightarrow \infty}\left\{\int_{-R}^{+R}\frac{dq_{\wh+}}{(2\pi)}\frac{{\rm e}^{-iq_{\wh+}x^{\wh+}}}{\mathbb C\left(q_{\wh+} - {\cal A}_{\wh+} + i\varepsilon^{\prime}\right)\left( q_{\wh+} + {\cal B}_{\wh+} - i\varepsilon^{\prime}\right)} \right.\notag\\
		&\left.+\int_{{C}_{ R}}\frac{dq_{\wh+}}{(2\pi)}\frac{{\rm e}^{-iq_{\wh+}x^{\wh+}}}{\mathbb C\left(q_{\wh+} - {\cal A}_{\wh+} + i\varepsilon^{\prime}\right)\left( q_{\wh+} + {\cal B}_{\wh+} - i\varepsilon^{\prime}\right)} \right\}.
		\end{align}
		
		
		The left-hand side of Eq.~(\ref{CauchyA}) is (by Cauchy's theorem) equal to $-i{\rm Res}({\mathcal A}_{\wh+}-i\varepsilon^{\prime})$, where the minus sign is due to the clockwise direction of the closed contour. Since the arc contribution in the limit $R \rightarrow \infty$ goes to zero, in this limit we have
		\begin{align}
		& \int_{-\infty}^{+\infty}\frac{dq_{\wh+}}{(2\pi)}\frac{{\rm e}^{-iq_{\wh+}x^{\wh+}}}{\mathbb C\left(q_{\wh+} - {\cal A}_{\wh+} + i\varepsilon^{\prime}\right)\left( q_{\wh+} + {\cal B}_{\wh+} - i\varepsilon^{\prime}\right)}\notag\\
		&= -i\frac{{\rm e}^{-i{\mathcal A}_{\wh+}x^{\wh+}}}{\mathbb C\left({\mathcal A}_{\wh+}+{\mathcal B}_{\wh+}\right)}, \qquad (x^{\wh+}>0). \label{CauchyresultA}
		\end{align}
		
		
		For $x^{\wh+}<0$, this implies that in order to have a converging exponential factor in the integrand, we must have ${\mathfrak {Im}} q_{\wh+}>0$. This means that the semi-circle ${C}_{ R}$ that closes the contour must now be located in the upper half of the complex $q_{\wh+}$ plane, in a counterclockwise direction. A closed contour in this sense encloses now the pole $q_{\wh+}=-{\mathcal B}_{\wh+}+i\varepsilon^{\prime}$. We thus have for this case:
		\begin{align}
		\label{CauchyB}
		& \oint \frac{dq_{\wh+}}{(2\pi)}\frac{{\rm e}^{-iq_{\wh+}x^{\wh+}}}{\mathbb C\left(q_{\wh+} - {\cal A}_{\wh+} + i\varepsilon^{\prime}\right)\left( p_{\wh+} + {\cal B}_{\wh+} - i\varepsilon^{\prime}\right)}  \nonumber \\
		= & \lim_{R\rightarrow \infty}\left\{\int_{-R}^{+R}\frac{dq_{\wh+}}{(2\pi)}\frac{{\rm e}^{-iq_{\wh+}x^{\wh+}}}{\mathbb C\left(q_{\wh+} - {\cal A}_{\wh+} + i\varepsilon^{\prime}\right)\left( q_{\wh+} + {\cal B}_{\wh+} - i\varepsilon^{\prime}\right)}\right.\notag\\
		&\left.+\int_{{C}_{ R}}\frac{dq_{\wh+}}{(2\pi)}\frac{{\rm e}^{-iq_{\wh+}x^{\wh+}}}{\mathbb C\left(q_{\wh+} - {\cal A}_{\wh+} + i\varepsilon^{\prime}\right)\left( q_{\wh+} + {\cal B}_{\wh+} - i\varepsilon^{\prime}\right)} \right\}.
		\end{align}
		
		The left-hand side of Eq.~(\ref{CauchyB}) is (by Cauchy's theorem) equal to $+i{\rm Res}(-{\mathcal B}_{\wh+}+i\varepsilon^{\prime})$, where the plus sign now is due to the counterclockwise direction of the closed contour. Since the arc contribution in the limit $R \rightarrow \infty$ goes to zero, in this limit we now have
		\begin{align}
		&\int_{-\infty}^{+\infty}\frac{dq_{\wh+}}{(2\pi)}\frac{{\rm e}^{-iq_{\wh+}x^{\wh+}}}{\mathbb C\left(q_{\wh+} - {\cal A}_{\wh+} + i\varepsilon^{\prime}\right)\left( q_{\wh+} + {\cal B}_{\wh+} - i\varepsilon^{\prime}\right)}\notag\\
		&= +i\frac{{\rm e}^{i{\mathcal B}_{\wh+}x^{\wh+}}}{\mathbb C\left(-{\mathcal B}_{\wh+}-{\mathcal A}_{\wh+}\right)}, \qquad (x^{\wh+}<0). \label{CauchyresultB}
		\end{align}
		In this last expression, we have already dropped the $\varepsilon^{\prime}$ in the result after the $q_{\wh+}$ integration, and put an reminder that this result is now valid for the specific case of $x^{\wh +}<0$.
		
		For $x^{\wh+}=0$, the main converging factor in the integrand becomes one, that is, ${\rm e}^0 =1$. We have therefore
				\begin{align}
				\label{CauchyC}
				& \oint \frac{dq_{\wh+}}{(2\pi)}\frac{1}{\mathbb C\left(q_{\wh+} - {\cal A}_{\wh+} + i\varepsilon^{\prime}\right)\left( q_{\wh+} + {\cal B}_{\wh+} - i\varepsilon^{\prime}\right)}  \nonumber \\
				= & \lim_{R\rightarrow \infty}\left\{\int_{-R}^{+R}\frac{dq_{\wh+}}{(2\pi)}\frac{1}{\mathbb C\left(q_{\wh+} - {\cal A}_{\wh+} + i\varepsilon^{\prime}\right)\left( q_{\wh+} + {\cal B}_{\wh+} - i\varepsilon^{\prime}\right)}\right. \nonumber\\
				& \left.+ \int_{{C}_{ R}}\frac{dq_{\wh+}}{(2\pi)}\frac{1}{\mathbb C\left(q_{\wh+} - {\cal A}_{\wh+} + i\varepsilon^{\prime}\right)\left( q_{\wh+} + {\cal B}_{\wh+} - i\varepsilon^{\prime}\right)} \right\}.
				\end{align}
		
		Although for this case the exponential factor in the integrand is absent, the denominator of the integrand has enough powers in $q_{\wh +}$ to make the arc contribution go to zero when $R\to \infty$. Therefore, closing the contour from below, that is, with $C_R$ in the clockwise direction. This encloses the pole $q_{\wh+}={\mathcal A}_{\wh+}-i\varepsilon^{\prime}$ and we get
		\begin{align}
		&\int\limits_{-\infty}^{+\infty}\frac{dq_{\wh+}}{(2\pi)}\frac{1}{\mathbb C\left(q_{\wh+} - {\cal A}_{\wh+} + i\varepsilon^{\prime}\right)\left( q_{\wh+} + {\cal B}_{\wh+} - i\varepsilon^{\prime}\right)}\notag\\
		&=\frac{-i}{\mathbb C\left({\mathcal A}_{\wh+}+{\mathcal B}_{\wh+}\right)} =-\frac{i}{2Q^{\wh+}}, \quad (x^{\wh+}=0).
		\end{align}
		
		Closing the contour in the counterclockwise direction, we enclose the other pole, $q_{\wh+}={\mathcal B}_{\wh+}+i\varepsilon^{\prime}$, and we obtain
		\begin{align}
		&\int\limits_{-\infty}^{+\infty}\frac{dq_{\wh+}}{(2\pi)}\frac{1}{2Q^{\wh+}\left(q_{\wh+} - {\cal A}_{\wh+} + i\varepsilon^{\prime}\right)\left( q_{\wh+} + {\cal B}_{\wh+} - i\varepsilon^{\prime}\right)}\notag\\
		&=\frac{+i}{\mathbb C\left(-{\mathcal B}_{\wh+}-{\mathcal A}_{\wh+}\right)}
		= -\frac{i}{2Q^{\wh+}}, \quad (x^{\wh+}=0). 
		\end{align}
		Thus, both circulations yield the same answer, as it should and serve as a check for our results. 
		
		Finally, the overall result for the Feynman propagator is given by 
		\begin{widetext}
			\begin{align}
			\label{Feynman result final}
			\Delta_{\rm F}(x)
			&=\int\frac{d^2{\bf q}_{\perp}}{(2\pi)^2} \int\limits_{-\infty}^{+\infty}\frac{dq_{\wh-}}{(2\pi)} \frac{1}{2Q^{\wh+}}\left\{{\Theta(x^{\wh+}){\rm e}^{-i\left({\mathcal A}_{\wh+}x^{\wh+}+q_{\wh-}x^{\wh-}+{\bf q}_{\perp}\cdot{\bf x}^{\perp}\right)}}+{\Theta(-x^{\wh+}){\rm e}^{i\left({\mathcal B}_{\wh+}x^{\wh+}+q_{\wh-}x^{\wh-}+{\bf q}_{\perp}\cdot{\bf x}^{\perp}\right)}}\right\},
			\end{align}
		\end{widetext}
		where we have made the variable shifts
		\begin{align}
		{\bf q}_{\perp} & \rightarrow  -{\bf q}_{\perp}, \label{shift1}\\
		q_{\wh-} & \rightarrow  - q_{\wh-}, \label{shift2}
		\end{align}
		in the second term, which are possible because the integration ranges for these variables are from $ -\infty $ to $ +\infty $.
		
		When $ \mathbb{C}=0 $, however, the denominator of the Feynman propagator is a linear expression in $ q_{\wh{+}}=q_+=q^- $ instead of a quadratic one, thus it has only one pole
		\begin{equation}
		q_{\mathrm{on}}^-=\frac{\mathbf{q}_{\perp}^2+m^2}{2q^+}-i\frac{\varepsilon}{2q^+}.
		\end{equation}
		Thus when $ q^+>0 $, the pole is located in the fourth quadrant of the $ q^- $ complex plane, and to make sure the arc contribution is zero, when $ x^+>0 $, one has to close the contour from below, while when $ x^+<0 $, one needs to close the contour from above, and it gives no contribution since there is no pole in the upper half plane. Similarly, when $ q^+<0 $, the pole is located in the second quadrant of the $ q^- $ complex plane, and to make sure the arc contribution is zero, when $ x^+>0 $, one has to close the contour from below, which again gives no contribution because there is no pole in the lower half plane, while when $ x^+<0 $, one needs to close the contour from above, catching the pole there. Thus, the light-front-time- ($ x^+ $-) ordering imposes clear cut on the signs of $ q^+ $ and consequently on $ q^- $ due to the sign correlation between them, so that when $ x^+>0 $, $ q^+ $ and $ q^- $ must both be positive, on the other hand when $ x^+<0 $, $ q^+ $ and $ q^- $ must both be negative. As a result, the integration ranges of the momentum variables in the two time-orderings are not both $ (-\infty,+\infty) $ , but are $ (0,+\infty) $ for the forward time and $ (-\infty,0) $ for the backward time.
		
		Doing the pole integration, we get in the light-front,
		\begin{widetext}
			\begin{align}
			\label{Feynman result LF}
			\Delta_{\rm F}(x)
			&=\int\frac{d^2{\bf q}_{\perp}}{(2\pi)^2}  \left\{{\Theta(x^{+})\int\limits_{0}^{+\infty}\frac{dq^{+}}{(2\pi)|2q^+|}{\rm e}^{-i\left( q_{\rm on}^-x^{+}+q^{+}x^{-}+{\bf q}_{\perp}\cdot{\bf x}^{\perp}\right)}}+{\Theta(-x^{+})\int\limits_{-\infty}^{0}\frac{dq^{+}}{(2\pi)|2q^+|}{\rm e}^{-i\left( q_{\rm on}^-x^{+}+q^{+}x^{-}+{\bf q}_{\perp}\cdot{\bf x}^{\perp}\right)}}\right\}\notag\\
			&=\int\frac{d^2{\bf q}_{\perp}}{(2\pi)^2}  \left\{{\Theta(x^{+})\int\limits_{0}^{+\infty}\frac{dq^{+}}{(2\pi)|2q^+|}{\rm e}^{-i\left( q_{\rm on}^-x^{+}+q^{+}x^{-}+{\bf q}_{\perp}\cdot{\bf x}^{\perp}\right)}}+{\Theta(-x^{+})\int\limits_{0}^{+\infty}\frac{dq^{+}}{(2\pi)|2q^+|}{\rm e}^{i\left( q_{\rm on}^-x^{+}+q^{+}x^{-}+{\bf q}_{\perp}\cdot{\bf x}^{\perp}\right)}}\right\}.
			\end{align}
		\end{widetext}		
		
		Using the interpolating step function given in Eq.~(\ref{eqn:Interpolating_Theta_Function_QED}) in Sec.~(\ref{ssb:propagator_decomposition}), we can combine the results and write as follows
		\begin{widetext}
			\begin{align}
			\label{Feynman result final_combine}
			\Delta_{\rm F}(x)
			&=\int\frac{d^2{\bf q}_{\perp}}{(2\pi)^2} \int\limits_{-\infty}^{+\infty}\frac{dq_{\wh-}}{(2\pi)} \frac{1}{2Q^{\wh+}}\wh{\Theta}(q_{\wh{-}})\left\{{\Theta(x^{\wh+}){\rm e}^{-i{\mathcal A}_{\wh\mu}x^{\wh\mu}}}+{\Theta(-x^{\wh+}){\rm e}^{i{\mathcal B}_{\wh\mu}x^{\wh\mu}}}\right\},
			\end{align}
		\end{widetext}
		where we have introduced the shorthand notation
		\begin{align}
		\mathcal{A}_{\wh{\mu}}\equiv\left(\mathcal{A}_{\wh{+}}, q_1,q_2,q_{\wh{-}} \right) \overset{\mathbb{C}\to0}{\to}\left( q_{\rm on}^-,q_1,q_2,q^+\right)\label{eqn:Amu} \\
		\mathcal{B}_{\wh{\mu}}\equiv\left(\mathcal{B}_{\wh{+}}, q_1,q_2,q_{\wh{-}} \right)\overset{\mathbb{C}\to0}{\to}\left( q_{\rm on}^-,q_1,q_2,q^+\right)\label{eqn:Bmu}
		\end{align}

		The explicit form of the Feynman propagator is given by \cite{Bogo}:
		\begin{widetext}
			\begin{align}
			\Delta_{\rm F}(x) & =  -\frac{1}{4\pi}\delta(x^2)+\frac{m}{8\pi \sqrt{x^2}}\;\Theta\left(x^2\right)\left\{ J_1\left(m\sqrt{x^2}\right)-iN_1\left(m\sqrt{x^2}\right) \right\}-\frac{im}{4\pi^2\sqrt{-x^2}}\;\Theta\left(-x^2\right)K_1\left(m\sqrt{-x^2}\right) \nonumber \\
			&=-\frac{1}{4\pi}\left\{\delta\left(x^2\right)+\frac{im}{\pi}\frac{K_1\left(m\sqrt {-x^2+i\varepsilon}\right)}{\sqrt{-x^2+i\varepsilon}}\right\},
			\end{align}
		\end{widetext}
		where $J_1(z)$, $N_1(z)$ and $K_1(z)$ are respectively the Bessel, Neumann and Hankel functions of order 1, and $\sqrt{-x^2+i\varepsilon}=i\sqrt{x^2}$, for $x^2>0$. Note here that the argument of the Hankel function is imaginary.

		To derive the fermion propagator we need to apply the Dirac operator on it, 
		\begin{align}
		S_{\rm F} (x)
		& =  \left(i\gamma^{\wh +}\partial_{\wh +}+i\gamma^{\wh -}\partial_{\wh -}+i{\boldsymbol{ \gamma}}^{\perp}\cdot\boldsymbol{\partial}_{\perp}+m\right)\Delta_{\rm F}(x),
		\end{align}
		where $\Delta_{\rm F}(x)$ is given by Eq.~(\ref{Feynman result final_combine}).
		We obtain
		\begin{widetext}
			\begin{align}
			\label{eqn:Fermion_interpolation}
			S_{\rm F}(x) = \int\frac{d^2{\bf q}_{\perp}}{(2\pi)^2} \int\limits_{-\infty}^{+\infty}\frac{dq_{\wh-}}{(2\pi)(2Q^{\wh+})}\wh{\Theta}(q_{\wh{-}}) &\left\{\Theta(x^{\wh+})\left(\slashed{\cal A}+m\right){\rm e}^{-i{\mathcal A}_{\wh\mu}x^{\wh \mu}}+{\Theta(-x^{\wh+})\left(-\slashed{\cal B}+m\right){\rm e}^{i{\mathcal B}_{\wh \mu}x^{\wh \mu}}}\right.\notag\\
			&\left.+i\gamma^{\wh +}\delta(x^{\wh +})\left( {\rm e}^{-i{\mathcal A}_{\wh\mu}x^{\wh\mu}}-{\rm e}^{i{\mathcal B}_{\wh\mu}x^{\wh\mu}}\right) \right\}.
			\end{align}
		\end{widetext}			
		Then, going back to the the momentum space, we note that when $ \wh{\Theta}(q_{\wh-})=1 $, i.e. the integration of $ q_{\wh-} $ goes from $ -\infty $ to $ +\infty $, the two $ \delta(x^{\wh+}) $ terms cancel each other exactly when a spatial integration is performed, while for $ \wh{\Theta}(q_{\wh-})=\Theta(q^+) $, they don't cancel and an ``instantaneous contribution" is leftover. We finally get
		\begin{align}
	&\Sigma_{\rm F}(q)\equiv iS_{\rm F}(q)=i\int d^4 x \ S_{\rm F}(x)\ {\rm e}^{iq_{\wh\mu}x^{\wh\mu}}\notag\\
		&=
		\begin{dcases}
		\frac{1}{2Q^{\wh+}}\left( \frac{\slashed{Q}_a+m}{q_{\wh+}-Q_{a\wh+}}+\frac{-\slashed{Q}_b+m}{-q_{\wh+}-Q_{b\wh+}}\right) , \qquad (\mathbb{C}\neq0),\\
		\frac{1}{2q^+}\frac{\slashed{q}_{\rm on}+m}{q^--q_{\rm on}^-}+\frac{\gamma^+}{2q^+},\qquad \qquad(\mathbb{C}=0),
		\end{dcases}
		\end{align}
		where $ \slashed{Q}_a\equiv\slashed{\cal A} $ and $ \slashed{Q}_b\equiv\slashed{\cal B} $ are defined in Eqs.~(\ref{eqn:Amu}) and (\ref{eqn:Bmu}), respectively, while $ Q_{a\wh+}\equiv{\cal A}_{\wh+} $ and $ Q_{b\wh+}\equiv{\cal B}_{\wh+} $ are defined in Eqs.~(\ref{eqn:poleA}) and (\ref{eqn:poleB}), respectively. Thus, we get the time-ordered propagators given in the main text.
		
	\section{Derivation of Interpolating QED Hamiltonian}
	\label{app:Hamiltonian}	
	
	In this Appendix, we show how the Hamiltonian in subsection (\ref{ssb:hamiltonian}) is derived, and how the consistency with the LFD formulation presented by Kogut and Soper \cite{KS} can be seen.
	
	We start from the interpolating QED Hamiltonian density, as given in Eq.~(\ref{eqn:Hamiltonian_T_+_+}),
	\begin{align}
	\mathcal{H}&=\bar{\psi}\left(-i\gamma^{j}\partial_{j}-i \gamma^{\wh{-}}\partial_{\wh{-}}+m \right)\psi+ eA_{\wh{\mu}}\bar{\psi}\gamma^{\wh{\mu}}\psi \notag\\
	& +\dfrac{1}{4}F^{\wh{\mu}\wh{\nu}}F_{\wh{\mu}\wh{\nu}}-F^{\wh{+} j}\partial_{\wh{+}}A_{j}-F^{\wh{+} \wh{-}}\partial_{\wh{+}}A_{\wh{-}},
	\label{eqn:Hamiltonian_T_+_+2}
	\end{align}
	Consider the first two terms of Eq.~(\ref{eqn:Hamiltonian_T_+_+2}), i.e. fermion and fermion---gauge boson interaction terms. According to the definition of free and constrained photon fields, Eqs.~(\ref{eqn:tilde{A}}) and (\ref{eqn:phi}), the first two terms can be written as
	\begin{align}
	\mathcal{H}_{\mathrm{f}}&=\bar{\psi}\left( -i\partial_{\wh{-}}\gamma^{\wh{-}}-i\partial_{j}\gamma^{j}+m\right)\psi+e\tilde{A}_{\wh{\mu}}\bar{\psi}\gamma^{\wh{\mu}} \psi+e\phi J^{\wh{+}}.
	\label{eqn:Hamiltonian_f}
	\end{align}
	
	Separating $ \psi=\tilde{\psi}+\delta_{\mathbb{C}0}\Upsilon $ for any general interpolation angle, we write
	\begin{align}
	\mathcal{H}_{\mathrm{f}}&=\left( \bar{\tilde{\psi}}+\delta_{\mathbb{C}0}\bar{\Upsilon}\right) \left( -i\partial_{\wh{-}}\gamma^{\wh{-}}-i\partial_{j}\gamma^{j}+m\right)\left( \tilde{\psi}+\delta_{\mathbb{C}0}\Upsilon\right) \notag\\
	&+e\tilde{A}_{\wh{\mu}}\left( \bar{\tilde{\psi}}+\delta_{\mathbb{C}0}\bar{\Upsilon}\right) \gamma^{\wh{\mu}} \left( \tilde{\psi}+\delta_{\mathbb{C}0}\Upsilon\right) +e\phi J^{\wh{+}}.
	\label{eqn:Hamiltonian_f_tildeandUpsilon}
	\end{align}
	
	The $ \Upsilon $ field exists only in the exact light-front, where we can make use of the identity given in Ref. ~\cite{KS}
	\begin{equation}
	\bar{\psi}\left[ \left( i\partial_j-eA_j\right)\gamma^j-m \right] \psi=-2\bar{\psi}\left( i\partial_-\gamma^-\right) \psi.
	\label{eqn:full_identity}
	\end{equation}
	
	Recalling in the light front we can separate the fermion field into the free one and constrained one $ \psi=\psi_++\psi_- =\tilde{\psi}_++\psi_-$ with $ \gamma^+\psi_-=\gamma^-\psi_+=0 $, and $ \psi_-=\tilde{\psi}_-+\Upsilon $ with $ \tilde{\psi}_- $ and $ \Upsilon $ given by Eqs.~(\ref{eqn:psi_minus_tilde}) and (\ref{eqn:Upsilon}), respectively, one realizes that identity (\ref{eqn:full_identity}) consists of four different identities
	\begin{equation}
	\bar{\tilde{\psi}}\left(i\partial_{j}\gamma^{j}-m \right) \tilde{\psi}=-2\bar{\tilde{\psi}}\left( i\partial_-\gamma^-\right) \tilde{\psi},\label{eqn:first_identity}
	\end{equation}
	\begin{equation}
	-eA_{j}\left( \bar{\tilde{\psi}}\gamma^{j}\Upsilon+\bar{\Upsilon}\gamma^{j}\tilde{\psi}\right) =-2\bar{\Upsilon}\left( i\partial_-\gamma^-\right) \Upsilon,\label{eqn:second_identity}
	\end{equation}
	\begin{align}
	&\bar{\tilde{\psi}}\left(i\partial_{j}\gamma^{j}-m \right)\Upsilon+\bar{\Upsilon} \left(i\partial_{j}\gamma^{j}-m \right)\tilde{\psi}\notag\\
	=&-\bar{\tilde{\psi}}\left(i\partial_-\gamma^- \right)\tilde{\Upsilon}-\bar{\Upsilon}\left(i\partial_-\gamma^- \right) \tilde{\psi},\label{eqn:third_identity}
	\end{align}
	and
	\begin{equation}
	-eA_{j}\bar{\tilde{\psi}}\gamma^{j}\tilde{\psi}=-\bar{\tilde{\psi}}\left(i\partial_-\gamma^- \right)\tilde{\Upsilon}-\bar{\Upsilon}\left(i\partial_-\gamma^- \right) \tilde{\psi}.\label{eqn:fourth_identity}
	\end{equation}
	The term $ \bar{\Upsilon}\left[ \left( i\partial_j-eA_j\right)\gamma^j-m \right]\Upsilon $ on the left hand side vanishes due to $ \gamma^{+2}=0 $.
	
	Noticing the fact that the transverse and mass components of the $ \bar{\Upsilon}-\Upsilon $ terms vanish, we can expand Eq.~(\ref{eqn:Hamiltonian_f_tildeandUpsilon}) as
	\begin{align}
	\mathcal{H}_{\mathrm{f}}&= \bar{\tilde{\psi}}\left( -i\partial_{\wh{-}}\gamma^{\wh{-}}-i\partial_{j}\gamma^{j}+m\right) \tilde{\psi} \notag\\
	&+\delta_{\mathbb{C}0}\left[ \bar{\tilde{\psi}}\left( -i\partial_-\gamma^-\right)\Upsilon+\bar{\Upsilon}\left( -i\partial_-\gamma^-\right)\tilde{\psi}\right.\notag\\
	&+\bar{\tilde{\psi}}\left(-i\partial_{j}\gamma^{j}+m\right)\Upsilon+\bar{\Upsilon}\left(-i\partial_{j}\gamma^{j}+m\right)\tilde{\psi}\notag\\
	&+\bar{\Upsilon}\left( -i\partial_-\gamma^-\right)\Upsilon \notag\\
	&\left.+e\tilde{A}_{j} \bar{\tilde{\psi}}\gamma^{j} \Upsilon+e\tilde{A}_{j} \bar{\Upsilon} \gamma^{j}  \tilde{\psi}\right]\notag\\
	&+e\tilde{A}_{\wh{\mu}} \bar{\tilde{\psi}}\gamma^{\wh{\mu}}\tilde{\psi} +e\phi J^{\wh{+}}.
	\label{eqn:Hamiltonian_f_termbyterm}
	\end{align}
	The first term is the free Hamiltonian $ \mathcal{H}_{\mathrm{f,0}}=\bar{\tilde{\psi}}\left( -i\partial_{\wh{-}}\gamma^{\wh{-}}-i\partial_{j}\gamma^{j}+m\right) \tilde{\psi}$, which can be reduced in the LF to  $\mathcal{H}_{\mathrm{f,0}}=\bar{\tilde{\psi}}\left( i\partial_-\gamma^-\right) \tilde{\psi} $ due to identity (\ref{eqn:first_identity}).
	The second and third lines of Eq.~(\ref{eqn:Hamiltonian_f_termbyterm}) cancel each other due to identity (\ref{eqn:third_identity}).
	The fifth line of Eq.~(\ref{eqn:Hamiltonian_f_termbyterm}) is equal to $ -2 $ times the fourth line due to identity (\ref{eqn:second_identity}). Thus Eq.~(\ref{eqn:Hamiltonian_f_termbyterm}) reduces to
	\begin{align}
	\mathcal{H}_{\mathrm{f}}&= \bar{\tilde{\psi}}\left( -i\partial_{\wh{-}}\gamma^{\wh{-}}-i\partial_{j}\gamma^{j}+m\right) \tilde{\psi}+e\tilde{A}_{\wh{\mu}} \bar{\tilde{\psi}}\gamma^{\wh{\mu}}  \tilde{\psi}\notag\\
	&+\delta_{\mathbb{C}0}\bar{\Upsilon}\left( i\partial_-\gamma^-\right)\Upsilon +e\phi J^{\wh{+}}.
	\label{eqn:Hamiltonian_f_final}
	\end{align}
	
	The rest of the Hamiltonian is the gauge boson part
	\begin{equation}
	\mathcal{H}_{\mathrm{g}}=\dfrac{1}{4}F^{\wh{\mu}\wh{\nu}}F_{\wh{\mu}\wh{\nu}}-F^{\wh{+} j}\partial_{\wh{+}}A_{j}-F^{\wh{+} \wh{-}}\partial_{\wh{+}}A_{\wh{-}},
	\end{equation}
	and similarly we want to separate it into the free part and the constraint part.
	\begin{equation}
	\mathcal{H}_{\mathrm{g}}=\mathcal{H}_{\mathrm{g}}^{\mathrm{free}}+\mathcal{H}_{\mathrm{g}}^{\mathrm{constraint}}
	\label{free_plus_constraint},
	\end{equation}
	where
	\begin{equation}
	\mathcal{H}_{\mathrm{g}}^{\mathrm{free}}=\dfrac{1}{4}\tilde{F}^{\wh{\mu}\wh{\nu}}\tilde{F}_{\wh{\mu}\wh{\nu}}-\tilde{F}^{\wh{+} j}\partial_{\wh{+}}\tilde{A}_{j}-\tilde{F}^{\wh{+} \wh{-}}\partial_{\wh{+}}\tilde{A}_{\wh{-}},\label{eqn:H_g_free}
	\end{equation}
	and $ \tilde{F}^{\wh{\mu}\wh{\nu}} $ is defined in terms of the free photon fields as given in Eq.~(\ref{eqn:tilde{A}}).
	
	Using $ A_{\wh{\mu}}=\tilde{A}_{\wh{\mu}}+{g_{\wh{\mu}}}^{\wh{+}}\phi $ and $ A^{\wh{\mu}}=\tilde{A}^{\wh{\mu}}+g^{\wh{\mu}\wh{+}}\phi $, we find 
	\begin{align}
	&\mathcal{H}_{\mathrm{g}}^{\mathrm{constraint}}=\mathcal{H}_{\mathrm{g}}-\mathcal{H}_{\mathrm{g}}^{\mathrm{free}}\notag\\
	&=\frac{1}{2}\left( -\mathbb{S}\partial^{\wh{+}}\phi\partial_{\wh{-}}\phi+\mathbb{C}\partial^{\wh{-}}\phi\partial_{\wh{-}}\phi+\mathbb{C}\partial^{j}\phi\partial_{j}\phi\right) 
	\end{align}
	with all other terms vanish upon applying the interpolation gauge condition $ \partial_{j}A_{j}=-\frac{1}{\mathbb{C}}\partial_{\wh{-}}A_{\wh{-}} $.
	
	Using integration by parts, 
	\begin{align} \mathcal{H}_{\mathrm{g}}^{\mathrm{constraint}}&=\frac{1}{2}\phi \left(\mathbb{S}\partial^{\wh{+}}\partial_{\wh{-}}-\mathbb{C}\partial^{\wh{-}}\partial_{\wh{-}}+\mathbb{C}\partial_{j}\partial_{j} \right) \phi\notag\\
	&=\frac{1}{2}\phi\left( \mathbb{C}\boldsymbol{\partial}_{\perp}^2+\partial_{\wh{-}}^2\right) \phi\notag\\
	&=-\frac{1}{2}e\phi J^{\wh{+}},
	\end{align}
	according to the definition of the constraint photon field $ \phi $ in Eq.~(\ref{eqn:phi}).
	
	Thus 
	\begin{equation}
	\mathcal{H}_{\mathrm{g}}=\mathcal{H}_{\mathrm{g}}^{\mathrm{free}}-\frac{1}{2}e\phi J^{\wh{+}},\label{eqn:H_g_final}
	\end{equation}
	with $ \mathcal{H}_{\mathrm{g}}^{\mathrm{free}} $ given by Eq.~(\ref{eqn:H_g_free}).
	
	Adding two pieces together, we can identify the free and interaction Hamiltonian
	\begin{equation}
	\mathcal{H}=\mathcal{H}_{\mathrm{f}}+\mathcal{H}_{\mathrm{g}}=\mathcal{H}_{0}+\mathcal{V},
	\end{equation}
	where 
	\begin{align}
	\mathcal{H}_{0}&=\bar{\tilde{\psi}}\left( -i\partial_{\wh{-}}\gamma^{\wh{-}}-i\partial_{j}\gamma^{j}+m\right) \tilde{\psi}\notag\\
	&+\dfrac{1}{4}\tilde{F}^{\wh{\mu}\wh{\nu}}\tilde{F}_{\wh{\mu}\wh{\nu}}-\tilde{F}^{\wh{+} j}\partial_{\wh{+}}\tilde{A}_{j}-\tilde{F}^{\wh{+} \wh{-}}\partial_{\wh{+}}\tilde{A}_{\wh{-}},
	\end{align}
	and
	\begin{align}
	\mathcal{V}&=e\tilde{A}_{\wh{\mu}} \bar{\tilde{\psi}}\gamma^{\wh{\mu}}  \tilde{\psi}
	+\delta_{\mathbb{C}0}\bar{\Upsilon}\left( i\partial_-\gamma^-\right)\Upsilon 
	+\frac{1}{2}e\phi J^{\wh{+}}.
	\end{align}
	
	Thus, we get the interpolating QED Hamiltonian density as shown in the main text Eqs.~(\ref{eqn:H0_QED}) and (\ref{eqn:V_QED}).
	
	\section{Sum of the Interpolating Time-Ordered Fermion Propagators}
	\label{app:Propagator}
	
	In this Appendix, we show how the addition of the two time-ordered propagators gives correctly the covariant one. We start with the expressions given in Eqs.~(\ref{eqn:propagator_TOa}) and (\ref{eqn:propagator_TOb}).
		\begin{align}
		&\Sigma_a+\Sigma_b\notag\\
		&=\frac{1}{2Q^{\wh{+}}}\left( \frac{\slashed{Q}_a+m}{q_{\wh{+}}-Q_{a\wh{+}}}-\frac{-\slashed{Q}_b+m}{q_{\wh{+}}+Q_{b\wh{+}}}\right)\notag\\
		&=\frac{1}{2Q^{\wh{+}}}\left(\frac{\mathbb{C}\slashed{Q}_a+\mathbb{C}m}{\mathbb{C}q_{\wh{+}}+\mathbb{S}q_{\wh{-}}-Q^{\wh{+}}}-\frac{-\mathbb{C}\slashed{Q}_b+\mathbb{C}m}{\mathbb{C}q_{\wh{+}}+\mathbb{S}q_{\wh{-}}+Q^{\wh{+}}}\right) \notag\\
		\end{align}
		where we have used (\ref{eqn:qanew+}) and (\ref{eqn:qbnew+}).
		
		Using the relationship between superscripts and subscripts it can furthermore be written as:
		\begin{widetext}
		\begin{align}
		\Sigma_a+\Sigma_b
		&=\frac{1}{2Q^{\wh{+}}}\left( \frac{\mathbb{C}\slashed{Q}_a+\mathbb{C}m}{q^{\wh{+}}-Q^{\wh{+}}}-\frac{-\mathbb{C}\slashed{Q}_b+\mathbb{C}m}{q^{\wh{+}}+Q^{\wh{+}}}\right)\notag\\
		&=\frac{\mathbb{C}}{2Q^{\wh{+}}}\left(\frac{\gamma^{\wh{+}}Q_{a\wh{+}}+\gamma^{\wh{-}}q_{\wh{-}}+\gamma^{\perp}.q_{\perp}+m }{q^{\wh{+}}-Q^{\wh{+}}}
		-\frac{-\gamma^{\wh{+}}Q_{b\wh{+}}+\gamma^{\wh{-}}q_{\wh{-}}+\gamma^{\perp}.q_{\perp}+m}{q^{\wh{+}}+Q^{\wh{+}}}\right) \notag\\
		\end{align}
		\end{widetext}
		where it is worth paying attention to the fact that the sign is different between $ q $ and $ q_b $ and we have replaced all $ q_b $'s with $ q $'s in the second line.
		
		The above equation can be further simplified:
		\begin{widetext}
		\begin{align}
		&\Sigma_a+\Sigma_b\notag\\
		&=\frac{\mathbb{C}}{2Q^{\wh{+}}}\; \frac{\gamma^{\wh{+}}\left( Q_{a\wh{+}}(q^{\wh{+}}+Q^{\wh{+}})+Q_{b\wh{+}}(q^{\wh{+}}-Q^{\wh{+}})\right)+\left( \gamma^{\wh{-}}q_{\wh{-}}+\gamma^{\perp}.q_{\perp}+m\right) \left( (q^{\wh{+}}+Q^{\wh{+}})-(q^{\wh{+}}-Q^{\wh{+}})\right)  }{(q^{\wh{+}})^2-(Q^{\wh{+}})^2}\notag\\
		&=\frac{\mathbb{C}}{2Q^{\wh{+}}}\;\frac{\gamma^{\wh{+}}\left( q^{\wh{+}}(Q_{a\wh{+}}+Q_{b\wh{+}})+Q^{\wh{+}}(Q_{a\wh{+}}-Q_{b\wh{+}})\right) +2Q^{\wh{+}}\left( \gamma^{\wh{-}}q_{\wh{-}}+\gamma^{\perp}.q_{\perp}+m\right)}{(q^{\wh{+}})^2-\left( q_{\wh{-}}^2+\mathbb{C}q_{\perp}^2+\mathbb{C}m^2\right) }\notag\\
		&=\frac{\mathbb{C}}{2Q^{\wh{+}}}\;\frac{\gamma^{\wh{+}}\left( \mathbb{C}q_{\wh{+}}+\mathbb{S}q_{\wh{-}}\right) \left( \frac{2Q^{\wh{+}}}{\mathbb{C}}\right) +\gamma^{\wh{+}}Q^{\wh{+}}\left( \frac{-2\mathbb{S}q_{\wh{-}}}{\mathbb{C}}\right) +2Q^{\wh{+}}\left( \gamma^{\wh{-}}q_{\wh{-}}+\gamma^{\perp}.q_{\perp}+m\right)}{q^{\wh{+}}\left( \mathbb{C}q_{\wh{+}}+\mathbb{S}q_{\wh{-}}\right) -q_{\wh{-}}\left( \mathbb{S}q^{\wh{+}}-\mathbb{C}q^{\wh{-}}\right) -\mathbb{C}q_{\perp}^2-\mathbb{C}m^2}\notag\\
		&=\frac{\mathbb{C}\gamma^{\wh{+}} q_{\wh{+}} +\mathbb{C}\left( \gamma^{\wh{-}}q_{\wh{-}}+\gamma^{\perp}\cdot q_{\perp}+m\right) }{\mathbb{C}q^{\wh{+}}q_{\wh{+}}+\mathbb{C}q_{\wh{-}}q^{\wh{-}}+\mathbb{C}q^{\perp}.q_{\perp}-\mathbb{C}m^2}\notag\\
		&=\frac{\slashed{q}+m}{q^2-m^2}
		\end{align}
	\end{widetext}
	Thus, the total result is proved to be consistent with the Feynman propagator.

	\section{Apparent Angle Distribution of Interpolating Helicity Amplitudes for the two scalar particle production in 
	$e^+ e^ -$ Annihilation Process}
	\label{app:ApprantAngle}
	
	\begin{figure}
		\centering
		\subfloat[]{
			\includegraphics[width=0.48\columnwidth]{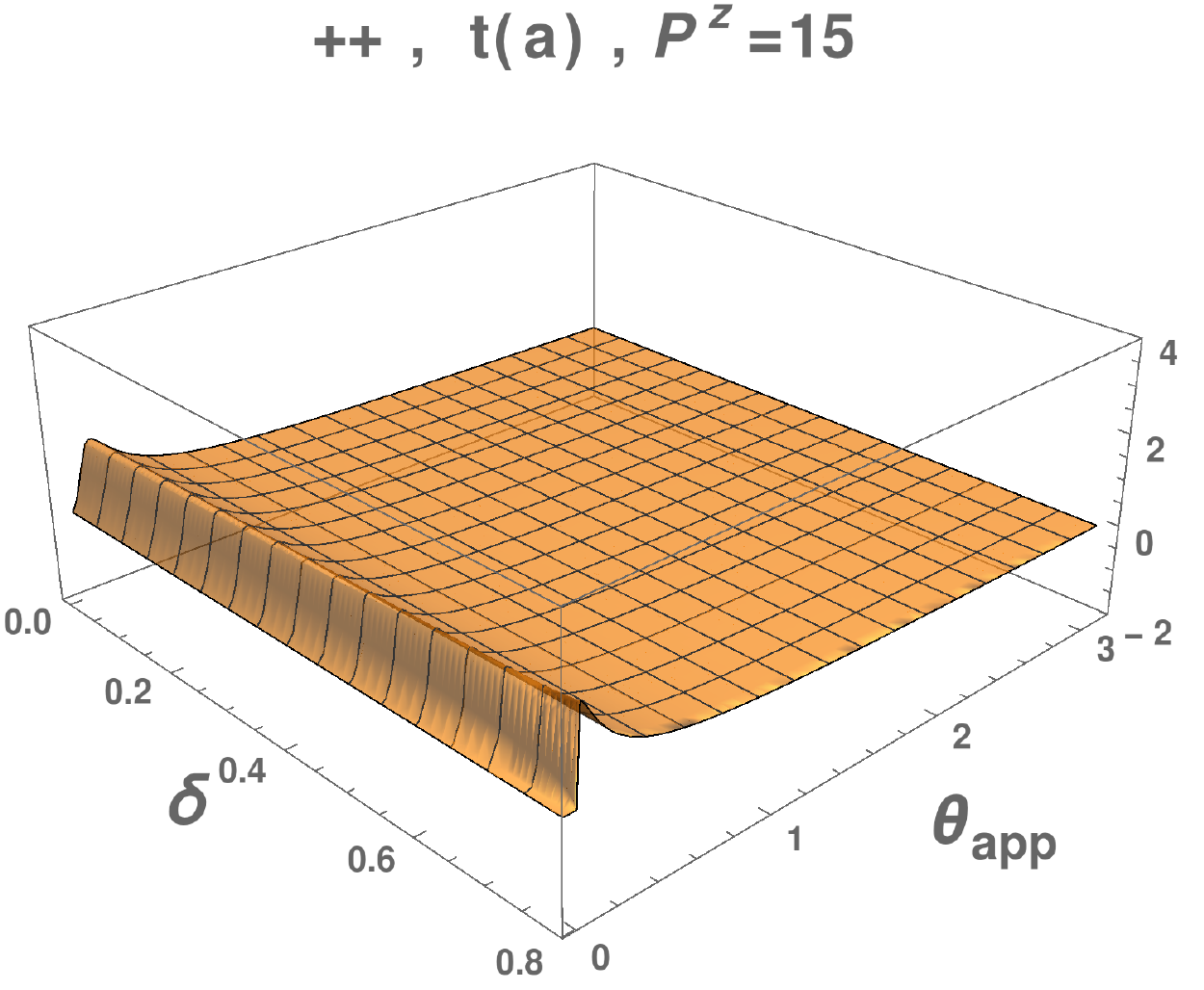}
			\label{fig:eessAppAngDistRRtaPzp}}
		\centering
		\subfloat[]{
			\includegraphics[width=0.48\columnwidth]{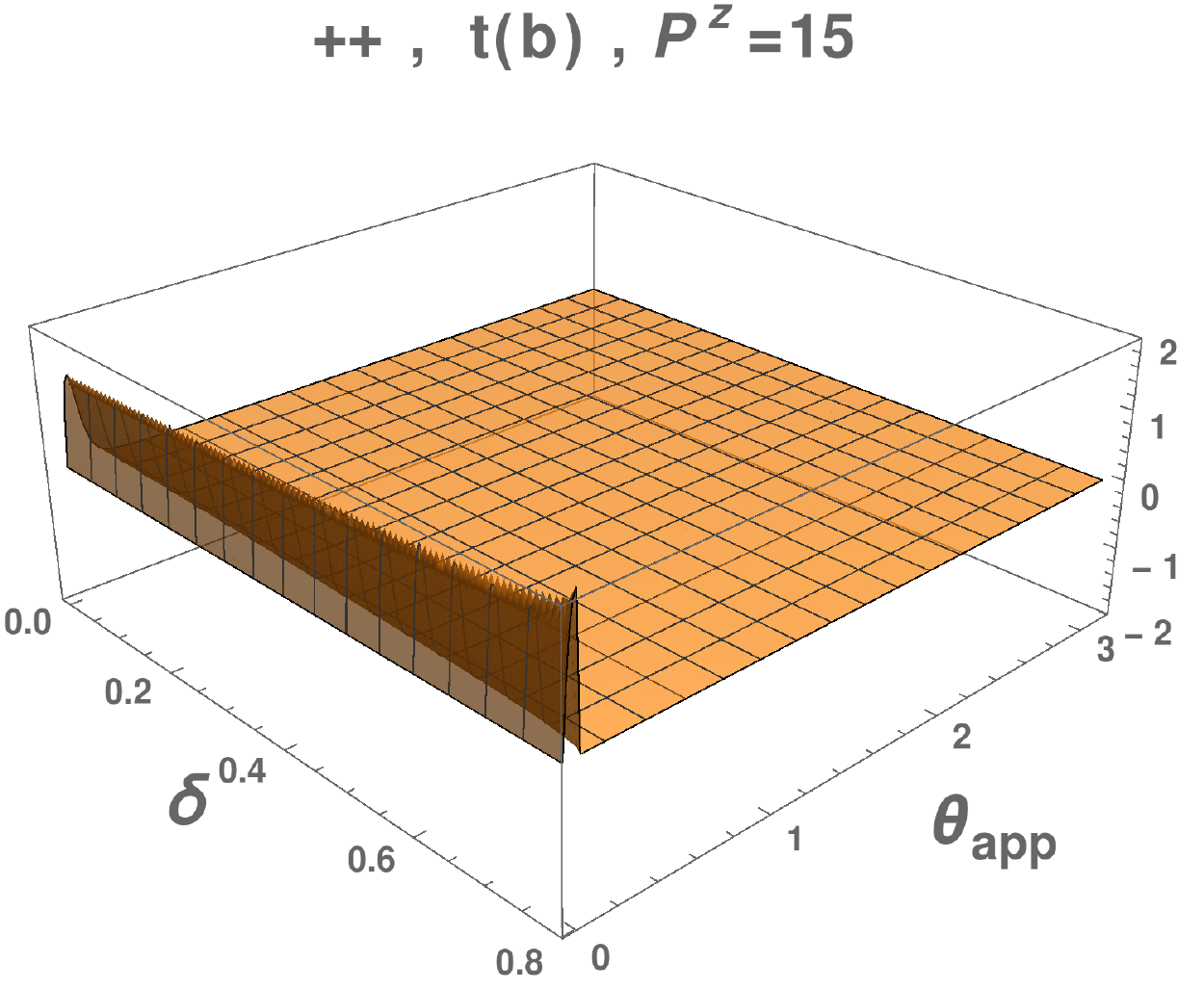}
			\label{fig:eessAppAngDistRRtbPzp}}
		\\
		\centering
		\subfloat[]{
			\includegraphics[width=0.48\columnwidth]{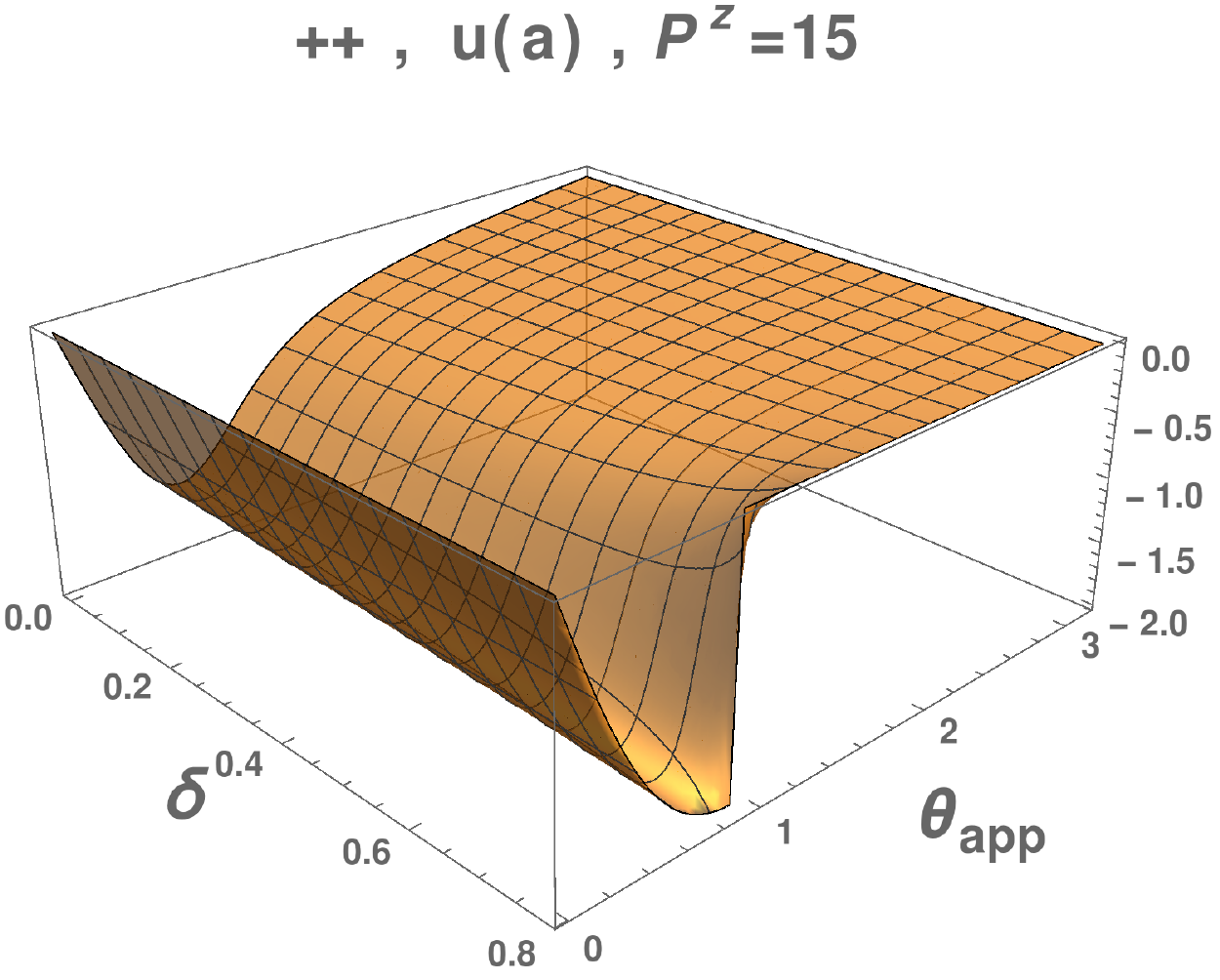}
			\label{fig:eessAppAngDistRRuaPzp}}
		\centering
		\subfloat[]{
			\includegraphics[width=0.48\columnwidth]{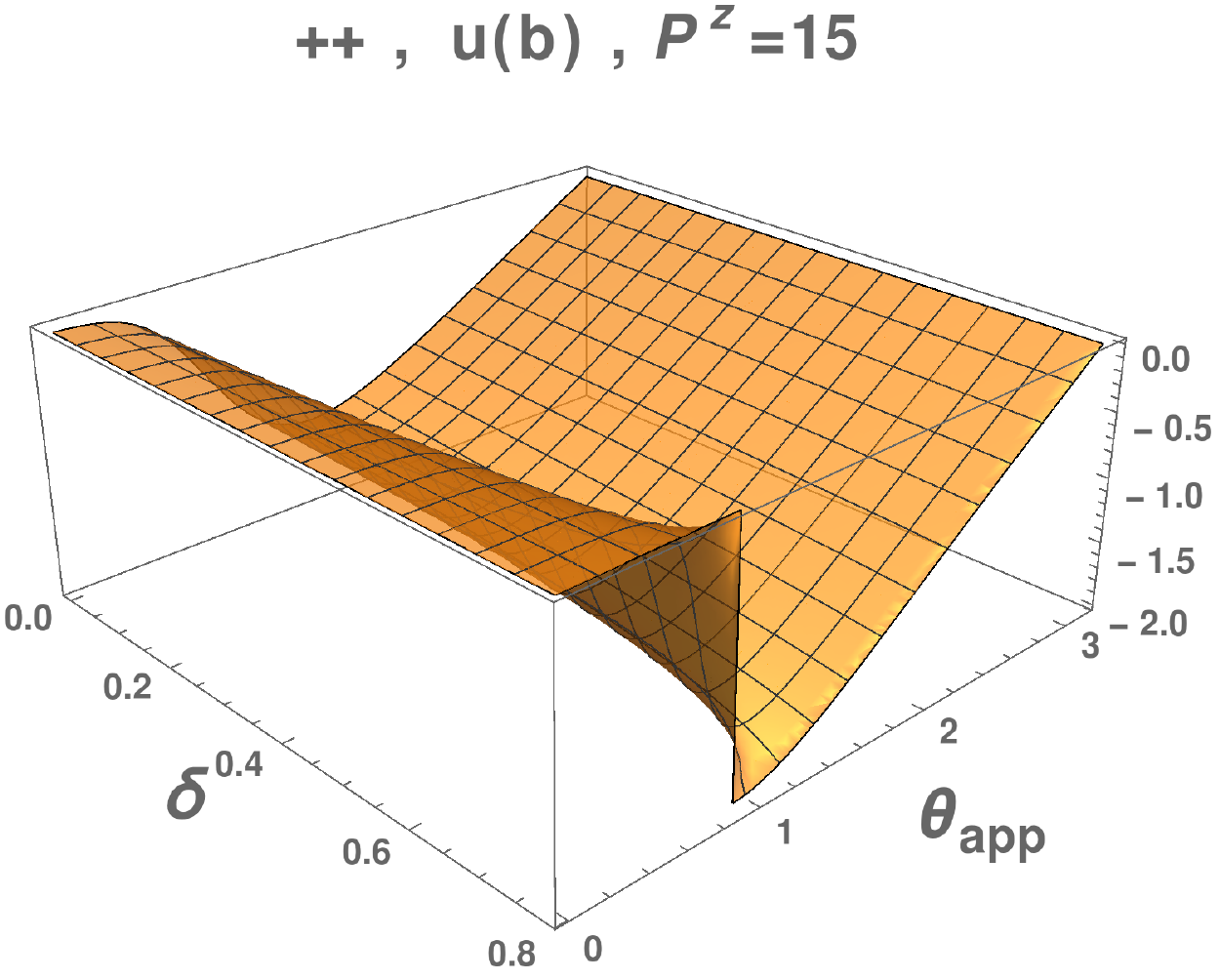}
			\label{fig:eessAppAngDistRRubPzp}}
		\caption{\label{fig:eessAppAngDistRRPzp}Apparent Angular distribution of the helicity amplitude $ ++ $ 
			for (a) t-channel a time-ordering, $\mathcal{M}_{a,t}^{+,+}$  (b) t-channel b time-ordering, $\mathcal{M}_{b,t}^{+,+}$ 
			(c) u-channel a time-ordering, $\mathcal{M}_{a,u}^{+,+}$ (d) u-channel b time-ordering, $\mathcal{M}_{b,u}^{+,+}$.}
	\end{figure}
	
	\begin{figure}
		\centering
		\subfloat[]{
			\includegraphics[width=0.48\columnwidth]{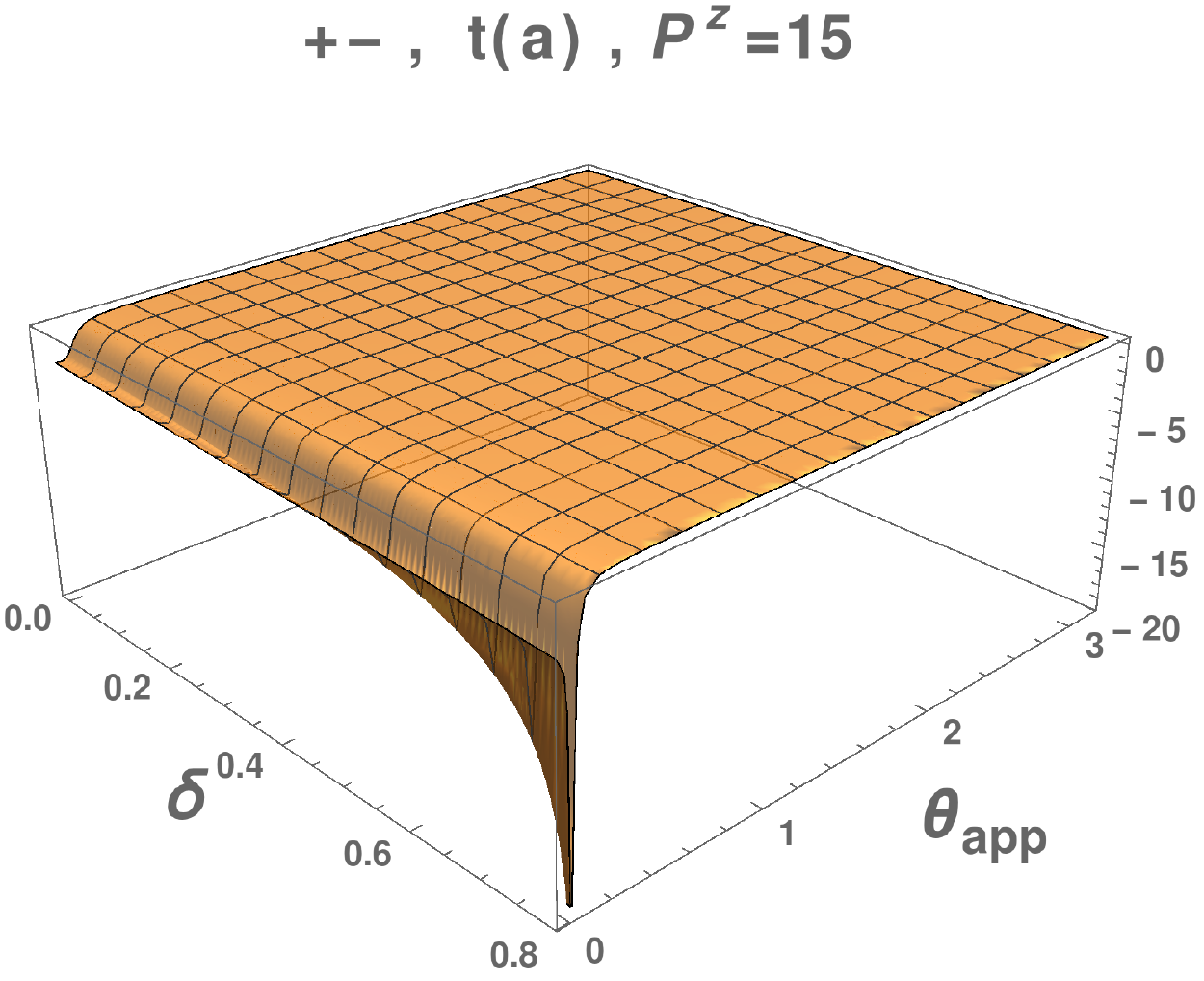}
			\label{fig:eessAppAngDistRLtaPzp}}
		\centering
		\subfloat[]{
			\includegraphics[width=0.48\columnwidth]{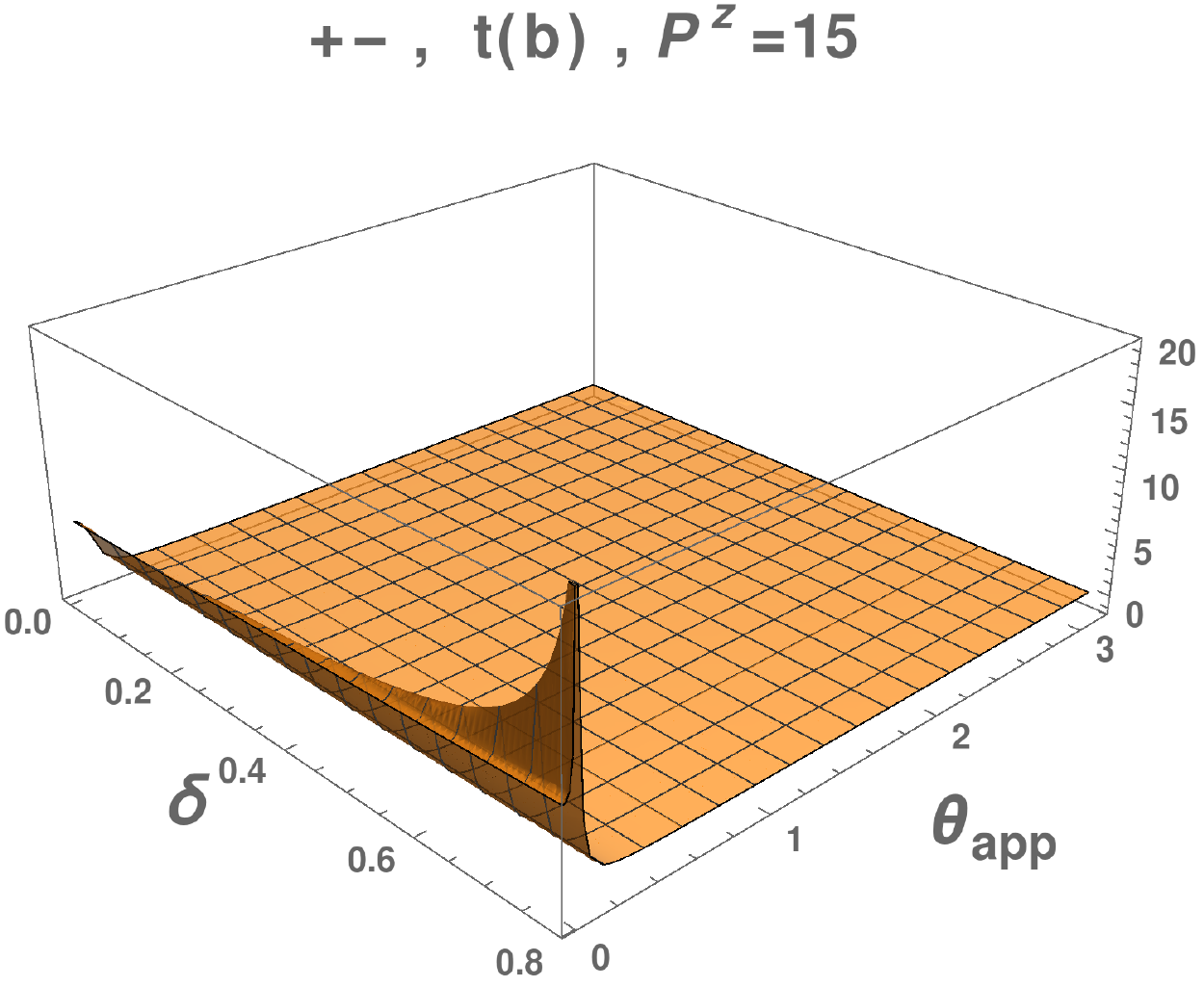}
			\label{fig:eessAppAngDistRLtbPzp}}
		\\
		\centering
		\subfloat[]{
			\includegraphics[width=0.48\columnwidth]{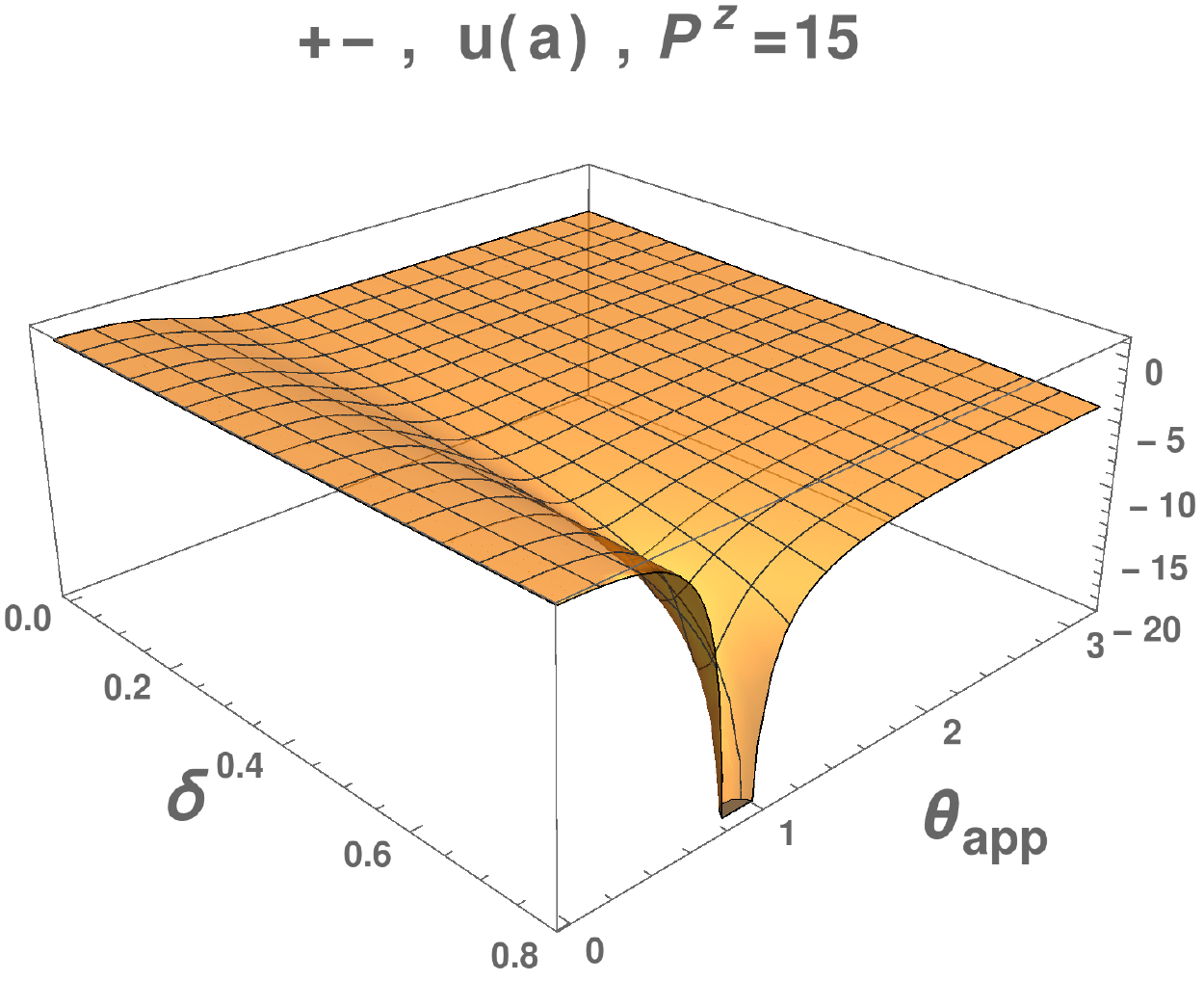}
			\label{fig:eessAppAngDistRLuaPzp}}
		\centering
		\subfloat[]{
			\includegraphics[width=0.48\columnwidth]{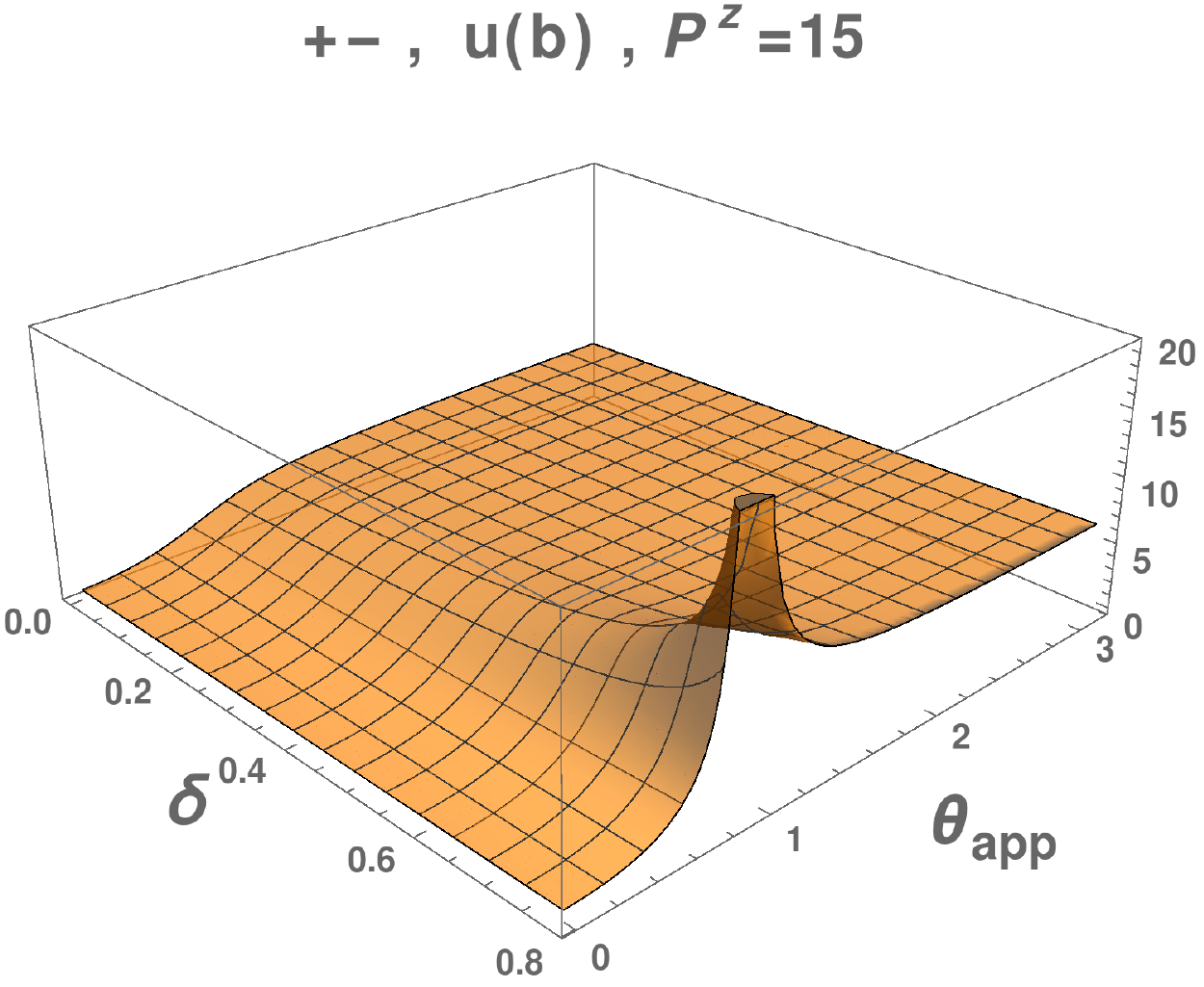}
			\label{fig:eessAppAngDistRLubPzp}}
		\caption{\label{fig:eessAppAngDistRLPzp}Apparent Angular distribution of the helicity amplitude $ +- $ 
			for (a) t-channel a time-ordering, $\mathcal{M}_{a,t}^{+,-}$  (b) t-channel b time-ordering, $\mathcal{M}_{b,t}^{+,-}$ 
			(c) u-channel a time-ordering, $\mathcal{M}_{a,u}^{+,-}$ (d) u-channel b time-ordering, $\mathcal{M}_{b,u}^{+,-}$.}
	\end{figure}
	
	\begin{figure}
		\centering
		\subfloat[]{
			\includegraphics[width=0.48\columnwidth]{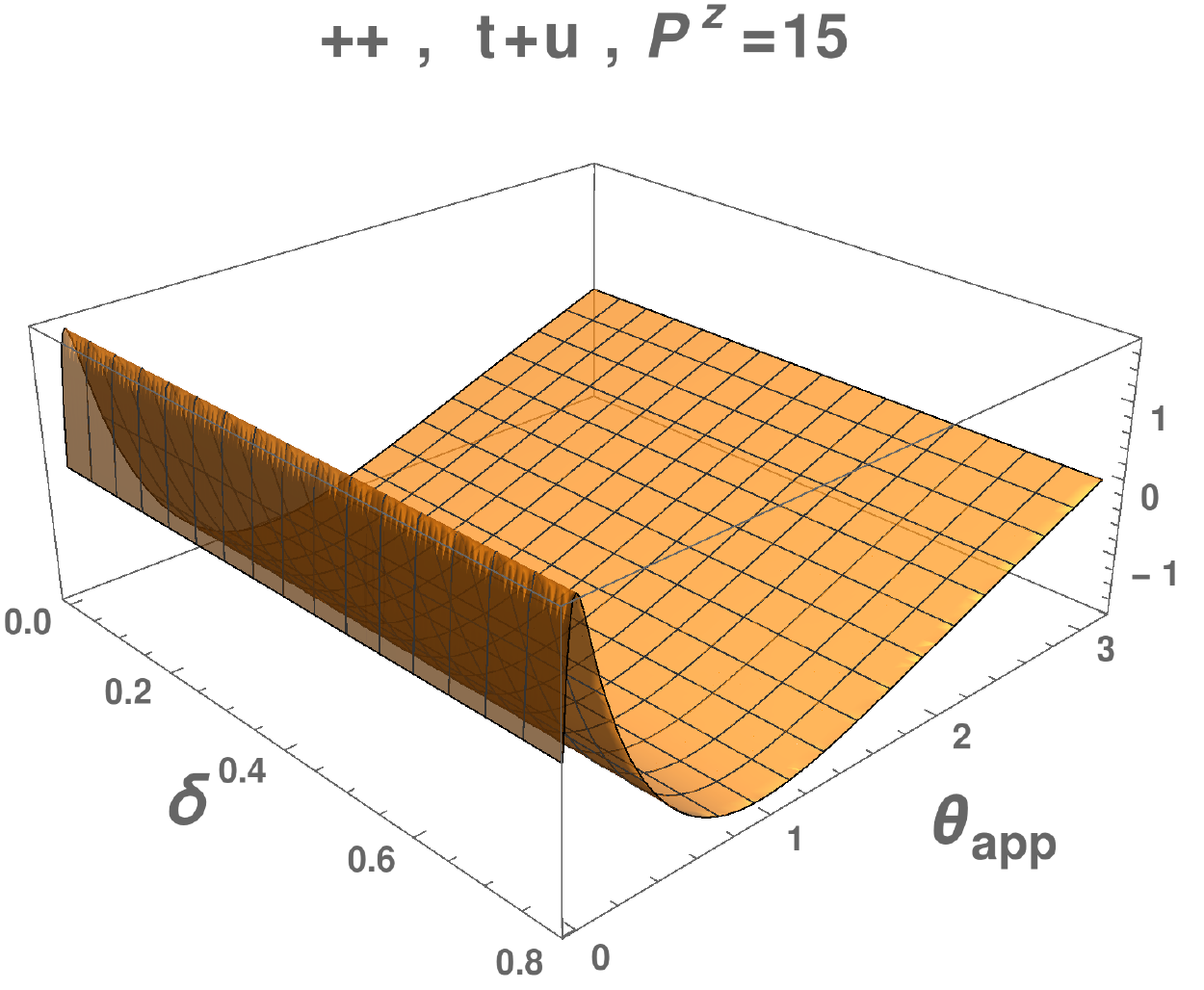}
			\label{fig:eessAppAngDistRRtuPzp}}
		\centering
		\subfloat[]{
			\includegraphics[width=0.48\columnwidth]{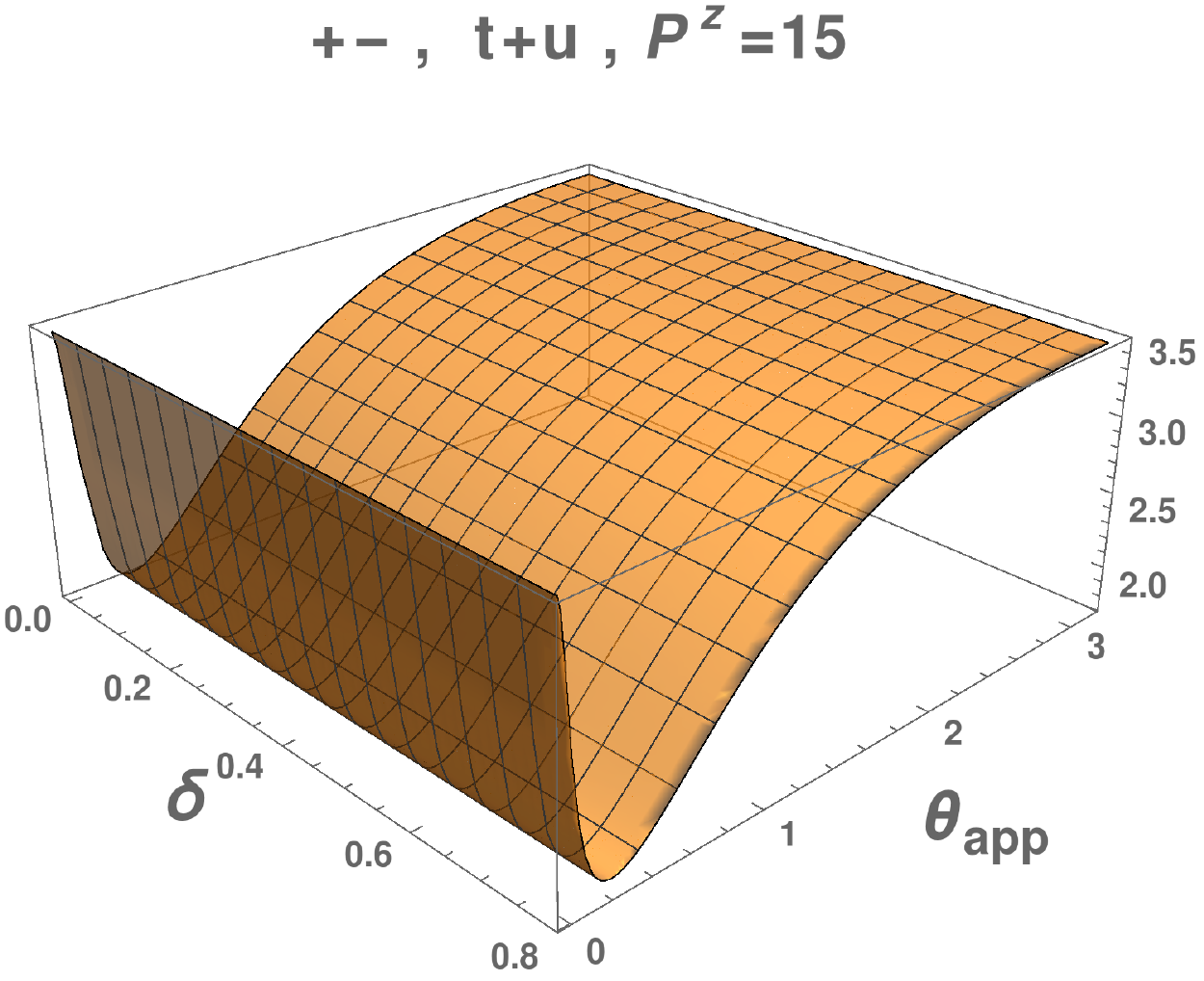}
			\label{fig:eessAppAngDistRLtuPzp}}
		\\
		\centering
		\subfloat[]{
			\includegraphics[width=0.48\columnwidth]{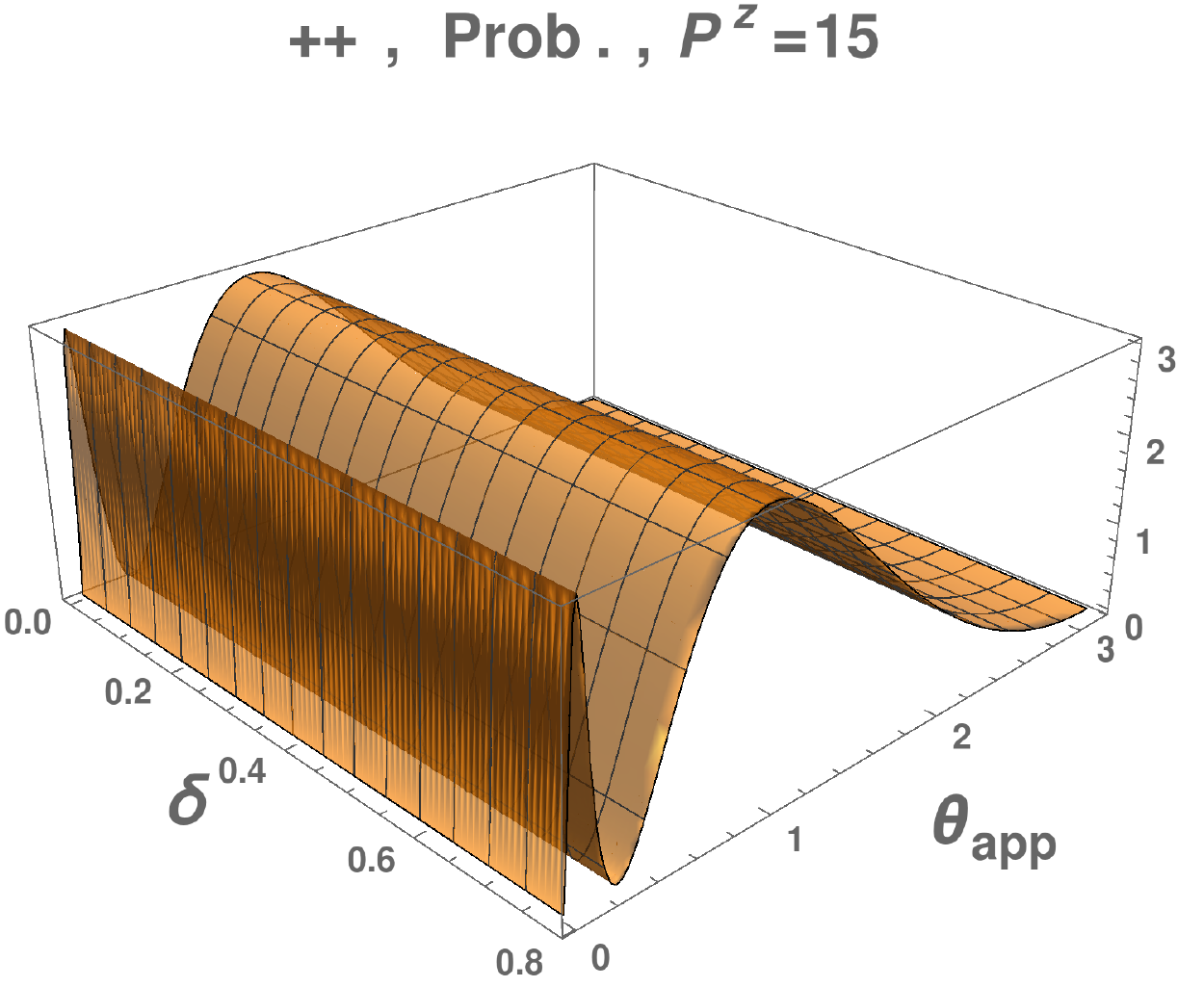}
			\label{fig:eessAppAngDistRRprobPzp}}
		\centering
		\subfloat[]{
			\includegraphics[width=0.48\columnwidth]{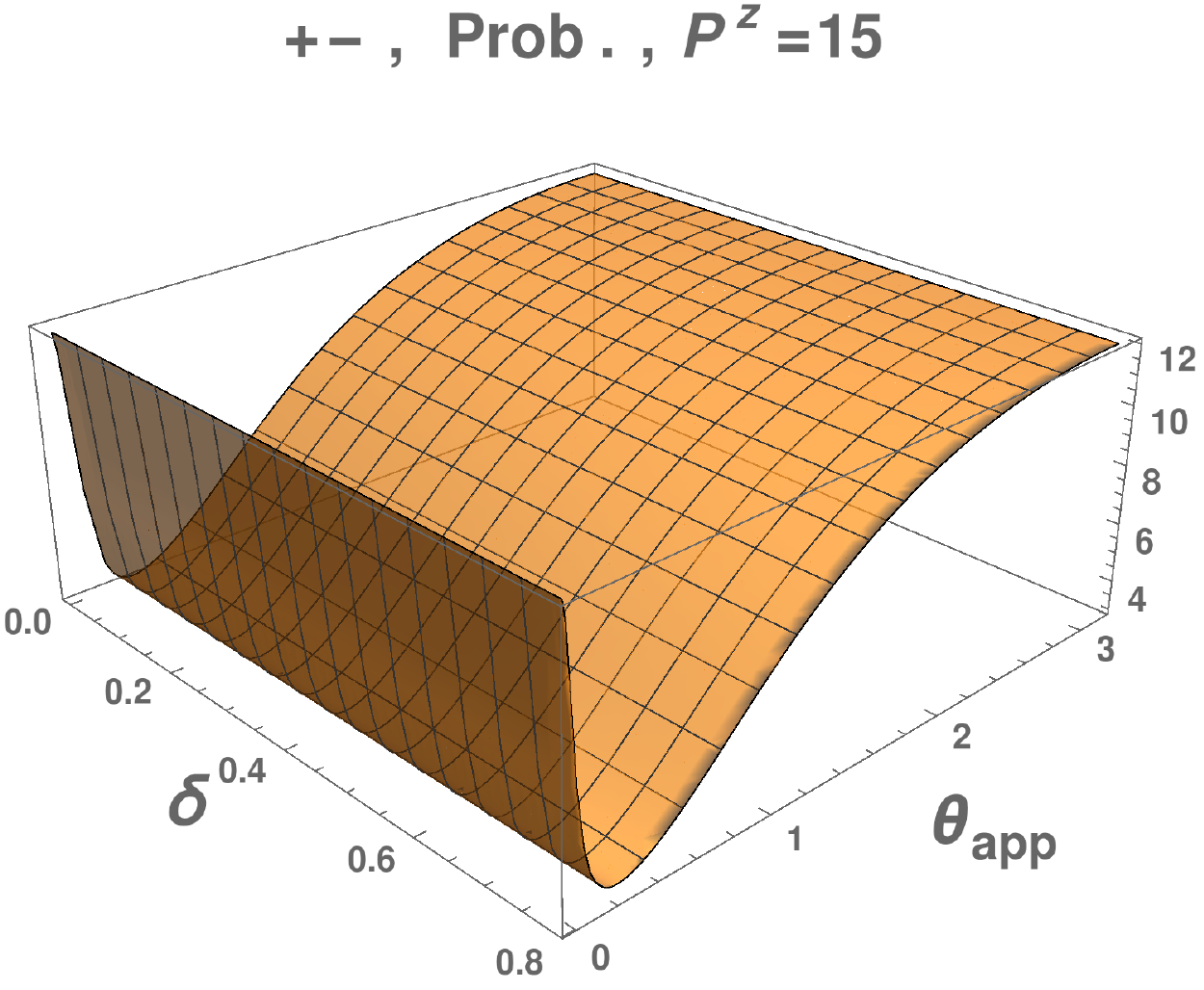}
			\label{fig:eessAppAngDistRLprobPzp}}
		\caption{\label{fig:eessAppAngDistAmpAndProbPzp} (a) $\mathcal{M}_{a,t}^{+,+}+\mathcal{M}_{b,t}^{+,+}+\mathcal{M}_{a,u}^{+,+}+\mathcal{M}_{a,u}^{+,+}$ 
			(b) $\mathcal{M}_{a,t}^{+,-}+\mathcal{M}_{b,t}^{+,-}+\mathcal{M}_{a,u}^{+,-}+\mathcal{M}_{a,u}^{+,-}$ 
			(c) $|\mathcal{M}_{a,t}^{+,+}+\mathcal{M}_{b,t}^{+,+}+\mathcal{M}_{a,u}^{+,+}+\mathcal{M}_{a,u}^{+,+}|^2$  
			(d) $|\mathcal{M}_{a,t}^{+,-}+\mathcal{M}_{b,t}^{+,-}+\mathcal{M}_{a,u}^{+,-}+\mathcal{M}_{a,u}^{+,-}|^2$ }
	\end{figure}
	\begin{figure}
		\centering
		\subfloat[]{
			\includegraphics[width=0.48\columnwidth]{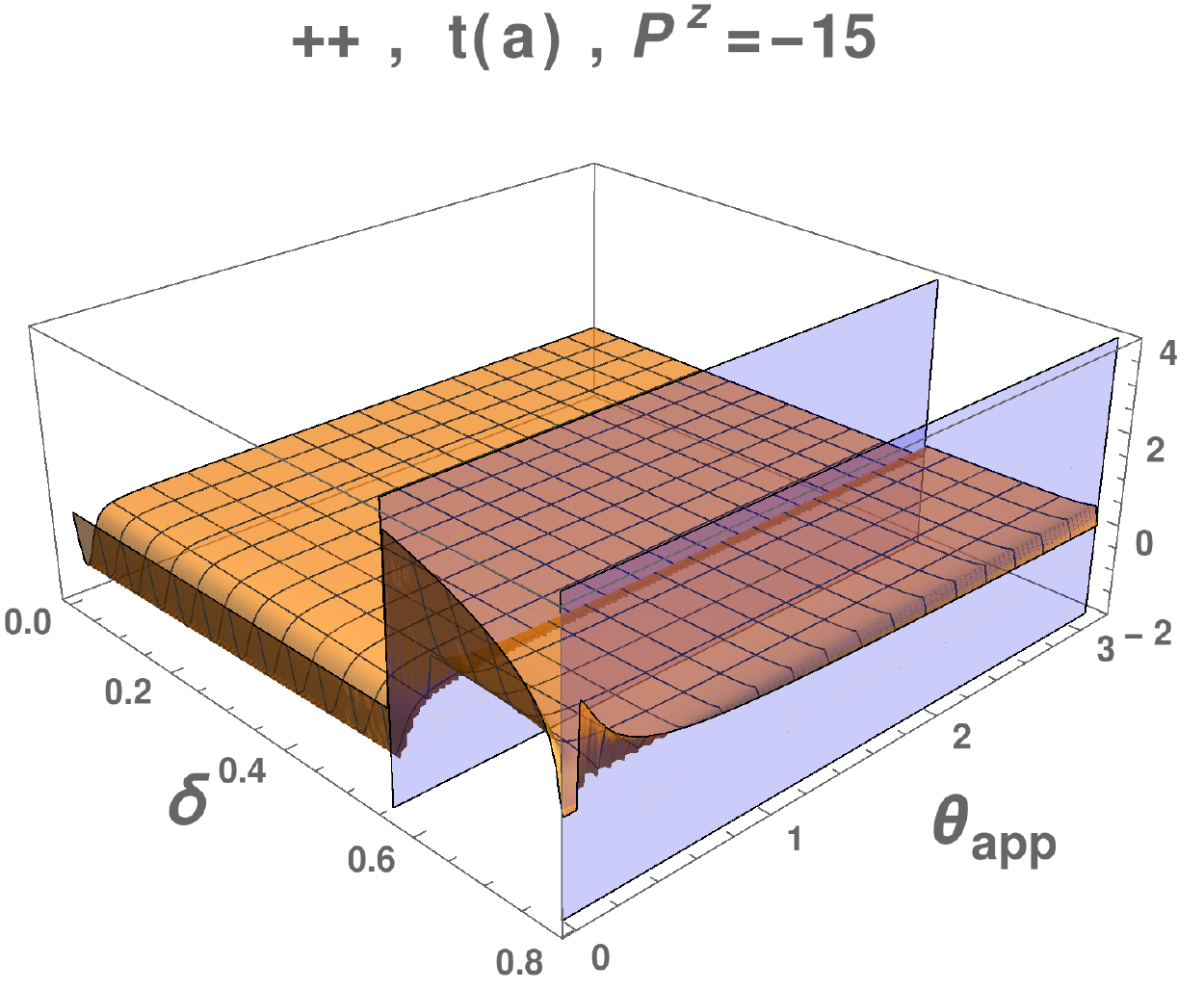}
			\label{fig:eessAppAngDistRRtaPzm}}
		\centering
		\subfloat[]{
			\includegraphics[width=0.48\columnwidth]{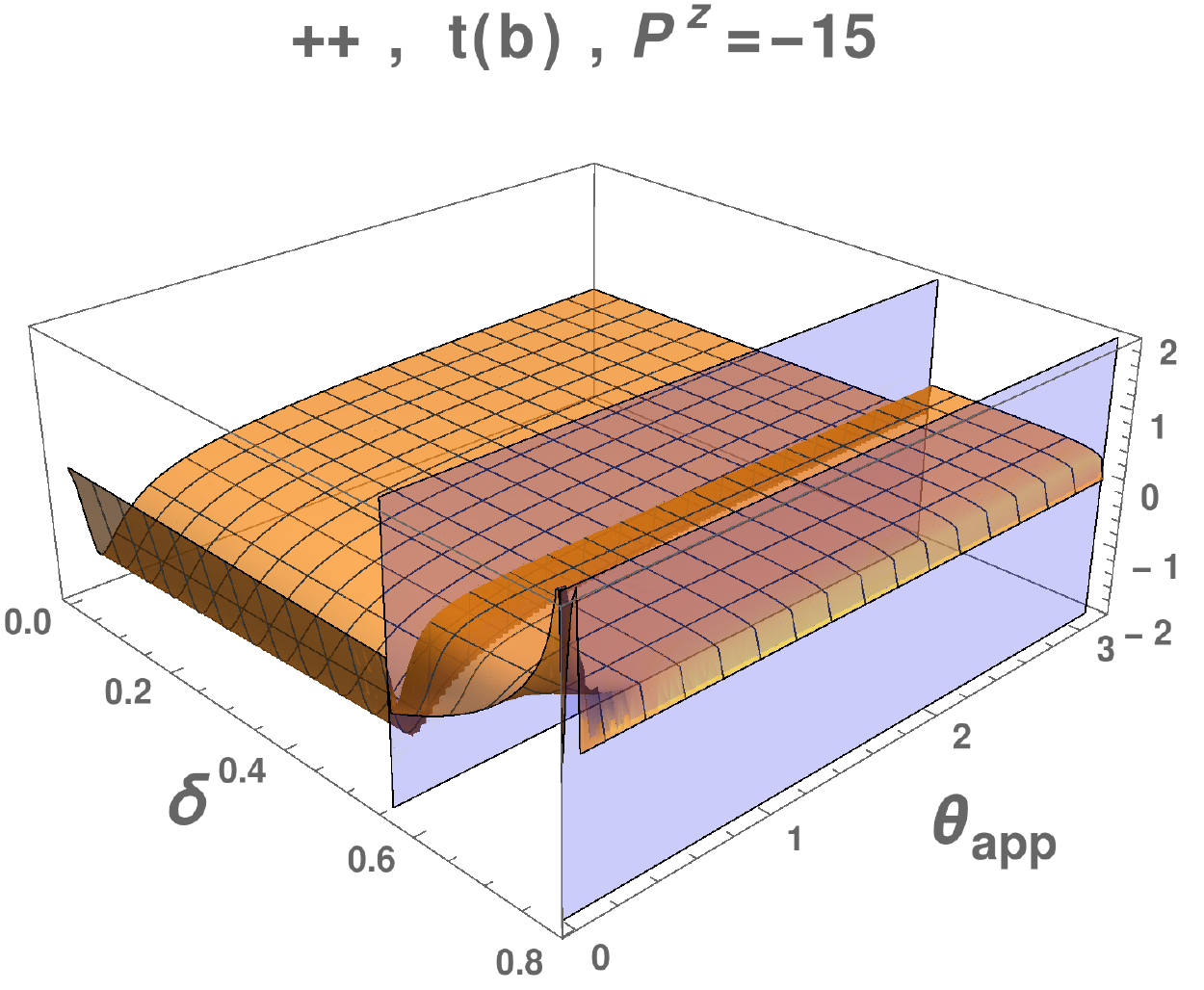}
			\label{fig:eessAppAngDistRRtbPzm}}
		\\
		\centering
		\subfloat[]{
			\includegraphics[width=0.48\columnwidth]{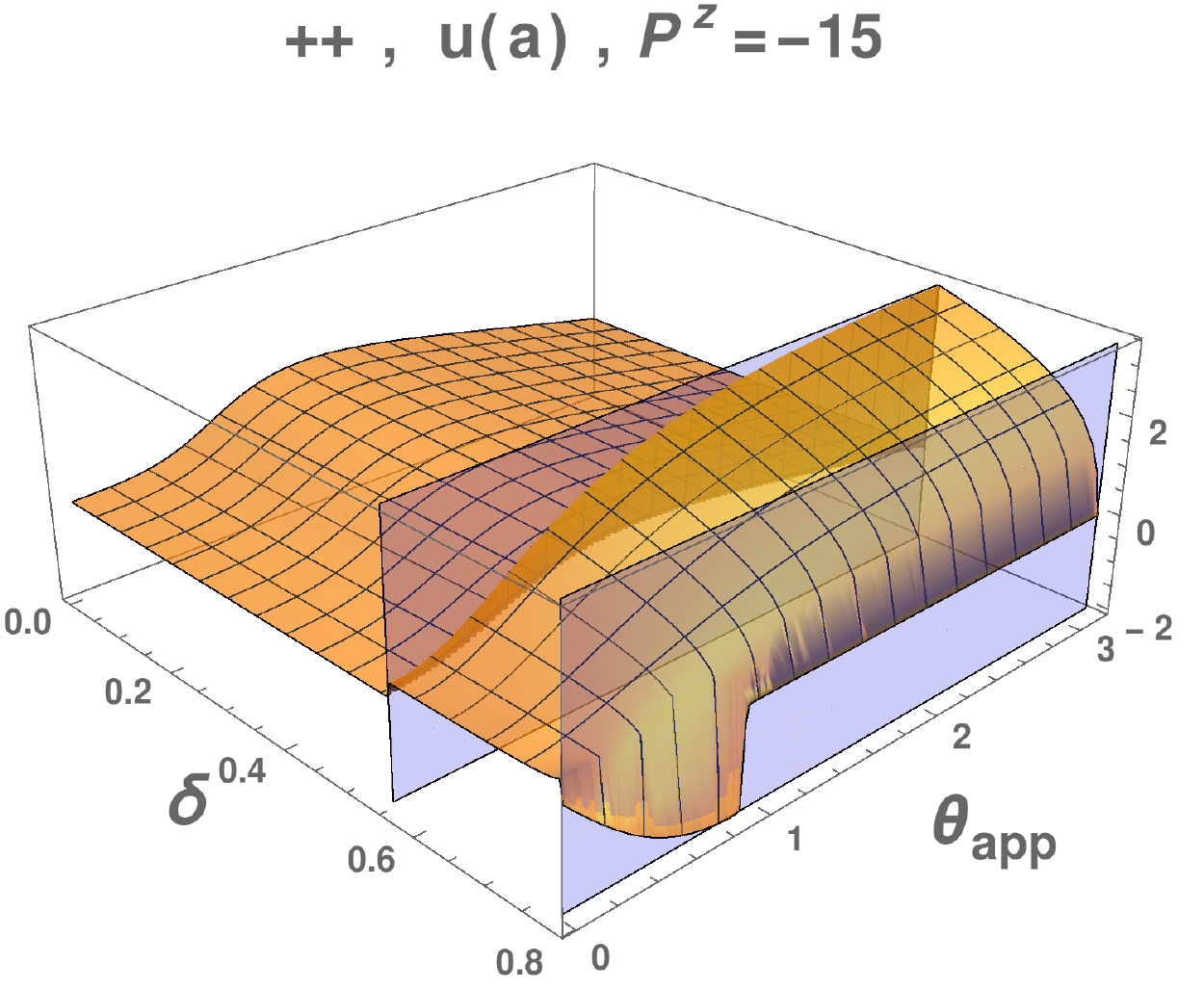}
			\label{fig:eessAppAngDistRRuaPzm}}
		\centering
		\subfloat[]{
			\includegraphics[width=0.48\columnwidth]{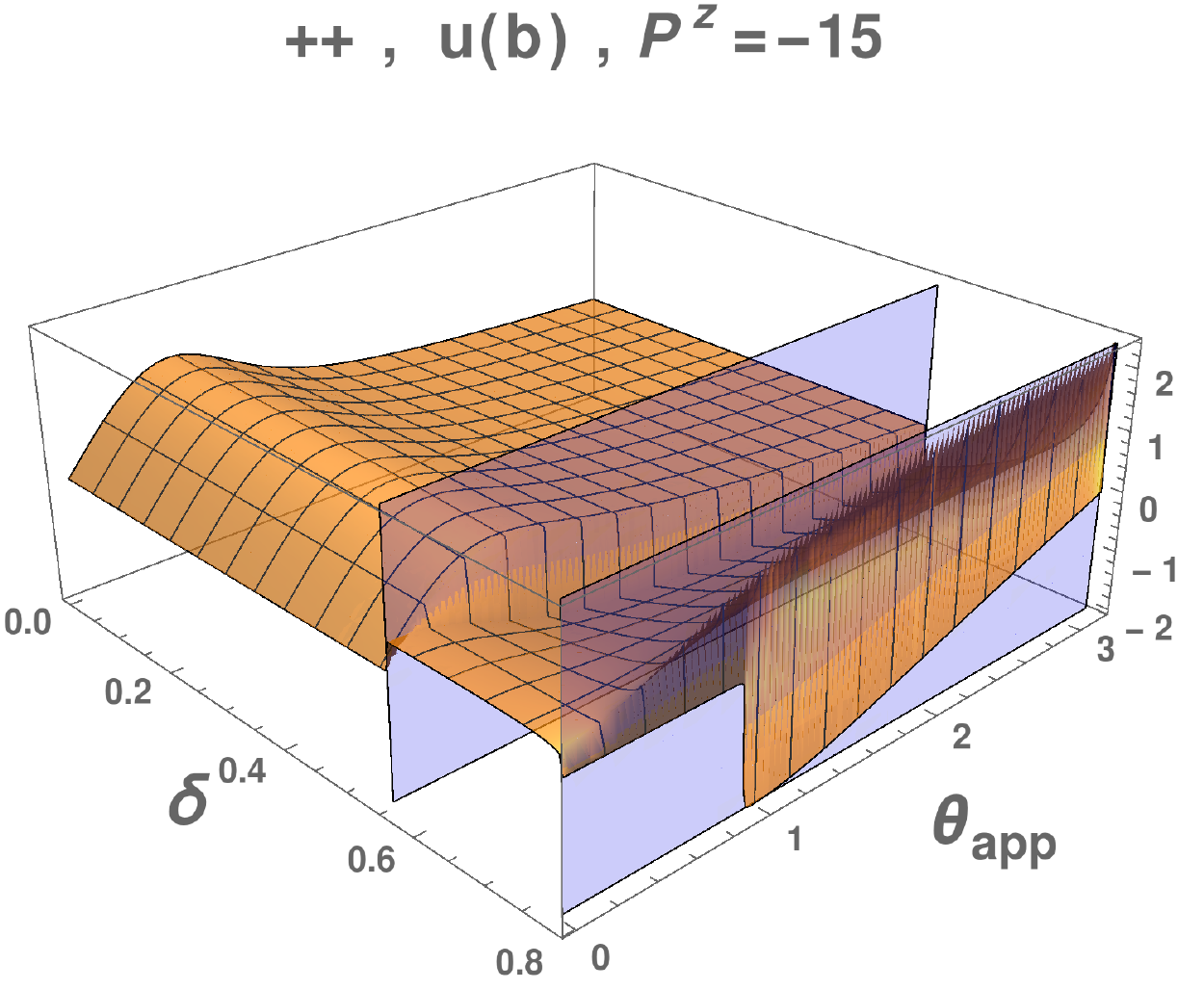}
			\label{fig:eessAppAngDistRRubPzm}}
		\caption{\label{fig:eessAppAngDistRRPzm}Apparent Angular distribution of the helicity amplitude $ ++ $ 
			for (a) t-channel a time-ordering, $\mathcal{M}_{a,t}^{+,+}$  (b) t-channel b time-ordering, $\mathcal{M}_{b,t}^{+,+}$ 
			(c) u-channel a time-ordering, $\mathcal{M}_{a,u}^{+,+}$ (d) u-channel b time-ordering, $\mathcal{M}_{b,u}^{+,+}$.}
	\end{figure}
	
	\begin{figure}
		\centering
		\subfloat[]{
			\includegraphics[width=0.48\columnwidth]{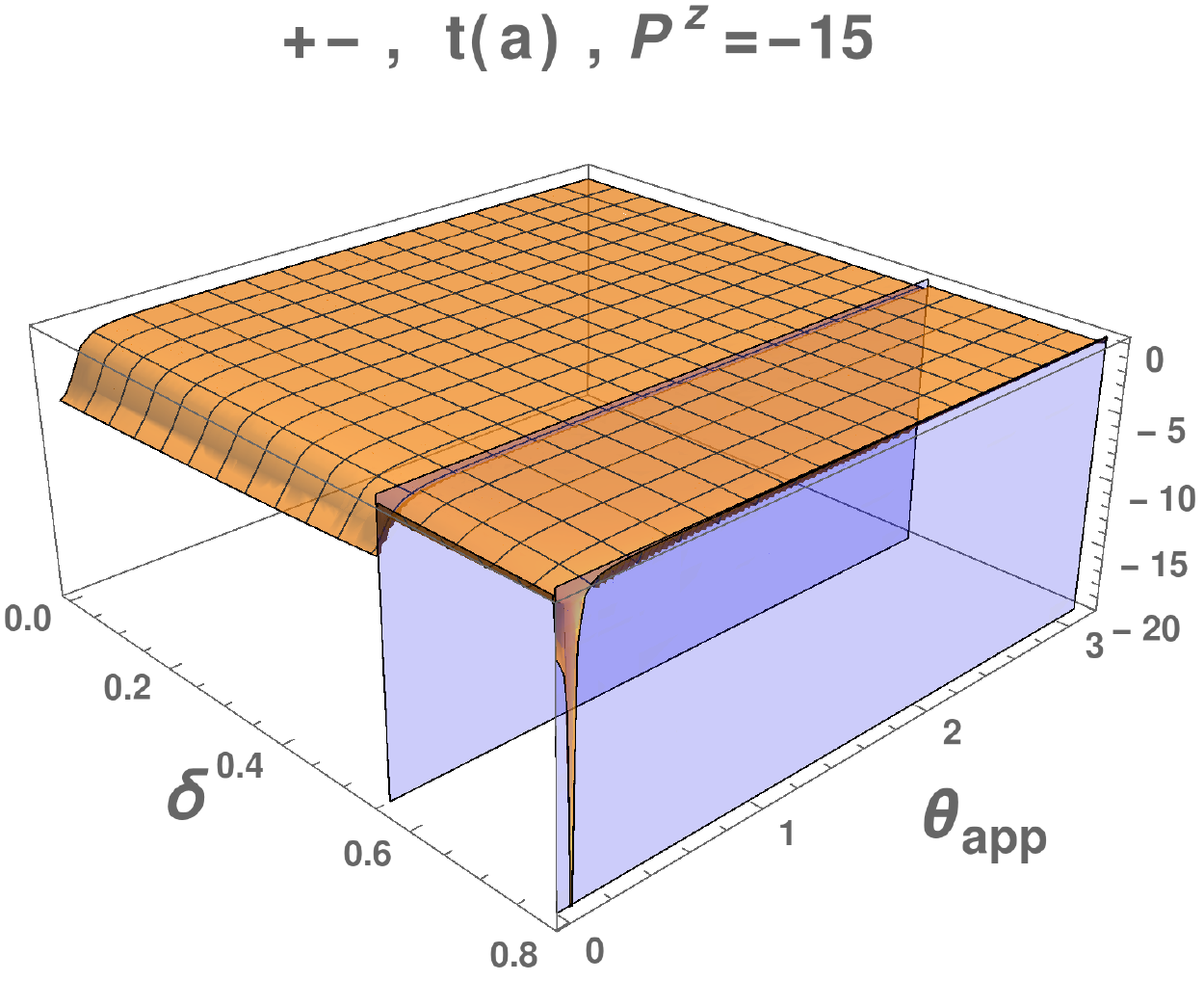}
			\label{fig:eessAppAngDistRLtaPzm}}
		\centering
		\subfloat[]{
			\includegraphics[width=0.48\columnwidth]{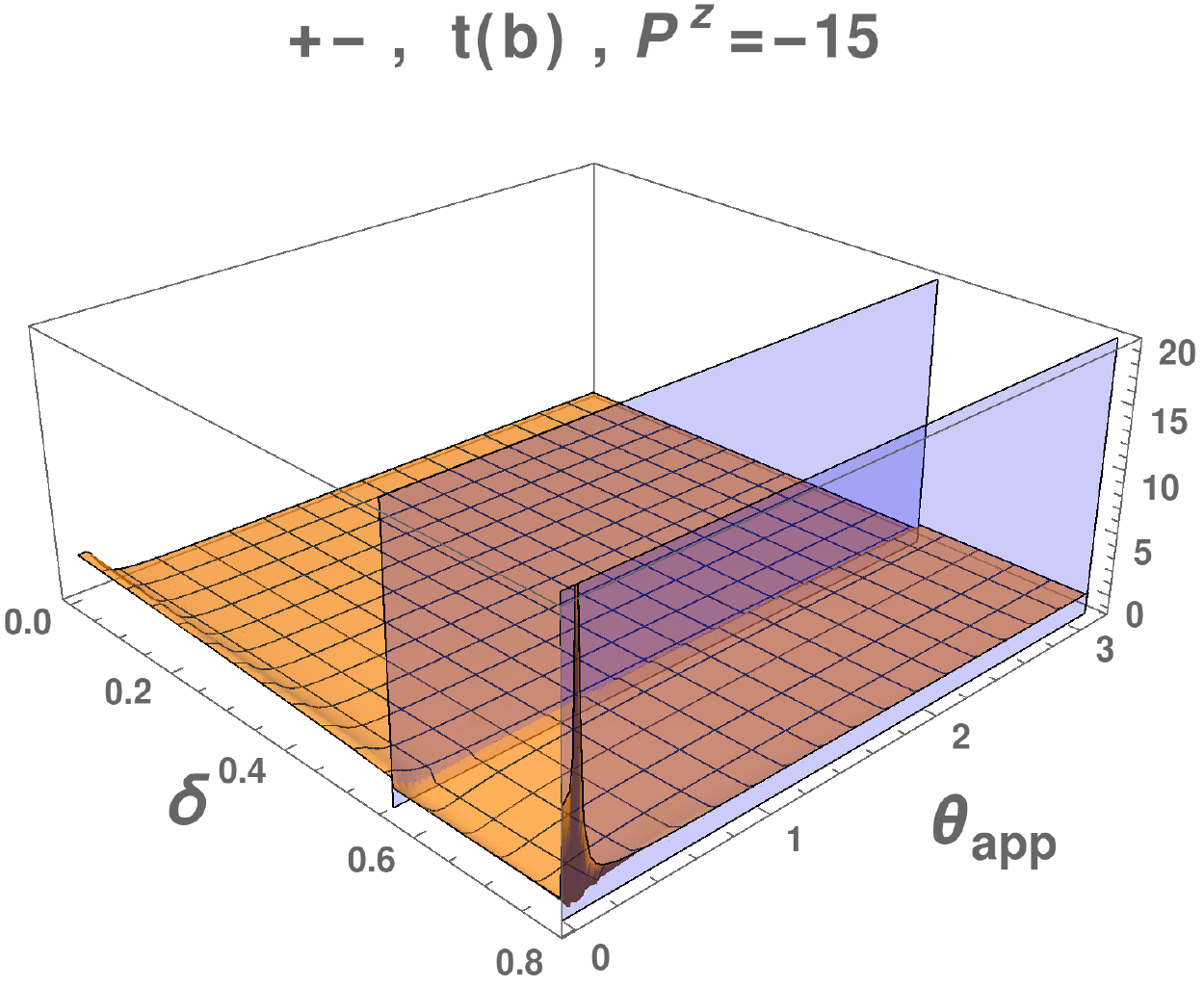}
			\label{fig:eessAppAngDistRLtbPzm}}
		\\
		\centering
		\subfloat[]{
			\includegraphics[width=0.48\columnwidth]{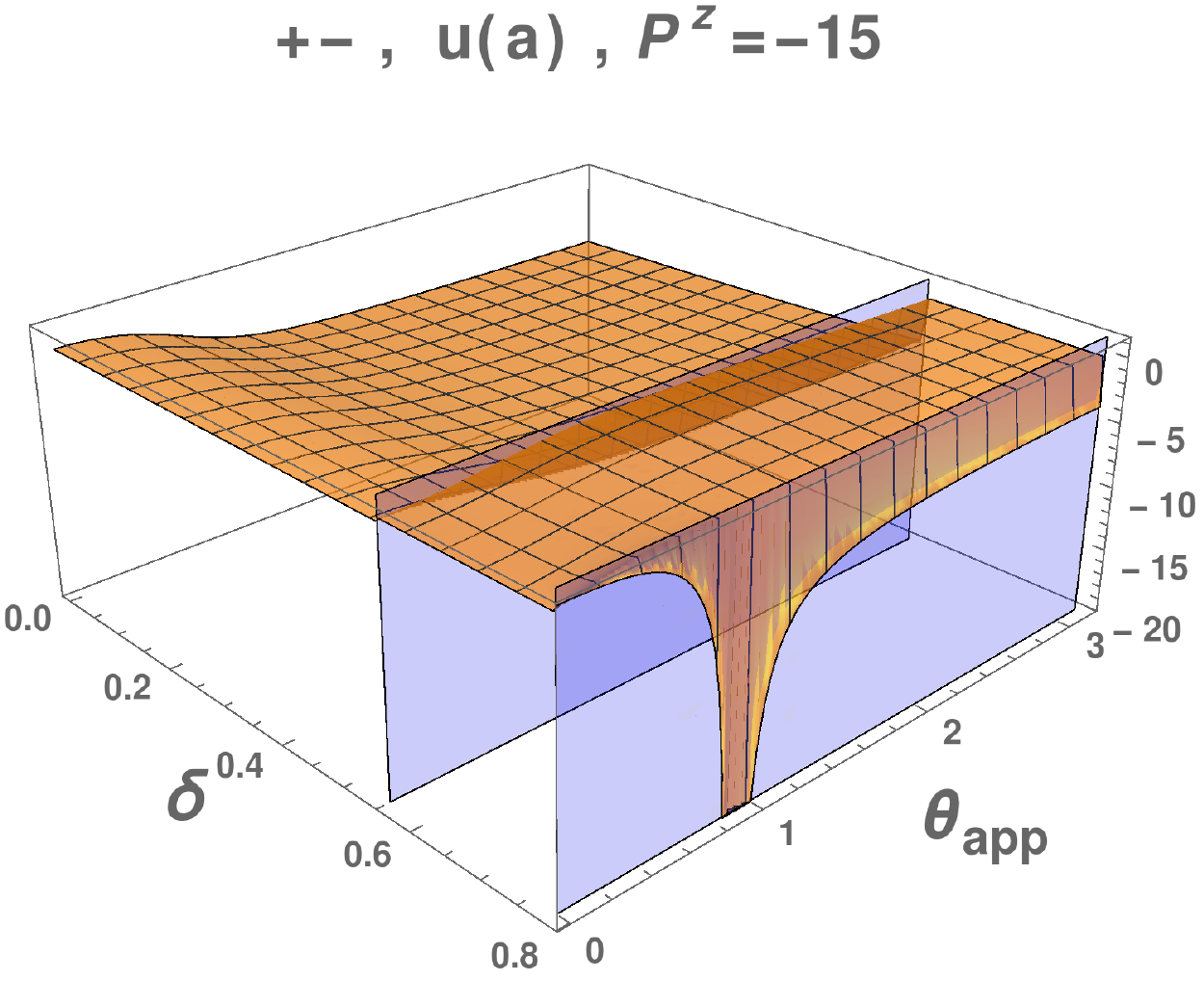}
			\label{fig:eessAppAngDistRLuaPzm}}
		\centering
		\subfloat[]{
			\includegraphics[width=0.48\columnwidth]{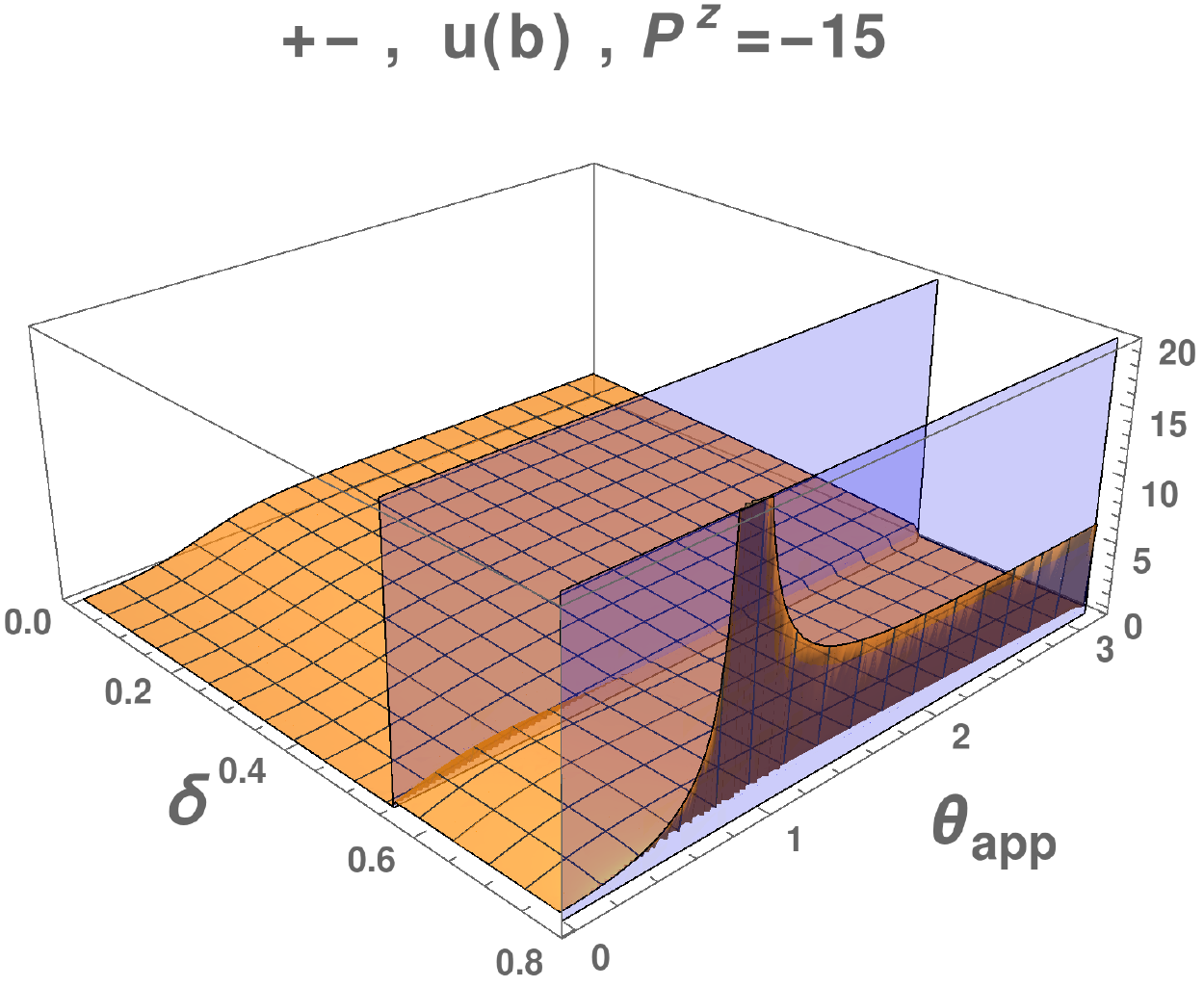}
			\label{fig:eessAppAngDistRLubPzm}}
		\caption{\label{fig:eessAppAngDistRLPzm}Apparent Angular distribution of the helicity amplitude $ +- $ 
			for (a) t-channel a time-ordering, $\mathcal{M}_{a,t}^{+,-}$  (b) t-channel b time-ordering, $\mathcal{M}_{b,t}^{+,-}$ 
			(c) u-channel a time-ordering, $\mathcal{M}_{a,u}^{+,-}$ (d) u-channel b time-ordering, $\mathcal{M}_{b,u}^{+,-}$.}
	\end{figure}
	
	\begin{figure}
		\centering
		\subfloat[]{
			\includegraphics[width=0.48\columnwidth]{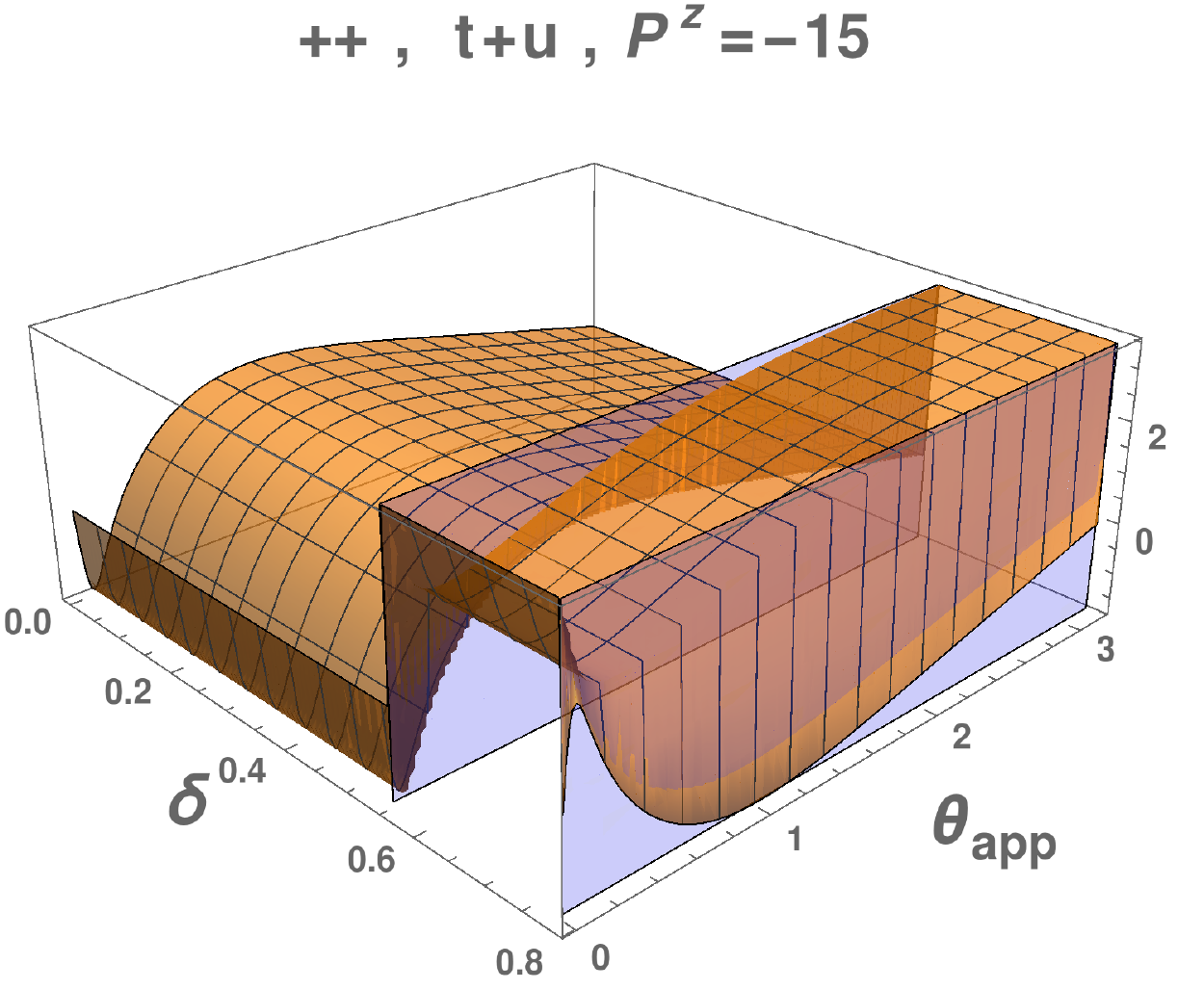}
			\label{fig:eessAppAngDistRRtuPzm}}
		\centering
		\subfloat[]{
			\includegraphics[width=0.48\columnwidth]{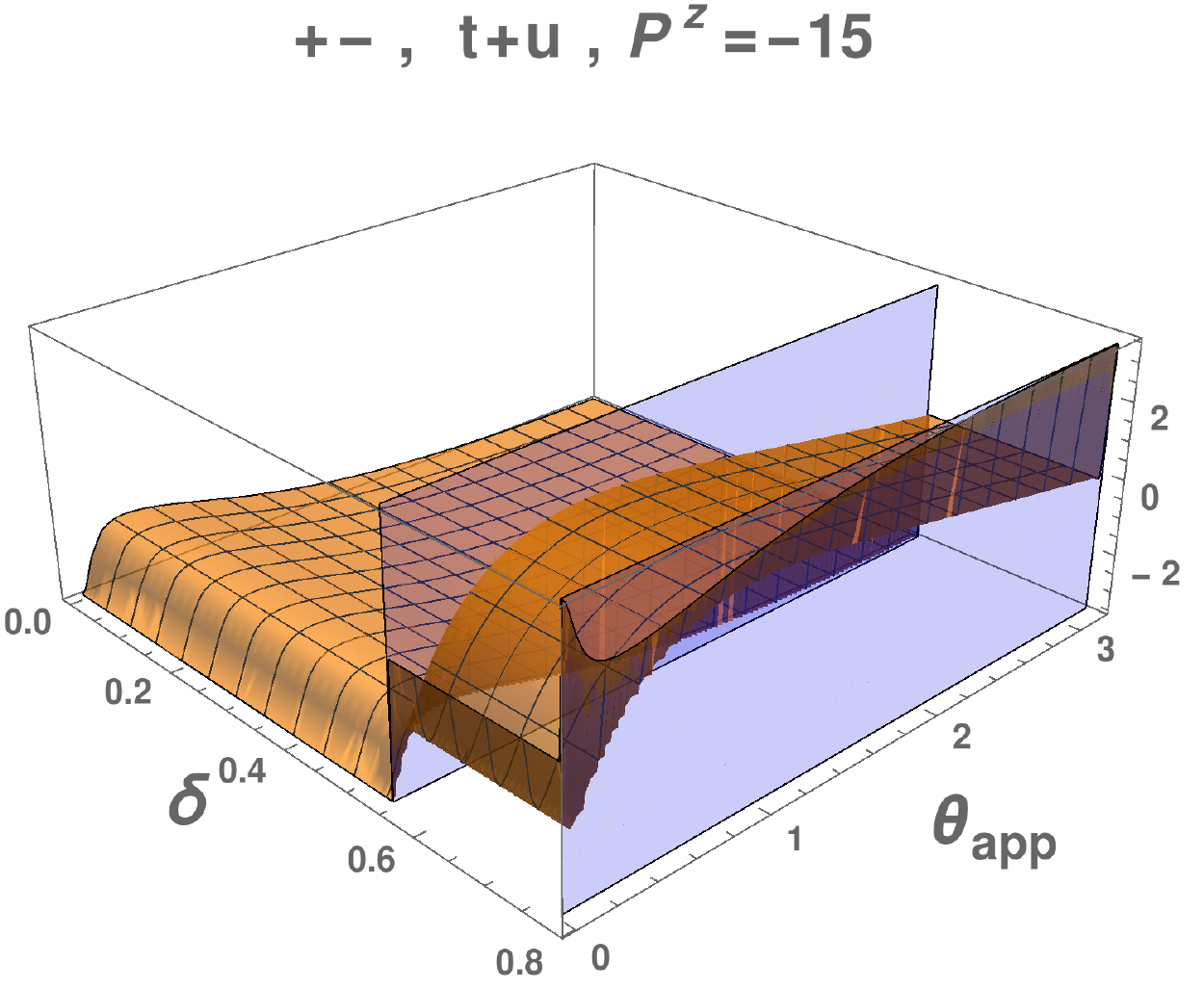}
			\label{fig:eessAppAngDistRLtuPzm}}
		\\
		\centering
		\subfloat[]{
			\includegraphics[width=0.48\columnwidth]{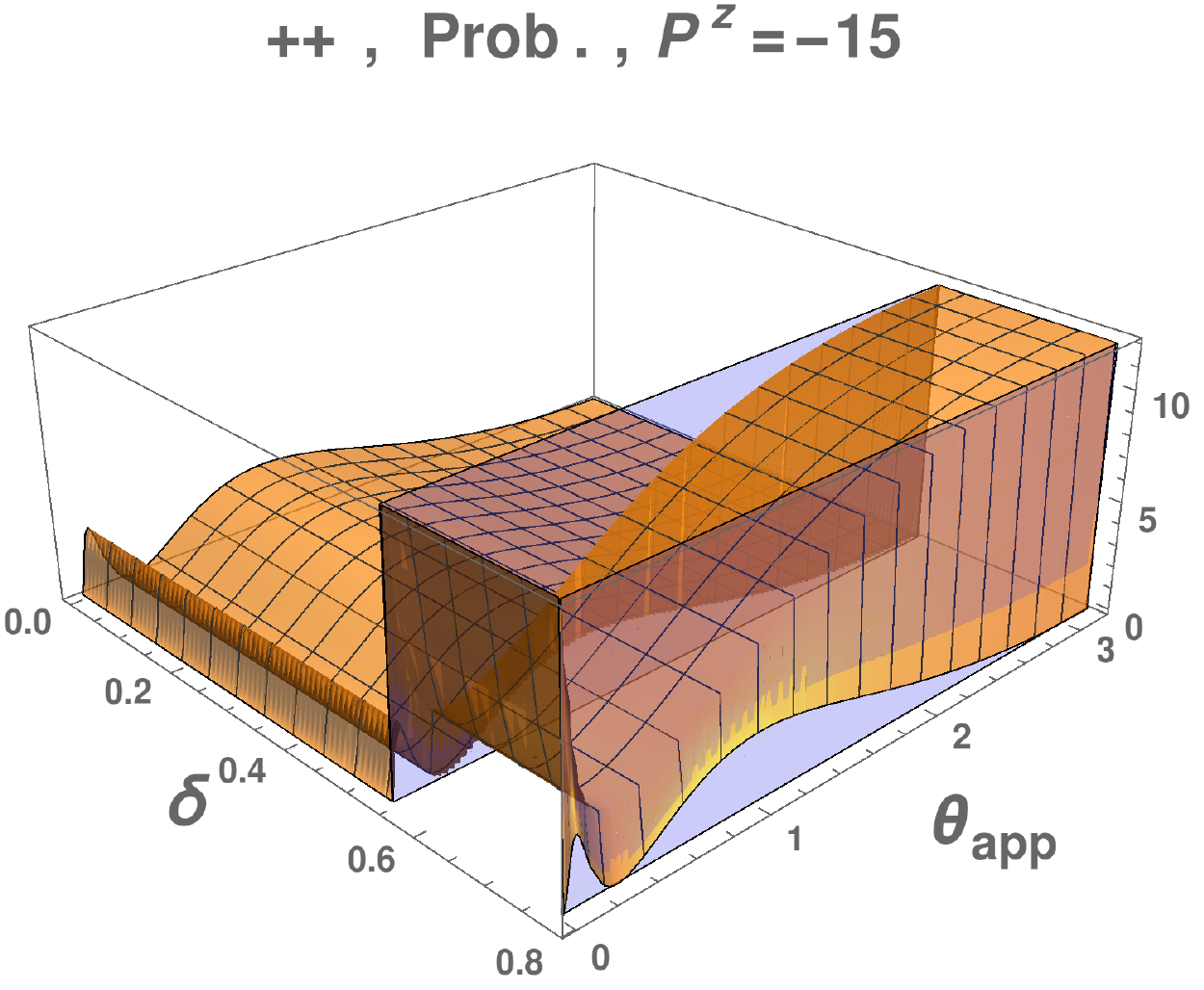}
			\label{fig:eessAppAngDistRRprobPzm}}
		\centering
		\subfloat[]{
			\includegraphics[width=0.48\columnwidth]{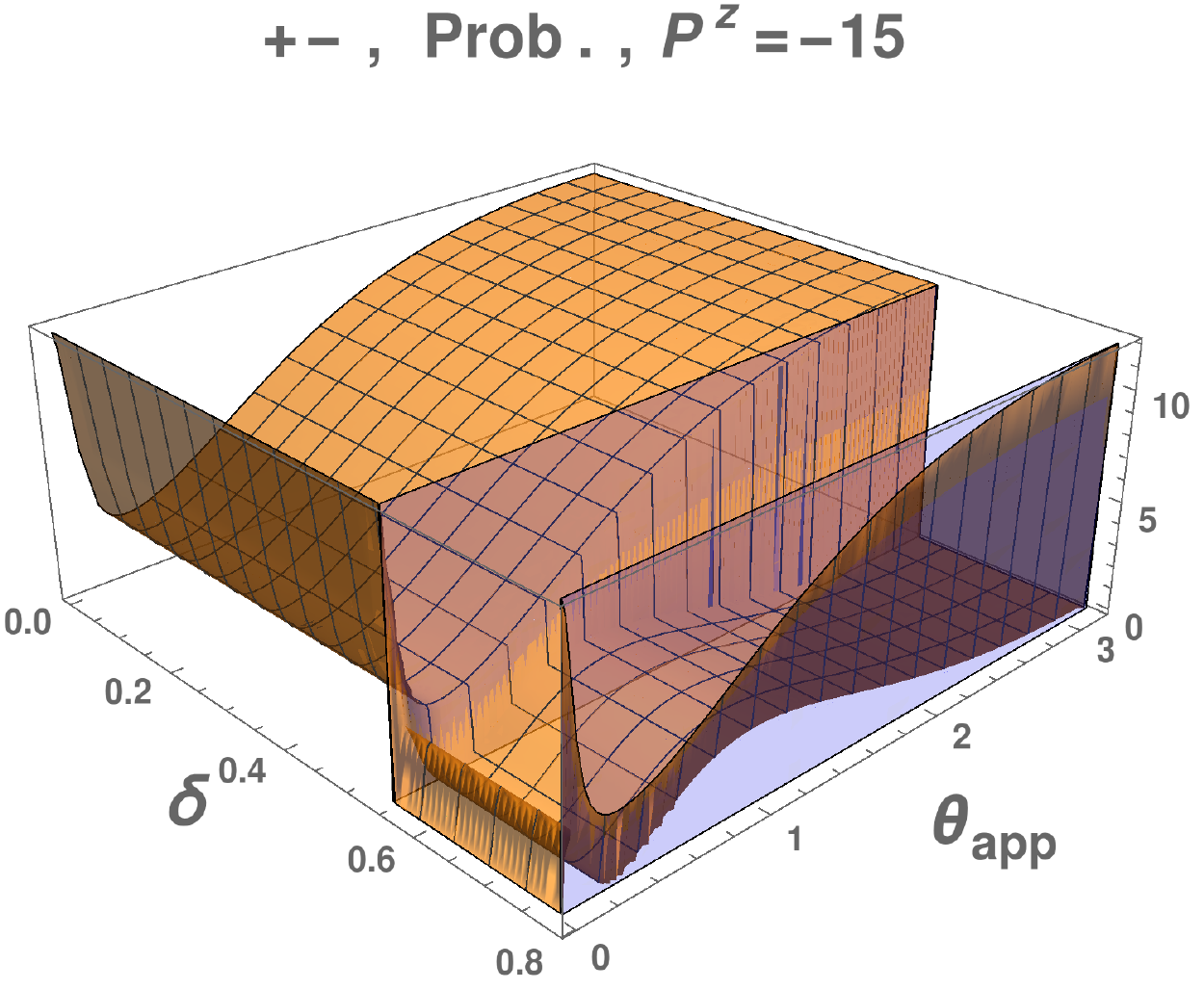}
			\label{fig:eessAppAngDistRLprobPzm}}
		\caption{\label{fig:eessAppAngDistAmpAndProbPzm} (a) $\mathcal{M}_{a,t}^{+,+}+\mathcal{M}_{b,t}^{+,+}+\mathcal{M}_{a,u}^{+,+}+\mathcal{M}_{a,u}^{+,+}$ 
			(b) $\mathcal{M}_{a,t}^{+,-}+\mathcal{M}_{b,t}^{+,-}+\mathcal{M}_{a,u}^{+,-}+\mathcal{M}_{a,u}^{+,-}$ 
			(c) $|\mathcal{M}_{a,t}^{+,+}+\mathcal{M}_{b,t}^{+,+}+\mathcal{M}_{a,u}^{+,+}+\mathcal{M}_{a,u}^{+,+}|^2$  
			(d) $|\mathcal{M}_{a,t}^{+,-}+\mathcal{M}_{b,t}^{+,-}+\mathcal{M}_{a,u}^{+,-}+\mathcal{M}_{a,u}^{+,-}|^2$ }
	\end{figure}
	
	\begin{figure}
		\centering
		\includegraphics[width=0.7\linewidth]{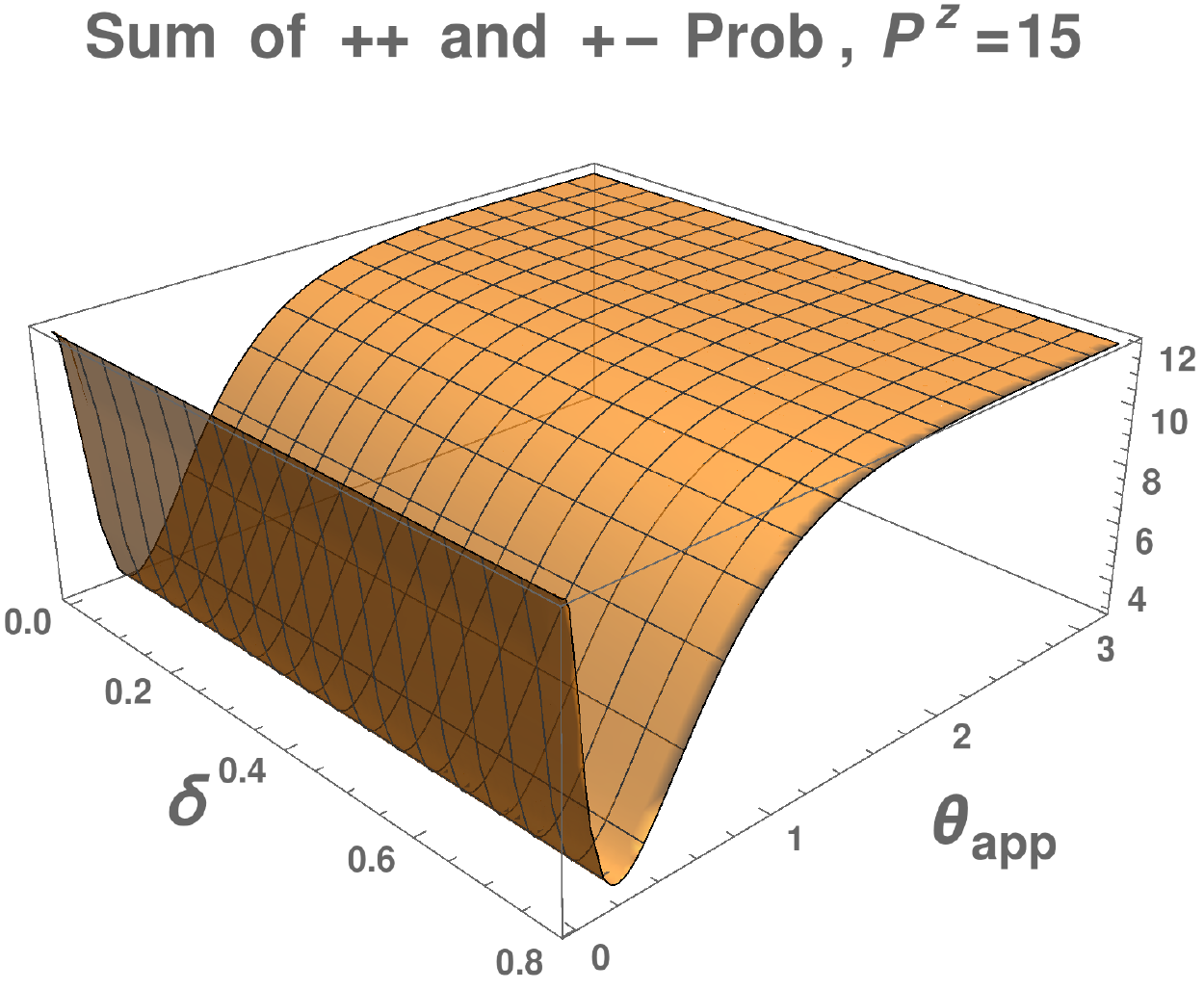}
		\caption{Sum of $++$ and $+-$ Helicity Probabilities}
		\label{fig:eessAppAngDistSumPzp}
	\end{figure}
	
	\begin{figure}
		\centering
		\includegraphics[width=0.7\linewidth]{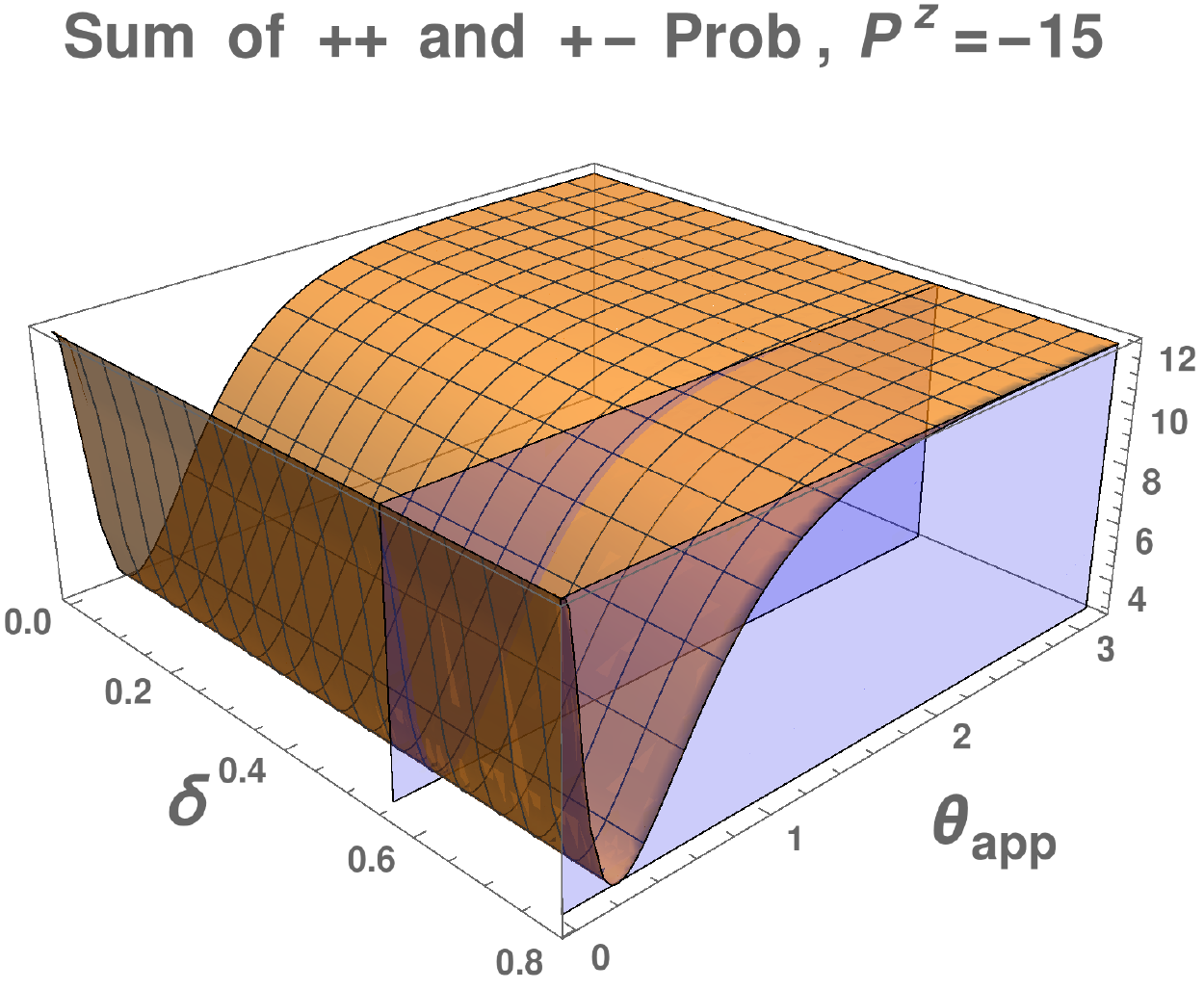}
		\caption{Sum of $++$ and $+-$ Helicity Probabilities}
		\label{fig:eessAppAngDistSumPzn}
	\end{figure}
			
	In this Appendix, we present the angular distribution shown in Sec.~\ref{sec:eess} re-plotted in terms of the apparent angle of the scattering/annihilation process in a moving frame viewed from the lab frame, $ \theta_{\mathrm{app}} $, as well as the interpolation angle $ \delta $. 
		
	By boosting the system with total momentum $ P^z $, we get 
	\begin{equation}
	\gamma=\sqrt{1+\left( \frac{P^z}{2E_0}\right)^2 },\ \beta=\frac{\left( \frac{P^z}{2E_0}\right) }{\sqrt{1+\left( \frac{P^z}{2E_0}\right)^2 }},
	\end{equation}
	where $ E_0 $ is the energy of the initial particle in the original center-of-mass frame.
	
	Then, by writing the 4-momentum of the boosted photon (particle 3), we can find its moving direction to be
	\begin{align}
	\tan\theta_{\mathrm{app}}&=\frac{E_0\sin\theta}{\gamma E_0\left( \beta+\cos\theta\right) }\notag\\
	&=\frac{\tan\theta}{\left( \frac{P^z}{2E_0}\right)\sqrt{1+\tan^2\theta}+\sqrt{1+\left( \frac{P^z}{2E_0}\right)^2} },\label{eqn:thetaapp}
	\end{align}
	when viewed from the lab frame.
	
	Reversing Eq.~(\ref{eqn:thetaapp}), we get
	\begin{equation}
	\tan\theta=\frac{\sqrt{1+\left( \frac{P^z}{2E_0}\right)^2}\tan\theta_{\mathrm{app}}+ \frac{\left|P^z\right|}{2E_0}\tan\theta_{\mathrm{app}}\sqrt{1+\tan^2\theta_{\mathrm{app}}}}{1-\left( \frac{P^z}{2E_0}\right)^2\tan^2\theta_{\mathrm{app}}}.
	\end{equation}

	Figs.~\ref{fig:eessAppAngDistRRPzp}-\ref{fig:eessAppAngDistAmpAndProbPzm} 
	plotted in the apparent angle given by Eq.~(\ref{eqn:thetaapp}) correspond to Figs.~\ref{fig:eessAngDistRRPzp}-\ref{fig:eessAngDistAmpAndProbPzm} plotted in the CMF scattering angle $\theta$. 
Figs.~\ref{fig:eessAppAngDistSumPzp} and ~\ref{fig:eessAppAngDistSumPzn} correspond to 
Fig.~\ref{fig:HelicityProbabilitySumPzp} and ~\ref{fig:HelicityProbabilitySumPzm}, respectively.

	\section{Boosted $e^+ e^- \to \gamma \gamma$ Interpolating Helicity Amplitudes}
	\label{app:BoostedAnnihilation}
				\begin{figure*}
					\centering
					\subfloat[]{\includegraphics[width=0.49\textwidth]{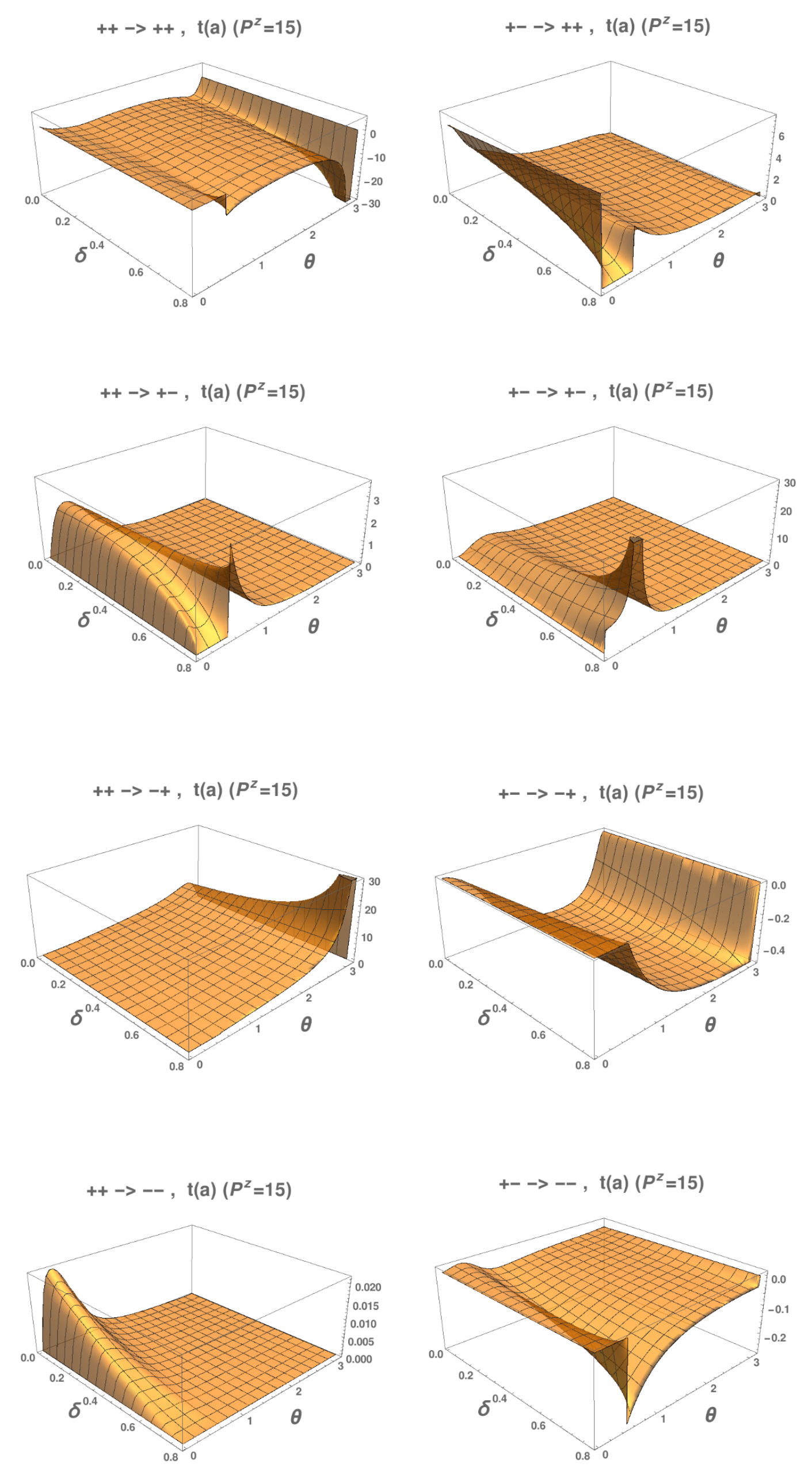}\label{fig:JoinedFigmtaPzp}}
					\centering
					\subfloat[]{\includegraphics[width=0.49\textwidth]{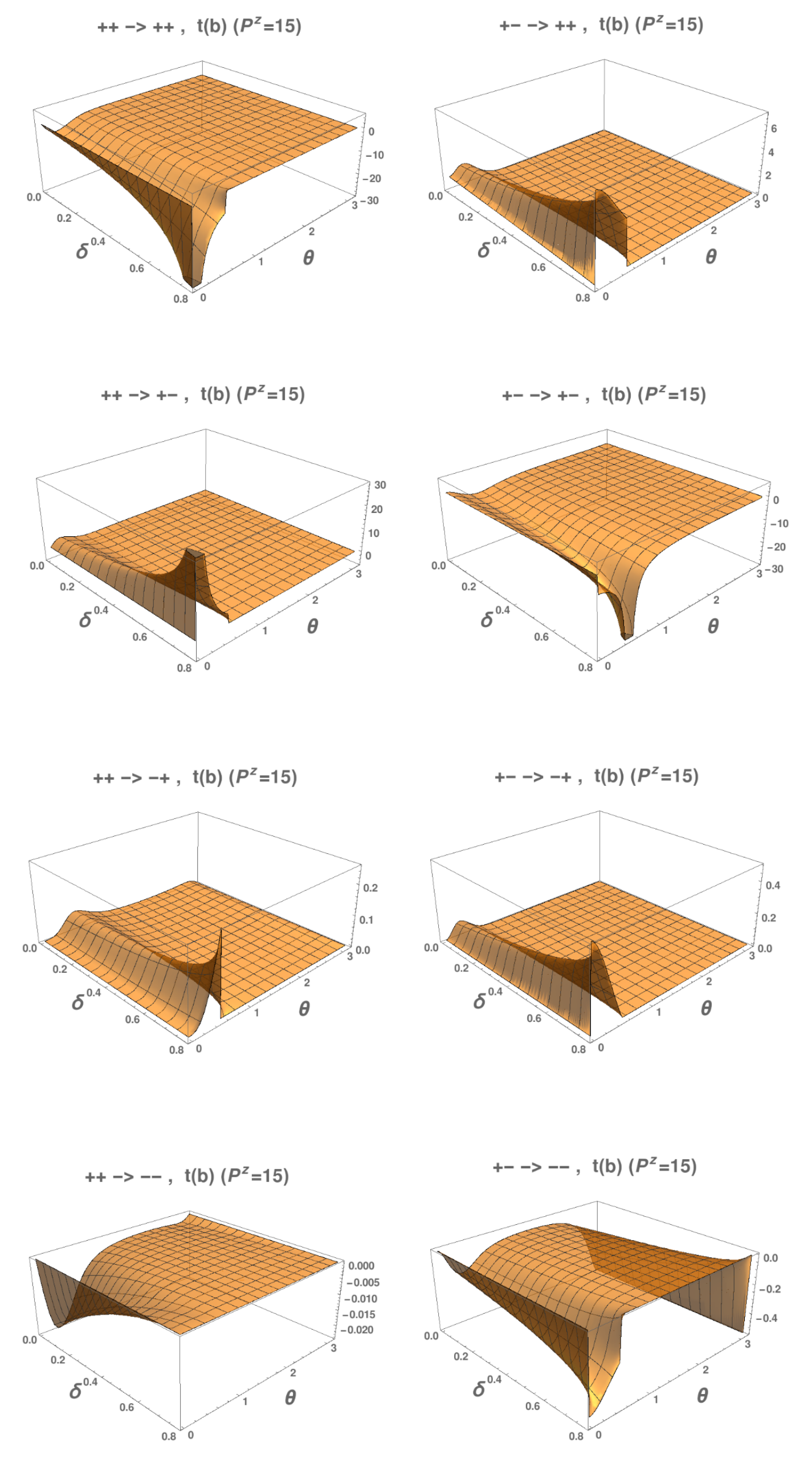}\label{fig:JoinedFigmtbPzp}}
					\caption{\label{fig:JoinedFigmtaandtbPzp}Angular distribution of the helicity amplitudes for (a) t-channel a time-ordering and (b) t-channel b time-ordering
					}
				\end{figure*}
				\begin{figure*}
					\centering
					\subfloat[]{\includegraphics[width=0.49\textwidth]{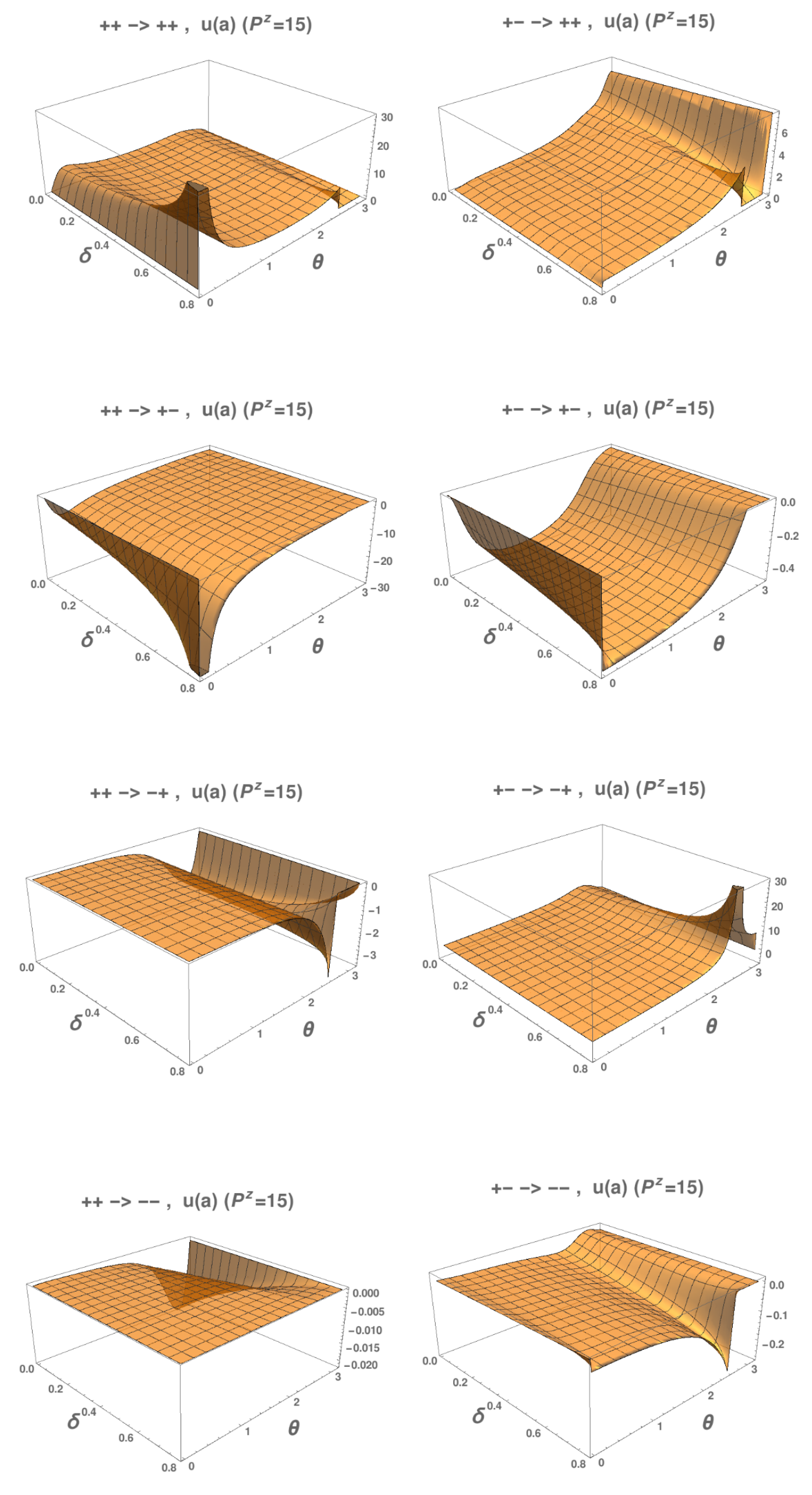}\label{fig:JoinedFigmuaPzp}}
					\centering
					\subfloat[]{\includegraphics[width=0.49\textwidth]{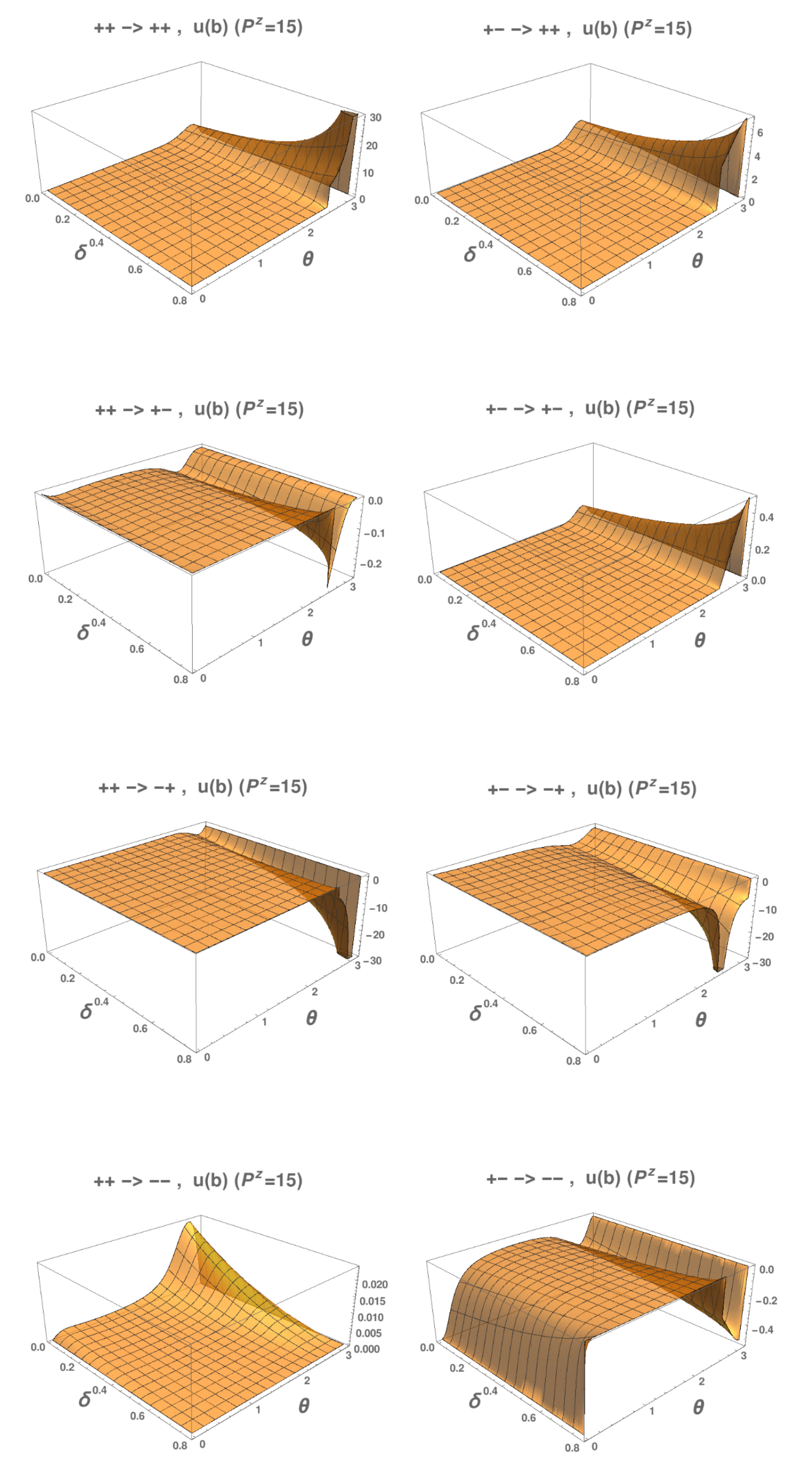}\label{fig:JoinedFigmubPzp}}
					\caption{\label{fig:JoinedFigmuaandubPzp}Angular distribution of the helicity amplitudes for (a) u-channel a time-ordering and (b) u-channel b time-ordering}
				\end{figure*}
				\begin{figure*}
					\centering
					\subfloat[]{\includegraphics[width=0.49\textwidth]{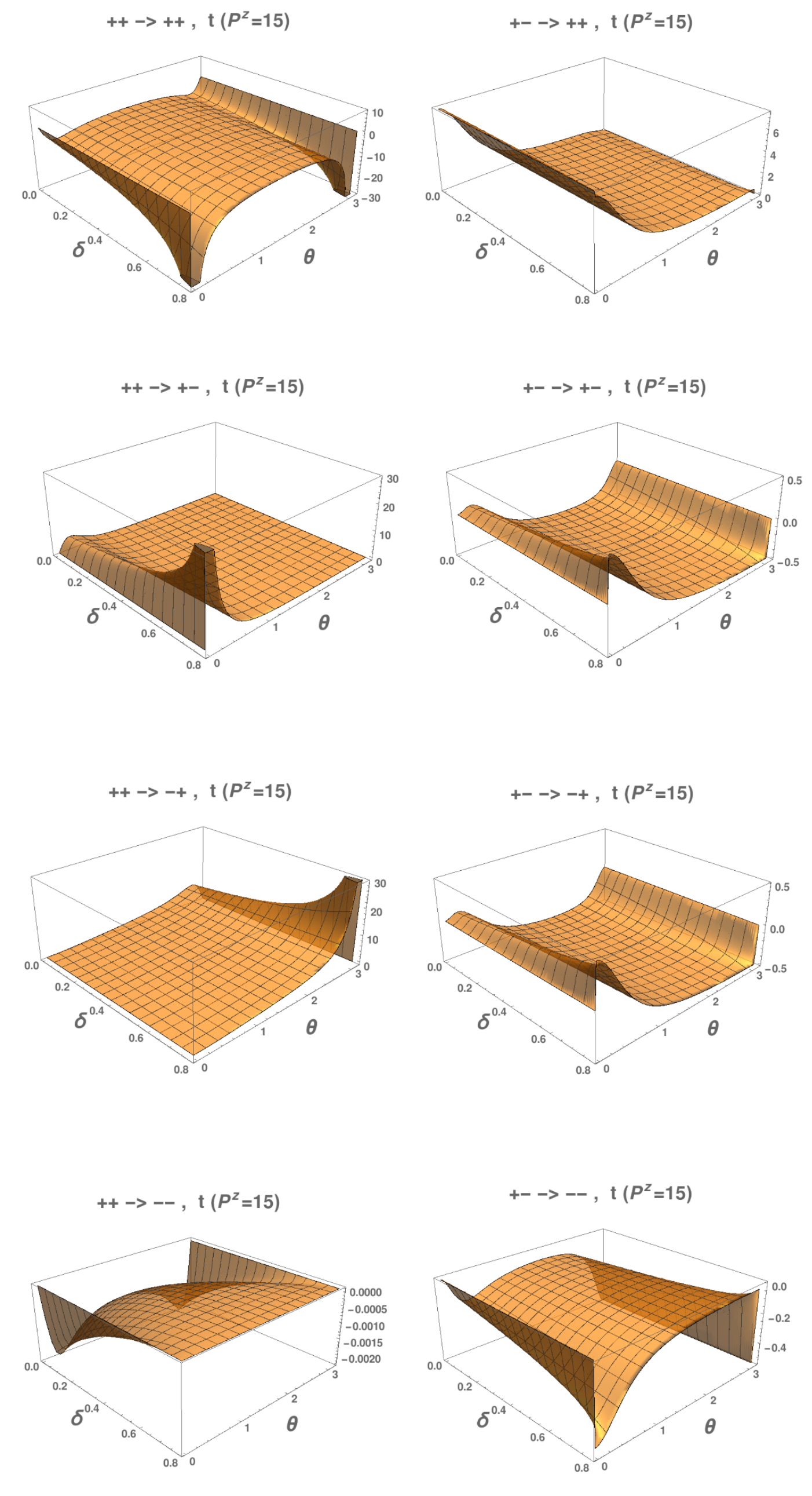}\label{fig:JoinedFigmtoPzp}}
					\centering
					\subfloat[]{\includegraphics[width=0.49\textwidth]{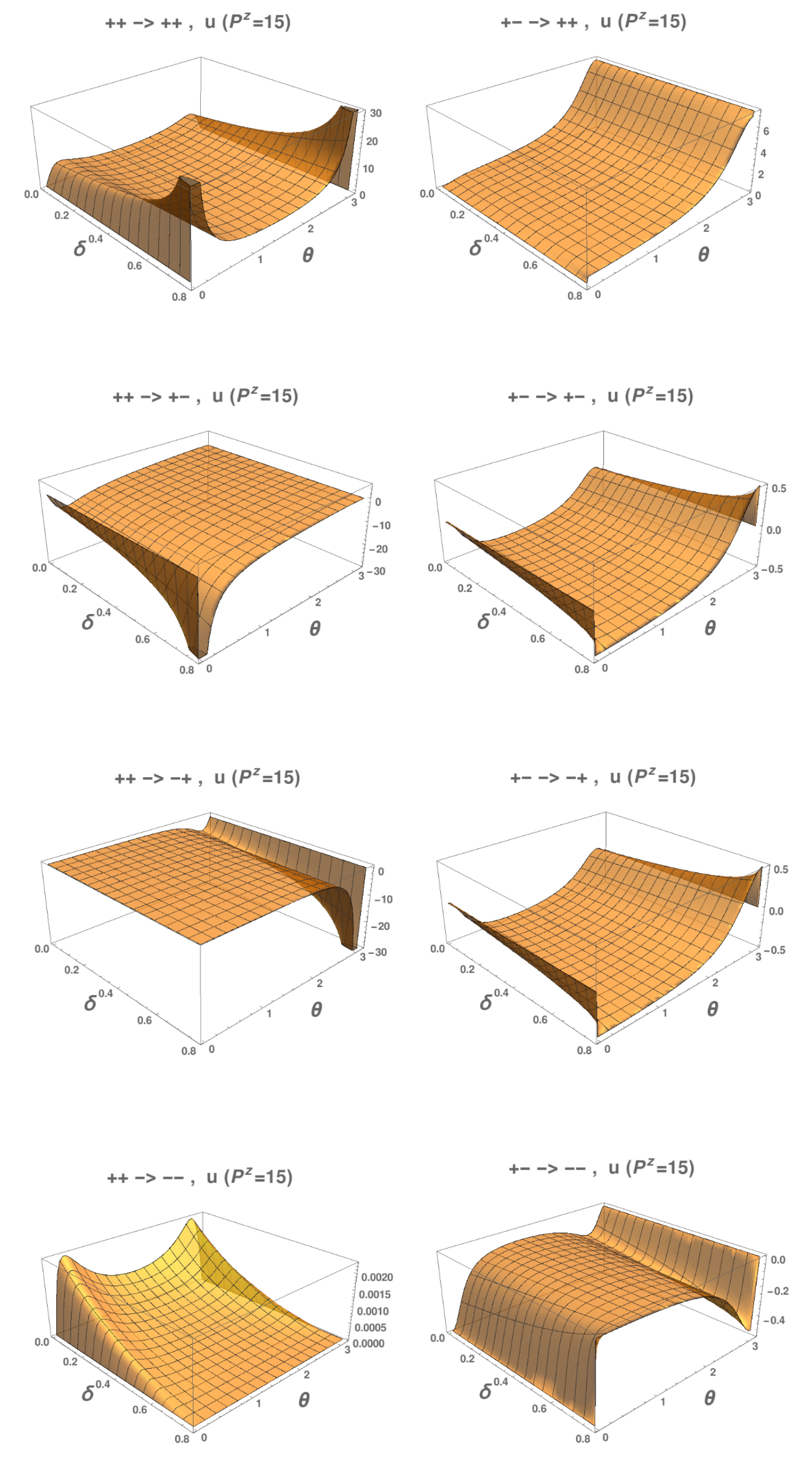}\label{fig:JoinedFigmuoPzp}}
					\caption{\label{fig:JoinedFigmtanduPzp}Angular distribution of the helicity amplitudes for (a) t-channel and (b) u-channel
					}
				\end{figure*}
				\begin{figure}
					\centering
					\includegraphics[width=0.49\textwidth]{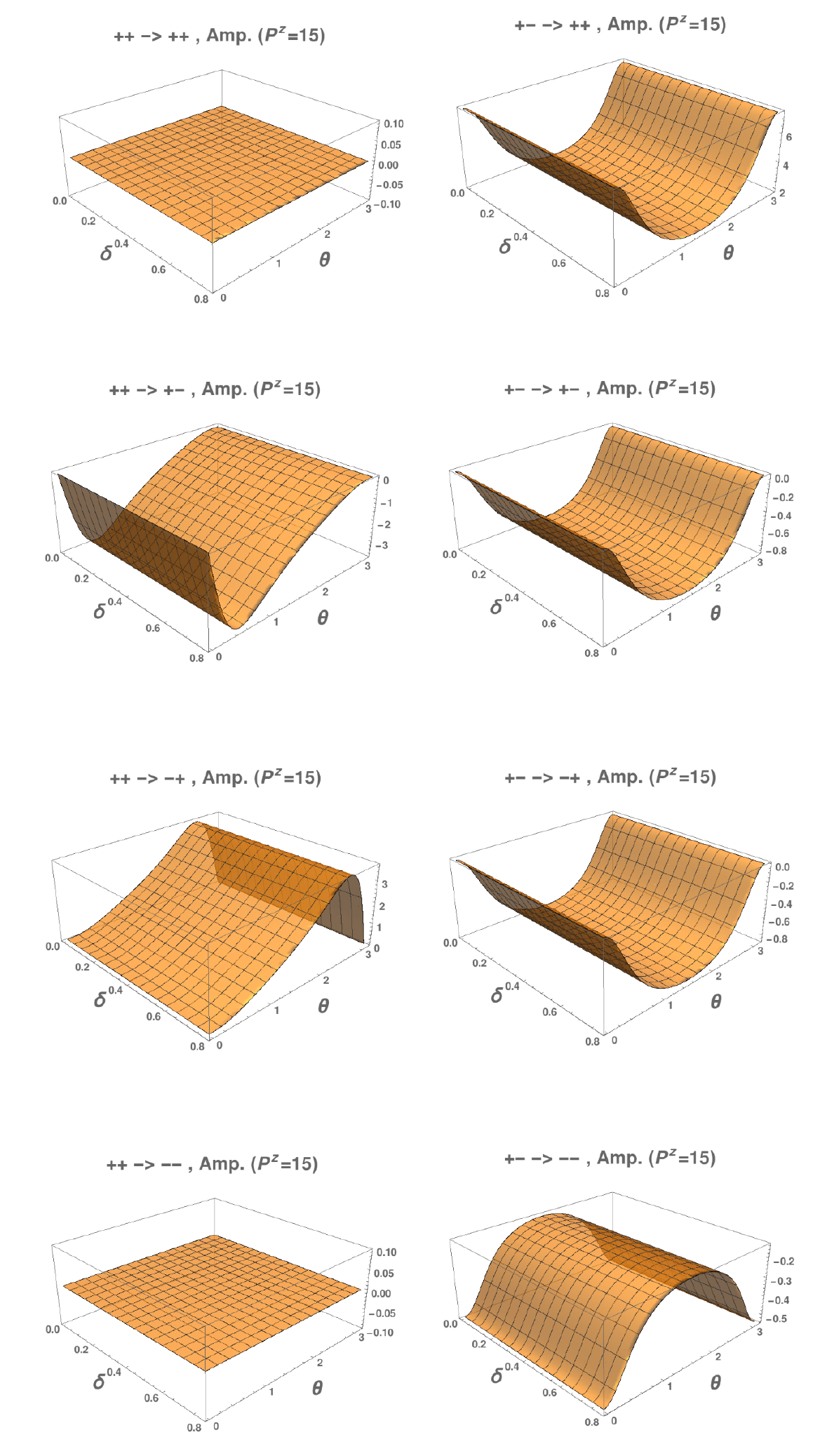}
					\caption{\label{fig:JoinedFigampPzp}Angular distribution of the helicity amplitudes for t+u amplitudes.
					}
				\end{figure}
				\begin{figure}
					\centering
					\includegraphics[width=0.49\textwidth]{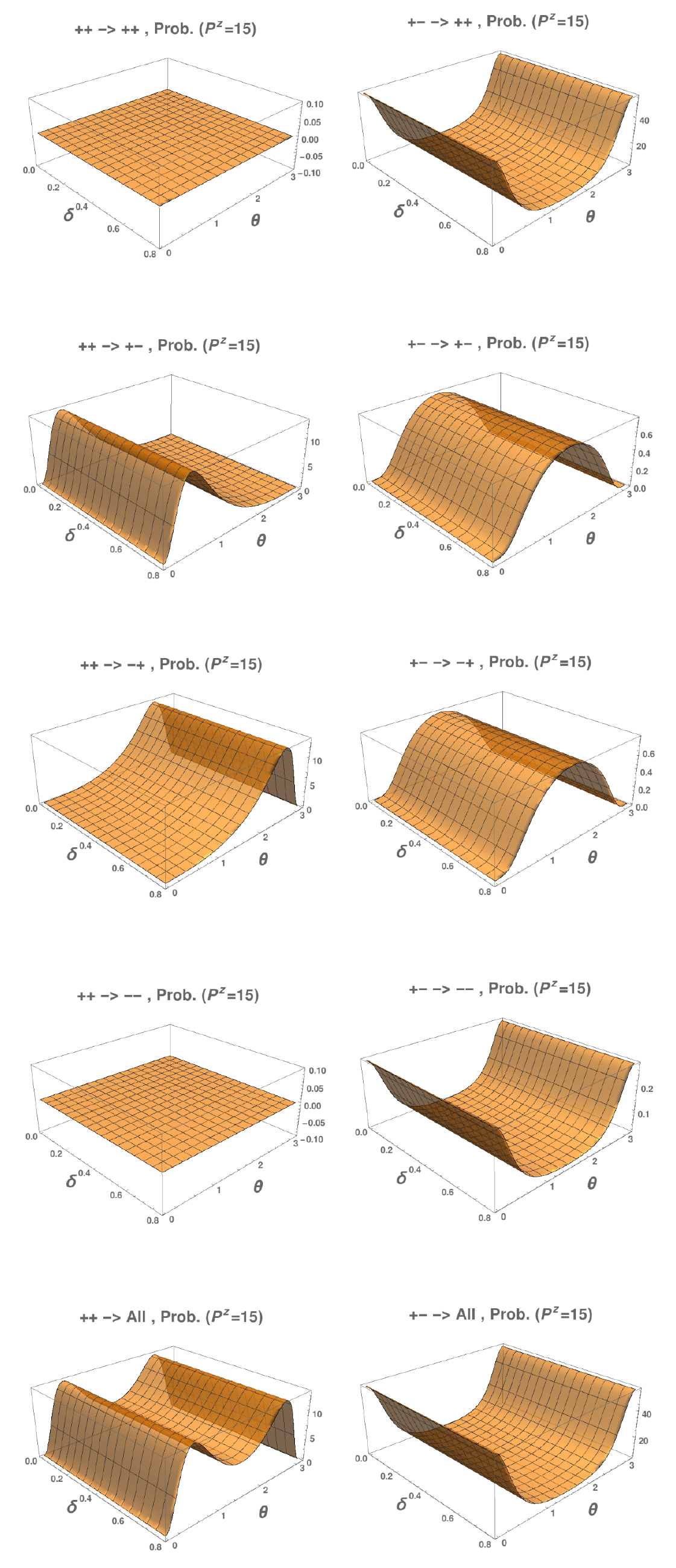}
					\caption{\label{fig:JoinedFigprobPzp}Angular distribution of the helicity amplitudes for the probabilities ,the figures in the last row is result of summing over all figures above it.
					}
				\end{figure}	
				\begin{figure*}
					\centering
					\subfloat[]{\includegraphics[width=0.49\textwidth]{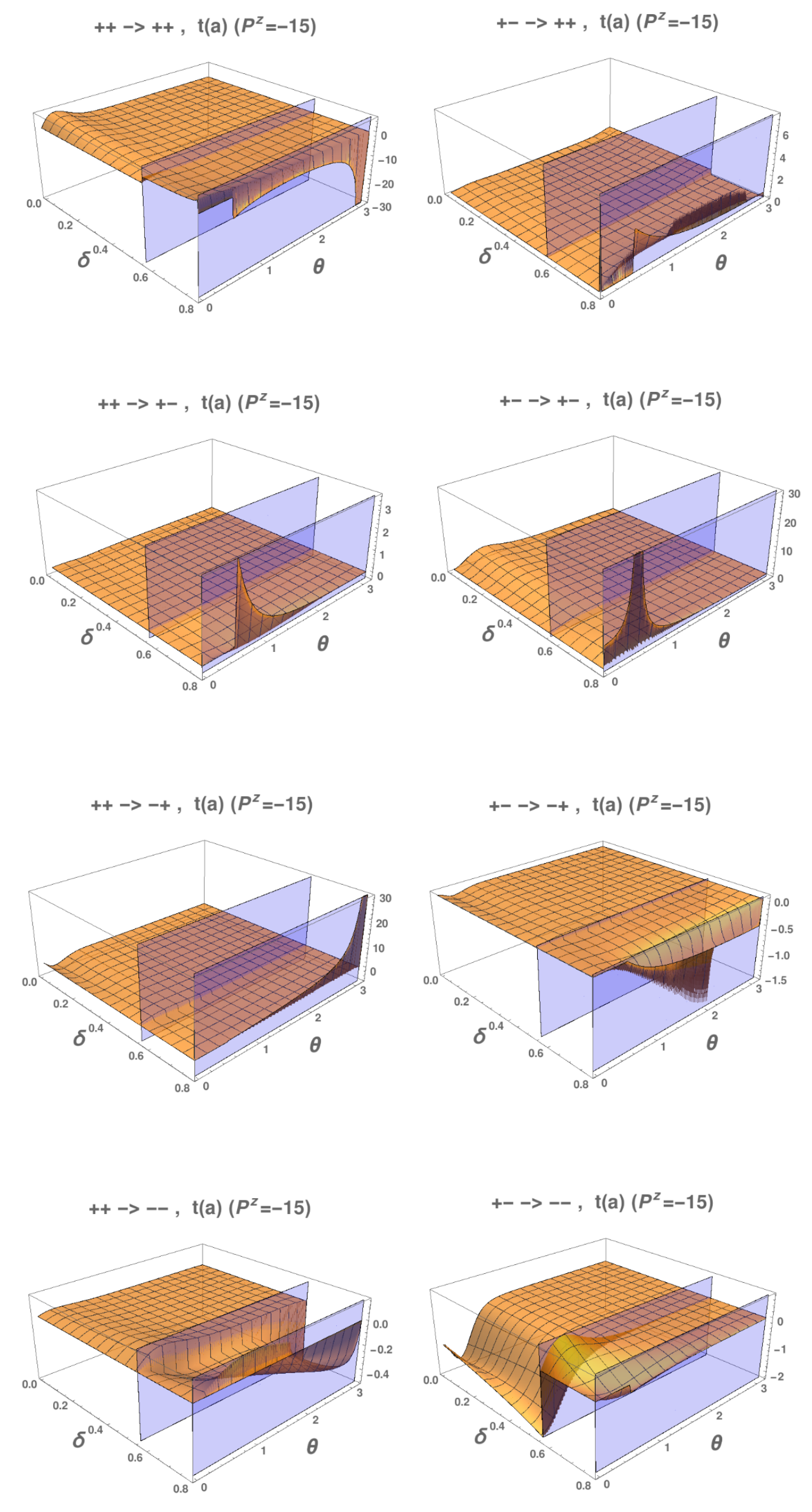}\label{fig:JoinedFigmtaPzm}}
					\centering
					\subfloat[]{\includegraphics[width=0.49\textwidth]{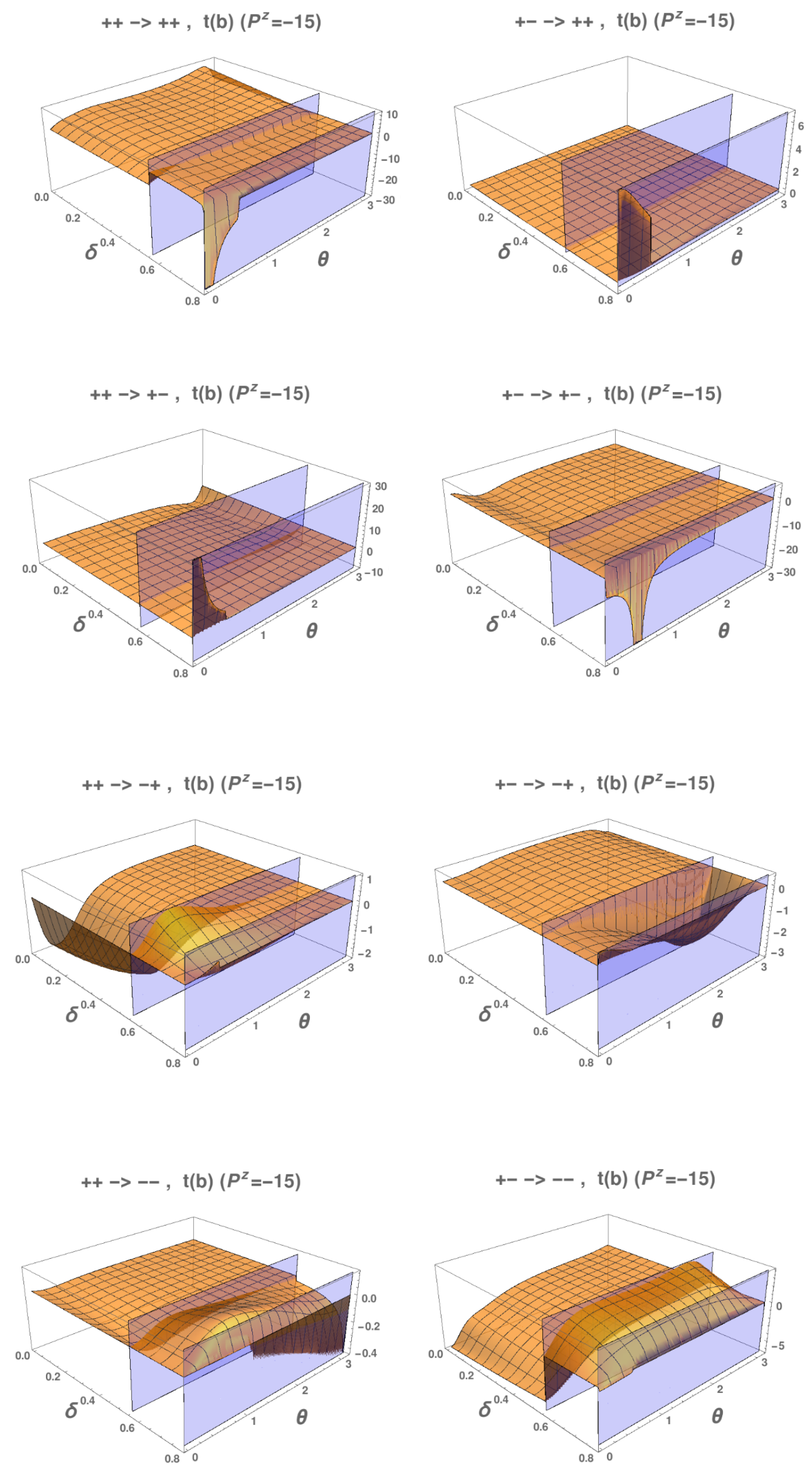}\label{fig:JoinedFigmtbPzm}}
					\caption{\label{fig:JoinedFigmtaandtbPzm}Angular distribution of the helicity amplitudes for (a) t-channel a time-ordering and (b) t-channel b time-ordering
					}
				\end{figure*}
				\begin{figure*}
					\centering
					\subfloat[]{\includegraphics[width=0.49\textwidth]{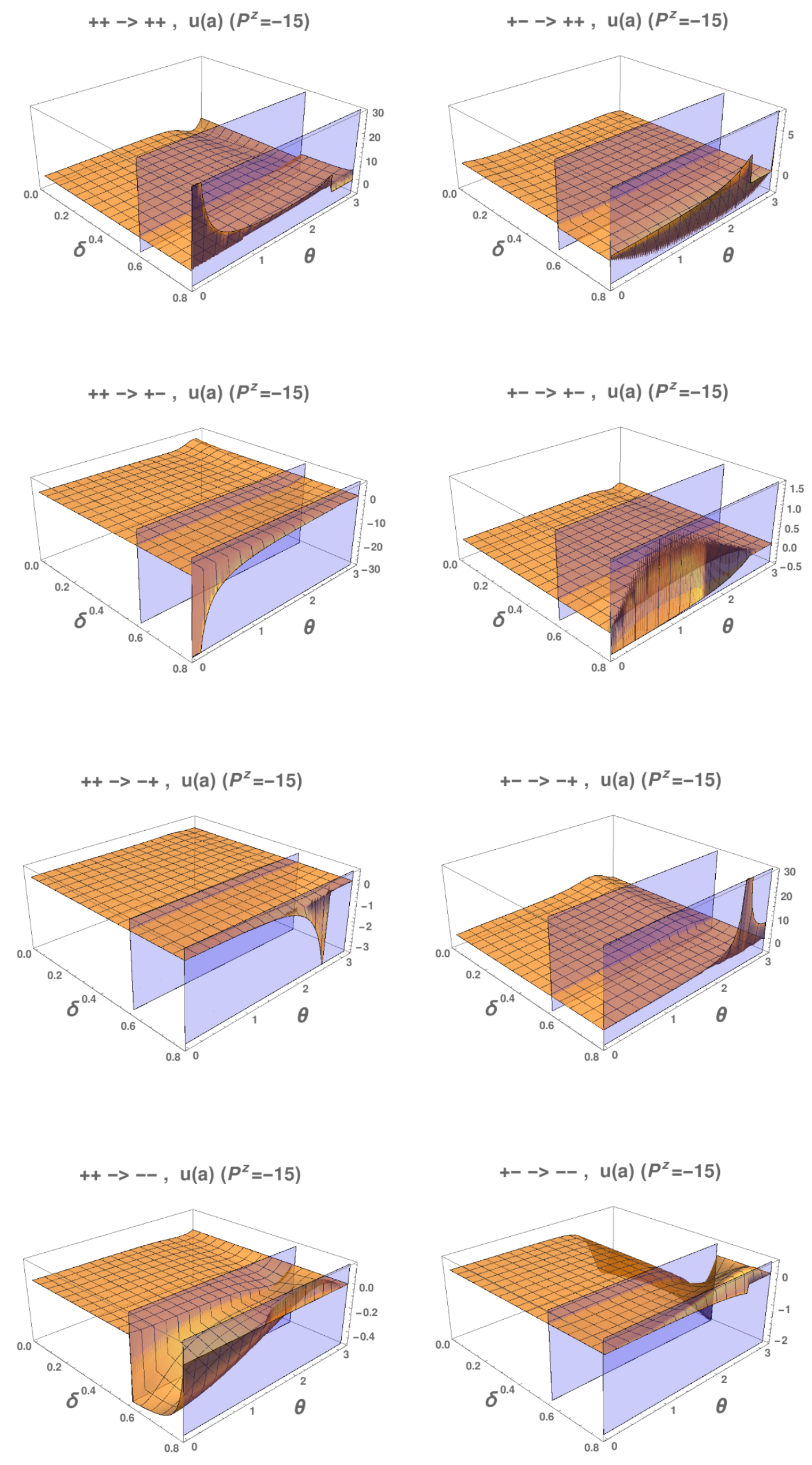}\label{fig:JoinedFigmuaPzm}}
					\centering
					\subfloat[]{\includegraphics[width=0.49\textwidth]{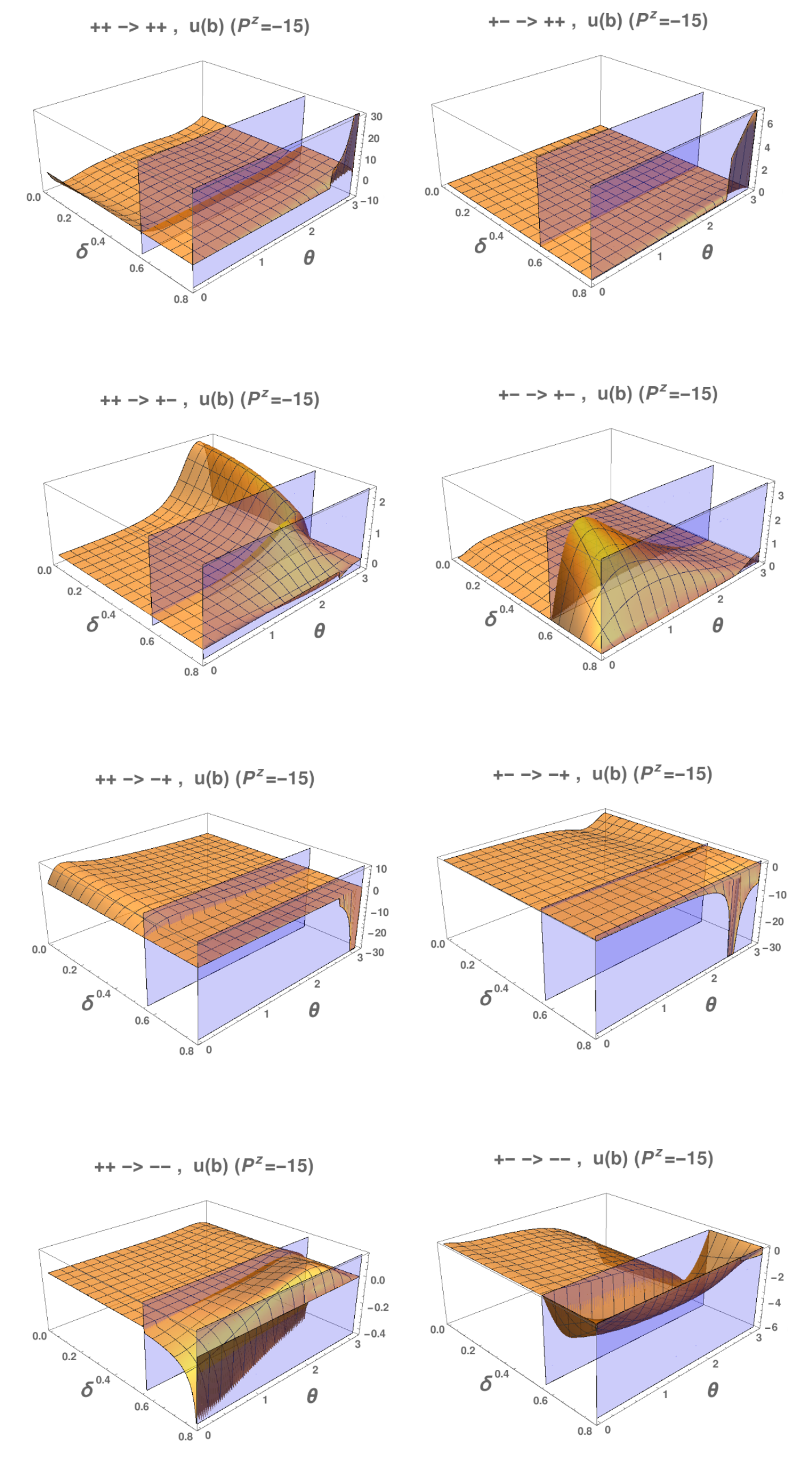}\label{fig:JoinedFigmubPzm}}
					\caption{\label{fig:JoinedFigmuaandubPzm}Angular distribution of the helicity amplitudes for (a) u-channel a time-ordering and (b) u-channel b time-ordering
					}
				\end{figure*}
				\begin{figure*}
					\centering
					\subfloat[]{\includegraphics[width=0.49\textwidth]{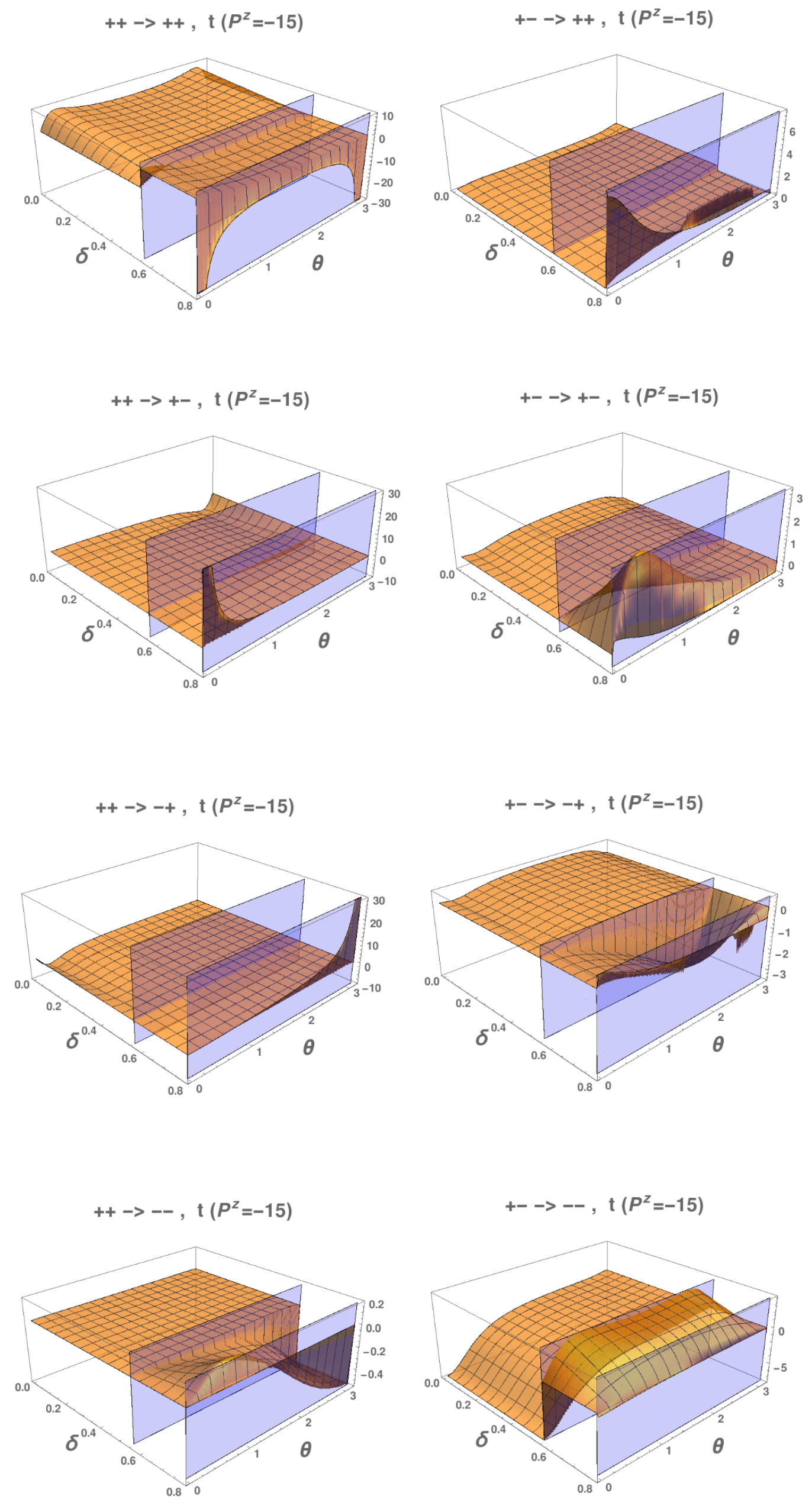}\label{fig:JoinedFigmtoPzm}}
					\centering
					\subfloat[]{\includegraphics[width=0.49\textwidth]{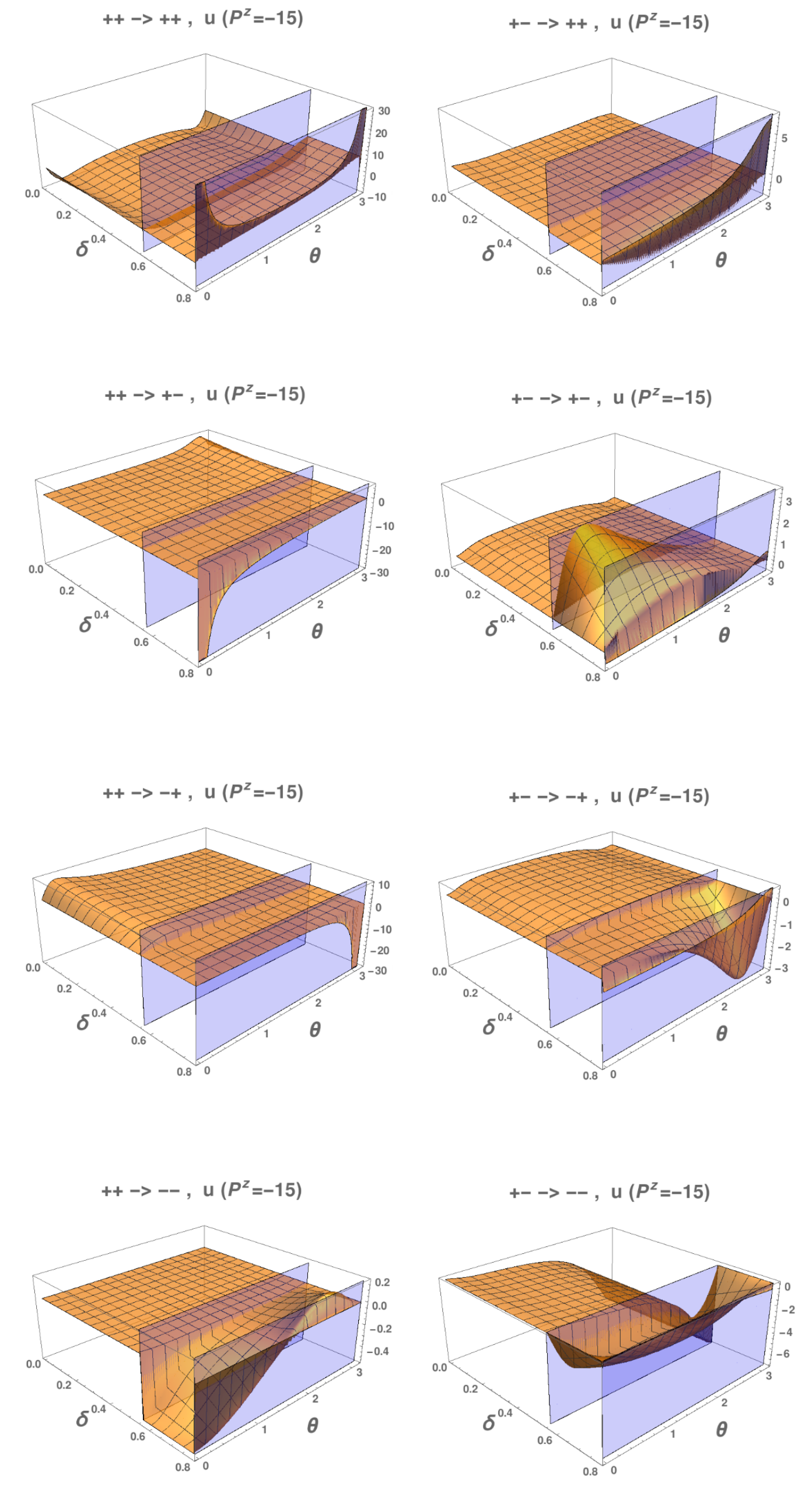}\label{fig:JoinedFigmuoPzm}}
					\caption{\label{fig:JoinedFigmtanduPzm}Angular distribution of the helicity amplitudes for (a) t-channel  and (b) u-channel
					}
				\end{figure*}
				\begin{figure}
					\centering
					\includegraphics[width=0.49\textwidth]{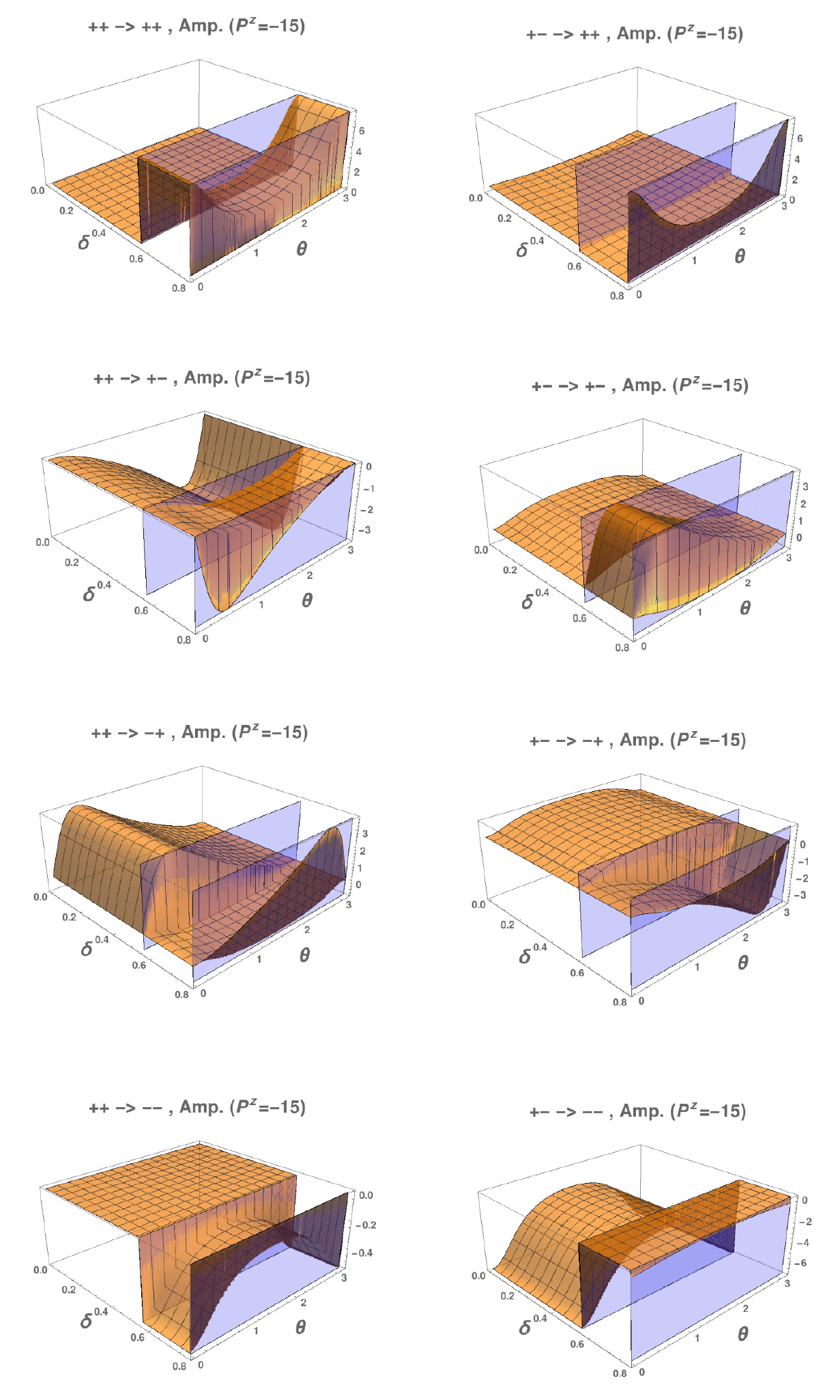}
					\caption{\label{fig:JoinedFigampPzm}Angular distribution of the helicity amplitudes for t+u amplitudes.
					}
				\end{figure}
				\begin{figure}
					\centering
					\includegraphics[width=0.49\textwidth]{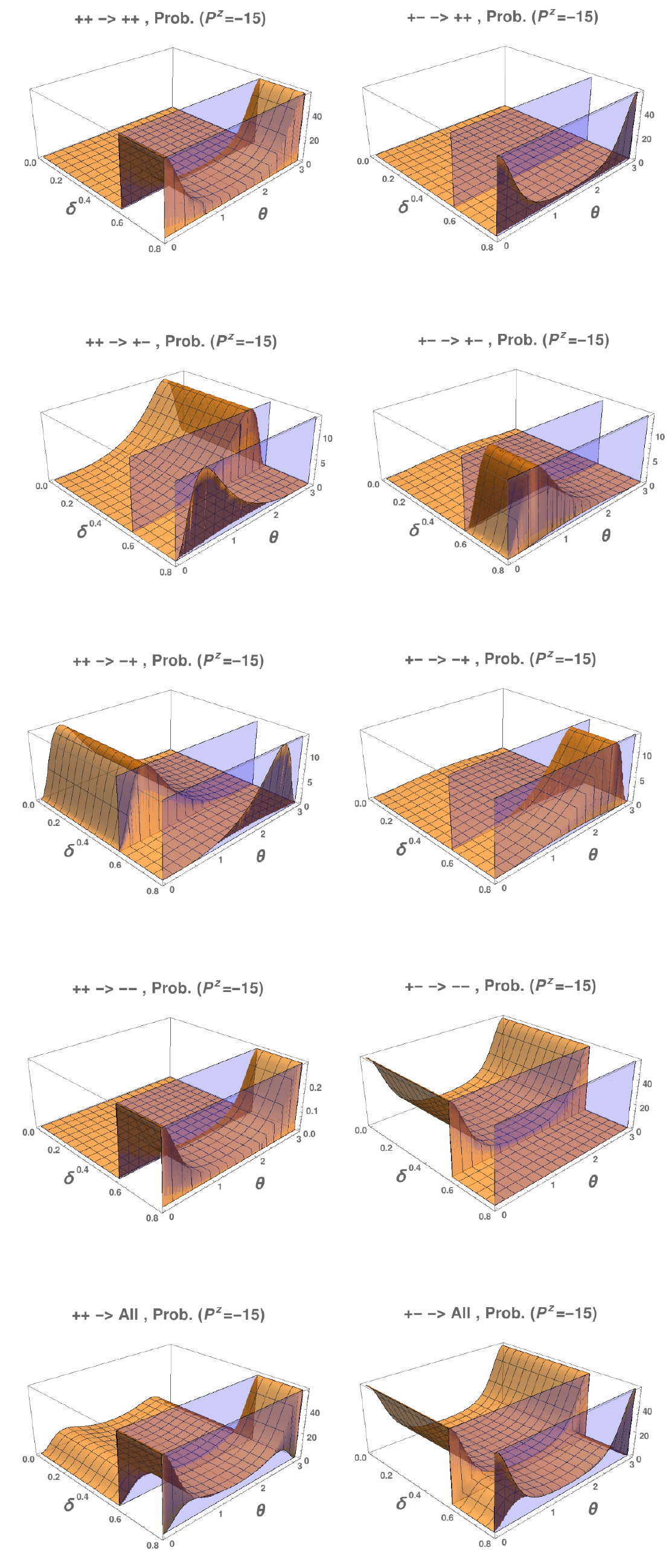}
					\caption{\label{fig:JoinedFigprobPzm}Angular distribution of the helicity amplitudes for the probabilities, the figures in the last row is result of summing over all figures above it.
					}
				\end{figure}
				\begin{figure*}
					\centering
					\subfloat[]{
						\includegraphics[width=0.48\textwidth]{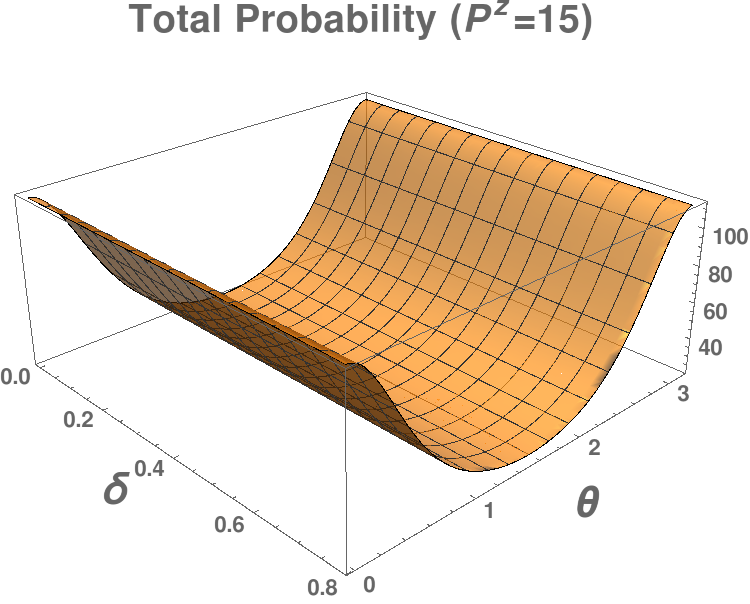}
						\label{fig:ProbPzp}}
					\centering
					\subfloat[]{
						\includegraphics[width=0.48\textwidth]{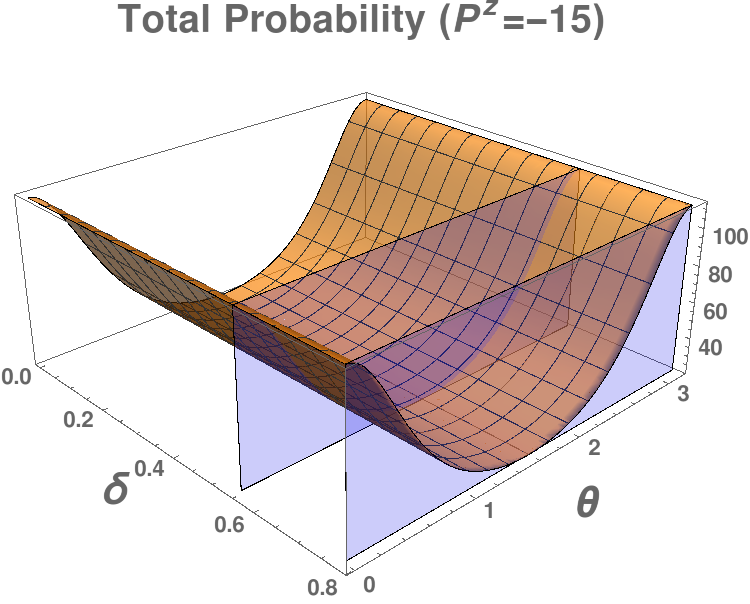}
						\label{fig:ProbPzm}}
					\caption{\label{fig:TwoGammaHelicityProbabilitySum}$e^+ e^- \to \gamma \gamma$ Sum of Helicity Probabilities for: (a) $P^z = +15 m_e$ and (b) $P^z = -15 m_e$.}
				\end{figure*}

In this Appendix, similar to what was done in Sec.~\ref{subsec:noncollnear}, we examine the frame dependence of the whole landscape of all the angular distributions of the helicity amplitudes discussed in Sec.~\ref{sub:annihilation} by computing them with non-zero center of momentum ($P^z=+15m_e $ and $ p^z=-15m_e $).
In Figs.~\ref{fig:JoinedFigmtaandtbPzp}, ~\ref{fig:JoinedFigmuaandubPzp},~\ref{fig:JoinedFigmtanduPzp},
~\ref{fig:JoinedFigampPzp} and ~\ref{fig:JoinedFigprobPzp},
we show the results for $P^z = +15 m_e$ while we do for $P^z = -15 m_e$ in Figs.~\ref{fig:JoinedFigmtaandtbPzm}, ~\ref{fig:JoinedFigmuaandubPzm},~\ref{fig:JoinedFigmtanduPzm},~\ref{fig:JoinedFigampPzm} and ~\ref{fig:JoinedFigprobPzm}. 
As we have seen in Sec.~\ref{subsec:noncollnear}, no helicity boundaries exist between IFD and LFD in the frame with $P^z = +15 m_e$ while there are two distinct helicity boundaries, one from electron and the other from positron
(see Eqs.~(\ref{eq:electron_boundary}) and (\ref{eq:positron_boundary}), respectively) between IFD and LFD for $ P^z=-15m_e $. The sum of the 16 helicity probabilities for $ P^z=+15m_e $ and $ P^z=-15m_e $ are shown in Fig.~\ref{fig:TwoGammaHelicityProbabilitySum}, and comparing with Fig.~\ref{fig:Prob} shown in Sec.~\ref{sub:annihilation}, we can see that the total probability is independent of reference frame, as well as the interpolation angle, as it should be.

	\section{Boost Dependence in $e^+ e^- \to \gamma \gamma$ Interpolating Helicity Amplitudes}
	\label{app:PzDepThetaPiOver3}
	
	\begin{figure*}
		\centering
		\includegraphics[width=1.0\linewidth]{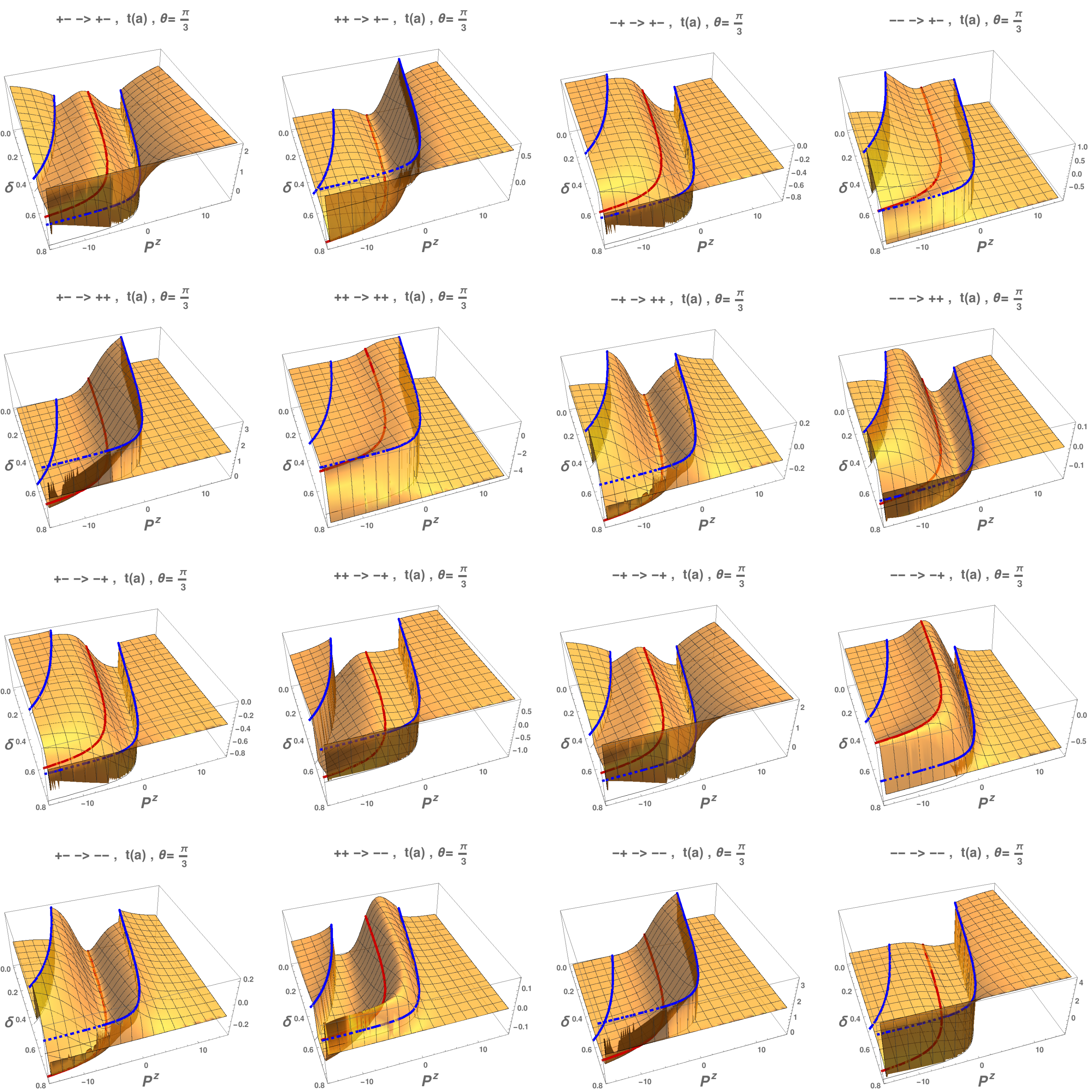}
		\caption{Annihilation Amplitudes --- t channel, time-ordering a}
		\label{fig:Annihilation_JoinedFigmta}
	\end{figure*}
	\begin{figure*}
		\centering
		\includegraphics[width=1.0\linewidth]{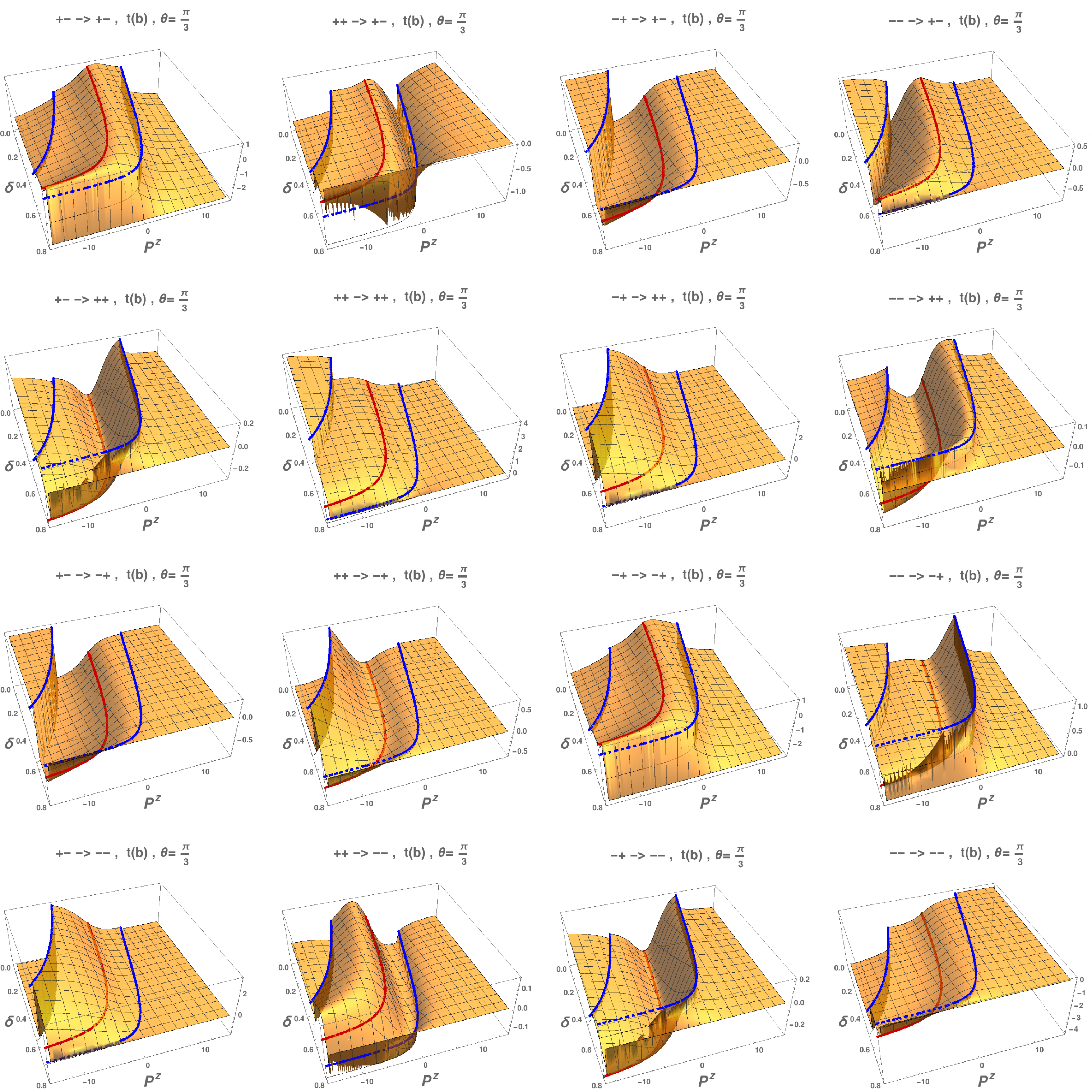}
		\caption{Annihilation Amplitudes --- t channel, time-ordering b}
		\label{fig:Annihilation_JoinedFigmtb}
	\end{figure*}
	\begin{figure*}
		\centering
		\includegraphics[width=1.0\linewidth]{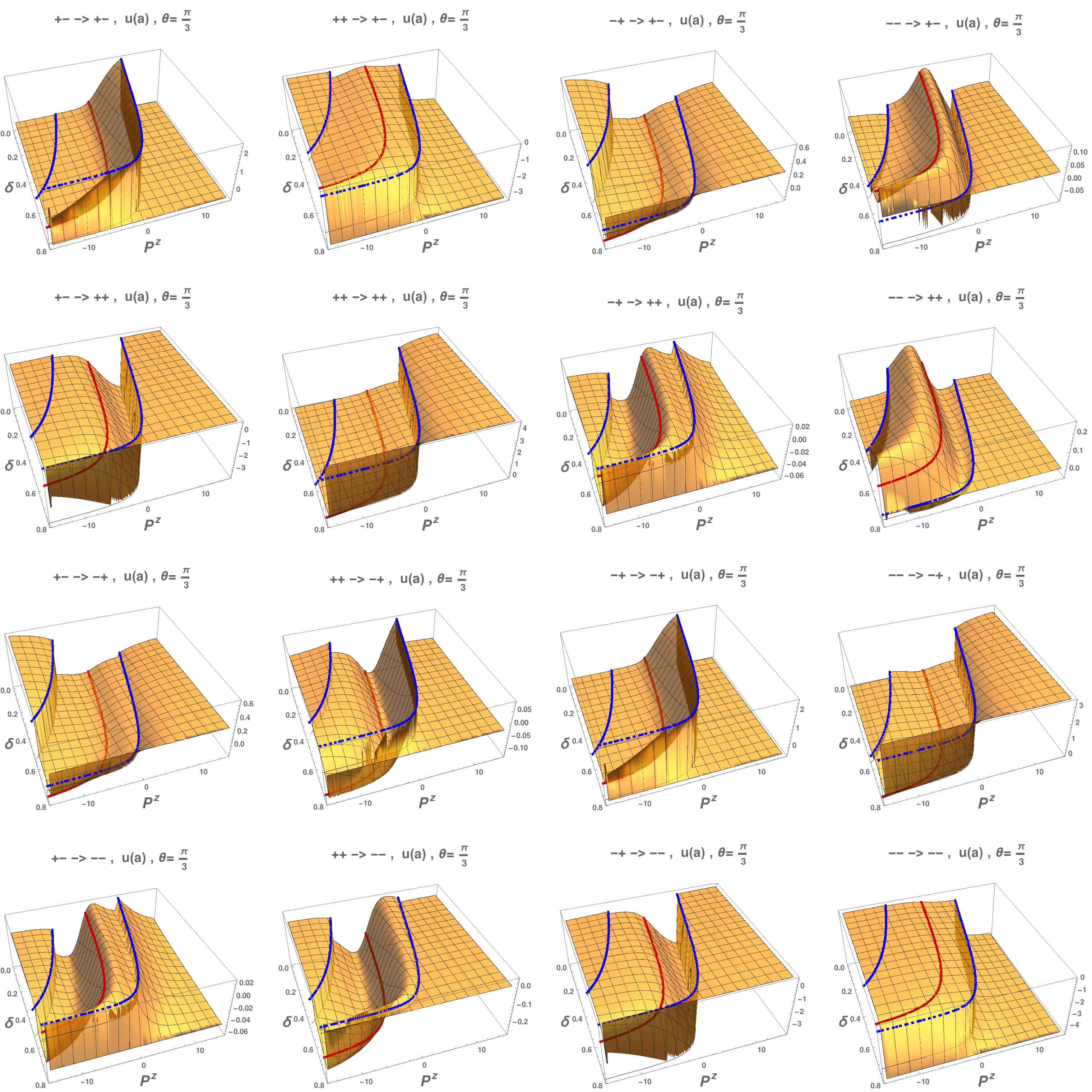}
		\caption{Annihilation Amplitudes --- u channel, time-ordering a}
		\label{fig:Annihilation_JoinedFigmua}
	\end{figure*}
	\begin{figure*}
		\centering
		\includegraphics[width=1.0\linewidth]{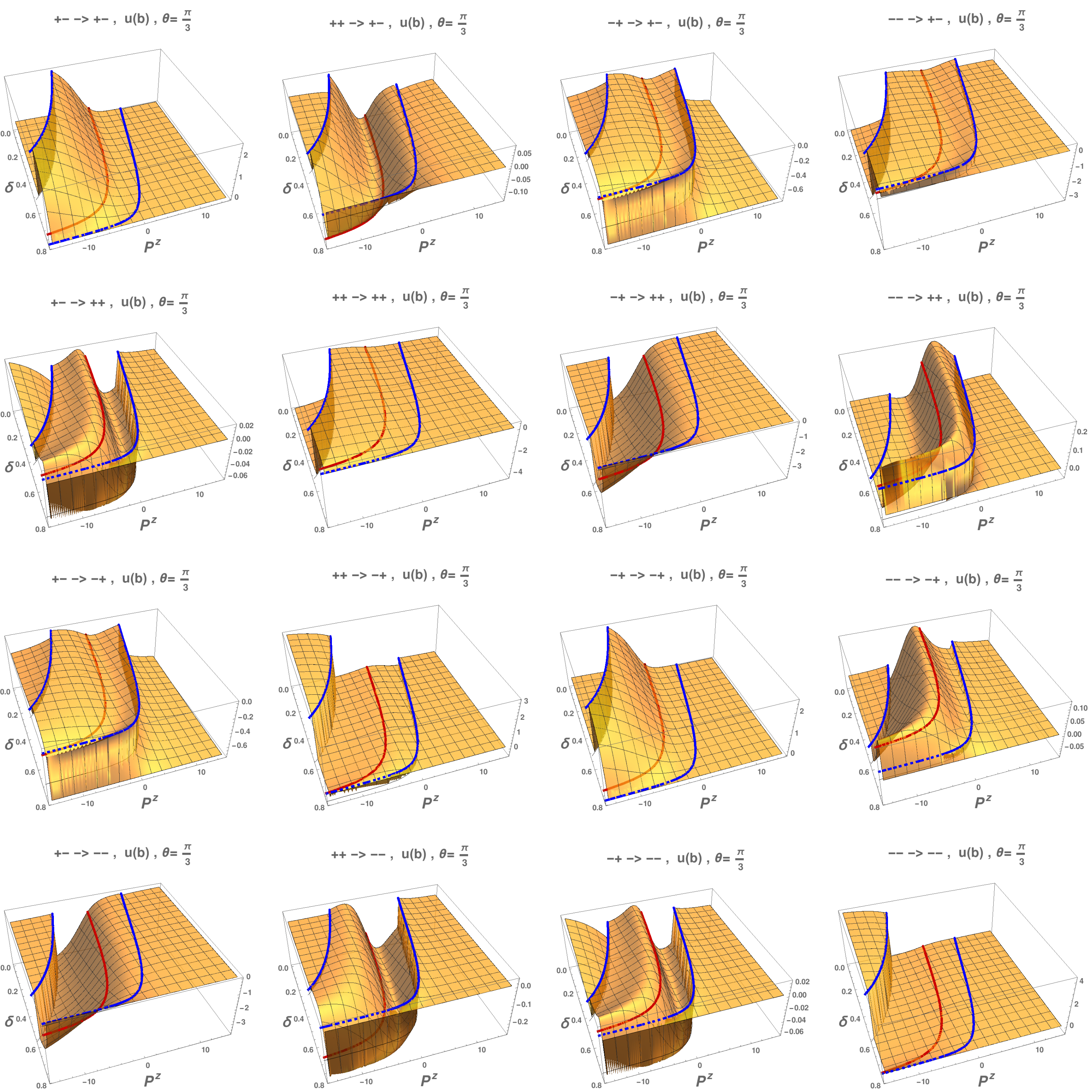}
		\caption{Annihilation Amplitudes --- u channel, time-ordering b}
		\label{fig:Annihilation_JoinedFigmub}
	\end{figure*}
	\begin{figure*}
		\centering
		\includegraphics[width=1.0\linewidth]{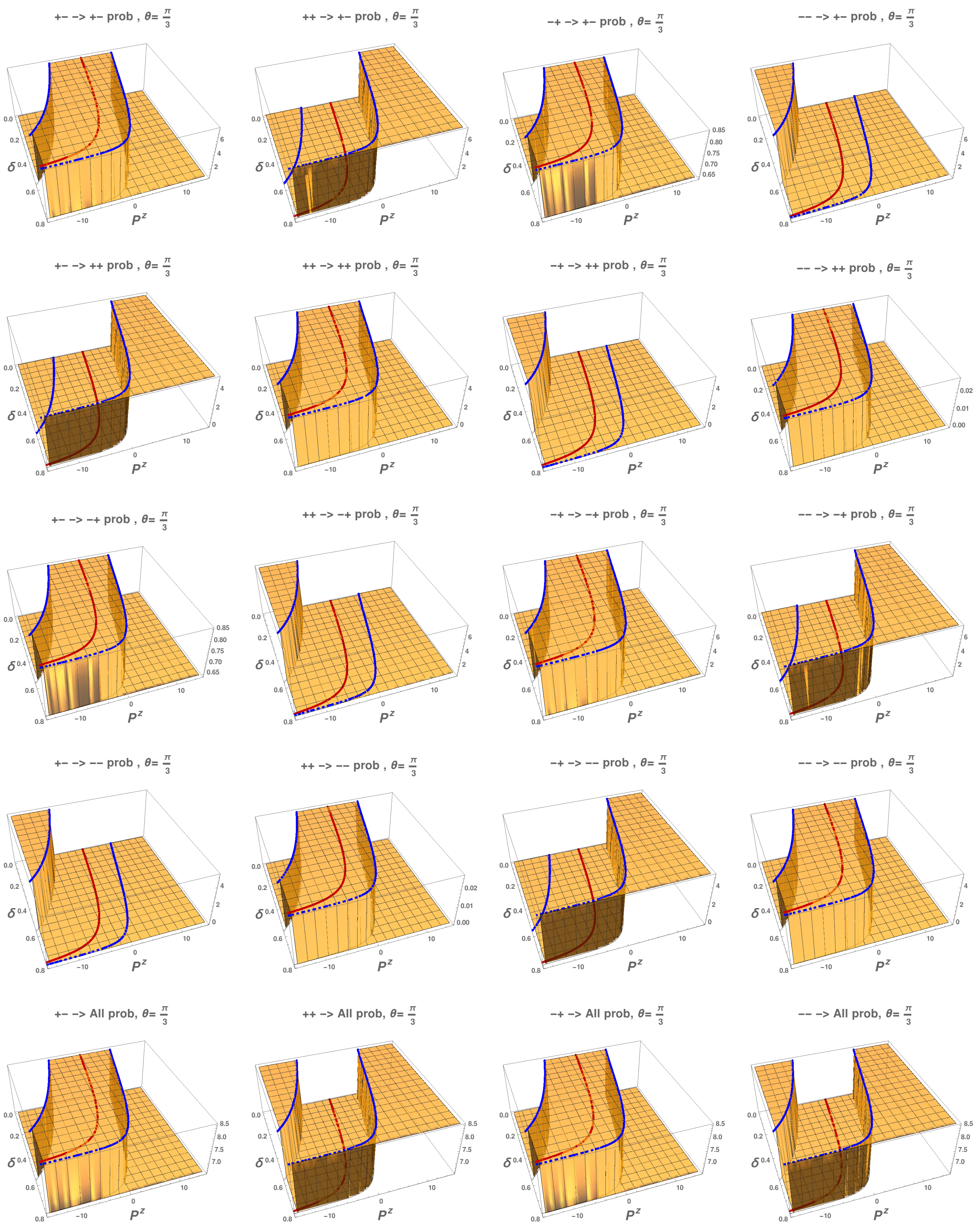}
		\caption{Annihilation probabilities}
		\label{fig:Annihilation_JoinedFigprob20}
	\end{figure*}
		\begin{figure}
			\centering
			\includegraphics[width=0.85\linewidth]{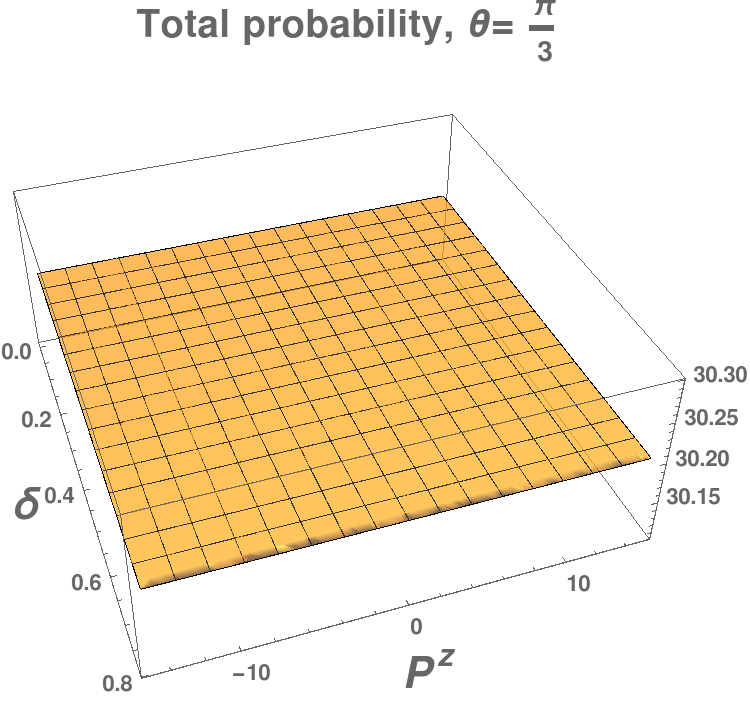}
			\caption{Total probability for $ e^+ e^-\to \gamma \gamma $ annihilation process}
			\label{fig:Annihilation_FigProb}
		\end{figure}
In this Appendix, we plot the helicity amplitudes of $e^+ e^- \to \gamma \gamma$, as given by Eq.~(\ref{eqn:TwoPhotonAmpsT}) and (\ref{eqn:TwoPhotonAmpsU}), in terms of both the interpolation angle $ \delta $ and the total momentum $ P^z $. As was done in Sec.~\ref{sub:annihilation}, we take $ m=m_e $, $ E_0=2m_e $, and instead of looking at the angular distribution, we fix the angle $ \theta $ to be $ \pi/3 $. The helicity amplitudes for the t-channel, corresponding to Feynman diagram Fig.~\ref{fig:PairAnnihilationto2s_tchannel}, with two time-orderings shown in Fig.~\ref{fig:PairAnnihilationto2s_tchannel_TO}, are presented in Figs. \ref{fig:Annihilation_JoinedFigmta} and  \ref{fig:Annihilation_JoinedFigmtb}, while the u-channel helicity amplitudes are shown in Figs.~\ref{fig:Annihilation_JoinedFigmua} and \ref{fig:Annihilation_JoinedFigmub}. The probabilities, after summing both time-orderings of both channels, are shown in Fig.~\ref{fig:Annihilation_JoinedFigprob20}, where the last row is the summation over all four final helicity states for each initial state. The total probability, obtained after summing all initial and final helicities, is shown in Fig.~\ref{fig:Annihilation_FigProb}, and is independent of boost momentum and interpolation angle. 

In these figures, the boundaries of bifurcated helicity branches between IFD and LFD due to the initial electron and positron moving in
$+\hat{z}$ and $-\hat{z}$ directions given by Eqs.~(\ref{eq:electron_boundary}) and (\ref{eq:positron_boundary}) are denoted by the blue curves, while the 
characteristic ``J-curve" given by Eq.~(\ref{J-curve}) is depicted as the red curve.
It is also apparent that the relationship between different helicity amplitudes given by 
Eq.~(\ref{eqn:amplitude_relation_Annihilation}) is satisfied, where $ \lambda_3 $ and $ \lambda_4 $ are the helicities of the outgoing photons while 
$ \lambda_1 $ and $ \lambda_2 $ are the incoming electron and positron helicities, respectively. 
This relationship holds as one can see in Figs.~\ref{fig:Annihilation_JoinedFigmta} through \ref{fig:Annihilation_JoinedFigmub}.  
Up to an overall sign difference, the upper left 2 by 2 block is the same with the lower right 2 by 2 block, while the upper right block is the same with the lower left block. In the right most column, however, all figures have their signs flipped from their counterparts. For the square of amplitudes shown in Fig.~\ref{fig:Annihilation_JoinedFigprob20}, the same correspondence holds without any sign difference as it should be.

\end{document}